\documentclass[a4paper,fleqn,usenatbib,useAMS]{mnras}

\usepackage{caption}

\usepackage{rotating}
\usepackage{lscape}
\usepackage{longtable}

\usepackage{subfigure}
\usepackage{graphicx}

\usepackage{dcolumn}
\newcolumntype{d}[1]{D{.}{.}{#1}}

\usepackage{multirow}

\usepackage{amsmath}

\usepackage{color}
\definecolor{highlight}{rgb}{1,0,0}
\definecolor{red}{rgb}{1,0,0}
\definecolor{green}{rgb}{0,1,0}
\definecolor{blue}{rgb}{0,0,1}

\newcommand{\kms}{km\,s$^{-1}$} 
\newcommand{\malt}{\mbox{MALT-45}}

\newcommand{\cs}{CS}
\newcommand{\ctfs}{C$^{34}$S}
\newcommand{\choh}{CH$_3$OH}
\newcommand{\sio}{SiO}
\newcommand{\nh}{NH$_3$}
\newcommand{\water}{H$_2$O}

\newcommand{\oh}{OH}

\newcommand{\hii}{H\,\textsc{ii}}

\newcommand{\hfoa}{H51$\alpha$}
\newcommand{\hfta}{H53$\alpha$}

\newcommand*\rfrac[2]{{}^{#1}\!/_{#2}}

\usepackage[T1]{fontenc}
\usepackage{ae,aecompl}

\usepackage{newtxtext,newtxmath}

\title[MALT-45 - II. Class I methanol maser follow-up]{MALT-45: A 7\,mm survey of the southern Galaxy - II. ATCA follow-up observations of 44\,GHz class I methanol masers}

\author[C. H. Jordan et al.]{Christopher H. Jordan,$^{1,2,3,4}$\thanks{E-mail: \href{mailto:christopher.jordan@curtin.edu.au}{christopher.jordan@curtin.edu.au}}
  Andrew~J.~Walsh,$^1$
  Shari~L.~Breen,$^{5,4}$\newauthor
  Simon~P.~Ellingsen,$^3$
  Maxim~A.~Voronkov,$^{3,4,6}$
  and Lucas~J.~Hyland$^{3,4}$\\
\\
$^1$International Centre for Radio Astronomy Research, Curtin University, Bentley, WA 6845, Australia\\
$^2$ARC Centre of Excellence for All-sky Astrophysics (CAASTRO)\\
$^3$School of Physical Sciences, Private Bag 37, University of Tasmania, Hobart, TAS 7001, Australia\\
$^4$CSIRO Astronomy and Space Science, PO Box 76, Epping, NSW 1710, Australia\\
$^5$Sydney Institute for Astronomy (SIfA), School of Physics, University of Sydney, Sydney, NSW 2006, Australia\\
$^6$Astro Space Centre, Profsouznaya st. 84/32, 117997 Moscow, Russia}

\date{Last updated yyyy mmm dd; in original form yyyy mmmmmmmmm d}

\pubyear{2016}

\begin{document}
\label{firstpage}
\pagerange{\pageref{firstpage}--\pageref{lastpage}}
\maketitle

\newcolumntype{d}[1]{D{.}{.}{#1} }

\begin{abstract}
  We detail interferometric observations of 44\,GHz class~I methanol masers detected by \malt{} (a 7\,mm unbiased auto-correlated spectral-line Galactic-plane survey) using the Australia Telescope Compact Array. We detect 238 maser spots across 77 maser sites. Using high-resolution positions, we compare the class~I~\choh{} masers to other star formation maser species, including \cs{}~(1--0), \sio{}~$v=0$ and the \hfta{} radio-recombination line. Comparison between the cross- and auto-correlated data has allowed us to also identify quasi-thermal emission in the 44\,GHz class~I~methanol maser line. We find that the majority of class~I methanol masers have small spatial and velocity ranges ($<$0.5\,pc and $<$5\,\kms{}), and closely trace the systemic velocities of associated clouds. Using 870\,$\mu$m dust continuum emission from the ATLASGAL survey, we determine clump masses associated with class~I masers, and find they are generally associated with clumps between 1000 and 3000\,$M_\odot$. For each class~I methanol maser site, we use the presence of \oh{} masers and radio recombination lines to identify relatively evolved regions of high-mass star formation; we find that maser sites without these associations have lower luminosities and preferentially appear toward dark infrared regions.
\end{abstract}




\section{Introduction}
Methanol masers are excellent tracers of high-mass star-formation (e.g. \citealt{walsh03}). Since their discovery \citep{menten91}, 6.7\,GHz methanol masers have been heralded as one of the most important spectral lines in astronomy; each of the transitions of the class~II~\choh{} maser family, of which the 6.7\,GHz transition belongs, occurs only towards regions of high-mass star formation (HMSF; \citealt{walsh01,minier03,green12,breen13}). These class~II transitions are powered by mid-infrared emission from a nearby young stellar object (YSO), and dissipate as the \hii{} region resulting from HMSF evolves \citep{walsh98}. Consequently, class~II~\choh{} masers signpost a specific evolutionary phase of the HMSF timeline.

On the other hand, class~I~\choh{} maser transitions have a relatively uncertain connection to HMSF. Observations have shown that class~I~\choh{} masers can be found towards many star-forming regions and evolutionary stages, but not consistently, and not necessarily with or without the presence of class~II masers \citep{ellingsen06,voronkov10a,breen10,ellingsen13}. Additionally, class~I maser emission has been found towards low-mass star formation \citep{kalenskii10}, supernova remnants \citep{pihlstrom14,mcewen14} and the centres of other galaxies \citep{ellingsen14,chen15}. In contrast to class~II masers, class~I masers are collisionally excited in dense molecular gas \citep{cragg92,voronkov10a,voronkov10b,voronkov14}.

Another collisionally-excited transition commonly found toward star-forming regions is thermal silicon monoxide (\sio{}~$v=0$), which traces a wide range of shocks \citep{nguyen-luong13,widmann16}, and has a rest frequency close to the class~I~\choh{} maser. Thus, it is prudent to compare these spectral lines and learn more about class~I~\choh{} masers. Additionally, high-density gas tracers such as carbon monosulfide (\cs{}) highlight regions with potentially sufficient molecular gas abundance for masing. Investigations of the interstellar medium utilising these spectral lines, which all occur in the 7\,mm waveband, form the primary motivation for the Millimetre Astronomer's Legacy Team - 45\,GHz survey (\malt{}; \citealt{jordan13,jordan15}). \malt{} is an unbiased, sensitivity-limited auto-correlation survey of spectral lines in the 7\,mm waveband, primarily surveying \cs{}~$J$=(1--0), class~I~\choh{} masers (the 44\,GHz 7(0,7)--6(1,6) A$^+$ transition) and \sio{}~$J$=(1--0); a table of all surveyed spectral lines is contained in Table~\ref{tab:meth_targets}. Prior to the \malt{} survey, class~I~\choh{} masers were primarily found towards regions containing other masers or regions of shocked gas, such as extended green objects (EGOs; \citealt{cyganowski08}); very little was known about regions containing only class~I maser emission.

The extent of the \malt{} survey is currently detailed in a single paper (\citealt{jordan15}; hereafter Paper~I). \malt{} is unique in that it utilises the Australia Telescope Compact Array (ATCA) by processing auto-correlated data (although cross-correlated data is simultaneously collected, it is not used due to extremely poor \emph{uv}-coverage). Within the mapped region $330^{\circ} \leq l \leq 335^{\circ}$, $b=\pm0.5^{\circ}$, Paper~I details the detection of 77 class~I methanol masers, 58 of which were new detections. For the first time, with a flux-density limited sample of class~I sources, we are able to assess the bulk-properties of these masers. Paper~I in this series of \malt{} publications briefly investigated the association of methanol masers with other detected spectral lines, but was ultimately limited by the spatial resolution of the auto-correlation survey ($\sim$1~arcmin). In this paper, we detail the results of interferometric follow-up observations towards each of the class~I~\choh{} masers observed in Paper~I, in order to derive accurate positions.

\section{Observations}

\begin{table*}
  \begin{center}
    \caption{Bright spectral lines between 42.2 to 49.2\,GHz, targeted by \malt{} and these observations. Column 1 lists the spectral line. Column 2 lists the rest frequency of the line. Column 3 classifies the line as either a maser or thermal line. Column 4 gives the ATCA half power beam width at the corresponding rest frequency. Column 5 indicates whether this line is catalogued in this paper (`Y') or not (`N'); this refers to the inclusion of Gaussian fits to the spectral emission from these lines. Column 6 lists the median RMS noise level for auto-correlation data per 32\,kHz spectral channel, with errors representing the standard deviation. Radio recombination line (RRL) frequencies are taken from \citet{lilley68}. All other rest frequencies are taken from the Cologne Database for Molecular Spectroscopy (CDMS; \citealt{muller05,mueller13}). Note that cross-correlation noise levels are given in Table~\ref{tab:meth_targets}.}
  \label{tab:spectral_lines}
  \begin{tabular}{ lcccc c }
    \hline
    Spectral line                & Rest frequency & Maser or        & Beam size & Catalogued in & Median RMS  \\
                                 & (GHz)          & thermal?        & (arcsec)  & this paper?   & noise level \\
    \hline
    \sio{}~(1--0) $v=3$          & 42.51938       & Maser           & 66 & N & \\
    \sio{}~(1--0) $v=2$          & 42.82059       & Maser           & 66 & N & \\
    \hfta{} (RRL)                & 42.95197       & Thermal         & 65 & Y & 6.5$\pm$1.0\,mK \\
    \sio{}~(1--0) $v=1$          & 43.12207       & Maser           & 65 & N & \\
    \sio{}~(1--0) $v=0$          & 43.42385       & Thermal         & 65 & Y & 5.9$\pm$0.67\,mK \\
    \choh{}~7(0,7)--6(1,6) A$^+$ & 44.06941       & Maser (Class~I) & 64 & Y & 170$\pm$62\,mJy \\
    \hfoa{} (RRL)                & 48.15360       & Thermal         & 58 & N & \\
    \ctfs{}~(1--0)               & 48.20694       & Thermal         & 58 & N & 8.1$\pm$0.85\,mK \\
    \choh{}~1$_0$--0$_0$ A$^+$   & 48.37246       & Thermal         & 58 & Y & 7.7$\pm$1.8\,mK \\
    \choh{}~1$_0$--0$_0$ E       & 48.37689       & Thermal         & 58 & N & \\
    OCS~(4--3)                   & 48.65160       & Thermal         & 58 & N & \\
    \cs{}~(1--0)                 & 48.99095       & Thermal         & 57 & Y & 8.9$\pm$1.3\,mK \\
    \hline
  \end{tabular}
  \end{center}
\end{table*}

Observations were conducted on 2013 September 7 and 8 using the ATCA. The array configuration was 1.5A, which has 5 antennas distributed in the East-West direction with baselines ranging from 153\,m to 1.5\,km; the maximum baseline is 4.5\,km including antenna 6 (CA06), however, due to poor weather, baselines including CA06 had bad phase stability. Consequently, data from CA06 is not included in this paper.

The correlator was programmed in the 64M-32k mode, which provides `zoom windows` for enhanced spectral resolution. In this mode, each zoom window has a channel resolution of 32\,kHz over a bandwidth of 64\,MHz. This same correlator mode was used in the observations detailed by Paper~I; however, unlike Paper~I, these observations made use of `stitched zooms': rather than using individual zoom windows, multiple zooms may be joined to increase the bandwidth over a spectral line, while maintaining the same spectral resolution. Each spectral line in these observations was observed with stitched zoom windows covering 96\,MHz (except for the window containing \hfoa{} and \ctfs{}~(1--0), which covers 224\,MHz). For observations at 7\,mm, the 32\,kHz channel resolution corresponds to a spectral resolution of approximately 0.21\,\kms{} per channel. The spectral lines observed are identical to that of Paper~I; the list of observed lines as well as velocity coverage information and noise statistics are contained in Table~\ref{tab:spectral_lines}. Since Paper~I, the rest frequencies of the \sio{}~(1--0) molecular lines have been updated \citep{mueller13}.

The class~I~\choh{} masers targeted in these observations are listed in Table~\ref{tab:meth_targets}. This target list was derived from a preliminary reduction of the \malt{} survey and hence do not exactly match the maser sites presented in Paper~I, however, there are only six sites that are not common to both papers. We include four sites in the current observations (G331.44$-$0.14, G331.44$-$0.16, G333.01$-$0.46 and G333.12$-$0.43) that were not listed in Paper~I, and we have not targeted two that are listed in Paper~I (G331.36$-$0.02 and G334.64$+$0.44), because they are tenuous detections that became apparent only in an improved processing of the \malt{} survey, after the observations detailed by this paper were conducted. Hence, a total of 79 targets were observed over the two days of observations. For more details, refer to Section~\ref{sec:results}.

To obtain good \emph{uv}-coverage, each target was observed multiple times over a single observing session. Each individual observation of a target lasted one minute (one `cut'), and each target has at least seven cuts, but some have up to ten; the noise levels for each target at the 44\,GHz class~I~\choh{} maser transition frequency are also listed in Table~\ref{tab:meth_targets}. Bandpass calibration was derived from PKS B1253$-$055, phase calibration from PMN J1646$-$5044 and flux-density calibration from PKS B1934$-$638. \citet{partridge16} give a flux density scale for B1934$-$638, which yields a flux density of 0.31\,Jy at the rest frequency of the 44\,GHz class~I~\choh{} maser spectral line. However, we used the recommended flux density of 0.39\,Jy derived from internal work with the ATCA. In addition to the cross-correlation data, auto-correlation data was simultaneously recorded and used; see Section~\ref{subsec:auto_correlation}.

\subsection{Data reduction}
\subsubsection{Cross-correlation}
\label{subsec:cross_correlation}
Class~I~\choh{} maser data were reduced using standard interferometric techniques with \textsc{MIRIAD} \citep{sault95}. For all data cubes, the synthesised beam is approximately 0.5$\times$1~arcsec in right ascension and declination, respectively. Similar to \citet{voronkov14}, the results presented in this paper make use of self-calibrated data cubes. Self-calibration affects the absolute positional accuracy of data; to remove this effect, a reference maser spot was chosen from data with no self calibration applied. The position of this reference spot was determined before and after self-calibration and the measured difference was used to determine the absolute positions for all spectral features in the self-calibrated data cube. Ideally, the reference position before and after self-calibration has not shifted significantly ($<$0.5~arcsec), but some significant offsets were observed in these data due to poor weather conditions during the observations, which leads to large corrections in the self-calibration model. Models were selected from the brightest channel of emission from CLEANed data, which is generally effective, but occasionally the quality of calibration was so poor that the resulting self-calibrated data experiences a dramatic spatial shift. Despite this, we are encouraged by the accuracy of positions once a reference position has been subtracted; for example, G333.23$-$0.06 experienced poor calibration, but the difference in position of the brightest maser spot of G333.23$-$0.06 between this work and that of \citet{voronkov14} is 1.2~arcsec. This difference may result from a combination of genuine maser variability and calibration uncertainties. Prior experience with the ATCA in this waveband suggests that, in the best case scenario, the standard phase transfer calibration procedure on the absolute positional accuracy results in an error of approximately 0.5~arcsec (1$\sigma$) \citep{voronkov14}. However, given that we experience non-ideal self-calibration offsets, the absolute positional error for any maser spot should be considered to be 1~arcsec.

Data cubes were produced for each of the observed targets, which were then used to identify maser spots by visual inspection. In this paper, a maser spot refers to a spatial and spectral peak of emission. Spots were identified if the spectrum contained at least three consecutive channels of 3$\sigma$ emission, with a peak of at least 5$\sigma$ within the same pixel (0.2~arcsec). We employ this conservative approach of maser identification so the reader can be confident that all identified maser spots are real. A small number of weaker, narrow features may not satisfy these criteria and so are excluded; however, from past experience and comparison of our cubes with previous observations (where available), we suggest that this is not common. The requirement of three significant consecutive channels has the effect of limiting our maser detections to a velocity width of at least 0.65\,\kms{}. Similar to \citet{voronkov14}, dynamic range limitations exist for channels containing extremely bright maser emission. This reduces the ability to identify real maser emission at a similar velocity to the bright feature; maser spots were excluded where they are believed to have been caused by such dynamic range artefacts. These exclusions were determined by visual inspection.

The position of each maser spot was fitted with the \textsc{MIRIAD} task \textsc{imfit} using each channel of significant emission ($>$3$\sigma$). \textsc{imfit} used a 3$\times$3~arcsec box around each maser spot to generate a position, except for those with nearby bright maser spots. To accurately position weaker maser spots close to bright ones, either the velocity range of emission was altered for position fitting, or the size of the box for fitting was decreased. Decreasing the size of the box for \textsc{imfit} was found to not significantly affect resultant positions, so we do not highlight the maser spots using these modified procedures. Each channel-fitted position is then flux density-weighted and is listed in Appendix~\ref{app:meth_detail}. The uncertainty reflects the 1$\sigma$ positional uncertainty of each channel-fitted position. In some cases, the maser spot positions have uncertainties of a few tens of milliarcsec. We caution that the ATCA is likely not able to determine relative positions to this accuracy for these observations, as dynamic range and calibration uncertainties become a limiting factor. In general, we found the class~I maser spots were spatially distinguished by our observations, with separations of 1~arcsec or greater, but occasionally spots were very near to a bright neighbour. In these cases, positions were manually assigned by using the brightest pixel, with a cautious position uncertainty of 100 milliarcsec in Right Ascension and 200 milliarcsec in Declination. Such maser spots are designated with an asterisk ($*$) in Appendix~\ref{app:meth_detail}.

Spectral information for each maser spot was derived from the image cubes, which were then characterised with Gaussian fits. All residuals are within 10 per cent of the fitted peak intensity. Coupled with uncertainties in calibration, we expect a 20 per cent uncertainty for all quoted flux-density values.

\subsubsection{Auto-correlation}
\label{subsec:auto_correlation}
Each observation target was also searched for \cs{}~(1--0), \sio{}~(1--0) $v=0$ and \choh{}~1$_0$--0$_0$~A$^+$. However, due to their thermal nature and the \emph{uv}-coverage of the observations (minimum baseline length of 153\,m), these thermal emission lines were not readily detectable in the cross-correlation data. To report on these properties, a \textsc{Ruby} script was written to assist with the analysis of the auto-correlated data. Unlike the observations detailed by Paper~I, no dedicated reference (i.e. off) position was observed in this work. Instead, data collected at a location where the emission is known to be weak and have a simple spectral profile was used as the reference spectrum for auto-correlations. The script routine pipeline was as follows:

(i) For each source, a raw auto-correlation spectrum was produced for each antenna and each cut;

(ii) For each antenna source cut, a quotient spectrum (i.e.~$\rfrac{\text{on}-\text{off}}{\text{off}}$) was formed from the source and reference cut data;

(iii) A first-order polynomial was fitted to and removed from the spectra. The velocity range of the fit was restricted between $-$160 and 20\,\kms{}, as this contains all of the emission in this section of the Galaxy, and constraining the fit improves the quality of the resulting spectra;

(iv) All quotient spectra from an antenna are averaged together, before averaging all antennas together;

(v) The flux-density calibration is achieved with antenna efficiencies and point-source sensitivities taken from \citet{urquhart10}.

\begin{figure}
  \centering
  \includegraphics[width=0.9\columnwidth]{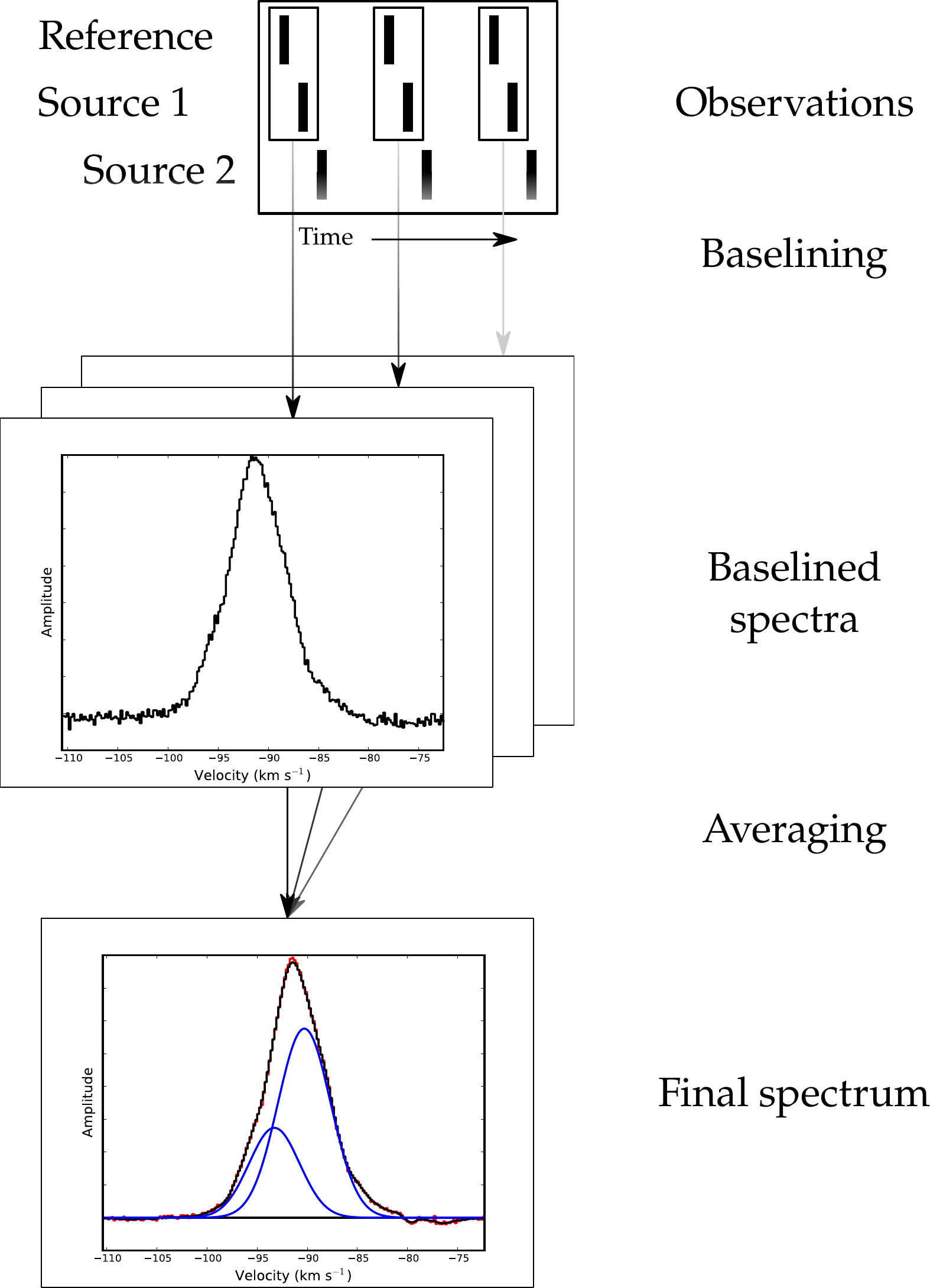}
  \caption{A graphical representation of the reduction procedure for auto-correlated data in this paper. The data from each source cut is combined with the nearest reference cut in time to produce a baselined spectrum. All baselined spectra are then averaged to produce an auto-correlation product. This procedure is performed for each antenna of the ATCA, and all products are then averaged together for a final product. The final panel contains this final product spectrum (red) and the same spectrum after being Hanning-smoothed (black). This example uses two Gaussians (blue) to characterise the data.}
  \label{fig:autocorrelation_procedure}
\end{figure}

This procedure is graphically demonstrated with the \cs{} emission of G330.95$-$0.18 in Fig.~\ref{fig:autocorrelation_procedure}.

For the spectral line data processed in this way, spectra were smoothed with a Hanning window of 9 spectral channels, and up to two Gaussians were fitted to the frequency data. This smoothing results in a velocity resolution of $\sim$0.39\,\kms{} for \cs{}, $\sim$0.44\,\kms{} for \sio{} $v=0$ and $\sim$0.40\,\kms{} for \choh{} 1$_0$--0$_0$ A$^+$. As with the cross-correlation data, Gaussians are fitted to minimise residuals; however, occasionally 10 per cent residuals are unavoidable. With additional flux-density calibration uncertainty applied to auto-correlation data, we place a 20 per cent uncertainty on all intensity values derived.

This auto-correlation procedure was also used for class~I~\choh{} masers, which we compare with the cross-correlation data in Section~\ref{sec:cross_vs_auto}.

\section{Results}
\label{sec:results}
The basic properties of each observed target are listed in Table~\ref{tab:meth_targets}. Note that five sources were not detected in cross-correlation by these observations, but three of these five were detected in auto-correlation (G331.44$-$0.14, G331.72$-$0.20 and G333.24$+$0.02). The reason that three of the sites were detected in auto-correlation data, but not cross-correlation data, could either be due to poor weather (causing significant decorrelation) or, alternatively, it may indicate that the emission region is large, and is therefore resolved out in the cross-correlation data. The fact that the spectral profile of these three sources is consistent with maser emission (although relatively weak maser emission; each of the three sources is less than $\sim$2\,Jy) suggests that the most likely explanation for their non-detection is due to poor weather. Further discussion of cross-correlation spectra compared to the auto-correlation data for the same source is contained in Section~\ref{sec:cross_vs_auto}.

The remaining two sources, G330.83$+$0.18 and G331.21$+$0.10, were detected in October~2013 with peak-flux densities of 3.1 and 3.5\,Jy. The median RMS noise of the survey observations was 0.90\,Jy, meaning that they were detected a 3.4 and 3.8-$\sigma$ level, respectively. It is therefore possible that these were spurious detections, or, alternatively, that their peak-flux densities have varied below the 0.85\,Jy 5$\sigma$ detection limit for the current observations. A variation factor of four seems reasonable for 44\,GHz masers (e.g. variations larger than four were found in two sources detailed by \citealt{kurtz04}), although there have been no dedicated studies. Given the current data, and the low significance of the original detections, these two sources cannot be considered reliable detections and as such we exclude them from further analysis. For the remainder of this paper, we discuss the 77 regions containing 44\,GHz class~I~\choh{} emission.

We have classified our class~I~\choh{} masers as associated with other maser transitions or selected HMSF tracers if they fall within 60~arcsec of each other, to be consistent with \citet{voronkov14}. The associations are detailed in Table~\ref{tab:meth_assoc}. Each site is compared with published positions of class~II~\choh{} \citep{caswell11}, water (\water{}) \citep{breen10,walsh11,walsh14} and hydroxyl (\oh{}) masers \citep{sevenster97,caswell98}, as well as ATLASGAL clumps \citep{contreras13} and EGOs \citep{cyganowski08}. The presence of each thermal line is determined by the method described in Section~\ref{subsec:auto_correlation}, using data from these observations. \cs{} is detected towards every source, and because of this is not listed in Table~\ref{tab:meth_assoc}. The Gaussian parameters determined for each of the thermal lines are listed in Appendix~\ref{app:thermal_detail}. Uncertainties for each of the fitted parameters are quoted, however, we remind the reader to consider the 20 per cent uncertainty adopted for the absolute flux-density scale.

Note that \citet{walsh11} details single-dish positions of \water{} masers, while \citet{walsh14} provides high-resolution follow-up positions of these same masers. Hence, we use positions from \citet{walsh14} wherever possible; however, two \water{} masers (G331.86$+$0.06 and G333.46$-$0.16) were not detected in follow-up observations, and thus we use their single-dish positions for comparisons in this paper. The non-detection of these masers is reasoned as being most likely due to intrinsic variability.

Kinematic distances are also contained in Table~\ref{tab:meth_assoc}. Distances were calculated using the program supplied by \citet{reid09}. All kinematic distance ambiguities were assumed to be at the `near' distance, except for sources which have been prescribed as `far' by \citet{green11}. All distances presented in this paper use the same parameters as Paper~I, specifically: $\Theta_0=246$\,\kms{}, $R_\odot=8.4$\,kpc, $U_\odot=11.1$\,\kms{}, $V_\odot=12.2$\,\kms{}, $W_\odot=7.25$\,\kms{}, $U_s=0$\,\kms{}, $V_s=-15.0$\,\kms{}, $W_s=0$\,\kms{}, $\sigma(V_{LSR})=3.32$\,\kms{}.

We consider all maser spots detected toward a single target to be part of the same maser site. There is potential for this to artificially to combine multiple maser sites into a single site, however, Apprendix~\ref{app:glimpse} shows that this is uncommon. To illustrate the derived properties of masers, a sample of Gaussian fits to spectra is included in Table~\ref{tab:meth_sample}. The complete collection of Gaussian spectral fits is included in Appendix~\ref{app:meth_detail}. Maser positions overlaid on infrared images from the \emph{Spitzer} Galactic Legacy Infrared Mid-Plane Survey Extraordinaire (GLIMPSE; \citealt{benjamin03}) can be seen in Appendix~\ref{app:glimpse}.

\begin{table*}
  \begin{center}
    \caption{Observational targets for class~I~\choh{} masers. Column 1 lists the site name taken from Paper~I. Column 2 lists the refined site name, which is determined from the mean position of maser spots within a single maser site. The refined site names are used throughout this paper. A refined name is not supplied if no maser emission was detected in this work. Columns 3 and 4 list the interferometric phase centre for these observations (central maser positions are listed in Table~\ref{tab:meth_assoc}). Column 5 gives the date on which the site was observed (7 or 8 September 2013). Column 6 lists the off-source image-plane median RMS noise value for the self-calibrated cube. Column 7 lists the smallest radius of a circle to encompass all emission of the site. Column 8 lists the velocity range of emission detected in auto-correlation data. All maser sites were observed with an approximate local standard of rest velocity coverage between $-$353 and 195\,\kms{}. Note that sites with a radius of $<1$~arcsec contain either only one maser spot, or spots very close together. Sites without a specified radius were not detected in this work. Sites without a velocity range were not detected in cross- or auto-correlation (see Section~\ref{sec:results}). Sites labelled with an asterisk ($*$) were not listed by Paper~I, but have class~I~\choh{} maser emission detected in these observations. Sites labelled with a dagger ($\dagger$) are designated as `young'; see Section~\ref{sec:evolutionary}. Sites labelled with a double dagger ($\ddagger$) have large shifts induced by self-calibration; see Section~\ref{subsec:cross_correlation}.}
    \label{tab:meth_targets}
    \begin{tabular}{ lcccc ccc }
      \hline
      \multicolumn{1}{c}{Paper~I}     & Refined     & \multicolumn{2}{c}{Interferometric phase centre}         & Obs.        & Median RMS   & Radius   & Velocity \\
      \multicolumn{1}{c}{site name}   & site name   & $\alpha_{2000}$ & $\delta_{2000}$                        & date        & noise level  & (arcsec) & range    \\
                                      &             & (h:m:s)         & ($^\circ$:$^\prime$:$^{\prime\prime}$) & (Sep. 2013) & (mJy)        &          & (\kms{})   \\
      \hline
      G330.30$-$0.39           & G330.294$-$0.393 & 16:07:37.0 & $-$52:30:59 & 8 & 46 & 4  & $-$82 to $-$76 \\
      G330.67$-$0.40           & G330.678$-$0.402 & 16:09:30.6 & $-$52:16:08 & 8 & 42 &$<1$& $-$69 to $-$60 \\
      G330.78$+$0.24$^\dagger$ & G330.779$+$0.249 & 16:07:12.2 & $-$51:43:07 & 8 & 42 &$<1$& $-$46 to $-$41 \\
      G330.83$+$0.18           &                  & 16:07:40.6 & $-$51:43:54 & 8 & 44 &    & \\
      G330.87$-$0.36           & G330.876$-$0.362 & 16:10:16.6 & $-$52:05:50 & 8 & 45 & 22 & $-$66 to $-$57 \\
      G330.88$-$0.38           & G330.871$-$0.383 & 16:10:21.1 & $-$52:06:42 & 8 & 44 &$<1$& $-$67 to $-$57 \\
      G330.92$-$0.41$^\dagger$ & G330.927$-$0.408 & 16:10:44.2 & $-$52:05:56 & 8 & 45 & 3  & $-$44 to $-$40 \\
      G330.93$-$0.26$^\dagger$ & G330.931$-$0.260 & 16:10:06.6 & $-$51:59:23 & 8 & 43 & 1  & $-$91 to $-$87 \\
      G330.95$-$0.18           & G330.955$-$0.182 & 16:09:52.0 & $-$51:54:59 & 8 & 46 &$<1$& $-$98 to $-$84 \\
      G331.13$-$0.48$^\dagger$ & G331.131$-$0.470 & 16:11:59.8 & $-$52:00:32 & 8 & 44 &$<1$& $-$70 to $-$65 \\
      G331.13$-$0.25           & G331.132$-$0.244 & 16:10:59.7 & $-$51:50:25 & 8 & 46 & 10 & $-$93 to $-$78 \\
      G331.13$-$0.50           & G331.134$-$0.488 & 16:12:05.9 & $-$52:01:33 & 8 & 37 &$<1$& $-$68 to $-$68 \\
      G331.13$+$0.15$^\dagger$ & G331.134$+$0.156 & 16:09:14.8 & $-$51:32:47 & 8 & 46 & 4  & $-$79 to $-$73 \\
      G331.21$+$0.10           &                  & 16:09:50.8 & $-$51:32:39 & 8 & 44 &    & \\
      G331.29$-$0.20           & G331.279$-$0.189 & 16:11:27.0 & $-$51:41:54 & 8 & 45 & 21 & $-$95 to $-$84 \\
      G331.34$-$0.35           & G331.341$-$0.347 & 16:12:25.6 & $-$51:46:16 & 8 & 46 &$<1$& $-$67 to $-$64 \\
      G331.37$-$0.40$^\dagger$ & G331.370$-$0.399 & 16:12:48.2 & $-$51:47:26 & 8 & 44 &$<1$& $-$66 to $-$64 \\
      G331.37$-$0.13$^\dagger$ & G331.371$-$0.145 & 16:11:40.3 & $-$51:35:52 & 8 & 47 &$<1$& $-$89 to $-$86 \\
      G331.39$+$0.15$^\dagger$ & G331.380$+$0.149 & 16:10:26.2 & $-$51:22:52 & 8 & 46 &$<1$& $-$47 to $-$43 \\
      G331.41$-$0.17$^\dagger$ & G331.409$-$0.164 & 16:11:59.2 & $-$51:35:33 & 8 & 45 &$<1$& $-$85 to $-$85 \\
      G331.44$-$0.14$^{*\dagger}$&                & 16:11:57.9 & $-$51:33:08 & 8 & 45 &    & $-$87 to $-$84 \\
      G331.44$-$0.19           & G331.440$-$0.187 & 16:12:11.5 & $-$51:35:02 & 8 & 44 & 5  & $-$92 to $-$85 \\
      G331.44$-$0.16$^{*\dagger}$&G331.442$-$0.158& 16:12:05.0 & $-$51:33:45 & 8 & 40 &$<1$& $-$87 to $-$85 \\
      G331.50$-$0.08           & G331.492$-$0.082 & 16:11:59.6 & $-$51:28:14 & 8 & 46 & 22 & $-$93 to $-$84 \\
      G331.50$-$0.10           & G331.503$-$0.109 & 16:12:10.5 & $-$51:29:23 & 8 & 44 & 17 & $-$101 to $-$86 \\
      G331.52$-$0.08           & G331.519$-$0.082 & 16:12:07.5 & $-$51:27:25 & 8 & 44 & 8  & $-$93 to $-$85 \\
      G331.54$-$0.10           & G331.530$-$0.099 & 16:12:14.5 & $-$51:27:34 & 8 & 45 & 19 & $-$95 to $-$85 \\
      G331.55$-$0.07           & G331.544$-$0.067 & 16:12:11.3 & $-$51:25:29 & 8 & 45 & 7  & $-$92 to $-$85 \\
      G331.56$-$0.12           & G331.555$-$0.122 & 16:12:26.7 & $-$51:27:41 & 8 & 44 & 3  & $-$104 to $-$97 \\
      G331.72$-$0.20$^\dagger$ &                  & 16:13:34.4 & $-$51:24:25 & 8 & 46 &    & $-$49 to $-$45 \\
      G331.86$-$0.13$^\dagger$ & G331.853$-$0.129 & 16:13:52.3 & $-$51:15:42 & 8 & 46 &$<1$& $-$52 to $-$48 \\
      G331.88$+$0.06$^\dagger$ & G331.887$+$0.063 & 16:13:10.5 & $-$51:06:09 & 8 & 44 & 12 & $-$91 to $-$84 \\
      G331.92$-$0.08$^\dagger$ & G331.921$-$0.083 & 16:13:58.0 & $-$51:10:56 & 8 & 46 &$<1$& $-$53 to $-$51 \\
      G332.09$-$0.42$^\dagger$ & G332.092$-$0.420 & 16:16:15.6 & $-$51:18:31 & 8 & 46 & 7  & $-$59 to $-$54 \\
      G332.24$-$0.05$^\dagger$ & G332.240$-$0.044 & 16:15:17.6 & $-$50:55:59 & 8 & 47 & 8  & $-$51 to $-$46 \\
      G332.30$-$0.09           & G332.295$-$0.094 & 16:15:46.0 & $-$50:55:54 & 8 & 46 & 8  & $-$55 to $-$45 \\
      G332.32$+$0.18$^\dagger$ & G332.318$+$0.179 & 16:14:40.0 & $-$50:43:11 & 8 & 45 & 10 & $-$50 to $-$44 \\
      G332.36$-$0.11           & G332.355$-$0.114 & 16:16:07.3 & $-$50:54:16 & 8 & 45 &$<1$& $-$51 to $-$49 \\
      G332.59$+$0.15$^\dagger$ & G332.583$+$0.147 & 16:15:58.5 & $-$50:33:22 & 8 & 45 &$<1$& $-$45 to $-$42 \\
      G332.60$-$0.17$^\dagger$ & G332.604$-$0.167 & 16:17:28.1 & $-$50:46:20 & 8 & 46 & 2  & $-$48 to $-$44 \\
      G332.72$-$0.05$^\dagger$ & G332.716$-$0.048 & 16:17:28.6 & $-$50:36:34 & 8 & 45 & 2  & $-$41 to $-$38 \\
      G333.00$-$0.43           & G333.002$-$0.437 & 16:20:27.7 & $-$50:41:06 & 8 & 45 &$<1$& $-$57 to $-$55 \\
      G333.01$-$0.46$^{* \ddagger}$ & G333.014$-$0.466 & 16:20:37.1 & $-$50:41:31 & 7 & 41 &$<1$& $-$55 to $-$52 \\
      G333.02$-$0.06$^\dagger$ & G333.029$-$0.063 & 16:18:55.9 & $-$50:24:03 & 8 & 43 & 1  & $-$43 to $-$39 \\
      G333.03$-$0.02$^\dagger$ & G333.029$-$0.024 & 16:18:45.9 & $-$50:22:21 & 7 & 41 &$<1$& $-$43 to $-$41 \\
      G333.07$-$0.44$^\ddagger$ & G333.068$-$0.446 & 16:20:49.6 & $-$50:38:51 & 7 & 46 &$<1$& $-$55 to $-$51 \\
      G333.07$-$0.40           & G333.071$-$0.399 & 16:20:37.0 & $-$50:36:33 & 7 & 42 & 6  & $-$54 to $-$51 \\
      G333.10$-$0.51           & G333.103$-$0.502 & 16:21:13.4 & $-$50:39:56 & 7 & 41 & 11 & $-$60 to $-$53 \\
      G333.12$-$0.43$^*$       & G333.121$-$0.433 & 16:20:58.5 & $-$50:35:41 & 7 & 54 & 3  & $-$57 to $-$44 \\
      G333.13$-$0.44           & G333.126$-$0.439 & 16:21:02.3 & $-$50:35:53 & 7 & 55 & 12 & $-$57 to $-$42 \\
      G333.14$-$0.42           & G333.137$-$0.427 & 16:21:03.0 & $-$50:34:59 & 7 & 49 & 17 & $-$57 to $-$43 \\
      G333.16$-$0.10           & G333.162$-$0.101 & 16:19:41.4 & $-$50:19:55 & 7 & 35 & 4  & $-$93 to $-$90 \\
      G333.18$-$0.09           & G333.184$-$0.090 & 16:19:45.2 & $-$50:18:45 & 7 & 37 & 10 & $-$89 to $-$84 \\
      G333.22$-$0.40           & G333.220$-$0.402 & 16:21:16.9 & $-$50:30:17 & 7 & 42 & 4  & $-$57 to $-$49 \\
      \hline
    \end{tabular}
  \end{center}
\end{table*}

\setcounter{table}{1}
\begin{table*}
  \begin{center}
    \caption{{\em - continued.}}
    \begin{tabular}{ lcccc ccc }
      \hline
      \multicolumn{1}{c}{Paper~I}   & Refined     & \multicolumn{2}{c}{Pointing centre}                      & Obs.        & Median RMS   & Radius   & Velocity \\
      \multicolumn{1}{c}{site name} & site name   & $\alpha_{2000}$ & $\delta_{2000}$                        & date        & noise level  & (arcsec) & range    \\
                                    &             & (h:m:s)         & ($^\circ$:$^\prime$:$^{\prime\prime}$) & (Sep. 2013) & (mJy)        &          & (\kms{})   \\
      \hline
      G333.23$-$0.06           & G333.233$-$0.061 & 16:19:49.4 & $-$50:15:17 & 7 & 42 & 14 & $-$96 to $-$81 \\
      G333.24$+$0.02$^\dagger$ &                  & 16:19:26.0 & $-$50:11:29 & 7 & 42 &    & $-$71 to $-$66 \\
      G333.29$-$0.38           & G333.284$-$0.373 & 16:21:28.1 & $-$50:26:35 & 7 & 43 &$<1$& $-$55 to $-$49 \\
      G333.30$-$0.35           & G333.301$-$0.352 & 16:21:24.5 & $-$50:24:47 & 7 & 40 &$<1$& $-$54 to $-$48 \\
      G333.31$+$0.10           & G333.313$+$0.106 & 16:19:27.6 & $-$50:04:45 & 7 & 55 & 8  & $-$51 to $-$43 \\
      G333.33$-$0.36           & G333.335$-$0.363 & 16:21:36.4 & $-$50:23:53 & 7 & 44 & 18 & $-$55 to $-$46 \\
      G333.37$-$0.20$^{\dagger \ddagger}$ & G333.376$-$0.202 & 16:21:05.2 & $-$50:15:18 & 7 & 44 & 5  & $-$63 to $-$56 \\
      G333.39$+$0.02           & G333.387$+$0.031 & 16:20:08.2 & $-$50:04:49 & 7 & 39 & 3  & $-$73 to $-$68 \\
      G333.47$-$0.16           & G333.467$-$0.163 & 16:21:19.5 & $-$50:09:42 & 7 & 40 & 12 & $-$49 to $-$38 \\
      G333.50$+$0.15$^\dagger$ & G333.497$+$0.143 & 16:20:06.5 & $-$49:55:25 & 7 & 42 & 2  & $-$115 to $-$111 \\
      G333.52$-$0.27$^\ddagger$ & G333.523$-$0.275 & 16:22:03.4 & $-$50:12:08 & 7 & 42 &$<1$& $-$52 to $-$49 \\
      G333.56$-$0.30           & G333.558$-$0.293 & 16:22:19.2 & $-$50:11:28 & 7 & 42 & 27 & $-$47 to $-$44 \\
      G333.56$-$0.02$^\dagger$ & G333.562$-$0.025 & 16:21:07.9 & $-$49:59:49 & 7 & 44 &$<1$& $-$43 to $-$37 \\
      G333.57$+$0.03$^\dagger$ & G333.569$+$0.028 & 16:20:55.4 & $-$49:57:25 & 7 & 42 &$<1$& $-$86 to $-$81 \\
      G333.59$-$0.21$^\ddagger$ & G333.595$-$0.211 & 16:22:06.2 & $-$50:06:30 & 7 & 45 & 4  & $-$52 to $-$44 \\
      G333.70$-$0.20           & G333.694$-$0.197 & 16:22:27.7 & $-$50:01:22 & 7 & 41 & 9  & $-$52 to $-$49 \\
      G333.71$-$0.12$^{\dagger \ddagger}$ & G333.711$-$0.115 & 16:22:10.8 & $-$49:57:21 & 7 & 41 & 5  & $-$32 to $-$30 \\
      G333.77$-$0.01$^{\dagger \ddagger}$ & G333.772$-$0.010 & 16:21:59.6 & $-$49:50:20 & 7 & 42 &$<1$& $-$91 to $-$88 \\
      G333.77$-$0.25$^{\dagger \ddagger}$ & G333.773$-$0.258 & 16:23:02.8 & $-$50:00:31 & 7 & 40 &$<1$& $-$50 to $-$47 \\
      G333.82$-$0.30$^\dagger$ & G333.818$-$0.303 & 16:23:29.5 & $-$50:00:40 & 7 & 41 & 1  & $-$50 to $-$46 \\
      G333.90$-$0.10$^\ddagger$ & G333.900$-$0.098 & 16:22:56.3 & $-$49:48:43 & 7 & 41 & 4  & $-$66 to $-$61 \\
      G333.94$-$0.14$^{\dagger \ddagger}$ & G333.930$-$0.133 & 16:23:14.1 & $-$49:48:40 & 7 & 44 & 5  & $-$43 to $-$39 \\
      G333.98$+$0.07$^\dagger$ & G333.974$+$0.074 & 16:22:31.0 & $-$49:38:08 & 7 & 40 &$<1$& $-$62 to $-$57 \\
      G334.03$-$0.04$^\dagger$ & G334.027$-$0.047 & 16:23:16.7 & $-$49:40:57 & 7 & 41 &$<1$& $-$85 to $-$83 \\
      G334.74$+$0.51$^\dagger$ & G334.746$+$0.506 & 16:23:57.9 & $-$48:46:40 & 7 & 38 & 4  & $-$66 to $-$59 \\
      \hline
    \end{tabular}
  \end{center}
\end{table*}

\begin{table*}
  \begin{center}
    \caption{Associations with each of the class~I~\choh{} maser sites. Note that the presence of \cs{}~(1--0) is not listed, because it is detected towards every source. Column 1 lists the refined region name. Columns 2 and 3 list the equatorial coordinates of the mean position of all maser spots within a single maser site. Column 4 lists the kinematic distance \citep{reid09}. Uncertainties are listed in units of the least significant figure. Column 5 lists the associations with other masers, EGOs and ATLASGAL sources within 1~arcmin of the centre position$^1$. If there is more than one of the same type of emission, it is listed as a subscript. Columns 6 through 9 lists whether thermal \sio{}~(1--0), \choh{} 1$_0$--0$_0$ A$^+$, \hfta{} or \ctfs{} (1--0) emission is detected (`Y') or not (`N'). Regions labelled with a dagger ($\dagger$) are designated as `young'; see Section~\ref{sec:evolutionary}. Regions labelled with a double dagger ($\ddagger$) have large shifts induced by self-calibration; see Section~\ref{subsec:cross_correlation}.}
    \label{tab:meth_assoc}
    \begin{tabular}{ lcccc cccc }
      \hline
      Region & \multicolumn{2}{c}{Centre of maser emission}             & Kinematic & Associations & Presence of   & Presence of         & Presence    & Presence     \\
      name   & $\alpha_{2000}$ & $\delta_{2000}$                        & distance  & within       & \sio{} (1--0) & \choh{}             & of \hfta{}? & of \ctfs{}   \\
             & (h:m:s)         & ($^\circ$:$^\prime$:$^{\prime\prime}$) & (kpc)     & 1 arcmin$^1$ & $v=0$?        & 1$_0$--0$_0$ A$^+$? &             & (1--0)?      \\
      \hline
      G330.294$-$0.393           & 16:07:37.9 & $-$52:30:58.29 & 4.9 (2) & WA             & N & Y & Y & Y \\
      G330.678$-$0.402           & 16:09:31.7 & $-$52:15:50.68 & 4.1 (2) & A              & Y & Y & Y & Y \\
      G330.779$+$0.249$^\dagger$ & 16:07:09.8 & $-$51:42:53.76 & 3.1 (2) & A              & Y & Y & N & Y \\
      G330.871$-$0.383           & 16:10:22.1 & $-$52:07:08.28 & 4.1 (2) & MA             & Y & Y & Y & Y \\
      G330.876$-$0.362           & 16:10:17.9 & $-$52:05:59.37 & 4.0 (2) & MWSCGA$_2$     & Y & Y & Y & Y \\
      G330.927$-$0.408$^\dagger$ & 16:10:44.7 & $-$52:05:57.33 & 3.1 (2) & A              & Y & Y & N & Y \\
      G330.931$-$0.260$^\dagger$ & 16:10:06.7 & $-$51:59:18.13 & 5.3 (2) & A              & N & Y & N & Y \\
      G330.955$-$0.182           & 16:09:53.0 & $-$51:54:52.84 & 5.5 (2) & MWCA           & Y & Y & Y & Y \\
      G331.131$-$0.470$^\dagger$ & 16:11:59.4 & $-$52:00:19.88 & 4.3 (2) & GA             & Y & Y & N & Y \\
      G331.132$-$0.244           & 16:10:59.8 & $-$51:50:21.78 & 5.4 (2) & MWCGA          & Y & Y & N & Y \\
      G331.134$-$0.488           & 16:12:05.1 & $-$52:01:01.37 & 4.3 (2) & A              & Y & Y & Y & Y \\
      G331.134$+$0.156$^\dagger$ & 16:09:15.1 & $-$51:32:39.79 & 4.6 (2) & MWA            & Y & Y & N & Y \\
      G331.279$-$0.189           & 16:11:27.2 & $-$51:41:57.04 & 5.3 (2) & MWCA           & Y & Y & Y & Y \\
      G331.341$-$0.347           & 16:12:26.4 & $-$51:46:20.50 & 4.2 (2) & MWCGA          & Y & Y & Y & Y \\
      G331.370$-$0.399$^\dagger$ & 16:12:48.5 & $-$51:47:24.39 & 4.2 (2) & GA             & N & Y & N & Y \\
      G331.371$-$0.145$^\dagger$ & 16:11:41.2 & $-$51:36:15.36 & 5.2 (2) &                & N & Y & N & Y \\
      G331.380$+$0.149$^\dagger$ & 16:10:26.8 & $-$51:22:58.46 & 3.2 (2) & A              & Y & Y & N & Y \\
      G331.409$-$0.164$^\dagger$ & 16:11:57.4 & $-$51:35:32.03 & 5.1 (2) & A$_3$          & Y & Y & N & Y \\
      G331.44$-$0.14$^\dagger$   & 16:11:57.9 & $-$51:33:07.81 & 5.1 (2) & A              & Y & Y & N & N \\
      G331.440$-$0.187           & 16:12:11.9 & $-$51:35:15.76 & 9.6 (2) & MWCA           & Y & Y & N & Y \\
      G331.442$-$0.158$^\dagger$ & 16:12:04.8 & $-$51:33:55.74 & 5.1 (2) &                & Y & Y & N & Y \\
      \hline
    \end{tabular}
    \flushleft
    $^1$M - 6.7\,GHz \choh{} maser from \citet{caswell11}; W - 22\,GHz \water{} maser from any of \citet{breen10,walsh11,walsh14}; S - 1612\,MHz \oh{} maser from \citet{sevenster97}; C - 1665 or 1667\,MHz \oh{} maser from \citet{caswell98}; G - EGO from \citet{cyganowski08}; A - ATLASGAL point source from \citet{contreras13}.
  \end{center}
\end{table*}

\setcounter{table}{2}
\begin{table*}
  \begin{center}
    \caption{{\em - continued.}}
    \begin{tabular}{ lcccc cccc }
      \hline
      Site   & \multicolumn{2}{c}{Centre of maser emission}             & Kinematic & Associations & Presence of   & Presence of         & Presence    & Presence     \\
      name   & $\alpha_{2000}$ & $\delta_{2000}$                        & distance  & within       & \sio{} (1--0) & \choh{}             & of \hfta{}? & of \ctfs{}   \\
             & (h:m:s)         & ($^\circ$:$^\prime$:$^{\prime\prime}$) & (kpc)     & 1 arcmin$^1$ & $v=0$?        & 1$_0$--0$_0$ A$^+$? &             & (1--0)?      \\
      \hline
      G331.492$-$0.082           & 16:11:59.0 & $-$51:28:34.01 & 5.3 (2) & A$_2$          & Y & Y & Y & Y \\
      G331.503$-$0.109           & 16:12:09.3 & $-$51:29:14.95 & 5.2 (2) & WSCA$_3$       & Y & Y & Y & Y \\
      G331.519$-$0.082           & 16:12:06.6 & $-$51:27:25.69 & 5.3 (2) & A              & Y & Y & Y & Y \\
      G331.530$-$0.099           & 16:12:14.3 & $-$51:27:44.11 & 5.4 (2) & A              & Y & Y & Y & Y \\
      G331.544$-$0.067           & 16:12:09.7 & $-$51:25:44.35 & 5.2 (2) & MCA            & Y & Y & Y & Y \\
      G331.555$-$0.122           & 16:12:27.1 & $-$51:27:42.69 & 5.8 (2) & MW$_2$CA$_2$   & Y & Y & Y & Y \\
      G331.72$-$0.20$^\dagger$   & 16:13:34.4 & $-$51:24:24.91 & 3.3 (2) & A              & Y & Y & N & Y \\
      G331.853$-$0.129$^\dagger$ & 16:13:52.5 & $-$51:15:46.63 & 3.4 (2) & A              & Y & Y & N & Y \\
      G331.887$+$0.063$^\dagger$ & 16:13:11.2 & $-$51:05:58.39 & 5.2 (2) & A              & Y & Y & N & Y \\
      G331.921$-$0.083$^\dagger$ & 16:13:59.2 & $-$51:10:56.00 & 3.5 (2) & A              & Y & Y & N & Y \\
      G332.092$-$0.420$^\dagger$ & 16:16:15.7 & $-$51:18:27.64 & 3.7 (2) & MWA            & Y & Y & N & Y \\
      G332.240$-$0.044$^\dagger$ & 16:15:17.2 & $-$50:56:01.32 & 3.4 (2) & A              & Y & Y & N & Y \\
      G332.295$-$0.094           & 16:15:45.3 & $-$50:55:54.87 & 3.5 (2) & MWGA           & Y & Y & Y & Y \\
      G332.318$+$0.179$^\dagger$ & 16:14:40.1 & $-$50:43:07.07 & 3.4 (2) & WA             & Y & Y & N & Y \\
      G332.355$-$0.114           & 16:16:07.2 & $-$50:54:19.92 & 3.5 (2) & MWCGA          & N & Y & N & Y \\
      G332.583$+$0.147$^\dagger$ & 16:16:01.0 & $-$50:33:30.96 & 11.8 (2) & MGA            & N & N & N & Y \\
      G332.604$-$0.167$^\dagger$ & 16:17:29.3 & $-$50:46:12.92 & 3.3 (2) & MWGA           & Y & Y & N & Y \\
      G332.716$-$0.048$^\dagger$ & 16:17:28.5 & $-$50:36:23.32 & 3.0 (2) & A              & N & Y & N & Y \\
      G333.002$-$0.437           & 16:20:28.7 & $-$50:41:00.91 & 3.8 (2) & WA$_3$         & Y & Y & Y & Y \\
      G333.014$-$0.466$^\ddagger$ & 16:20:39.5 & $-$50:41:47.45 & 3.6 (2) & A              & Y & Y & Y & Y \\
      G333.029$-$0.063$^\dagger$ & 16:18:56.7 & $-$50:23:54.64 & 3.0 (2) & MWA            & N & Y & N & Y \\
      G333.029$-$0.024$^\dagger$ & 16:18:46.6 & $-$50:22:14.57 & 3.1 (2) & MA$_2$         & Y & Y & N & Y \\
      G333.068$-$0.446$^\ddagger$& 16:20:48.7 & $-$50:38:38.68 & 3.7 (2) & MWA            & Y & Y & Y & Y \\
      G333.071$-$0.399           & 16:20:37.2 & $-$50:36:30.13 & 3.6 (2) & A              & Y & Y & Y & Y \\
      G333.103$-$0.502           & 16:21:13.1 & $-$50:39:31.37 & 3.8 (2) & MA             & Y & Y & Y & Y \\
      G333.121$-$0.433           & 16:20:59.5 & $-$50:35:51.03 & 3.4 (2) & M$_4$W$_6$C$_2$A$_2$&Y & Y & Y & Y \\
      G333.126$-$0.439           & 16:21:02.6 & $-$50:35:52.44 & 3.5 (2) & M$_4$W$_6$C$_2$A$_2$&Y & Y & Y & Y \\
      G333.137$-$0.427           & 16:21:02.2 & $-$50:34:56.03 & 3.5 (2) & MWC$_2$A       & Y & Y & Y & Y \\
      G333.162$-$0.101           & 16:19:42.5 & $-$50:19:56.29 & 5.3 (2) & MA$_2$         & N & Y & Y & Y \\
      G333.184$-$0.090           & 16:19:45.5 & $-$50:18:34.59 & 5.1 (1) & MGA            & Y & Y & Y & Y \\
      G333.220$-$0.402           & 16:21:17.9 & $-$50:30:19.97 & 3.6 (2) & A              & Y & Y & Y & Y \\
      G333.24$+$0.02$^\dagger$   & 16:19:26.0 & $-$50:11:29.26 & 4.4 (1) & A              & Y & Y & N & Y \\
      G333.233$-$0.061           & 16:19:50.8 & $-$50:15:15.94 & 5.1 (1) & M$_2$W$_3$SCA  & Y & Y & N & Y \\
      G333.284$-$0.373           & 16:21:27.3 & $-$50:26:22.48 & 3.6 (2) & A              & Y & Y & Y & Y \\
      G333.301$-$0.352           & 16:21:25.9 & $-$50:24:46.26 & 3.5 (2) & A$_2$          & Y & Y & Y & Y \\
      G333.313$+$0.106           & 16:19:28.5 & $-$50:04:45.61 & 11.8 (2) & MWCGA          & Y & Y & N & Y \\
      G333.335$-$0.363           & 16:21:38.0 & $-$50:23:46.78 & 3.6 (2) & A              & Y & Y & Y & Y \\
      G333.376$-$0.202$^{\dagger \ddagger}$ & 16:21:06.2 & $-$50:15:13.21 & 4.0 (2) & WA  & Y & Y & N & Y \\
      G333.387$+$0.031           & 16:20:07.5 & $-$50:04:49.51 & 10.6 (1) & MWCA           & Y & Y & N & Y \\
      G333.467$-$0.163           & 16:21:20.2 & $-$50:09:41.70 & 3.1 (2) & MWCGA          & Y & Y & Y & Y \\
      G333.497$+$0.143$^\dagger$ & 16:20:07.6 & $-$49:55:26.21 & 6.3 (2) & A$_2$          & Y & Y & N & Y \\
      G333.523$-$0.275$^\ddagger$ & 16:22:04.5 & $-$50:12:05.32 & 3.5 (2) & A             & Y & Y & Y & Y \\
      G333.558$-$0.293           & 16:22:18.7 & $-$50:11:23.38 & 3.1 (2) & A$_2$          & Y & Y & Y & Y \\
      G333.562$-$0.025$^\dagger$ & 16:21:08.7 & $-$49:59:48.85 & 12.0 (2) & MA             & Y & Y & N & Y \\
      G333.569$+$0.028$^\dagger$ & 16:20:56.6 & $-$49:57:15.91 & 5.0 (1) & A              & Y & Y & N & Y \\
      G333.595$-$0.211$^\ddagger$ & 16:22:06.7 & $-$50:06:21.28 & 3.5 (2) & WSCA          & Y & Y & Y & Y \\
      G333.694$-$0.197           & 16:22:29.1 & $-$50:01:31.84 & 3.5 (2) & A$_2$          & Y & Y & Y & Y \\
      G333.711$-$0.115$^{\dagger \ddagger}$ & 16:22:12.1 & $-$49:57:20.77 & 2.5 (2) & A   & N & N & N & Y \\
      G333.772$-$0.010$^{\dagger \ddagger}$ & 16:22:00.3 & $-$49:50:15.79 & 5.2 (1) &     & N & Y & N & Y \\
      G333.773$-$0.258$^{\dagger \ddagger}$ & 16:23:06.2 & $-$50:00:43.31 & 3.5 (2) & A   & Y & Y & N & Y \\
      G333.818$-$0.303$^\dagger$ & 16:23:30.0 & $-$50:00:41.97 & 3.4 (2) &                & N & Y & N & Y \\
      G333.900$-$0.098$^\ddagger$ & 16:22:57.3 & $-$49:48:35.60 & 10.9 (2) & MSA           & Y & Y & N & Y \\
      G333.930$-$0.133$^{\dagger \ddagger}$ & 16:23:14.5 & $-$49:48:45.94 & 3.1 (2) & MWA & Y & Y & N & Y \\
      G333.974$+$0.074$^\dagger$ & 16:22:31.4 & $-$49:38:08.31 & 3.9 (2) & A              & N & Y & N & Y \\
      G334.027$-$0.047$^\dagger$ & 16:23:16.7 & $-$49:41:00.19 & 5.0 (1) & A$_2$          & N & Y & N & Y \\
      G334.746$+$0.506$^\dagger$ & 16:23:58.0 & $-$48:46:59.12 & 4.2 (1) & A              & Y & Y & N & Y \\
      \hline
    \end{tabular}
  \end{center}
\end{table*}

\begin{table*}
  \begin{center}
    \caption{A sample of Gaussian spectral fits for class~I~\choh{} maser spots: only those in region G331.132$-$0.244. The remainder of fits are provided in Appendix~\ref{app:meth_detail}. Column 1 lists the Galactic coordinates of each distinct spectral feature. In the case where there is more than one velocity component at exactly the same location, the letter associated with the Galactic position is incremented. Spots labelled with an asterisk ($*$) have been manually fitted; see Section~\ref{subsec:cross_correlation}. Columns 2 and 3 list the fitted position for the Gaussian. Columns 4 through 6 list the fitted Gaussian parameters. Note that the uncertainty for each parameter is quoted in parentheses, in units of the least significant figure. Column 7 lists the integrated flux density of the Gaussian.}
    \label{tab:meth_sample}
    \begin{tabular}{ l ll ll d{1}l d{1}l d{1}l d{5} }
      \hline
      Spot name & \multicolumn{2}{c}{$\alpha_{2000}$} & \multicolumn{2}{c}{$\delta_{2000}$}                        & \multicolumn{2}{c}{Peak flux} & \multicolumn{2}{c}{Peak}      & \multicolumn{2}{c}{FWHM}     & \multicolumn{1}{c}{Integrated}   \\
                & \multicolumn{2}{c}{(h:m:s)}         & \multicolumn{2}{c}{($^\circ$:$^\prime$:$^{\prime\prime}$)} & \multicolumn{2}{c}{density}   & \multicolumn{2}{c}{velocity}  & \multicolumn{2}{c}{(\kms{})} & \multicolumn{1}{c}{flux density}    \\
                & &                                   & &                                                          & \multicolumn{2}{c}{(Jy)}      & \multicolumn{2}{c}{(\kms{})}  & &                            & \multicolumn{1}{c}{(Jy\,\kms{})} \\
      \hline
      G331.1308$-$0.2441A & 16:10:59.53 & (2) & $-$51:50:25.79 & (4) & 13.0 & (6) & -87.50 & (3) & 1.45 & (7) & 14.148 \\
      G331.1308$-$0.2441B & 16:10:59.54 & (6) & $-$51:50:25.72 & (5) & 21 & (4) & -90.1 & (1) & 1.6 & (3) & 24.901 \\
      G331.1308$-$0.2441C & 16:10:59.54 & (6) & $-$51:50:25.68 & (9) & 93 & (3) & -91.11 & (1) & 0.93 & (3) & 65.276 \\
      G331.1333$-$0.2458A & 16:11:00.7 & (5) & $-$51:50:23.97 & (6) & 45 & (1) & -88.531 & (8) & 0.70 & (2) & 23.769 \\
      G331.1333$-$0.2410A & 16:10:59.41 & (2) & $-$51:50:11.8 & (1) & 23.1 & (2) & -84.630 & (2) & 0.610 & (5) & 10.604 \\
      G331.1333$-$0.2410B & 16:10:59.45 & (2) & $-$51:50:11.64 & (2) & 2.9 & (1) & -86.19 & (1) & 0.53 & (3) & 1.151 \\
      G331.1322$-$0.2454A & 16:11:00.25 & (6) & $-$51:50:25.71 & (1) & 6.5 & (2) & -84.34 & (1) & 0.81 & (3) & 3.986 \\
      G331.1322$-$0.2454B & 16:11:00.29 & (4) & $-$51:50:26.0 & (1) & 6.3 & (2) & -87.667 & (8) & 0.47 & (2) & 2.244 \\
      G331.1332$-$0.2407A & 16:10:59.30 & (4) & $-$51:50:10.90 & (3) & 2.11 & (6) & -85.718 & (8) & 0.59 & (2) & 0.931 \\
      G331.1313$-$0.2434A & 16:10:59.499 & (9) & $-$51:50:22.76 & (5) & 4.19 & (7) & -86.097 & (3) & 0.412 & (7) & 1.300 \\
      G331.1315$-$0.2441A & 16:10:59.73 & (8) & $-$51:50:24.2 & (1) & 0.90 & (5) & -82.85 & (2) & 0.57 & (4) & 0.386 \\
      G331.1313$-$0.2451A & 16:10:59.9 & (2) & $-$51:50:27.2 & (2) & 0.89 & (7) & -82.53 & (4) & 0.91 & (8) & 0.610 \\
      \hline
    \end{tabular}
  \end{center}
\end{table*}

The auto-correlation sensitivity of these observations is approximately a factor of 5 better than that of Paper~I, and has improved the detection rate of thermal lines towards the class~I~\choh{} masers. Paper~I described detectable \cs{} emission towards 95 per cent of class~I~\choh{} masers, whereas these data have detectable \cs{} towards every maser. Similarly, the detection rate of thermal \sio{} has greatly improved; these data have detectable \sio{} towards 83 per cent of regions (64/77), which is a significant increase over Paper~I (30 per cent). The thermal 1$_0$--0$_0$ A$^+$ line of \choh{} was not characterised in Paper~I, but auto-correlations have been produced for this paper. The thermal line of \choh{} in this frequency band is detected towards almost all regions (75/77; 97 per cent). In addition, we list detections of \hfta{} and \ctfs{} in Table~\ref{tab:meth_assoc}, along with median RMS noise levels in Table~\ref{tab:spectral_lines}. The thermal lines are used for various comparisons in Section~\ref{discussion}. The Gaussian parameters determined for each of the thermal lines are listed in Appendix~\ref{app:thermal_detail}.

\section{Discussion}
\label{discussion}
With the first large and unbiased sample of class~I~\choh{} masers, we perform a number of analyses to determine any relationships between this maser transition and other star formation tracers. Given that we simultaneously collect useful thermal lines associated with HMSF, such as \cs{}~(1--0), we often compare these with the class~I masers to better understand their environments. In addition, we use other surveys pertaining to star formation, such as the Methanol MultiBeam (MMB) and ATLASGAL, in an attempt to identify what class~I~\choh{} maser emission can tell us about their host star-forming regions.

\subsection{Class~I~\choh{} masers on a HMSF evolutionary timeline}
\label{sec:evolutionary}
\citet{voronkov14} found that class~I~\choh{} masers likely trace multiple evolutionary phases of the HMSF timeline, perhaps ranging from very young sources tracing outflows to very evolved sources featuring expanding \hii{} regions. In this subsection, we attempt to divide class~I~\choh{} masers into `young' or `evolved' categories. Paper~I briefly discusses class~I~\choh{} masers with associated class~II~\choh{} or \oh{} masers; class~I~\choh{} masers associated with either or both of these types of maser were assumed to be in a relatively late stage of HMSF. Paper~I notes that the majority of class~I~\choh{} masers lack these associations and therefore seem more likely to be associated with earlier stages of HMSF. With the additional sensitivity provided by the follow-up observations discussed in this paper, we may also use radio recombination line (RRL) data to help discriminate HMSF regions between evolved and young stages. RRLs are associated with \hii{} regions, which typically signpost more evolved stages of HMSF than class~II~\choh{} masers \citep{walsh98}. As \oh{} masers also occur late in an evolutionary timeline (e.g. \citealt{caswell97}), in this paper, we categorise maser sites as `evolved' if \hfta{} emission was detected or an \oh{} maser is associated, or both. Otherwise, the class~I~\choh{} site is deemed `young'. The presence of \hfta{} emission as well as all maser associations with class~I~\choh{} maser sites is listed in Table~\ref{tab:meth_assoc}.

From the 77 observed class~I~\choh{} maser regions, we classify 41 as evolved (being associated with either an \oh{} maser or \hfta{} emission, 53 per cent). Curiously, when using catalogues of class~II~\choh{} and \oh{} masers, we find that a large number of maser sites (16/77, 21 per cent) featuring \hfta{} emission are without class~II~\choh{} or \oh{} masers. Perhaps these regions are too evolved to harbour class~II~\choh{} or \oh{} masers, or alternatively, it could be that the class~II~\choh{} or \oh{} maser emission is too weak to be detected. The population of \oh{} masers within the \malt{} survey region are primarily formed by the catalogue of \citet{caswell98}; this survey details detections to a limit of 0.16\,Jy, but \citet{caswell98} note that the survey is not complete at this level, requiring emission across several channels or corroboration with other data. Additional masers from the SPLASH survey may reveal new associations \citep{dawson14,qiao16}. Across the \malt{} survey region, class~II~\choh{} masers from the MMB survey are complete to a 5$\sigma$ detection limit at 1.0\,Jy \citep{caswell11}. Given the good sensitivity of these surveys, it seems unlikely that undiscovered class~II~\choh{} and \oh{} masers are associated with these class~I~\choh{} masers, but follow-up observations at these transitions may prove effective for finding new detections.

Another explanation for 21 per cent of our sites featuring \hfta{} emission but neither class~II~\choh{} nor \oh{} masers is that bright \hfta{} emission originates from nearby sources of HMSF, contaminating what we would otherwise deem as young regions of star formation. The G333 giant molecular cloud \citep{bains06,fujiyoshi06} powers bright \hfta{} emission, which may cause false-positive associations of \hfta{} at sites near to this complex. Thus, it can be difficult to accurately discriminate between young and evolved HMSF regions by using only RRL data; however, we suggest that this is generally uncommon, as the \water{} Galactic Plane survey (HOPS; \citealt{walsh11}) finds approximately 10 other examples of comparable RRL regions in 100 square-degrees of the Galactic plane. Therefore, we proceed with the presence of \hfta{} emission or an \oh{} maser to discriminate young and evolved regions of star formation.

Non-evolved class~I~\choh{} maser regions comprise 47 per cent of the total population, indicating that class~I~\choh{} masers can occur over a broad span of time in star-forming regions. In the following text, we qualitatively discuss infrared associations with young class~I~\choh{} masers. Images in Appendix~\ref{app:glimpse} show the infrared environment of the detected masers. It can be seen that many masers are cospatial with dark infrared regions, presumably IRDCs, which are dense regions of cold gas projected in front of a bright background. IRDCs are known to host HMSF, thus it is not surprising to find a large population of class~I~\choh{} masers towards these locations.

The three class~I~\choh{} maser sites detected in auto- but not cross-correlated data are also within the young population. G331.72$-$0.20 and G333.24$+$0.02 appear to be associated with dark infrared regions, but G331.44$-$0.14 lacks any obvious infrared feature; it may be originating in an IRDC behind foreground emission, or an example of a region lacking star formation. At present, we are unable to explain why these sites were not detected in cross-correlation data based on their infrared associations; with a larger population, an explanation may become apparent.

\begin{figure}
  \includegraphics[width=\columnwidth]{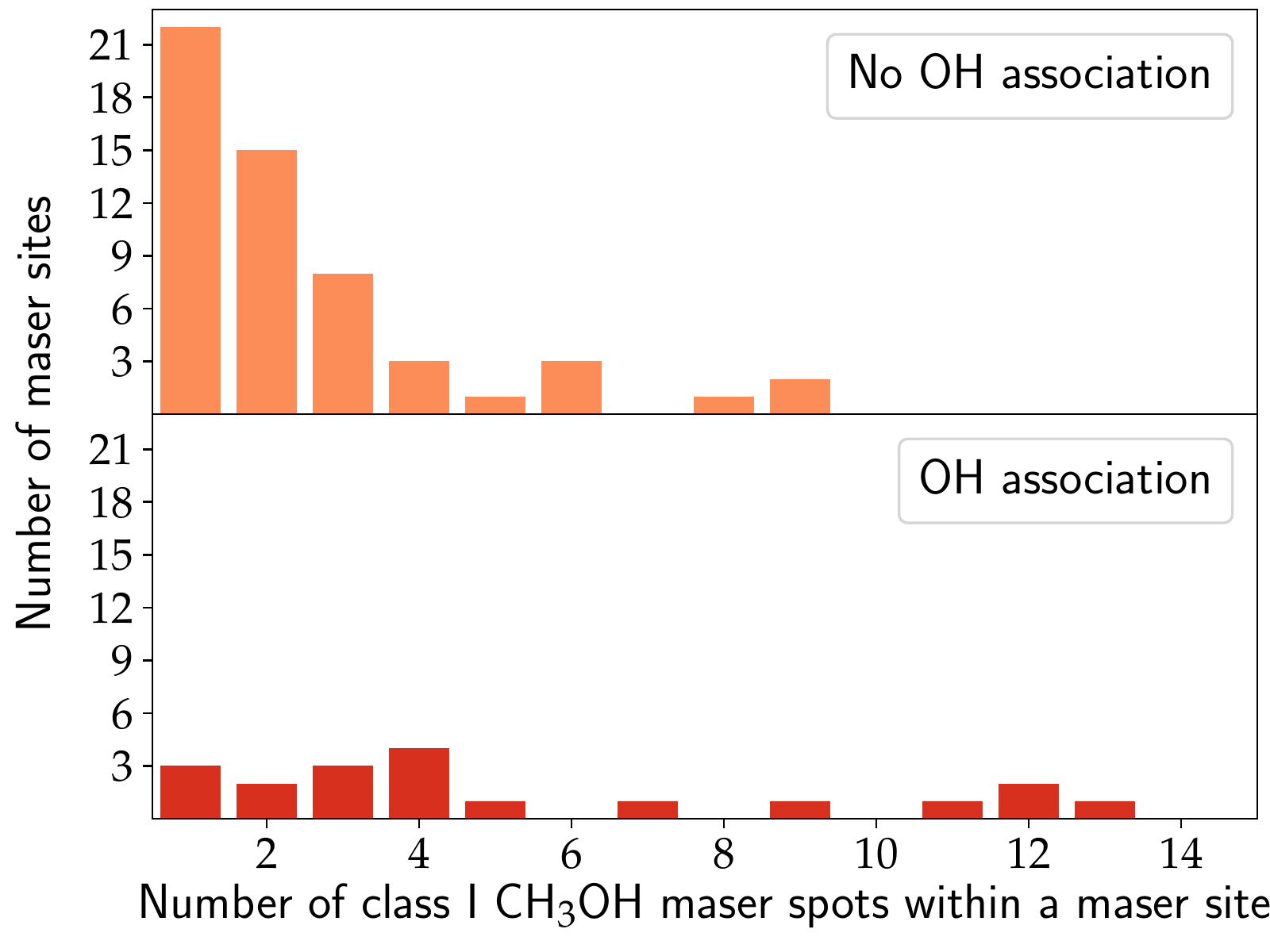}
  \includegraphics[width=\columnwidth]{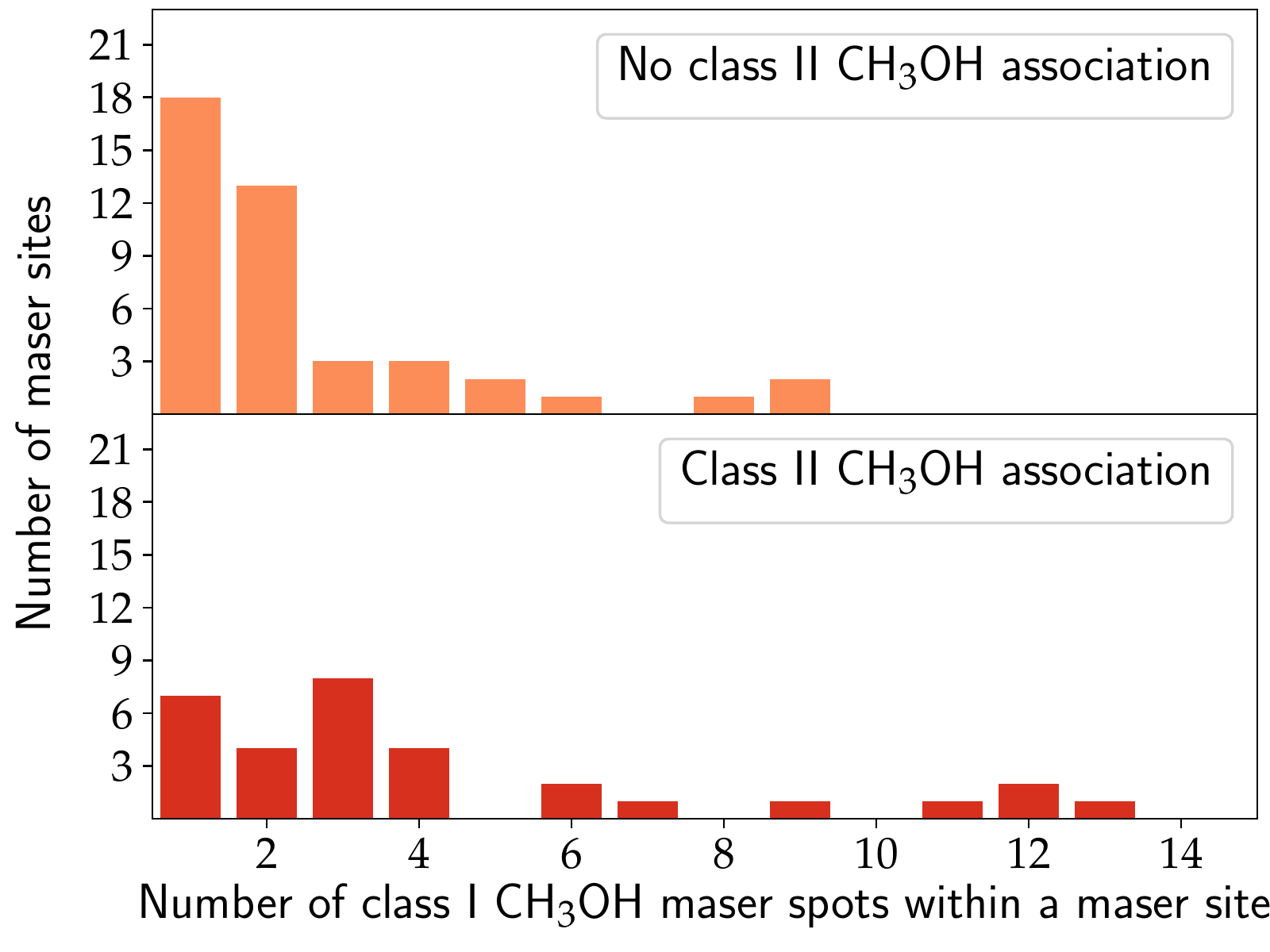}
  \includegraphics[width=\columnwidth]{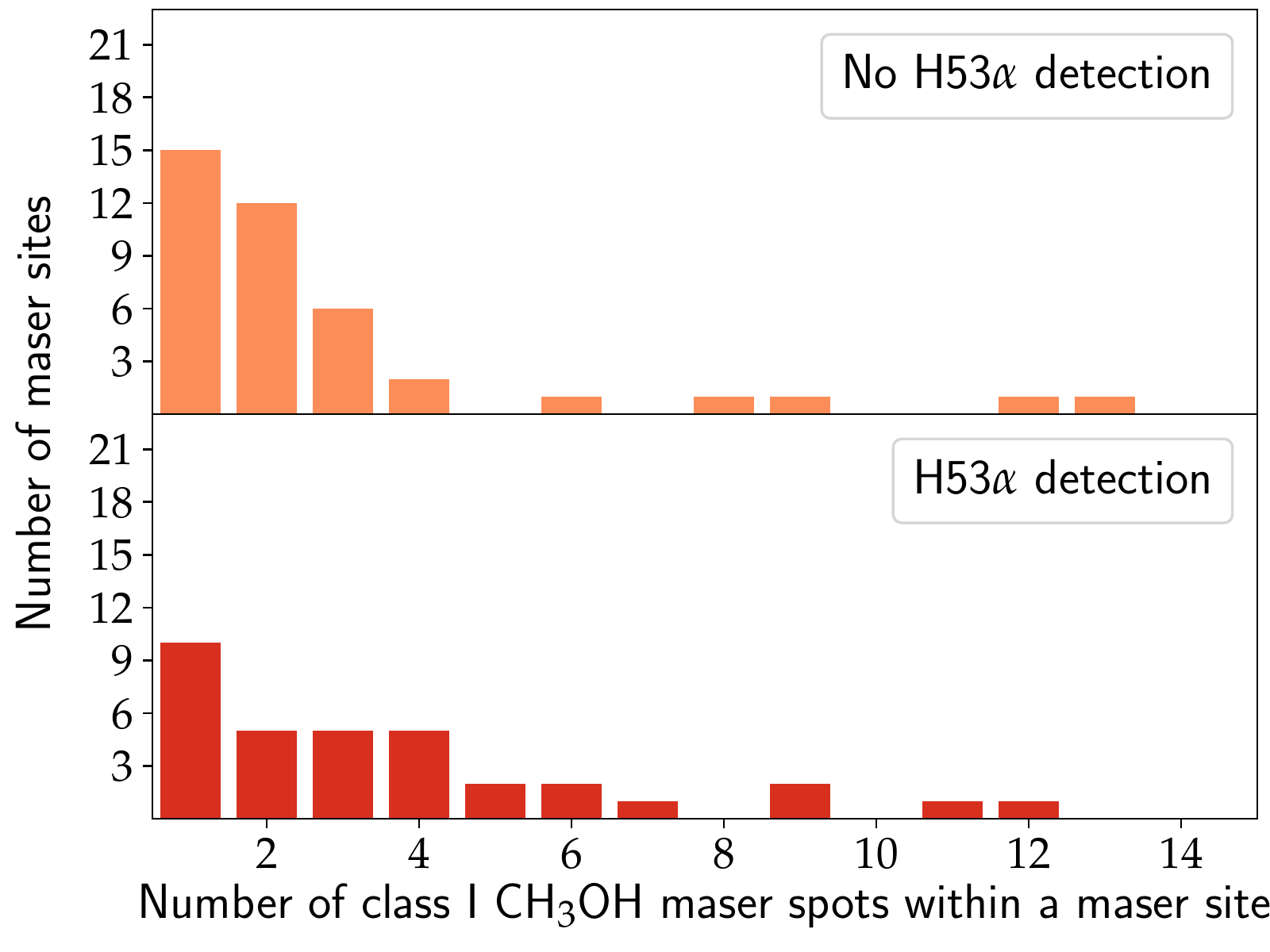}
  \caption{Histograms of class~I~\choh{} maser spot counts within each maser region, highlighting associations with and without an \oh{} maser, class~II~\choh{} maser or \hfta{} detection. The majority of class~I~\choh{} maser regions with only one or two spots do not have an associated \oh{} or class~II~\choh{} maser, but there is a tendency for more associations in regions with more class~I~\choh{} maser spots. A KS test on the samples with and without an \oh{} maser association shows a 1.2 per cent probability that the two samples are drawn from the same population. The same KS test performed on class~II~\choh{} maser and \hfta{} detection samples have a 1.1 and 22 per cent probability, respectively.}
  \label{fig:spot_counts}
\end{figure}

Fig.~\ref{fig:spot_counts} shows histograms of class~I~\choh{} maser site spot counts with and without other star-formation tracers. Perhaps most interestingly, the comparison with \oh{} masers suggests that class~I~\choh{} maser sites with few spots (less than five) are unlikely to be associated with an \oh{} maser. To check the significance of this claim, as well as the similarity between the other populations, we have performed Kolmogorov-Smirnov (KS) tests. KS tests on the populations associated with and without \oh{} masers, class~II~\choh{} masers and \hfta{} emission finds 1.2, 1.1 and 22 per cent probability that they are drawn from the same distribution, respectively. It seems that the number of detected class~I~\choh{} maser spots is a good indication for the presence of other masers, but not necessarily the presence of radio recombination line emission.

With only a few possible exceptions, each of the class~I~\choh{} masers identified by \malt{} appear to be associated with HMSF. As we find many maser sites without associated class~II~\choh{} or \oh{} masers or \hfta{} emission, we appear to be identifying early stages of HMSF. Thus, class~I~\choh{} masers can provide a useful means of identifying star formation, and further \malt{} observations will find more young star-forming regions purely through maser emission. In addition to isolating a young population, this work has shown that the number of class~I~\choh{} maser spots detected may be indicative of the evolutionary stage of the region, which further demonstrates the usefulness of this spectral line.

\subsection{Properties of detected class I \choh{} masers}
\label{sec:basic_properties}
\begin{figure}
  \subfigure{\includegraphics[width=\columnwidth]{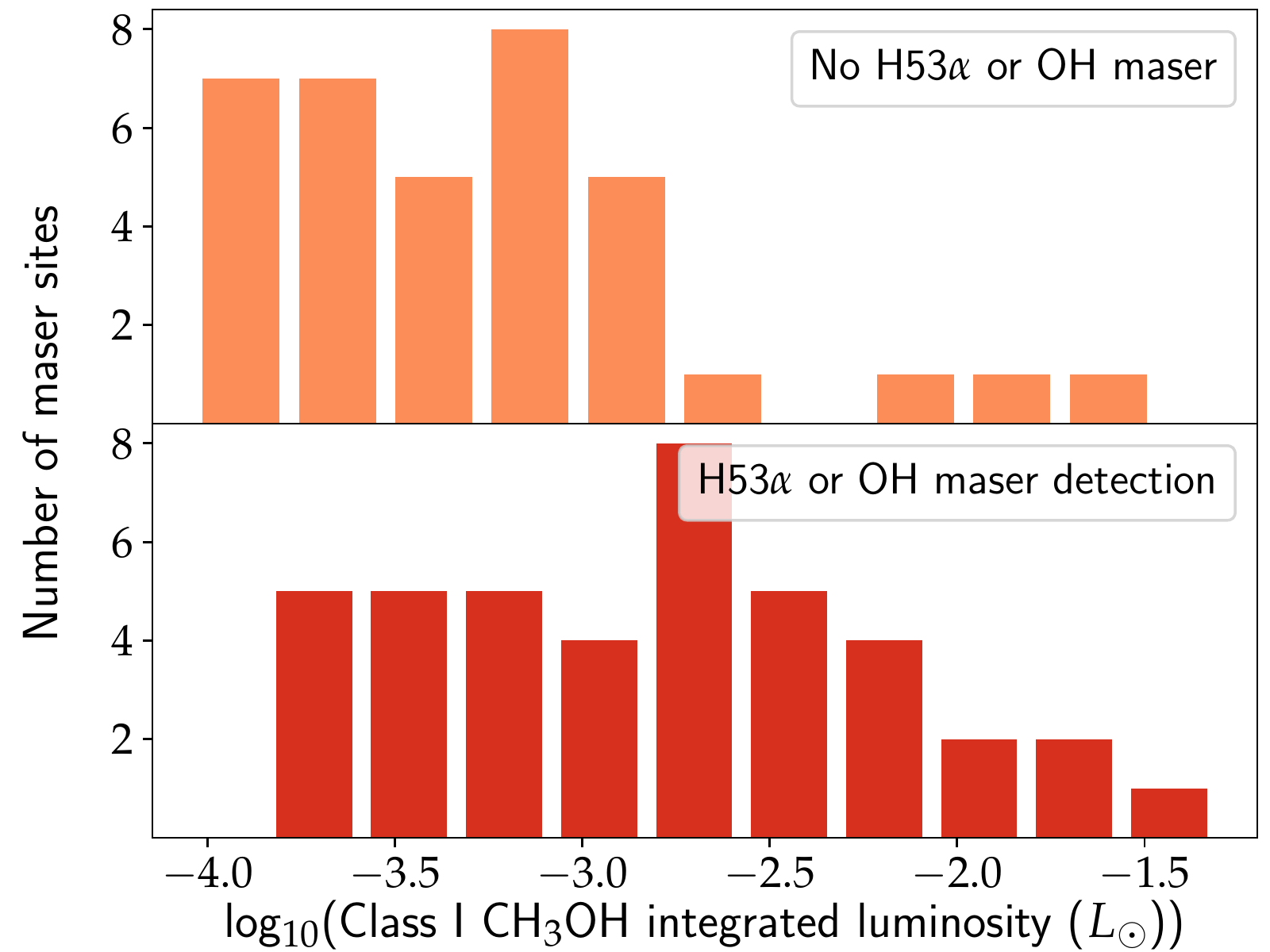}}
  \caption{Luminosities of class~I~\choh{} maser regions, associated with and without \hfta{} emission or an \oh{} maser. There are relatively few masers with high luminosities that are not associated with \hfta{} emission or an \oh{} maser. A KS test finds a 0.2 per cent chance that both samples are drawn from the same population.}
  \label{fig:lum_properties}
\end{figure}

\begin{figure*}
  \subfigure{\includegraphics[width=0.9\textwidth]{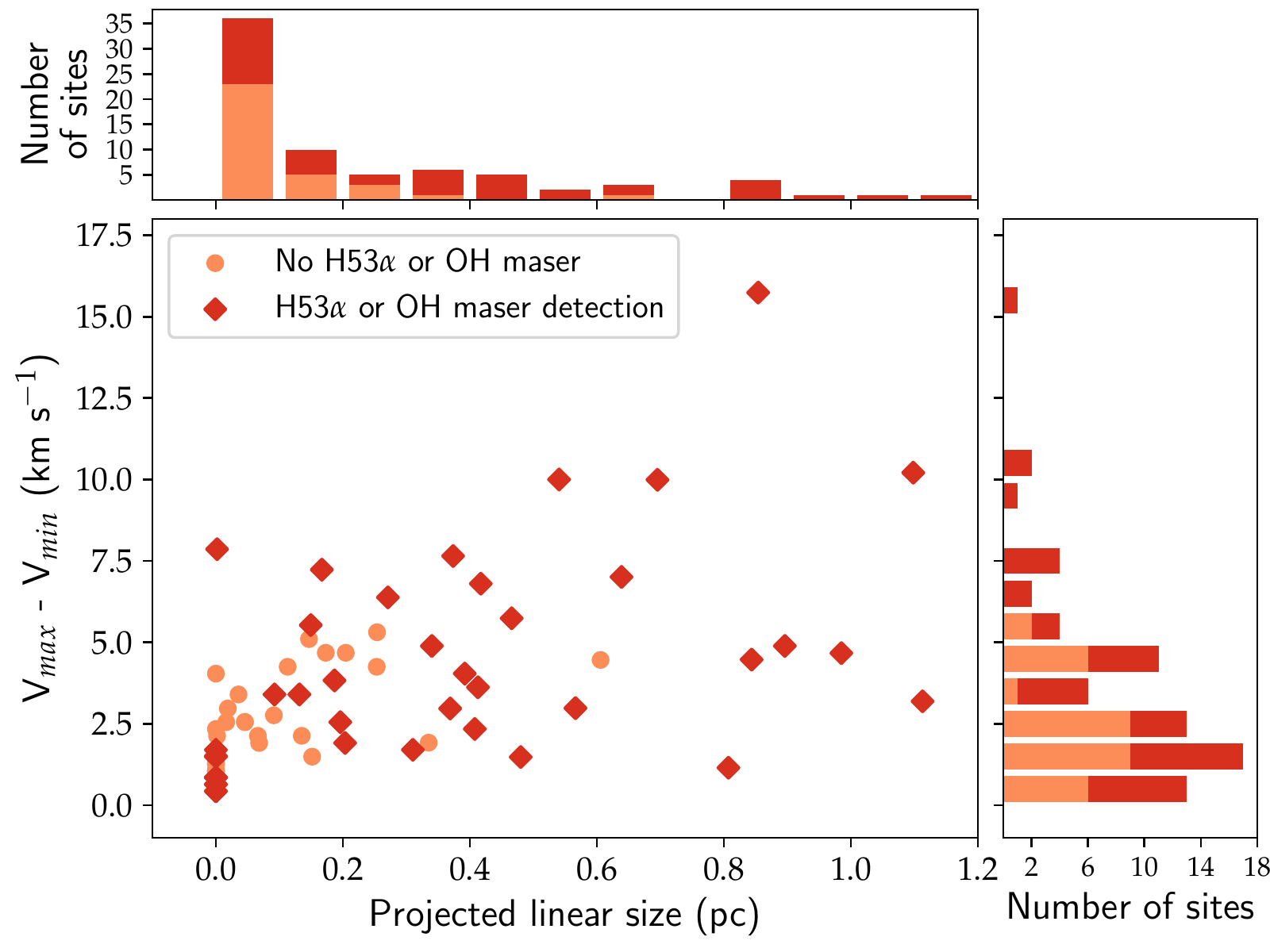}}
  \caption{Projected linear sizes and velocity ranges of class~I~\choh{} maser regions. The majority of maser regions have relatively small projected linear sizes and velocity ranges. Regions that are not evolved tend to have lower luminosities and smaller projected linear sizes and velocity ranges; KS tests suggest that the populations with and without \hfta{} emission and or an associated \oh{} maser are distinct (see Section~\ref{sec:sizes_vels}).}
  \label{fig:size_vel_properties}
\end{figure*}

\subsubsection{Luminosities}
Using the information gathered from these observations, we conduct analyses of the basic properties of class~I~\choh{} masers. Fig.~\ref{fig:lum_properties} shows the luminosities of these masers. Luminosities are determined from the integrated intensity of auto-correlated emission, calculated by Gaussian fits to maser spectra.

The histogram in Fig.~\ref{fig:lum_properties} presents luminosities of class~I~\choh{} masers, highlighting the populations associated with and without \hfta{} emission or an \oh{} maser. A KS test finds a 0.2 per cent chance that both samples are drawn from the same population. Using the presence of \hfta{} emission or an \oh{} maser as an indication for a relatively evolved region, this result suggests that the luminosity of class~I~\choh{} masers can indicate the evolutionary stage of its host star-forming region.

\subsubsection{Projected linear sizes and velocity ranges}
\label{sec:sizes_vels}
Fig.~\ref{fig:size_vel_properties} compares the projected linear sizes and velocity ranges of young and evolved class~I~\choh{} maser regions. The projected linear size is calculated using the angular size of maser emission across a region and its derived kinematic distance (see Table~\ref{tab:meth_assoc}). Velocity range simply refers to the difference between the most redshifted and blueshifted emission within a single region. The majority of sources are confined to small projected linear sizes ($<$0.5\,pc) and small velocity ranges ($<$5\,\kms{}). Of the maser sites that exceed these sizes and velocity ranges, relatively few are young. KS tests were performed on both young and evolved samples of projected linear sizes and velocities; the probability that each sample is drawn from the same distribution is $<$10$^{-1}$ per cent and 4.3 per cent, respectively, suggesting that these populations are distinct.

\citet{voronkov14} also analysed the projected linear sizes and velocity ranges of class~I~\choh{} masers, but between populations with and without an associated \oh{} maser. Here, we observe similarities between our two populations and theirs. Class~I~\choh{} masers observed by \citet{voronkov14} without an associated \oh{} maser are confined to relatively small projected linear sizes ($<$0.4\,pc) and velocity ranges ($<$10\,\kms{}) (fig.~5 of their paper). In addition, when associated with an \oh{} maser, \citet{voronkov14} find that the projected linear sizes and velocity ranges of class~I~\choh{} maser sites are typically larger.

While the projected linear sizes of our study and that of \citet{voronkov14} have similar values (up to ~1\,pc), the velocity ranges do not. The mean velocity ranges of our class~I~\choh{} masers are 4.1 and 2.4\,\kms{} for the evolved and young populations, respectively, while the \citet{voronkov14} with and without \oh{} maser association are approximately 10 and 5\,\kms{}, respectively. For comparison, the mean velocity range of our class~I~\choh{} masers associated with a class~II~\choh{} and \oh{} maser is 5.3\,\kms{}, whereas class~II~\choh{} without \oh{} is 3.2\,\kms{}. This difference is likely due to the sample of class~I~\choh{} masers used by \citet{voronkov14} being typically biased toward class~II~\choh{} masers. The difference between this value and that of \citet{voronkov14} might be due to our sample being more unbiased.

\subsubsection{Spatial distributions}
\label{sec:spatial_distributions}
The spatial distributions of class~I~\choh{} masers have been discussed in the literature. \citet{kurtz04} measured the distance between class~I~\choh{} masers and \hii{} regions, and \citet{voronkov14} compared the distance between class~II and class~I~\choh{} masers. For our class~I~\choh{} maser regions featuring another type of star-formation maser (\oh{}, class~II~\choh{} or \water{}), a projected linear distance comparison is presented in Section~\ref{sec:maser_separation}. As we do not have a common object to compare within each maser site (such as an \hii{} region or class~II~\choh{} maser), we instead compare the maximum distance between any two maser spots within a site. Here, we briefly discuss the maximum spatial offset (projected linear size) between any two maser spots within a site; see Fig.~\ref{fig:size_vel_properties}.

Seven maser sites have a projected linear size greater than 0.8\,pc; these sites are G330.876$-$0.362, G331.279$-$0.189, G331.492$-$0.082, G331.503$-$0.109, G331.530$-$0.099, G333.313$+$0.106 and G333.558$-$0.293. Upon inspection of the distribution of these masers with infrared maps, it is likely that not all the class~I~\choh{} masers are related to a single high-mass object; see Appendix~\ref{app:glimpse}. G330.876$-$0.362 has only two spots, very far from each other; one is closely associated with class~II~\choh{}, \water{} and \oh{} masers, the other with an infrared dark cloud (IRDC). G331.279$-$0.189 has a single maser spot very far from the rest of the relatively clustered spots, and may not be powered by a common source. G333.558$-$0.293 has only two maser spots; the infrared map shows what appears to be a different IRDC associated with each. G331.492$-$0.082, G331.503$-$0.109, G331.530$-$0.099 and G333.313$+$0.106 may be exceptions. There is no obvious infrared distinction between the maser spots of G331.492$-$0.082 to rule out a common source. The spots associated with G331.503$-$0.109 and G331.530$-$0.099 both may be powered by the same source located at G331.512$-$0.100, which also features \water{} and \oh{} masers, as there are no other apparent infrared sources that could be powering either maser spot. Despite the large projected linear size, the spots associated with G333.313$+$0.106 appear to be associated with the same object, which also features an EGO and a class~II~\choh{} maser.

While it is difficult to discern the origin of each maser spot from observations of maser emission alone, it seems that the large offsets between spots are not necessarily common, and may be erroneously generated by calculating a radius for each observed maser `site'. While these masers are collisionally-excited and therefore more likely to appear at large distances from their powering source, we suggest that genuine class~I~\choh{} maser associations over large distances are uncommon. Perhaps the class~I masers are tracing weak, continuous C-type shocks rather than powerful J-type shocks \citep{widmann16}; in addition, we might expect masers triggered by stronger shocks to have broader velocity ranges than those we observe.

\subsection{Comparing class I \choh{} masers with other masers and thermal lines}
\subsubsection{Separation from other maser species in star-forming regions}
\label{sec:maser_separation}
\begin{figure}
  \includegraphics[width=\columnwidth]{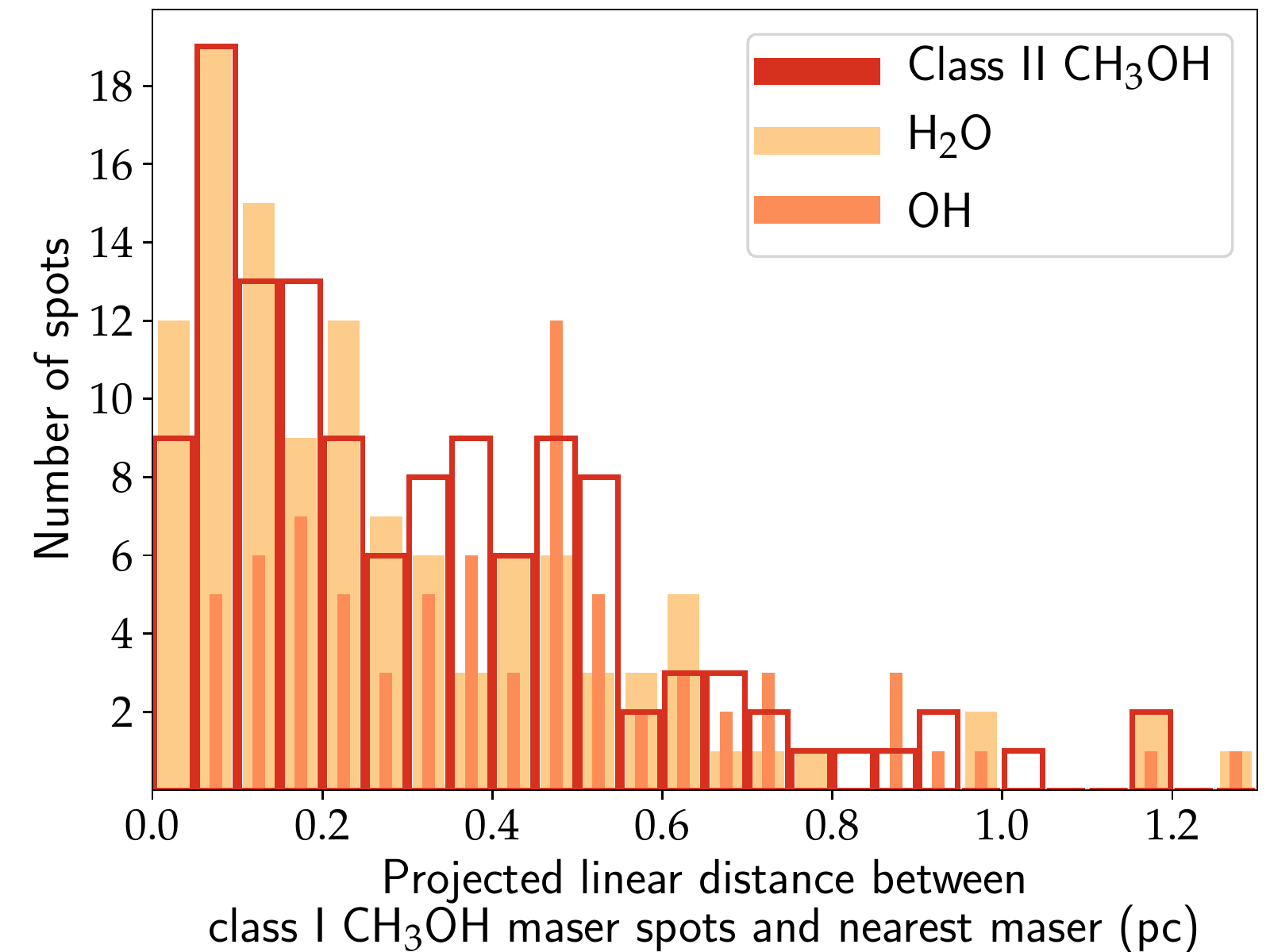}
  \caption{Histogram of the projected linear distance between every class~I~\choh{} maser spot position and class~II~\choh{}, \oh{} and \water{} masers. This histogram uses the projected distance between masers only when their angular offset is less than 60~arcsec. The majority of projected linear distances to masers are within 0.5\,pc.}
  \label{fig:maser_projected_hist}
\end{figure}

Section~\ref{sec:spatial_distributions} is able to infer the age and size of host star-forming regions by simply using the properties of class~I~\choh{} maser spots. Here, we compare the distances between class~I~\choh{} maser spots and other maser species to determine the spatial extents of star-forming regions; this is prudent in conjunction with Section~\ref{sec:spatial_distributions}, as class~II~\choh{} and \oh{} masers are relatively near to their powering source.

The projected linear distances between every spot of a class~I~\choh{} maser site and class~II~\choh{}, \oh{} and \water{} masers are presented in Fig.~\ref{fig:maser_projected_hist}. The other star-formation maser positions were gathered from \citet{caswell11,sevenster97,caswell98,breen10,walsh11,walsh14}. A maser site was considered to be associated with a secondary maser site if the two were less than 60~arcsec from each other. If more than one of the same maser species is within 60~arcsec of a class~I~\choh{} maser spot, only the closest was used for comparison.

Most projected linear distances are less than 0.5\,pc. \citet{kurtz04} find that the distance between class~I~\choh{} masers and \hii{} regions tend to be within 0.5\,pc, in agreement with our results. \citet{voronkov14} were able to model an exponential decay to the number of class~I~\choh{} maser spots with distance from class~II~\choh{} masers; with their larger sample size, they see many masers at projected linear distances beyond 0.5\,pc. However, the vast majority are within 0.5\,pc, also in agreement with our results. Given that our maser spots are derived from an unbiased sample, it seems that typical regions of HMSF power class~I~\choh{} maser activity within a distance of 0.5\,pc.

The majority of projected linear distances between class~I~\choh{} maser spots and class~II~\choh{} or \water{} masers are at small distances ($<$0.2\,pc), whereas \oh{} masers have a flat distribution out to approximately 0.5\,pc. The similarity of the projected linear distances in each comparison can be determined with KS tests; the probabilities of each distribution being drawn from the same sample are $<$10$^{-1}$, $<$10$^{-1}$ and 49.0 per cent for \oh{} and \water{}, \oh{} and class~II, and \water{} and class~II, respectively. Thus, \oh{} masers have significantly higher projected linear distances from class~I~\choh{} masers compared to class~II~\choh{} or \water{} masers.

In general, maser associations are well characterised within distances of 0.5\,pc. We believe that most associations beyond 0.5\,pc are also genuine, but we would require additional information to conclude otherwise.

\subsubsection{Velocities}
\begin{figure*}
  \includegraphics[width=\textwidth]{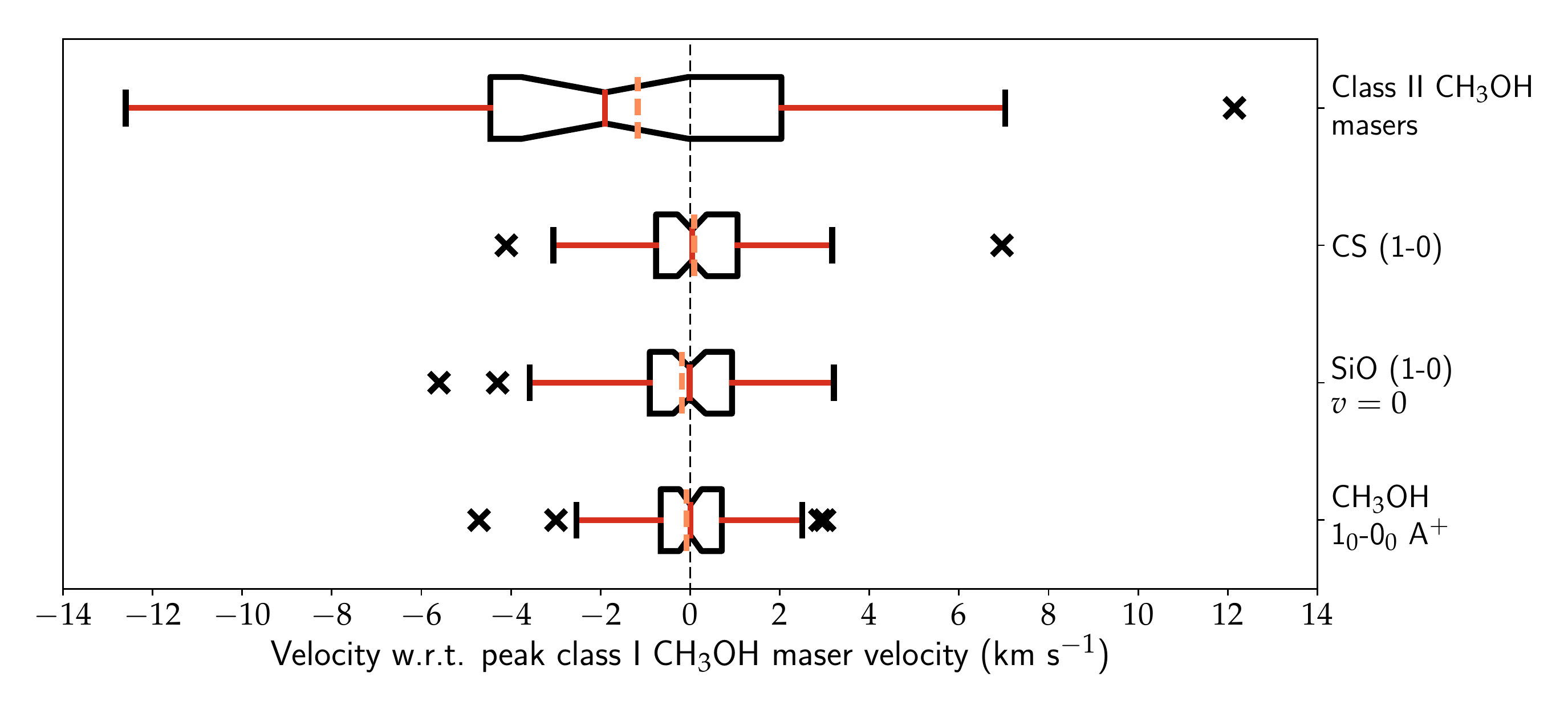}
  \caption{Notched box-and-whisker plot of peak velocity differences between class~I~\choh{} masers and various species associated with star formation. The red solid vertical line within the boxes indicates the median, the orange dashed lines the mean. The notched region about the median indicates the 95 per cent confidence interval on the median, and the box covers the interquartile region (IQR; middle 50 per cent of data). Whiskers extend past the IQR by up to 1.5$\times$IQR. Black crosses indicate outlying data. The dashed black line indicates zero peak velocity difference. The class~II~\choh{} maser box plot has a relatively wide range of velocity differences compared to the other lines. \cs{}, \sio{} and thermal \choh{} velocities are all closely associated with the maser velocity.}
  \label{fig:velocity_difference_box}
\end{figure*}

Paper~I finds that class~I~\choh{} masers are good tracers of systemic velocity; with these follow-up data, we repeat this type of analysis with more thermal lines, as well as class~II~\choh{} maser emission. The peak velocities of class~I~\choh{} masers in auto-correlated data are compared with class~II~\choh{} masers, thermal \cs{}, \sio{} and \choh{} in Fig.~\ref{fig:velocity_difference_box}. We discuss the resulting distributions in the following text.

\citet{voronkov14} used a large sample of class~I and class~II~\choh{} masers to investigate their relative velocities. Analysing the distribution of velocity differences between class~I and class~II~\choh{} masers, they find a peak velocity difference at $-0.57 \pm 0.07$\,\kms{} with a standard deviation of $3.32 \pm 0.07$\,\kms{} and a slight blueshifted asymmetry. The cause of the blueshift could not be attributed to either the class~I or class~II~\choh{} masers. As class~II~\choh{} masers occur near to a YSO, their velocities are thought to be tracers of systemic velocities, albeit with a large dispersion \citep{szymczak07,green11}. In the following discussion, we analyse the peak velocities of thermal lines observed in these data, and find that class~I~\choh{} masers are significantly better tracers of systemic velocities than class~II~\choh{} masers.

The distribution of velocity differences in Fig.~\ref{fig:velocity_difference_box} which compares class~I and class~II~\choh{} masers has the statistics $\mu = -1.17 \pm 0.87$\,\kms{}, $\tilde{x} = -1.90$\,\kms{} and $\sigma = 4.75 \pm 0.61$\,\kms{} ($\mu$ is the mean, $\tilde{x}$ the median and $\sigma$ the standard deviation). The parameters of this comparison are consistent with that of \citet{voronkov14}. The median and mean are also blueshifted, but unlike \citet{voronkov14}, are statistically insignificant.

Paper~I compared the peak velocities of \cs{}~(1--0) and class~I~\choh{} masers where each maser was detected. From that work, the resulting statistics were $\mu = 0.0 \pm 0.2$\,\kms{}, $\tilde{x} = -0.1$\,\kms{} and $\sigma = 1.5 \pm 0.1$\,\kms{}. The results of the same comparison with these data corroborate that of Paper~I: $\mu = 0.09 \pm 0.18$\,\kms{}, $\tilde{x} = 0.04$\,\kms{} and $\sigma = 1.56 \pm 0.17$\,\kms{}. Given that \cs{} traces very dense gas ($n_c > 10^5$\,cm$^{-3}$), the peak \cs{} velocity is likely closely related to the systemic velocity of a molecular cloud. Hence, the statistics suggest that class~I~\choh{} masers are also good tracers of systemic velocities. As \cs{} emission can be quite bright in these data, it is possible that optically thick emission, if present, is causing uncertainty in the systemic velocity, and hence skews the distribution in Fig.~\ref{fig:velocity_difference_box}. To help resolve this matter, we analysed the peak velocities of \ctfs{} emission. With only one exception, \ctfs{} was detected toward each observed class~I~\choh{} region, and the peak velocity of \ctfs{} agrees with the peak \cs{} velocity to within 0.5\,\kms{} in each region. Therefore, we consider peak \cs{} velocities as accurate systemic velocity tracers. \citet{green11} compared the peak velocity of \cs{} (2--1) and mid-velocity of class~II~\choh{} maser emission, finding a mean and median velocity difference of 3.6 and 3.2\,\kms, respectively; their relatively large velocity differences corroborate our results for class~II~\choh{} masers compared against class~I~\choh{} masers.

The measured difference between \sio{} emission and class~I~\choh{} maser velocities is similar to the difference between \cs{} emission and class~I~\choh{} maser velocities: $\mu = -0.19 \pm 0.21$\,\kms{}, $\tilde{x} = -0.01$\,\kms{} and $\sigma = 1.65 \pm 0.15$\,\kms{}. The distribution featuring \sio{} is slightly wider than that featuring \cs{}, but we place less emphasis on this difference in line width, since the difference is small. Overall, the peak velocities of class~I~\choh{} masers and \sio{} emission closely agree.

The differences in velocity between the peak velocity of the thermal 1$_0$--0$_0$ A$^+$ line of \choh{} and class~I~\choh{} masers also shows a tight association: $\mu = -0.08 \pm 0.15$\,\kms{}, $\tilde{x} = 0.01$\,\kms{} and $\sigma = 1.33 \pm 0.11$\,\kms{}. As a sufficient abundance of \choh{} gas is needed for maser emission, it is perhaps not surprising that the thermal velocity is closely matched to the maser velocity. Given the distribution relative to \cs{}, it seems that thermal \choh{} is also a good tracer of systemic velocities.

The differences in the peak velocity for the thermal \cs{}, \sio{} and \choh{} spectral lines all appear consistent with zero, with uncertainties of a couple of \kms{}. The distribution featuring velocity differences using class~II~\choh{} masers, however, is broad and has a relatively distinct mean and median blueshift. This hints that the peak velocity of class~II~\choh{} masers tend to be blueshifted from the systemic velocity, although the statistics presented above only provide tentative evidence of this. One explanation of the blueshift is that the strongest class~II~\choh{} masers in a region are preferentially detected in the foreground of star-forming regions rather than in the background. This preference may occur because at 6.7\,GHz radio continuum free-free emission may be optically thick. This would be especially true for younger sources. The preference to see blueshifted class~II~\choh{} masers could be explained by the masers occurring in an outflow or an expanding shell. But blueshifted masers are not easily explained if the masers occur in a circumstellar accretion disk.

\subsubsection{Brightness}
\label{sec:thermal_brightness}
\begin{figure*}
  \includegraphics[width=\textwidth]{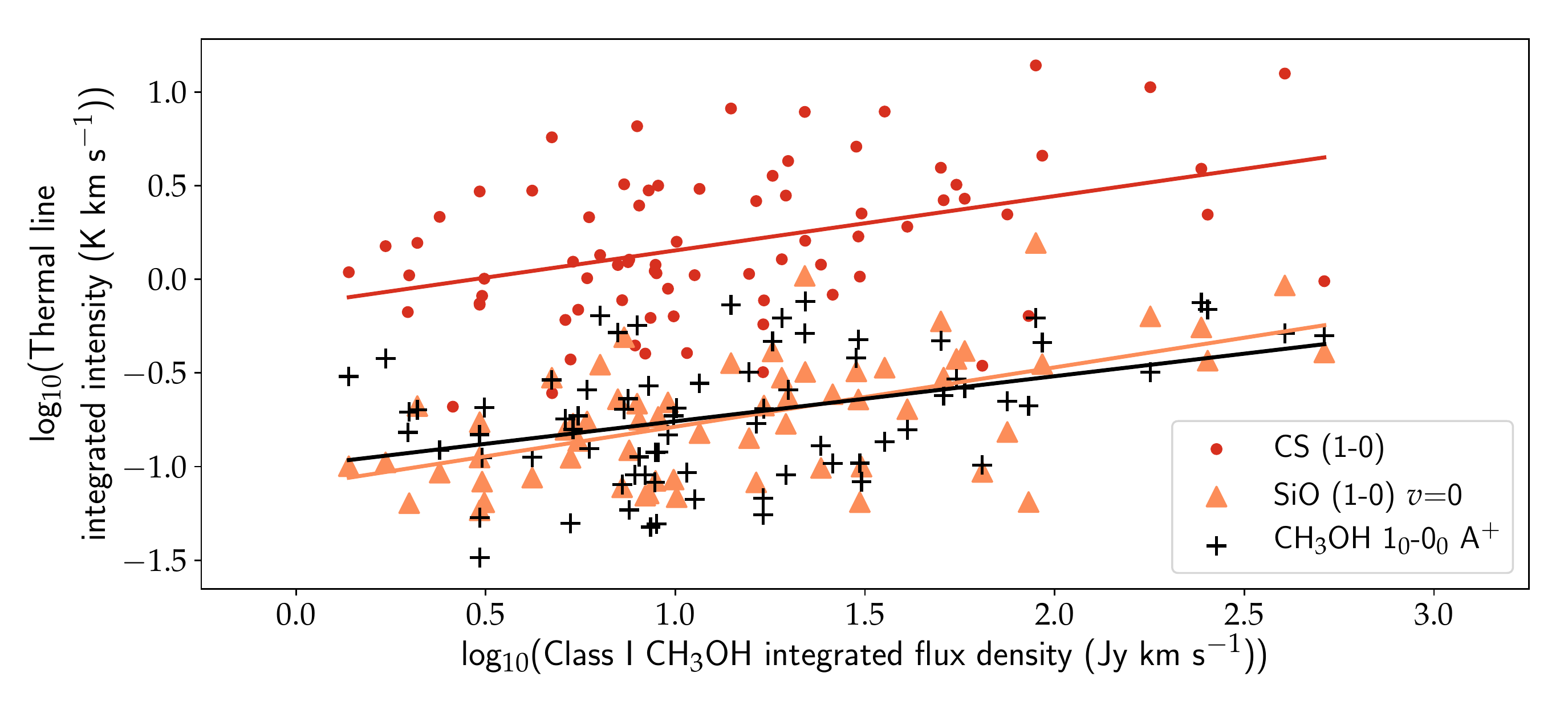}
  \caption{Integrated flux density scatter plots of \cs{}, \sio{} and thermal \choh{} against class~I~\choh{} masers. All three comparisons show a similar trend, albeit with a large degree of scatter; the $r$-values for the lines of best fit with \cs{}, \sio{} and thermal \choh{} are 0.41, 0.57 and 0.40, respectively.}
  \label{fig:thermal_scatter}
\end{figure*}

With a statistically-complete sample of class~I~\choh{} masers and enhanced sensitivity to auto-correlated emission, the observations of this work have the opportunity to identify relations, if any, between the luminosities of class~I~\choh{} masers and other species associated with star formation. In this subsection, we use the auto-correlated integrated flux densities of class~I~\choh{} maser emission to compare with the thermal \cs{}, \sio{} and \choh{} integrated intensities; see Fig.~\ref{fig:thermal_scatter}.

In each comparison, a similar positive trend exists, although the degree of scatter in each comparison is large. The $r$-value for the \cs{} comparison is 0.41.

\sio{}~(1--0)~$v=0$ is typically thought to trace strongly-shocked gas, particularly found in outflows. As class~I~\choh{} masers are collisionally excited by weak shocks, they may be excited in regions also containing \sio{} emission. As discussed earlier, analyses with \sio{} were not possible in Paper~I, due to the low detection rate and relatively weak intensities of \sio{}~(1--0)~$v=0$. Consequently, Paper~I speculates on the nature of the class~I~\choh{} maser sites, given that the \sio{} detection rate was quite low (30 per cent). As the majority of regions now have confirmed \sio{} detections, it seems that a large portion of class~I~\choh{} maser sites indeed have shocked gas associated with them. As bright \sio{} emission within these maser sites is rare, perhaps their faint intensities indicate weak shocks, similar to what was found by \citet{widmann16}. The correlation coefficient ($r$-value) between the integrated intensity of \sio{} and class~I~\choh{} maser integrated flux density (IFD) is 0.57.

Similar to \sio{}, comparisons with thermal \choh{} was not possible in Paper~I. The correlation here is the weakest, with an $r$-value of 0.40.

The correlation coefficient for each comparison indicates a moderate correlation. This combined with a significant positive slope indicates that brighter class~I~\choh{} masers are more likely to be associated with brighter thermal lines. We might expect that higher-mass regions of star formation to contain more molecular gas, and thus the thermal line emission to be brighter. Given the moderate correlations between the brightness of class~I~\choh{} masers and each of the thermal lines discussed in this section, class~I masers may in turn hint at the mass of their host star-forming regions.

Class II~\choh{} masers have been found to be more luminous in more evolved regions of HMSF \citep{breen10b}; if there was a relationship between class~I and class~II~\choh{} masers, then the same might be true for class~I~\choh{} masers. The luminosity of class~I and class~II~\choh{} masers was compared, but no correlation was observed, corroborating the results of Paper~I. Due to class~I~\choh{} masers being associated with more than one evolutionary phase on HMSF timeline, it is likely that a simple relationship between them and class~II~\choh{} masers does not exist.

\subsection{Comparing class I \choh{} masers with 870\,$\mu$m dust continuum from ATLASGAL}
\label{subsec:atlasgal}
\begin{figure}
  \includegraphics[width=\columnwidth]{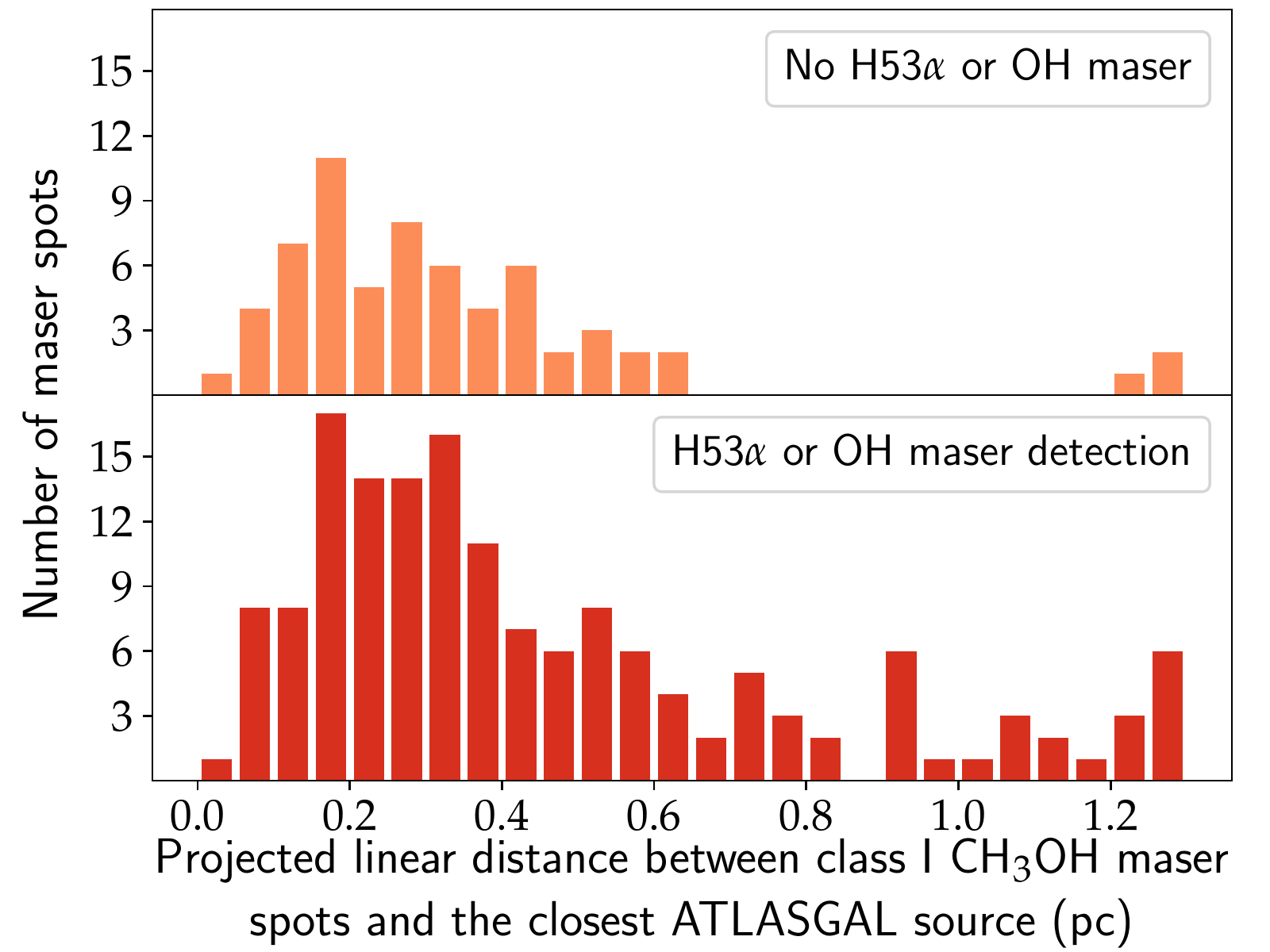}
  \caption{Histograms of offsets between ATLASGAL clumps and class~I~\choh{} maser spots, for young and evolved regions. The majority of associations have projected linear distances less than 0.4\,pc, and almost all are within 0.6\,pc. This indicates a strong relation between the region containing the ATLASGAL source and class~I~\choh{} maser emission. A small population also exists with projected linear distances larger than 1\,pc. Note that for each class~I~\choh{} maser spot, if multiple ATLASGAL sources are present within 60~arcsec, only the closest was used.}
  \label{fig:atlasgal_offset_hist}
\end{figure}

ATLASGAL surveyed 870\,$\mu$m dust continuum emission across a large part of the Galactic plane ($330^\circ \leq l \leq 60^\circ$; \citealt{schuller09}), including the \malt{} region. This sub-millimetre emission traces cold dust, which is optically thin, and can in turn be used to infer column densities and clump masses.

The catalogue of \citet{contreras13} provides the location and integrated flux densities of 870\,$\mu$m dust continuum emission point sources. Using these, we can investigate relationships between class~I~\choh{} masers and 870\,$\mu$m emission, such as the brightness of each and the typical clump mass containing a maser site. We associated any class~I~\choh{} masers and ATLASGAL clumps within an angular offset of 60~arcsec, to be consistent with the other maser associations, and list the associations in Table~\ref{tab:meth_assoc}. Only 4 of our 77 maser regions lack an association with an 870\,$\mu$m clump (5 per cent).

\subsubsection{Separation between ATLASGAL 870\,$\mu$m dust clumps and class~I~\choh{} maser spots}
Section~\ref{sec:maser_separation} analyses the distances between class~I~\choh{} masers and other masers associated with star formation; here, we analyse how class~I masers compare against point sources of dust continuum.

Histograms of projected linear distances between class~I~\choh{} maser spots and 870\,$\mu$m clumps are provided in Fig.~\ref{fig:atlasgal_offset_hist}. The majority of maser spots are within 0.4\,pc of an ATLASGAL source, but the more evolved maser distribution has a long tail down to approximately 1.2\,pc. Quantified with a KS test, we find a 2.4 per cent probability that both samples are drawn from the same population. This is similar to the projected linear distances between class~I~\choh{} maser spots and \oh{} masers shown in Fig.~\ref{fig:maser_projected_hist}, and may be due to the same effect: evolved regions of star formation are more likely to affect a greater volume than younger ones.

Note that Fig.~\ref{fig:atlasgal_offset_hist} includes any ATLASGAL clump within 60~arcsec of a maser spot. Maser sites without ATLASGAL associations are discussed in Section~\ref{sec:atlasgal_exceptions}.

\subsubsection{Masses}
\begin{figure}
  \includegraphics[width=\columnwidth]{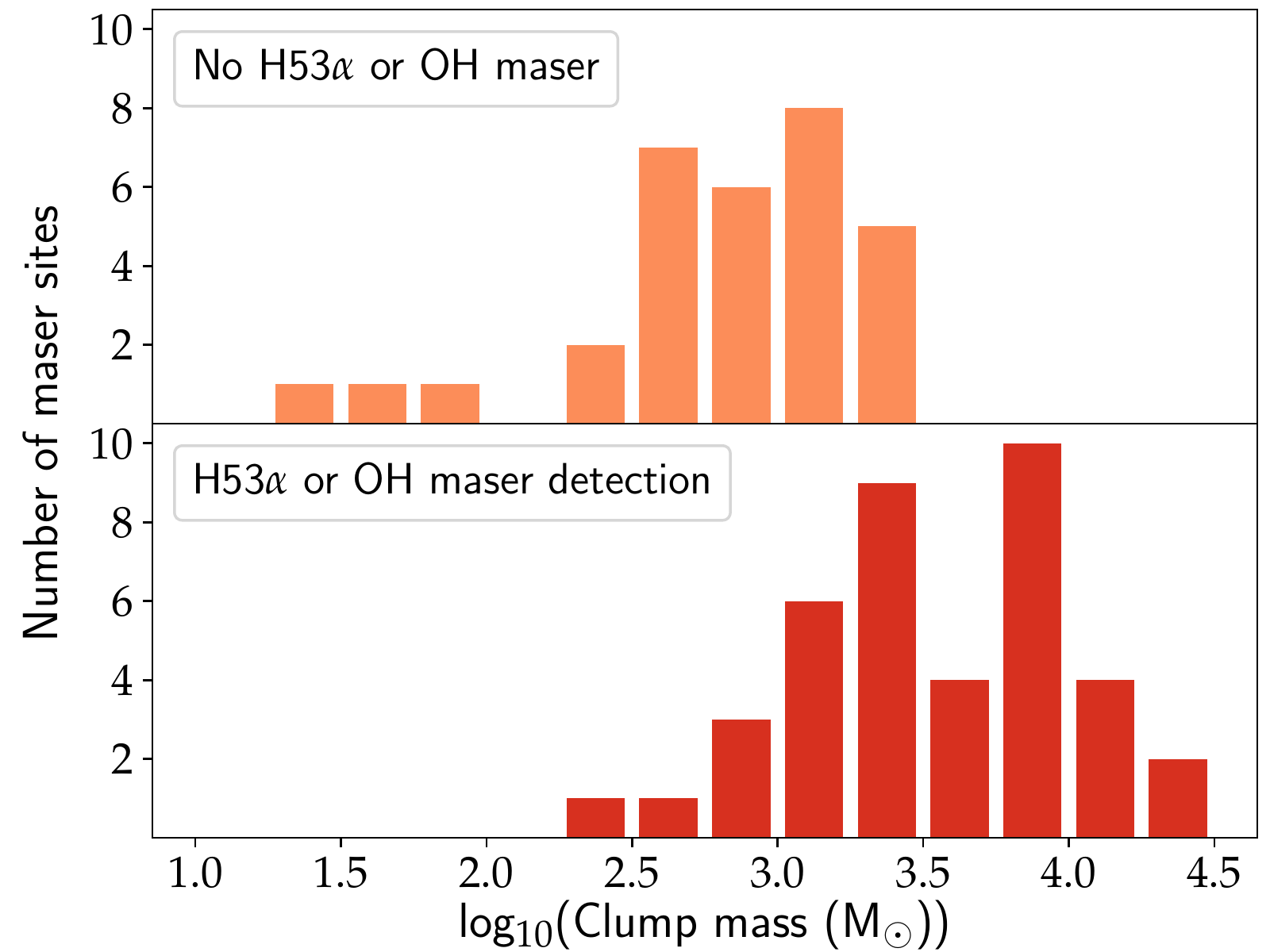}
  \caption{Histograms of clump masses associated with class~I~\choh{} masers, determined by using 870\,$\mu$m ATLASGAL dust continuum, for young and evolved populations. The mean mass of the young distribution is $10^{3.0}$\,$M_{\odot}$, whereas the mean mass of the evolved distribution is $10^{3.74}$\,$M_{\odot}$. The highest mass values are restricted to evolved masers regions, while the lowest masses are all young maser regions.}
  \label{fig:atlasgal_mass_hist}
\end{figure}

It is useful to compare the properties of class~I~\choh{} masers against the mass of their host star-forming regions. \citet{chen12} and \citet{urquhart13} estimate clump masses from the integrated intensity of millimetre-wavelength continuum emission, assuming the emission is optically thin. In the same manner, we use ATLASGAL data to estimate clump masses via:

\begin{subequations}\label{eq:mass}
  \begin{align}
    M_{\text{gas}} = \frac{S_{\nu} D^2 R_d}{\kappa_d B_\nu \left(T_{\text{dust}}\right)}
    \tag{1}
  \end{align}
\end{subequations}

where $M_{\text{gas}}$ is the mass of the gas, $S_{\nu}$ is the integrated flux density of ATLASGAL 870\,$\mu$m emission, $D$ is distance to the maser, $R_d$ is the ratio of gas and dust masses, $\kappa_d$ is the mass-absorption coefficient per unit mass of dust, and $B_\nu\left(T_{\text{dust}}\right)$ is the Planck function at temperature $T_{\text{dust}}$. \citet{urquhart13} justify the choice of $T_{\text{dust}} = 20$\,K and $\kappa_d = 1.85$\,cm$^2$\,g$^{-1}$, which we also use here. We also assume $R_d = 100$. The assumption of a single dust temperature is not realistic, but in the absence of temperature measurements toward every individual region, is necessary. \citet{pandian12} find the kinetic temperatures of gas in regions with class~II~\choh{} masers, determined by \nh{} observations, have mean and median values of 26 and 23.4\,K, respectively. \citet{urquhart11} also find mean and median temperatures toward high-mass YSOs to be 22.1 and 21.4\,K, respectively. If we instead assume a dust temperature of 25\,K, masses will decrease to 73 per cent of those determined with 20\,K. Using a temperature of 15\,K increases masses by 56 per cent relative to the 20\,K masses. These uncertainties are comparable to those given by the kinematic distances. The histogram of masses can be seen in Fig.~\ref{fig:atlasgal_mass_hist}.

Class~I~\choh{} masers appear to be associated with a wide range of clump masses ($10^{1.25} < \frac{M}{M_\odot} < 10^{4.5}$). \citet{urquhart13} associate 870\,$\mu$m dust continuum emission with class~II~\choh{} masers, and calculate clump masses. They find clump masses associated with class~II~\choh{} masers can range from $10^{-2} < \frac{M}{M_\odot} < 10^{6}$. As clump mass is consistent with that of \citet{urquhart13}, and all class~II~\choh{} masers are associated with HMSF, the lower end of these clump masses cannot necessarily serve as an indication of low-mass star formation regions.

Fig~\ref{fig:atlasgal_offset_hist} shows a difference in the calculated mass range for clumps associated with \hfta{} emission or an \oh{} maser, however, this difference is consistent with the sources we expect to be less evolved (clumps not associated with \hfta{} emission or an \oh{} maser) to be at a lower temperature than assumed 20\,K. Other small contributions to this mass discrepancy may be attributable to a slight bias in the detectability of \hfta{} whereby higher-mass objects are more likely to have detectable \hfta{} emission at slightly younger age, or the possibility of genuine associations with low-mass stars. A KS test shows that the probability that both histograms are drawn from the same distribution is $<$10$^{-1}$ per cent.

\subsubsection{Brightness}
\begin{figure}
  \includegraphics[width=\columnwidth]{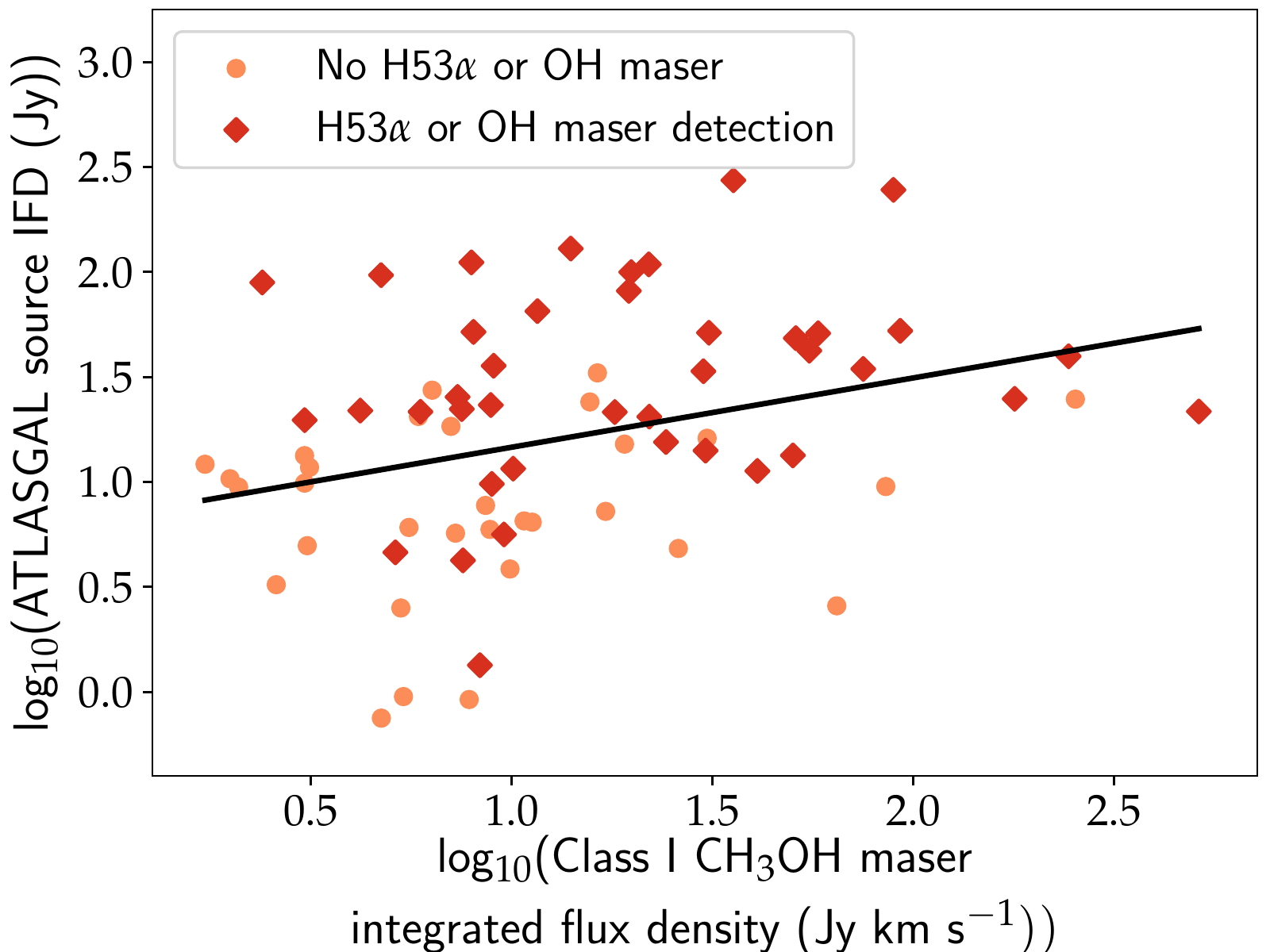}
  \caption{Scatter plot of ATLASGAL 870\,$\mu$m dust continuum emission and class~I~\choh{} maser integrated flux densities. The line of best fit for all data (young and evolved) is shown, and has an $r$-value of 0.33.}
  \label{fig:atlasgal_scatter}
\end{figure}

Fig.~\ref{fig:atlasgal_scatter} compares the IFD of 870\,$\mu$m ATLASGAL dust continuum clumps and the luminosity of class~I~\choh{} masers, but only a weak correlation exists. \citet{chen12} performed a similar comparison using 95\,GHz class~I~\choh{} masers towards Bolocam Galactic Plane Survey (BGPS) sources of 1.1\,mm thermal dust emission (fig.~6 of their paper). They find a strong correlation between the maser luminosity and clump mass, with a $r$-value of 0.84 and $p$-value of 8.1$\times$10$^{-13}$. Our line of best fit has the coefficients $r =$ 0.33, $p = $ 5.7$\times$10$^{-3}$, and is shown on Fig.~\ref{fig:atlasgal_scatter}. The large difference between correlation coefficients may be attributed to the \citet{chen12} sample of BGPS sources being selected based on their GLIMPSE colours, which is likely biasing their statistics. On the other hand, given that the relationship between 44\,GHz and 95\,GHz class~I~\choh{} masers is well established \citep{valtts00}, our lack of correlation observed may highlight differences between 870\,$\mu$m and 1.1\,mm emission, with 1.1\,mm emission originating from more evolved regions. More data is necessary to establish a connection between class~I~\choh{} masers (44 and 95\,GHz, to eliminate biases) and dust continuum emission. In particular, finding more 44\,GHz class~I~\choh{} masers toward sources with the same GLIMPSE colours selected by \citet{chen12} would help to eliminate biases.

Here, our correlation coefficient of 0.33 is not much weaker than those found in our comparisons of thermal lines to class~I~\choh{} masers (0.41, 0.57 and 0.40 for \cs{}, \sio{} and thermal \choh{}, respectively) in Section~\ref{sec:thermal_brightness}. In that section, we speculate that the brightness of the thermal lines is proportional to mass; indeed, a comparison between the integrated intensity of \cs{} and the IFD 870\,$\mu$ dust continuum emission has a correlation coefficient of 0.84. As the brightness of dust continuum emission is directly proportional to mass, the result in this section strengthens the notion that the brightness of class~I~\choh{} masers can hint at the mass of their host star-forming regions.

\subsubsection{Class~I~\choh{} maser sites without an ATLASGAL source}
\label{sec:atlasgal_exceptions}
Only four class~I~\choh{} maser regions are without an ATLASGAL point source classified by \citet{contreras13}: G331.371$-$0.145, G331.442$-$0.158, G333.772$-$0.010 and G333.818$-$0.303. Using the data provided by the ATLASGAL team, we are able to investigate the 870\,$\mu$m emission in each of these regions. Within $\sim$12~arcsec, G331.371$-$0.145, G331.442$-$0.158 and G333.818$-$0.303 each have peak 870\,$\mu$m flux densities of 0.41, 0.63 and 0.23\,Jy\,beam$^{-1}$, respectively. The ATLASGAL 1$\sigma$ sensitivity over the \malt{} area is approximately 60\,mJy\,beam$^{-1}$. Point sources were classified by the source extraction algorithm \textsc{SExtractor}. Despite the pixel values of 870\,$\mu$m dust emission toward these three maser sites being at least 3$\sigma$, \textsc{SExtractor} has most likely not been satisfied with a spatial morphology criterion. However, we do consider the 870\,$\mu$m emission at these locations to be genuine. For these three regions, we determine the IFD of 870\,$\mu$m emission to be 0.57, 0.28 and 0.24\,Jy, respectively. Using Equation~\ref{eq:mass}, these values correspond to clump masses of 85, 40 and 15\,$M_\odot$, respectively. These masses are relatively low compared to those discussed in Section~\ref{subsec:atlasgal}, but even lower masses were determined from integrated flux densities in the ATLASGAL catalogue.

G333.772$-$0.010 has a single pixel of 870\,$\mu$m dust continuum emission with a peak flux density of 0.25\,Jy\,beam$^{-1}$ at a projected linear distance of $\sim$27~arcsec. However, unlike the other regions of 870\,$\mu$m emission, nearby pixels are relatively dim ($<$0.14\,Jy\,beam$^{-1}$, $<\sim$2$\sigma$), indicating this pixel is likely a random noise spike. For sources at a similar distance ($\sim$5.2\,kpc), we might expect to see a similar distribution of 870\,$\mu$m emission, but this is not the case. The class~I~\choh{} maser emission is closely associated with compact infrared emission. Interestingly, the maser emission is strong (peak of 17.7\,Jy, IFD of 11.7\,Jy\,\kms{}), but the thermal lines are weak (the integrated intensity of \cs{} and thermal \choh{} are 0.319 and 0.057\,K\,\kms{}, respectively, and no \sio{} is detected). Since \citet{urquhart15} do not detect an ATLASGAL counterpart towards approximately 7 per cent of class~II~\choh{} masers from the MMB, we may have a similar example of star formation without a significant dust detection.

\subsection{Comparing class I \choh{} masers in cross- and auto-correlation}
\label{sec:cross_vs_auto}
\citet{minier02} and \citet{bartkiewicz14} discuss class~II~\choh{} maser emission detected by very-long-baseline interferometry and single-dish/auto-correlation observations. These comparisons between cross- and auto-correlation data reveal a significant amount of resolved-out flux density; \citet{bartkiewicz14} report between 24 and 86 per cent. Naturally, an interferometer will resolve out any extended emission relative to the synthesised beam. The authors elaborate that the `missing flux' is not dependant on distance, and that lower-resolution cross-correlation data have similar compact structures. This suggests that missing emission is quite diffuse, and is seemingly independent of the properties of compact maser emission. \citet{minier02} also attribute missing flux as being due to diffuse emission.

While these investigations were exclusively focused on class~II~\choh{} masers, the observations detailed in this paper are able to undertake a similar investigation for class~I~\choh{} masers. It is worth noting that the observations of \citet{minier02} and \citet{bartkiewicz14} have a resolution approximately two orders of magnitude better than ours. They find that class~II~\choh{} maser emission is partially resolved on 100\,km baselines, or longer. This is equivalent to a baseline length of approximately 15\,km at 44\,GHz. Since we note significantly reduced flux densities on \malt{} baselines of 6\,km or less, this implies that the 44\,GHz \choh{} emission regions are on typically larger scales than the 6.7\,GHz \choh{} regions.

\begin{figure}
  \centering
  \includegraphics[height=0.30\textheight]{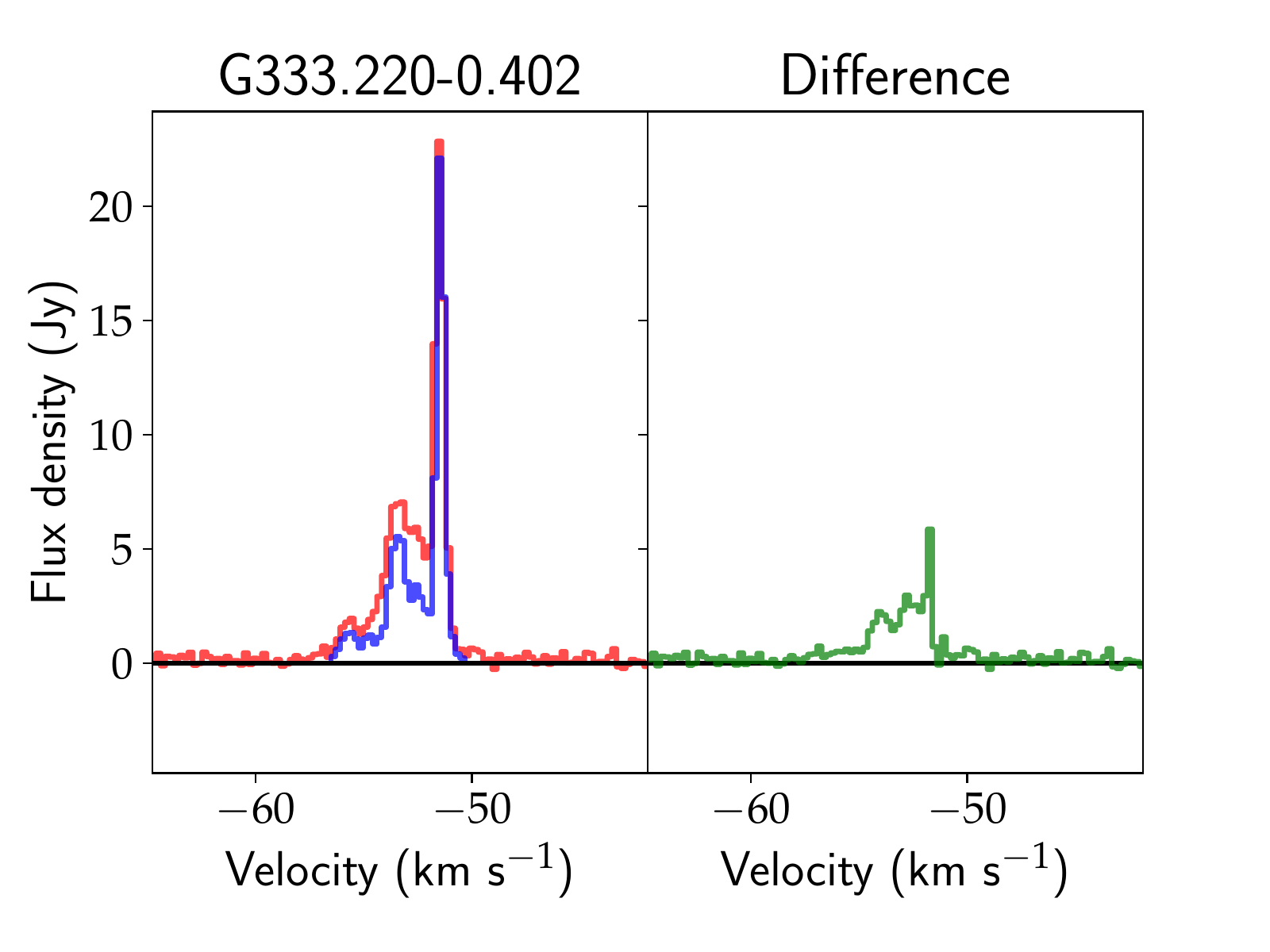}
  \includegraphics[height=0.30\textheight]{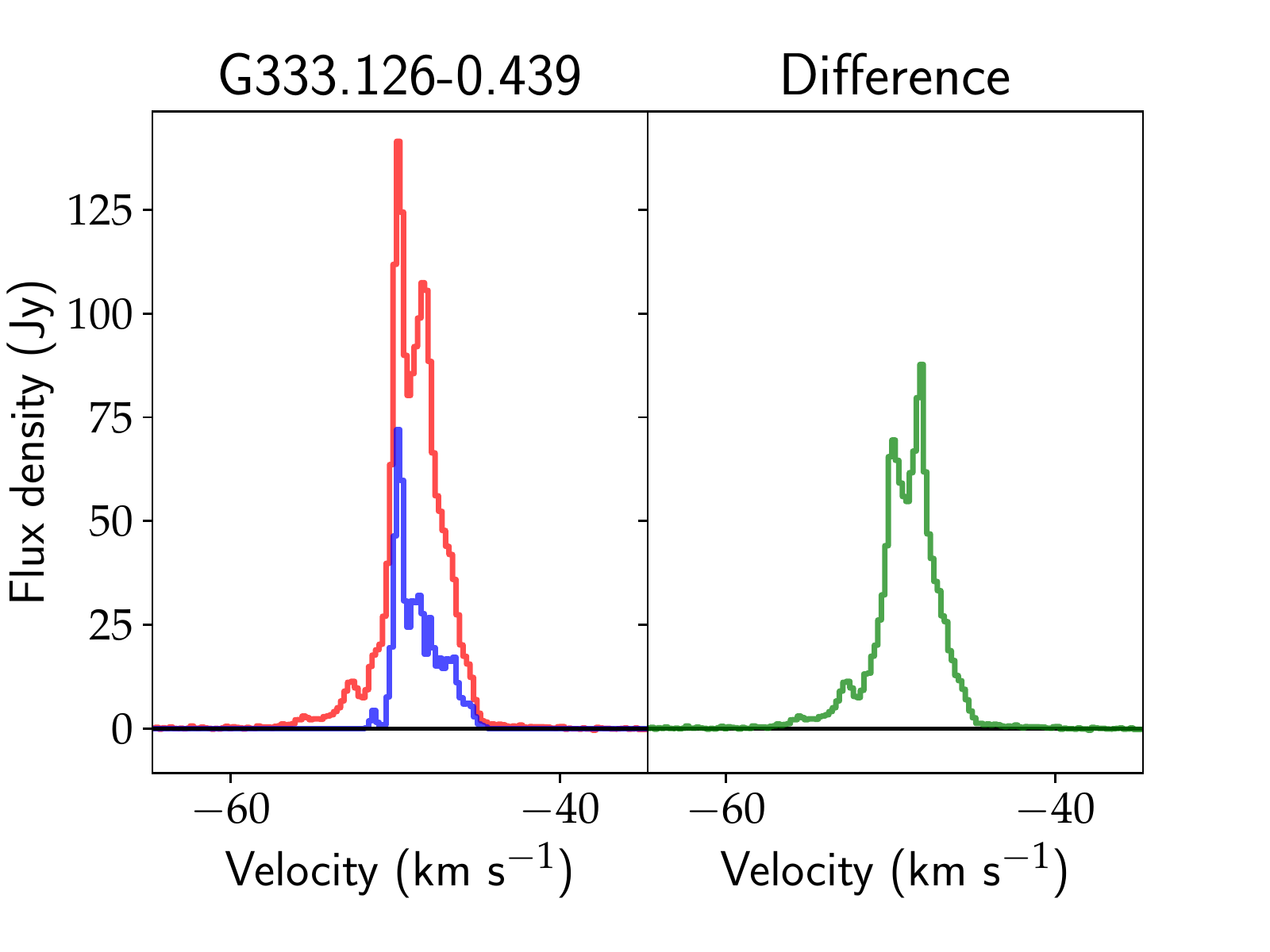}
  \includegraphics[height=0.30\textheight]{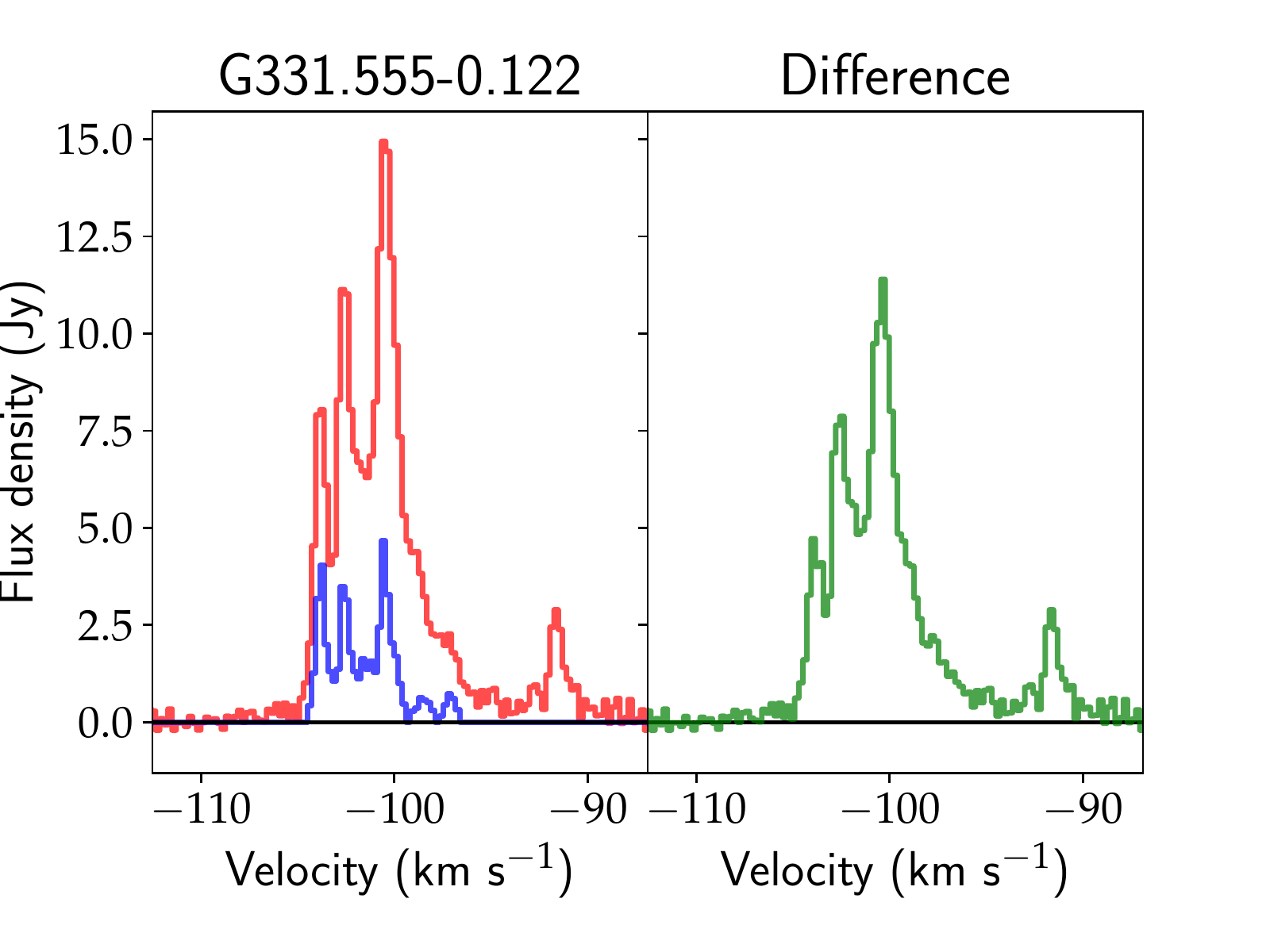}
  \caption{Cross- and auto-correlation spectra of three class~I~\choh{} maser regions, with strong, moderate and weak maser strengths. Auto-correlation spectra are plotted red, cross-correlation spectra blue and the difference green.}
  \label{fig:cross_auto_difference}
\end{figure}

Auto-correlation spectra for the \choh{} masers were extracted using the method discussed in Section~\ref{subsec:auto_correlation}. Cross-correlation spectra were then plotted with their corresponding auto-correlation spectrum; three example regions are shown in Fig.~\ref{fig:cross_auto_difference}. With the 1.5A array configuration for the ATCA, the maximum detectable scale of the cross-correlation data for these follow-up observations is approximately 4.6~arcsec.

The relative strength of compact maser emission to diffuse emission of a class~I~\choh{} maser region was assessed by computing the ratio of cross-correlation IFD to auto-correlation IFD, i.e., if the cross-correlation IFD closely matches the auto-correlation IFD, the emission is compact. Fig.~\ref{fig:cross_auto_difference} shows a large variation in missing flux density is across several regions. We analysed the ratios of cross-correlated IFDs to auto-correlated IFDs, but no trends are apparent for evolved, young or total populations. The ratio of cross-correlated IFD to auto-correlated IFD was also compared against the flux density of the masers, but no trend was observed. Class~I~\choh{} emission can be strongly confined to compact structures (cross-correlated emission $\approx$ auto-correlated emission), extended (auto-correlated emission $\gg$ cross-correlated emission), or a combination of the two. This may explain why three targeted sites were not detected in the cross-correlation dataset (G331.44$-$0.14, G331.72$-$0.20 and G333.24$+$0.02); their 44\,GHz emission is likely too extended and lacks any bright, compact maser components.

\begin{figure}
  \includegraphics[width=\columnwidth]{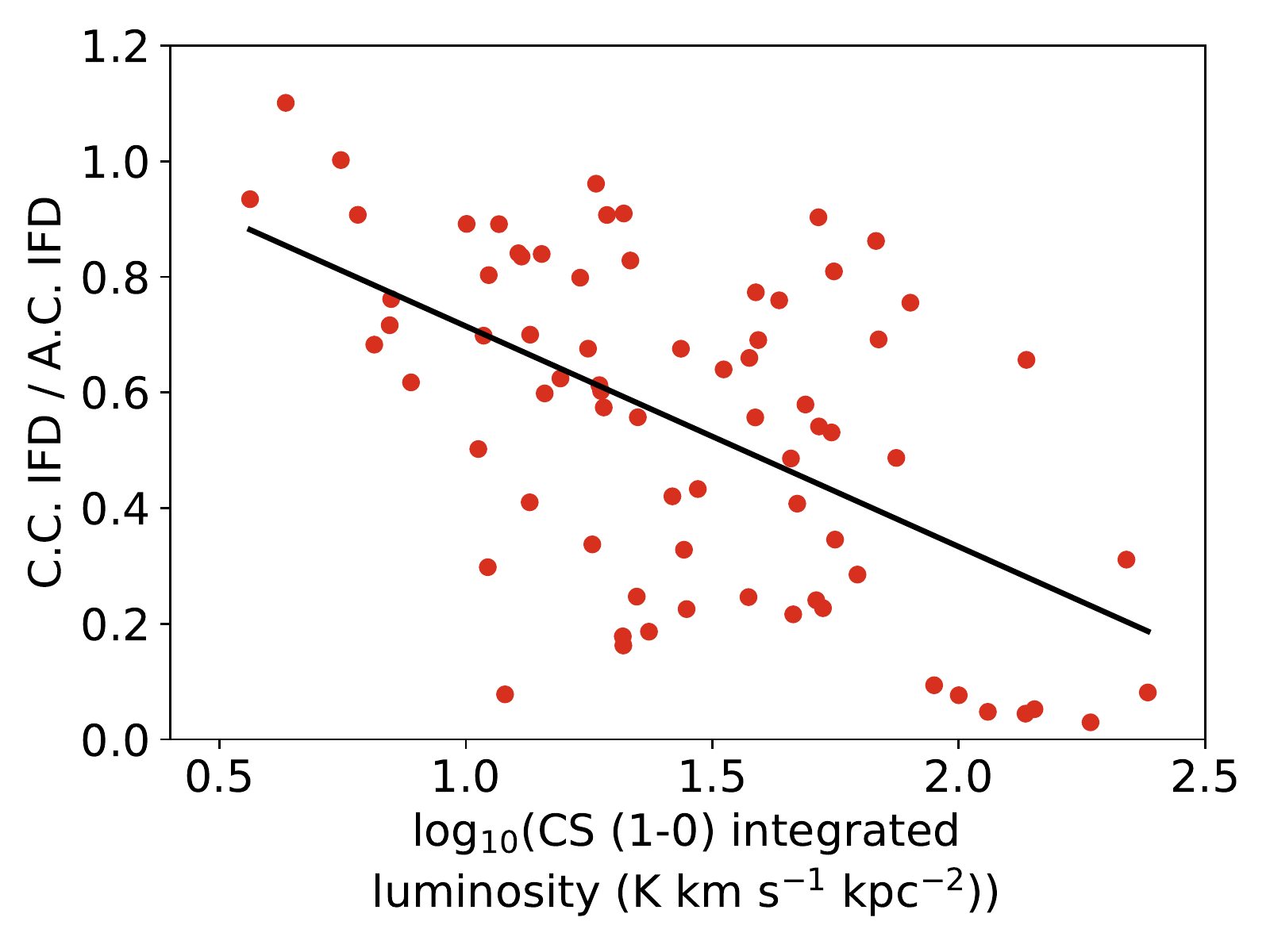}
  \includegraphics[width=\columnwidth]{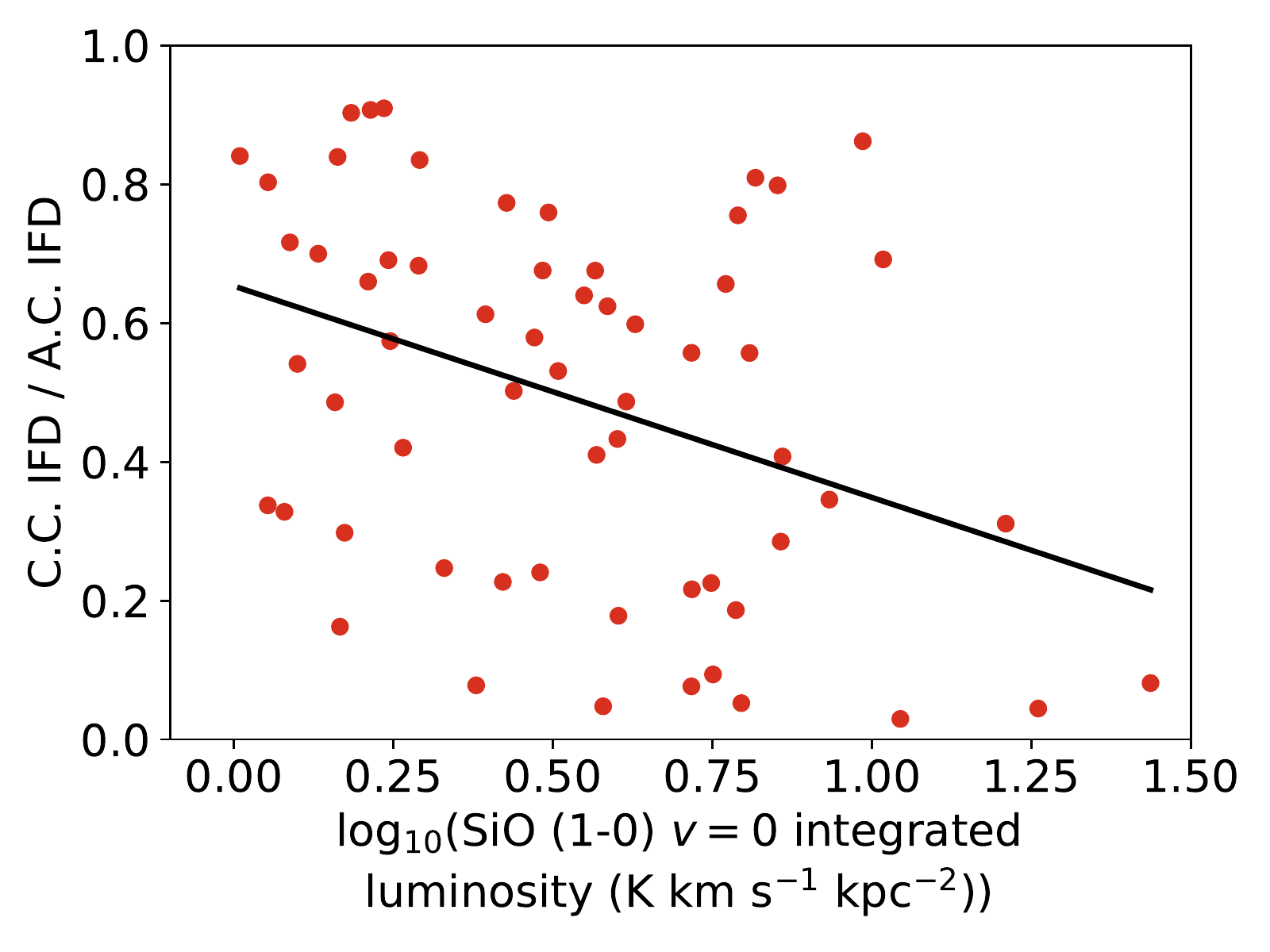}
  \includegraphics[width=\columnwidth]{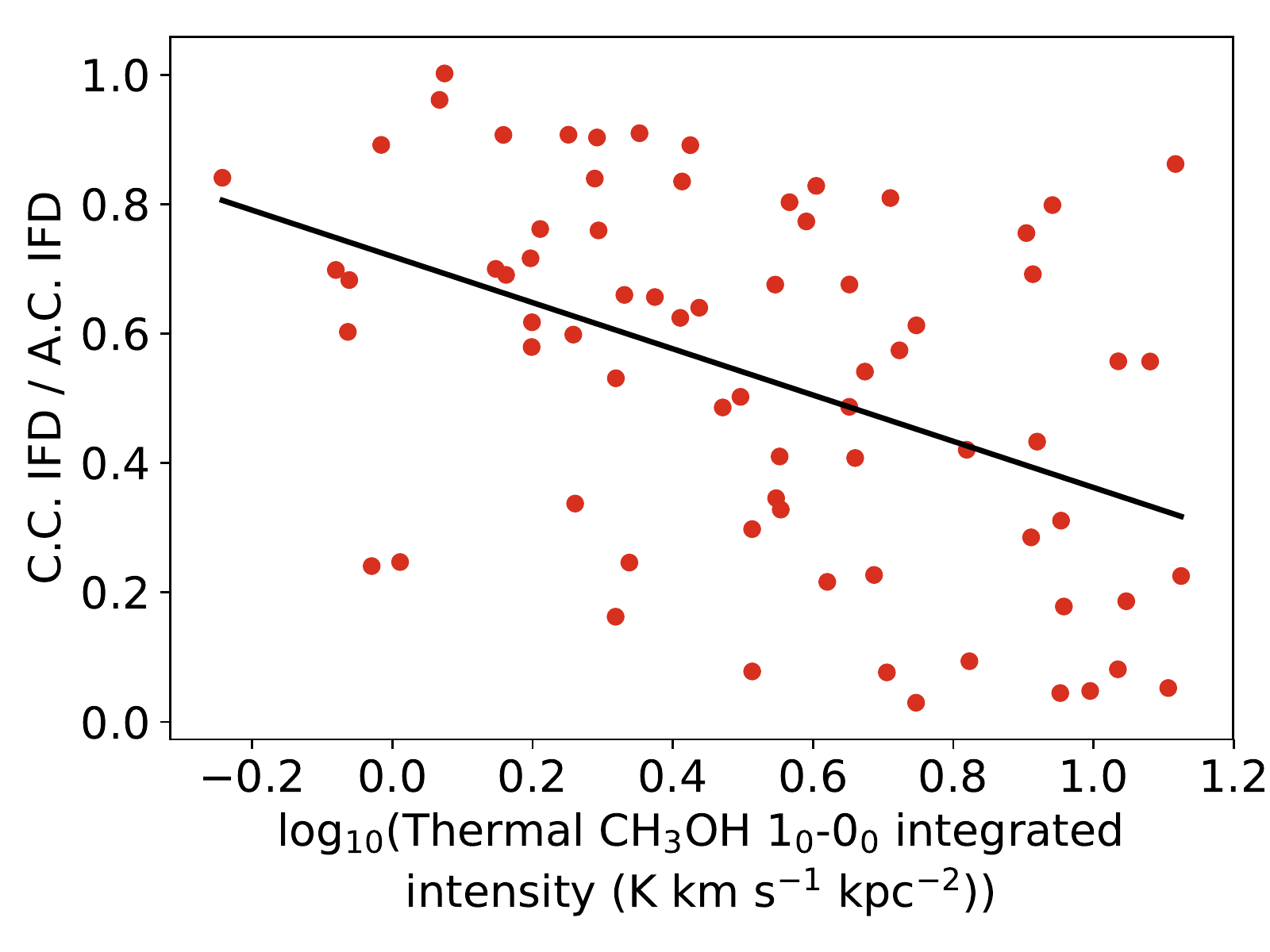}
  \caption{Scatter plots of the ratio of cross-correlation integrated flux density (C.C. IFD) to auto-correlation (A.C.) IFD against the thermal integrated intensity of \cs{}, \sio{} and \choh{}. The solid line represents the least squares fit. A similar, weak trend appears for all lines, but the large scatter hampers correlation ($r$-values of $-$0.55, $-$0.38 and $-$0.45, respectively). The single datum with a larger C.C. IFD than A.C. IFD has an over-contribution of cross-correlation emission, due to spectral blending in nearby maser spots.}
  \label{fig:maserness_scatter}
\end{figure}

To test if compact or diffuse regions can be separated, populations were investigated with and without various maser associations. These associations included class~II~\choh{}, \water{} and \oh{} masers, but no dependence was found for the proportion compact to diffuse maser emission with and without these associations. This proportion of compact to diffuse emission was also compared with thermal line integrated intensities in \cs{}~(1--0), \sio{}~(1--0)~$v=0$ and \choh{}~1$_0$--0$_0$; see Fig.~\ref{fig:maserness_scatter}. A similar trend can be seen in each comparison; the $r$-values for \cs{}, \sio{} and thermal \choh{} are $-$0.55, $-$0.38 and $-$0.45, respectively. The corresponding $p$-values are 3.3$\times$10$^{-7}$, 2.7$\times$10$^{-3}$ and 7.7$\times$10$^{-5}$, respectively. These statistics suggest that our lines of best fit are significant.

Any indication of class~I~\choh{} emission being dominated by compact components appears to be an intrinsic property of class~I masers, but with a significant scatter amongst the sources. Another star formation tracer may reveal a correlation with the property of maser strength discussed here. Further observations of class~I~\choh{} maser regions with better sensitivity and zero-spacing data will also help to identify weaker, more diffuse cross-correlation emission, in order to better understand the relationship between the diffuse and compact components of the \choh{} maser emission.

\section{Summary and Conclusions}
We have extracted high-resolution positions, flux densities and velocities for class~I methanol masers detected in the \malt{} survey, and presented various properties for each observed region. The unbiased population from \malt{} provides the first opportunity to assess class~I methanol masers in a sensitivity-limited sample which are free from target selection biases, such as class~II methanol masers or extended green objects. In addition, the thermal lines mapped by \malt{} were observed with better sensitivity toward each maser site, providing more detail about the regions containing the class~I methanol masers. We have determined:

(i) Class I methanol maser sites with fewer spots of emission are less likely to be associated with a class~II methanol or hydroxyl maser;

(ii) Class I methanol masers without an associated hydroxyl maser or radio recombination line emission have lower luminosities.

(iii) The spatial extent and velocity range of class~I methanol masers is typically small ($<$0.5\,pc and $<$5\,\kms{}, respectively), particularly for those without an associated class~II methanol or hydroxyl maser;

(iv) Class I methanol masers are generally located within 0.5\,pc of a class~II methanol or hydroxyl maser, but can be up to 0.8\,pc away;

(v) Class I methanol masers are reliable tracers of systemic velocities, and are better than class~II methanol masers;

(vi) The brightness of class~I methanol masers are weakly correlated with the brightness of the thermal \cs{}~(1--0), \sio{}~(1--0)~$v=0$ and \choh{}~1$_0$--0$_0$ lines, as well as 870\,$\mu$m dust continuum emission from ATLASGAL. These results suggest that the brightness of a class~I methanol maser is proportional to the mass of its host star-forming region;

(vii) Class I methanol masers have high association rates with 870\,$\mu$m dust continuum point sources catalogued by ATLASGAL, with typical offsets not exceeding 0.4\,pc;

(viii) Class I methanol masers are found towards a large range of clump masses (10$^{1.25}$ to 10$^{4.5}$\,$M_\odot$), but peak between 10$^{3.0}$ and 10$^{3.5}$\,$M_\odot$. Additionally, masers associated with clump masses between 10$^{3.25}$ and 10$^{4.5}$\,$M_\odot$ are almost all evolved regions of star formation;

(ix) The amount of diffuse emission in the 44\,GHz class~I methanol transition varies from source to source, but appears to increase as the brightness of thermal lines increase.

\section*{Acknowledgements}
Parts of this research were conducted by the Australian Research Council Centre of Excellence for All-sky Astrophysics (CAASTRO), through project number CE110001020. The Australia Telescope Compact Array is part of the Australia Telescope National Facility which is funded by the Commonwealth of Australia for operation as a National Facility managed by CSIRO. Shari Breen is the recipient of an Australian Research Council DECRA Fellowship (project number DE130101270). This research has made use of NASA's Astrophysics Data System Bibliographic Services, and the SIMBAD database,
operated at CDS, Strasbourg, France. Computation was aided by the \textsc{NumPy} \citep{numpy}, \textsc{SciPy} \citep{scipy} and \textsc{Astropy} \citep{astropy} libraries. Figures were generated with \textsc{matplotlib} \citep{matplotlib}. Some figures in this paper use colours from \url{www.ColorBrewer.org} \citep{colorbrewer}. The \textsc{MIRIAD}\footnote{http://www.atnf.csiro.au/computing/software/miriad/} software suite is managed and maintained by CSIRO Astronomy and Space Science.

\bibliographystyle{mnras}
\bibliography{references}

\newpage
\appendix

\begin{table*}
  \section{Gaussian spectral fits to cross-correlation class I \choh{} maser emission}
  \begin{center}
    \caption{Gaussian fits for the spectra of class~I~\choh{} maser spots. Column 1 lists the maser spot name, along with a letter designating the spot. Spots labelled with an asterisk ($*$) have been manually fitted as these spots appear close to a bright neighbour; see Section~\ref{subsec:cross_correlation}. Columns 2 and 3 list the fitted position for the Gaussian. Columns 4 through 6 list the fitted Gaussian parameters. Note that the uncertainty for each parameter is quoted in parentheses, in units of the least significant figure. Column 7 lists the integrated flux density of the Gaussian. Rows without spot names correspond to a second Gaussian fitted to the spot directly above.}
    \label{app:meth_detail}
    \begin{tabular}{ l ll ll d{1}l d{1}l d{1}l d{5} }
      \hline
      Spot name & \multicolumn{2}{c}{$\alpha_{2000}$} & \multicolumn{2}{c}{$\delta_{2000}$}                        & \multicolumn{2}{c}{Peak flux} & \multicolumn{2}{c}{Peak}     & \multicolumn{2}{c}{FWHM}   & \multicolumn{1}{c}{Integrated}   \\
                & \multicolumn{2}{c}{(h:m:s)}         & \multicolumn{2}{c}{($^\circ$:$^\prime$:$^{\prime\prime}$)} & \multicolumn{2}{c}{density}   & \multicolumn{2}{c}{velocity} & \multicolumn{2}{c}{(\kms{})} & \multicolumn{1}{c}{flux density} \\
                & &                                   & &                                                          & \multicolumn{2}{c}{(Jy)}      & \multicolumn{2}{c}{(\kms{})}   & &                          & \multicolumn{1}{c}{(Jy\,\kms{})}   \\
      \hline
      G330.2940$-$0.3921A & 16:07:37.51 & (5) & $-$52:30:56.1 & (2) & 1.00 & (4) & -80.13 & (2) & 0.90 & (4) & 0.679 \\
      G330.2945$-$0.3943A & 16:07:38.24 & (4) & $-$52:31:00.5 & (1) & 0.52 & (3) & -77.55 & (2) & 0.93 & (5) & 0.363 \\
      G330.6780$-$0.4023A & 16:09:31.70 & (5) & $-$52:15:50.7 & (1) & 1.73 & (6) & -64.78 & (1) & 0.78 & (3) & 1.015 \\
      G330.7790$+$0.2488A & 16:07:09.75 & (3) & $-$51:42:53.76 & (5) & 0.72 & (2) & -43.227 & (6) & 0.57 & (1) & 0.308 \\
      G330.8749$-$0.3557A & 16:10:15.99 & (6) & $-$52:05:46.6 & (1) & 14.3 & (1) & -61.284 & (2) & 0.476 & (5) & 5.120 \\
      G330.8774$-$0.3676A & 16:10:19.8 & (1) & $-$52:06:12.1 & (2) & 2.4 & (1) & -58.06 & (2) & 0.84 & (5) & 1.519 \\
      G330.8711$-$0.3830A & 16:10:22.13 & (8) & $-$52:07:08.23 & (9) & 0.85 & (6) & -58.10 & (4) & 1.08 & (8) & 0.690 \\
      G330.8711$-$0.3830B & 16:10:22.1 & (1) & $-$52:07:08.3 & (1) & 0.77 & (5) & -62.86 & (3) & 0.89 & (7) & 0.513 \\
      &  &  &  &  & 0.51 & (4) & -64.50 & (8) & 2.2 & (2) & 0.340 \\
      G330.9261$-$0.4083A & 16:10:44.6 & (2) & $-$52:06:00.37 & (4) & 11.8 & (4) & -42.36 & (1) & 0.66 & (2) & 5.877 \\
      &  &  &  &  & 1.2 & (2) & -41.37 & (5) & 0.7 & (1) & 0.598 \\
      G330.9274$-$0.4072A & 16:10:44.723 & (9) & $-$52:05:54.29 & (9) & 1.9 & (1) & -41.41 & (2) & 0.52 & (4) & 0.744 \\
      G330.9304$-$0.2602A & 16:10:06.61 & (4) & $-$51:59:18.5 & (1) & 3.9 & (1) & -90.00 & (1) & 0.91 & (3) & 2.675 \\
      G330.9311$-$0.2605A & 16:10:06.87 & (6) & $-$51:59:17.78 & (5) & 3.94 & (8) & -89.014 & (5) & 0.50 & (1) & 1.480 \\
      G330.9548$-$0.1823A & 16:09:52.99 & (5) & $-$51:54:52.8 & (2) & 1.55 & (6) & -93.66 & (1) & 0.57 & (3) & 0.670 \\
      G331.1308$-$0.4696A & 16:11:59.38 & (9) & $-$52:00:19.5 & (2) & 0.78 & (5) & -68.94 & (2) & 0.59 & (5) & 0.346 \\
      G331.1308$-$0.4698A & 16:11:59.4 & (1) & $-$52:00:20.24 & (5) & 2.6 & (1) & -67.45 & (3) & 1.47 & (8) & 2.876 \\
      G331.1336$-$0.4882A & 16:12:05.14 & (2) & $-$52:01:01.37 & (5) & 3.13 & (6) & -67.911 & (4) & 0.459 & (9) & 1.082 \\
      G331.1308$-$0.2441A & 16:10:59.53 & (2) & $-$51:50:25.79 & (4) & 13.0 & (6) & -87.50 & (3) & 1.45 & (7) & 14.148 \\
      G331.1308$-$0.2441B & 16:10:59.54 & (6) & $-$51:50:25.72 & (5) & 21 & (4) & -90.1 & (1) & 1.6 & (3) & 24.901 \\
      G331.1308$-$0.2441C & 16:10:59.54 & (6) & $-$51:50:25.68 & (9) & 93 & (3) & -91.11 & (1) & 0.93 & (3) & 65.276 \\
      G331.1333$-$0.2458A & 16:11:00.7 & (5) & $-$51:50:23.97 & (6) & 45 & (1) & -88.531 & (8) & 0.70 & (2) & 23.769 \\
      G331.1333$-$0.2410A & 16:10:59.41 & (2) & $-$51:50:11.8 & (1) & 23.1 & (2) & -84.630 & (2) & 0.610 & (5) & 10.604 \\
      G331.1333$-$0.2410B & 16:10:59.45 & (2) & $-$51:50:11.64 & (2) & 2.9 & (1) & -86.19 & (1) & 0.53 & (3) & 1.151 \\
      G331.1322$-$0.2454A & 16:11:00.25 & (6) & $-$51:50:25.71 & (1) & 6.5 & (2) & -84.34 & (1) & 0.81 & (3) & 3.986 \\
      G331.1322$-$0.2454B & 16:11:00.29 & (4) & $-$51:50:26.0 & (1) & 6.3 & (2) & -87.667 & (8) & 0.47 & (2) & 2.244 \\
      G331.1332$-$0.2407A & 16:10:59.30 & (4) & $-$51:50:10.90 & (3) & 2.11 & (6) & -85.718 & (8) & 0.59 & (2) & 0.931 \\
      G331.1313$-$0.2434A & 16:10:59.499 & (9) & $-$51:50:22.76 & (5) & 4.19 & (7) & -86.097 & (3) & 0.412 & (7) & 1.300 \\
      G331.1315$-$0.2441A & 16:10:59.73 & (8) & $-$51:50:24.2 & (1) & 0.90 & (5) & -82.85 & (2) & 0.57 & (4) & 0.386 \\
      G331.1313$-$0.2451A & 16:10:59.9 & (2) & $-$51:50:27.2 & (2) & 0.89 & (7) & -82.53 & (4) & 0.91 & (8) & 0.610 \\
      G331.1329$+$0.1560A & 16:09:15.03 & (5) & $-$51:32:42.42 & (7) & 1.03 & (9) & -78.36 & (3) & 0.66 & (7) & 0.509 \\
      G331.1339$+$0.1569A & 16:09:15.07 & (6) & $-$51:32:37.79 & (4) & 1.42 & (3) & -78.336 & (8) & 0.72 & (2) & 0.768 \\
      G331.1348$+$0.1562A & 16:09:15.50 & (2) & $-$51:32:37.3 & (1) & 10.0 & (3) & -77.19 & (1) & 0.77 & (3) & 5.760 \\
      G331.1332$+$0.1565A & 16:09:15.00 & (7) & $-$51:32:40.5 & (1) & 76 & (1) & -75.532 & (5) & 0.56 & (1) & 32.195 \\
      &  &  &  &  & 4.0 & (6) & -76.48 & (6) & 0.8 & (1) & 1.694 \\
      G331.1335$+$0.1562A & 16:09:15.16 & (2) & $-$51:32:40.63 & (9) & 10.4 & (2) & -76.139 & (5) & 0.60 & (1) & 4.664 \\
      G331.1331$+$0.1568A & 16:09:14.884 & (9) & $-$51:32:40.070 & (7) & 4.8 & (1) & -74.512 & (8) & 0.52 & (2) & 1.889 \\
      G331.2846$-$0.1953A & 16:11:30.32 & (4) & $-$51:41:59.9 & (1) & 1.08 & (4) & -94.03 & (3) & 1.44 & (6) & 1.168 \\
      G331.2800$-$0.1899A & 16:11:27.57 & (5) & $-$51:41:56.89 & (5) & 8.4 & (6) & -89.31 & (3) & 0.77 & (6) & 4.883 \\
      G331.2800$-$0.1899B & 16:11:28 & (0) & $-$51:41:57.02 & (1) & 6.3 & (9) & -89.94 & (6) & 0.9 & (2) & 4.333 \\
      G331.2800$-$0.1899C & 16:11:27.58 & (7) & $-$51:41:57.03 & (2) & 12.2 & (3) & -90.77 & (1) & 0.81 & (3) & 7.417 \\
      G331.2800$-$0.1899D & 16:11:27.55 & (6) & $-$51:41:57.07 & (1) & 6 & (5) & -91.31 & (5) & 0.2 & (2) & 1.021 \\
      &  &  &  &  & 1.0 & (1) & -92.17 & (8) & 1.2 & (2) & 0.170 \\
      G331.2802$-$0.1898A* & 16:11:27.6 & (2) & $-$51:41:56.4 & (2) & 5.4 & (6) & -89.92 & (5) & 0.8 & (1) & 3.427 \\
      G331.2803$-$0.1899A & 16:11:27.66 & (4) & $-$51:41:56.26 & (7) & 5.0 & (4) & -88.49 & (3) & 0.88 & (7) & 3.306 \\
      G331.2768$-$0.1870A & 16:11:25.91 & (5) & $-$51:41:57.21 & (5) & 3.0 & (3) & -86.83 & (4) & 0.8 & (1) & 1.861 \\
      G331.2793$-$0.1873A & 16:11:26.69 & (2) & $-$51:41:51.98 & (7) & 2.7 & (1) & -85.764 & (9) & 0.44 & (2) & 0.890 \\
      G331.2764$-$0.1874A & 16:11:25.91 & (3) & $-$51:41:59.38 & (4) & 3.78 & (7) & -85.259 & (6) & 0.67 & (1) & 1.896 \\
      G331.2783$-$0.1884A & 16:11:26.71 & (3) & $-$51:41:57.3 & (1) & 1.34 & (4) & -87.816 & (6) & 0.38 & (1) & 0.382 \\
      G331.2765$-$0.1871A & 16:11:25.85 & (2) & $-$51:41:58.05 & (3) & 2.0 & (1) & -87.76 & (1) & 0.50 & (3) & 0.752 \\
      G331.3409$-$0.3470A & 16:12:26.40 & (4) & $-$51:46:20.50 & (2) & 30.7 & (9) & -65.69 & (1) & 0.67 & (2) & 15.454 \\
      G331.3703$-$0.3990A & 16:12:48.54 & (5) & $-$51:47:24.39 & (5) & 5.15 & (9) & -65.132 & (4) & 0.471 & (9) & 1.826 \\
      G331.3705$-$0.1446A & 16:11:41.22 & (4) & $-$51:36:15.37 & (7) & 0.90 & (5) & -87.95 & (3) & 1.09 & (7) & 0.740 \\
      G331.3705$-$0.1446B & 16:11:41.23 & (5) & $-$51:36:15.3 & (1) & 0.90 & (7) & -89.07 & (3) & 0.76 & (7) & 0.517 \\
      G331.3798$+$0.1490A & 16:10:26.81 & (2) & $-$51:22:58.5 & (1) & 1.47 & (7) & -45.72 & (2) & 0.74 & (4) & 0.816 \\
      G331.4093$-$0.1643A & 16:11:57.37 & (4) & $-$51:35:32.033 & (9) & 1.27 & (2) & -85.050 & (5) & 0.54 & (1) & 0.518 \\
      G331.4407$-$0.1869A & 16:12:12.2 & (1) & $-$51:35:14.1 & (2) & 3.3 & (3) & -91.66 & (2) & 0.49 & (4) & 1.228 \\
      &  &  &  &  & 0.56 & (4) & -90.06 & (6) & 1.5 & (1) & 0.208 \\
      \hline
    \end{tabular}
  \end{center}
\end{table*}

\setcounter{table}{0}
\begin{table*}
  \begin{center}
    \caption{{\em - continued.}}
    \begin{tabular}{ l ll ll d{1}l d{1}l d{1}l d{5} }
      \hline
      Spot name & \multicolumn{2}{c}{$\alpha_{2000}$} & \multicolumn{2}{c}{$\delta_{2000}$}                        & \multicolumn{2}{c}{Peak flux} & \multicolumn{2}{c}{Peak}     & \multicolumn{2}{c}{FWHM}   & \multicolumn{1}{c}{Integrated} \\
                & \multicolumn{2}{c}{(h:m:s)}         & \multicolumn{2}{c}{($^\circ$:$^\prime$:$^{\prime\prime}$)} & \multicolumn{2}{c}{density}   & \multicolumn{2}{c}{velocity} & \multicolumn{2}{c}{(\kms{})} & \multicolumn{1}{c}{flux density}  \\
                & &                                   & &                                                          & \multicolumn{2}{c}{(Jy)}      & \multicolumn{2}{c}{(\kms{})}   & &                          & \multicolumn{1}{c}{(Jy\,\kms{})} \\
      \hline
      G331.4408$-$0.1869A & 16:12:12.25 & (7) & $-$51:35:13.88 & (6) & 1.10 & (6) & -87.94 & (2) & 0.65 & (4) & 0.536 \\
      G331.4381$-$0.1864A & 16:12:11.3 & (1) & $-$51:35:19.3 & (1) & 1.57 & (3) & -87.756 & (6) & 0.56 & (1) & 0.660 \\
      &  &  &  &  & 0.70 & (9) & -86.88 & (5) & 0.8 & (1) & 0.294 \\
      G331.4417$-$0.1580A & 16:12:04.85 & (5) & $-$51:33:55.74 & (3) & 0.73 & (3) & -85.98 & (2) & 1.05 & (5) & 0.574 \\
      G331.4876$-$0.0863A & 16:11:58.9 & (1) & $-$51:28:54.33 & (6) & 0.73 & (7) & -90.19 & (4) & 0.9 & (1) & 0.483 \\
      G331.4954$-$0.0786A & 16:11:59.05 & (2) & $-$51:28:14.81 & (6) & 31.7 & (4) & -89.001 & (4) & 0.673 & (9) & 16.069 \\
      G331.4962$-$0.0780A & 16:11:59.12 & (4) & $-$51:28:11.66 & (2) & 1.7 & (1) & -88.45 & (3) & 0.84 & (7) & 1.073 \\
      G331.4874$-$0.0865A & 16:11:58.87 & (3) & $-$51:28:55.23 & (3) & 1.41 & (4) & -87.906 & (7) & 0.54 & (2) & 0.575 \\
      G331.5046$-$0.1158A & 16:12:11.44 & (3) & $-$51:29:30.20 & (3) & 2.43 & (4) & -100.45 & (1) & 1.19 & (2) & 2.171 \\
      G331.5033$-$0.1066A & 16:12:08.64 & (8) & $-$51:29:09.1 & (1) & 0.70 & (4) & -88.39 & (2) & 0.73 & (5) & 0.386 \\
      G331.5028$-$0.1068A & 16:12:08.55 & (2) & $-$51:29:10.76 & (4) & 1.39 & (7) & -86.90 & (1) & 0.49 & (3) & 0.508 \\
      G331.5032$-$0.1067A & 16:12:08.65 & (9) & $-$51:29:09.71 & (6) & 0.87 & (8) & -86.18 & (4) & 1.0 & (1) & 0.625 \\
      G331.5219$-$0.0811A & 16:12:07.16 & (6) & $-$51:27:16.2 & (2) & 0.65 & (4) & -90.84 & (1) & 0.49 & (4) & 0.237 \\
      G331.5179$-$0.0823A & 16:12:06.36 & (3) & $-$51:27:29.28 & (3) & 3.7 & (2) & -88.04 & (3) & 0.97 & (6) & 2.708 \\
      &  &  &  &  & 2.5 & (3) & -89.36 & (4) & 0.8 & (1) & 1.830 \\
      G331.5179$-$0.0823B & 16:12:06.34 & (6) & $-$51:27:29.42 & (5) & 7.2 & (3) & -90.08 & (1) & 0.60 & (3) & 3.267 \\
      G331.5185$-$0.0823A & 16:12:06.52 & (7) & $-$51:27:27.81 & (5) & 0.53 & (2) & -87.063 & (8) & 0.50 & (2) & 0.199 \\
      G331.5290$-$0.0982A & 16:12:13.665 & (9) & $-$51:27:43.92 & (2) & 18.7 & (6) & -93.017 & (7) & 0.50 & (2) & 7.093 \\
      G331.5327$-$0.0993A & 16:12:15.00 & (9) & $-$51:27:37.6 & (2) & 79 & (2) & -91.827 & (9) & 0.58 & (2) & 34.306 \\
      &  &  &  &  & 10 & (1) & -91.06 & (3) & 0.45 & (6) & 4.343 \\
      G331.5237$-$0.1009A & 16:12:12.87 & (2) & $-$51:28:04.02 & (1) & 4.53 & (9) & -91.082 & (4) & 0.422 & (9) & 1.437 \\
      G331.5312$-$0.0987A & 16:12:14.40 & (2) & $-$51:27:39.71 & (6) & 1.21 & (3) & -90.662 & (5) & 0.42 & (1) & 0.380 \\
      G331.5341$-$0.0997A & 16:12:15.50 & (5) & $-$51:27:35.31 & (6) & 3.82 & (6) & -89.621 & (6) & 0.75 & (1) & 2.160 \\
      G331.5432$-$0.0660A & 16:12:09.17 & (2) & $-$51:25:44.2 & (1) & 1.6 & (1) & -88.77 & (3) & 0.74 & (8) & 0.893 \\
      G331.5432$-$0.0660B & 16:12:09.18 & (2) & $-$51:25:44.24 & (7) & 0.80 & (8) & -89.54 & (4) & 0.9 & (1) & 0.519 \\
      G331.5461$-$0.0688A & 16:12:10.73 & (6) & $-$51:25:44.6 & (2) & 0.62 & (3) & -87.74 & (1) & 0.59 & (3) & 0.274 \\
      &  &  &  &  & 0.44 & (5) & -87.02 & (3) & 0.52 & (7) & 0.194 \\
      G331.5548$-$0.1227A & 16:12:27.4 & (1) & $-$51:27:44.49 & (1) & 3.2 & (1) & -104.01 & (1) & 0.59 & (3) & 1.418 \\
      G331.5544$-$0.1221A & 16:12:27.1 & (1) & $-$51:27:44.1 & (1) & 3.8 & (3) & -100.68 & (3) & 0.95 & (8) & 2.714 \\
      G331.5544$-$0.1221B & 16:12:27.10 & (3) & $-$51:27:44.13 & (6) & 1.2 & (1) & -102.05 & (5) & 1.0 & (1) & 0.912 \\
      G331.5544$-$0.1221C & 16:12:27.09 & (2) & $-$51:27:44.1 & (1) & 0.68 & (7) & -103.75 & (6) & 1.2 & (1) & 0.595 \\
      G331.5547$-$0.1211A & 16:12:26.90 & (2) & $-$51:27:40.69 & (4) & 4.01 & (9) & -102.801 & (4) & 0.40 & (1) & 1.215 \\
      G331.5552$-$0.1215A & 16:12:27.16 & (3) & $-$51:27:40.6 & (2) & 0.74 & (2) & -97.311 & (8) & 0.62 & (2) & 0.348 \\
      G331.5552$-$0.1215B & 16:12:27.15 & (7) & $-$51:27:40.7 & (1) & 0.64 & (3) & -98.72 & (2) & 0.87 & (5) & 0.419 \\
      G331.8527$-$0.1295A & 16:13:52.5 & (1) & $-$51:15:46.63 & (6) & 1.15 & (8) & -49.77 & (4) & 1.05 & (9) & 0.913 \\
      G331.8855$+$0.0619A & 16:13:11.24 & (2) & $-$51:06:04.75 & (3) & 0.99 & (2) & -89.865 & (5) & 0.58 & (1) & 0.433 \\
      G331.8855$+$0.0627A & 16:13:11.05 & (6) & $-$51:06:02.6 & (2) & 4.3 & (3) & -88.88 & (3) & 0.86 & (6) & 2.774 \\
      G331.8854$+$0.0625A* & 16:13:11.1 & (2) & $-$51:06:03.5 & (2) & 3.7 & (2) & -88.81 & (2) & 0.76 & (5) & 2.112 \\
      G331.8847$+$0.0617A & 16:13:11.08 & (9) & $-$51:06:07.4 & (2) & 2.06 & (4) & -88.375 & (4) & 0.436 & (9) & 0.675 \\
      G331.8854$+$0.0630A & 16:13:10.96 & (2) & $-$51:06:02.21 & (2) & 0.82 & (2) & -87.135 & (5) & 0.405 & (9) & 0.250 \\
      G331.8892$+$0.0641A & 16:13:11.70 & (2) & $-$51:05:49.90 & (8) & 0.85 & (3) & -86.785 & (8) & 0.46 & (2) & 0.294 \\
      &  &  &  &  & 0.56 & (3) & -86.12 & (2) & 0.58 & (4) & 0.194 \\
      G331.8864$+$0.0618A & 16:13:11.522 & (9) & $-$51:06:02.7 & (1) & 0.73 & (3) & -86.378 & (9) & 0.42 & (2) & 0.230 \\
      G331.8898$+$0.0658A & 16:13:11.44 & (3) & $-$51:05:43.9 & (1) & 0.80 & (3) & -85.978 & (8) & 0.48 & (2) & 0.286 \\
      G331.8880$+$0.0658A & 16:13:10.94 & (3) & $-$51:05:48.5 & (1) & 1.51 & (1) & -86.863 & (1) & 0.330 & (2) & 0.375 \\
      G331.9209$-$0.0829A & 16:13:59.17 & (7) & $-$51:10:56.00 & (6) & 4.7 & (2) & -51.63 & (1) & 0.44 & (3) & 1.549 \\
      &  &  &  &  & 0.7 & (2) & -52.18 & (4) & 0.4 & (1) & 0.231 \\
      G332.0933$-$0.4213A & 16:16:16.59 & (7) & $-$51:18:27.90 & (8) & 3.6 & (2) & -58.82 & (1) & 0.53 & (3) & 1.442 \\
      &  &  &  &  & 1.7 & (2) & -58.16 & (3) & 0.67 & (8) & 0.681 \\
      G332.0911$-$0.4194A & 16:16:15.44 & (5) & $-$51:18:28.51 & (8) & 8.1 & (2) & -55.849 & (6) & 0.45 & (1) & 2.742 \\
      G332.0908$-$0.4184A & 16:16:15.12 & (3) & $-$51:18:26.51 & (5) & 0.68 & (3) & -58.05 & (1) & 0.53 & (3) & 0.270 \\
      G332.2420$-$0.0443A & 16:15:17.74 & (6) & $-$50:55:58.09 & (9) & 2.4 & (2) & -47.32 & (3) & 0.89 & (6) & 1.599 \\
      G332.2420$-$0.0443B & 16:15:17.78 & (8) & $-$50:55:58.0 & (1) & 121 & (3) & -49.178 & (7) & 0.63 & (2) & 57.263 \\
      &  &  &  &  & 15 & (4) & -49.91 & (5) & 0.36 & (9) & 7.099 \\
      G332.2417$-$0.0437A & 16:15:17.53 & (2) & $-$50:55:57.30 & (4) & 14.4 & (3) & -48.527 & (4) & 0.509 & (9) & 5.513 \\
      G332.2393$-$0.0425A & 16:15:16.55 & (2) & $-$50:55:60.0 & (1) & 6.8 & (5) & -48.51 & (3) & 0.61 & (6) & 3.146 \\
      G332.2401$-$0.0429A & 16:15:16.88 & (2) & $-$50:55:59.27 & (2) & 6.1 & (4) & -47.93 & (1) & 0.40 & (3) & 1.838 \\
      G332.2390$-$0.0451A & 16:15:17.2 & (2) & $-$50:56:07.5 & (4) & 34 & (1) & -47.31 & (1) & 0.86 & (3) & 21.996 \\
      G332.2387$-$0.0452A* & 16:15:17.1 & (2) & $-$50:56:08.7 & (2) & 6.5 & (8) & -46.78 & (3) & 0.50 & (7) & 2.442 \\
      G332.2390$-$0.0428A & 16:15:16.55 & (6) & $-$50:56:01.78 & (7) & 1.10 & (4) & -46.247 & (9) & 0.44 & (2) & 0.367 \\
      G332.2959$-$0.0945A & 16:15:45.8 & (2) & $-$50:55:54.5 & (3) & 20.2 & (8) & -49.70 & (2) & 0.91 & (4) & 13.820 \\
      \hline
    \end{tabular}
  \end{center}
\end{table*}

\setcounter{table}{0}
\begin{table*}
  \begin{center}
    \caption{{\em - continued.}}
    \begin{tabular}{ l ll ll d{1}l d{1}l d{1}l d{5} }
      \hline
      Spot name & \multicolumn{2}{c}{$\alpha_{2000}$} & \multicolumn{2}{c}{$\delta_{2000}$}                        & \multicolumn{2}{c}{Peak flux} & \multicolumn{2}{c}{Peak}     & \multicolumn{2}{c}{FWHM}   & \multicolumn{1}{c}{Integrated} \\
                & \multicolumn{2}{c}{(h:m:s)}         & \multicolumn{2}{c}{($^\circ$:$^\prime$:$^{\prime\prime}$)} & \multicolumn{2}{c}{density}   & \multicolumn{2}{c}{velocity} & \multicolumn{2}{c}{(\kms{})} & \multicolumn{1}{c}{flux density}  \\
                & &                                   & &                                                          & \multicolumn{2}{c}{(Jy)}      & \multicolumn{2}{c}{(\kms{})}   & &                          & \multicolumn{1}{c}{(Jy\,\kms{})} \\
      \hline
      G332.2959$-$0.0945B & 16:15:45.80 & (5) & $-$50:55:54.47 & (6) & 21.4 & (6) & -50.627 & (9) & 0.59 & (2) & 9.558 \\
      &  &  &  &  & 14.9 & (2) & -49.98 & (1) & 0.13 & (2) & 6.655 \\
      G332.2956$-$0.0945A* & 16:15:45.7 & (2) & $-$50:55:55.0 & (2) & 15.5 & (5) & -49.83 & (1) & 0.60 & (2) & 6.951 \\
      G332.2953$-$0.0944A & 16:15:45.61 & (8) & $-$50:55:55.62 & (5) & 9.0 & (2) & -48.355 & (7) & 0.58 & (2) & 3.924 \\
      G332.2941$-$0.0924A & 16:15:44.771 & (9) & $-$50:55:53.23 & (5) & 4.5 & (2) & -48.915 & (7) & 0.35 & (2) & 1.180 \\
      G332.2924$-$0.0919A & 16:15:44.15 & (2) & $-$50:55:56.39 & (6) & 2.69 & (7) & -46.107 & (9) & 0.74 & (2) & 1.492 \\
      G332.3191$+$0.1763A & 16:14:41.1 & (2) & $-$50:43:12.02 & (8) & 2.9 & (2) & -48.15 & (2) & 0.82 & (5) & 1.799 \\
      G332.3174$+$0.1817A & 16:14:39.18 & (5) & $-$50:43:02.11 & (2) & 0.55 & (3) & -49.28 & (1) & 0.54 & (3) & 0.222 \\
      G332.3546$-$0.1145A & 16:16:07.18 & (3) & $-$50:54:19.9 & (4) & 6.1 & (1) & -50.221 & (8) & 0.81 & (2) & 3.698 \\
      G332.5832$+$0.1469A & 16:16:00.97 & (5) & $-$50:33:30.96 & (9) & 3.01 & (5) & -42.631 & (6) & 0.76 & (1) & 1.718 \\
      G332.6036$-$0.1679A & 16:17:29.34 & (5) & $-$50:46:15.0 & (1) & 1.6 & (1) & -46.15 & (3) & 0.63 & (7) & 0.763 \\
      G332.6042$-$0.1668A & 16:17:29.21 & (4) & $-$50:46:10.9 & (1) & 10.5 & (3) & -45.836 & (9) & 0.72 & (2) & 5.714 \\
      G332.7163$-$0.0480A & 16:17:28.34 & (6) & $-$50:36:22.70 & (4) & 2.8 & (1) & -40.02 & (2) & 1.11 & (5) & 2.347 \\
      G332.7167$-$0.0488A & 16:17:28.65 & (5) & $-$50:36:23.9 & (1) & 1.10 & (5) & -39.71 & (1) & 0.58 & (3) & 0.480 \\
      &  &  &  &  & 0.82 & (5) & -38.66 & (2) & 0.77 & (5) & 0.358 \\
      G333.0024$-$0.4368A & 16:20:28.68 & (3) & $-$50:41:00.91 & (5) & 5.7 & (2) & -56.387 & (8) & 0.48 & (2) & 2.041 \\
      &  &  &  &  & 1.4 & (1) & -55.77 & (3) & 0.57 & (7) & 0.501 \\
      G333.0289$-$0.0623A & 16:18:56.54 & (6) & $-$50:23:53.6 & (1) & 0.755 & (5) & -42.125 & (2) & 0.513 & (5) & 0.292 \\
      &  &  &  &  & 0.768 & (3) & -43 & (8) & 0.14 & (2) & 0.297 \\
      G333.0290$-$0.0629A & 16:18:56.73 & (1) & $-$50:23:55.17 & (4) & 6.7 & (2) & -40.03 & (1) & 0.75 & (3) & 3.800 \\
      G333.0290$-$0.0629B & 16:18:56.70 & (4) & $-$50:23:55.09 & (3) & 7.4 & (4) & -40.88 & (2) & 0.63 & (4) & 3.502 \\
      G333.0294$-$0.0243A & 16:18:46.64 & (2) & $-$50:22:14.57 & (4) & 5.60 & (7) & -42.224 & (3) & 0.419 & (5) & 1.767 \\
      G333.0135$-$0.4660A & 16:20:39.46 & (2) & $-$50:41:47.45 & (3) & 1.21 & (5) & -52.65 & (1) & 0.58 & (3) & 0.528 \\
      G333.0678$-$0.4460A & 16:20:48.72 & (5) & $-$50:38:38.7 & (2) & 0.49 & (2) & -54.19 & (2) & 0.71 & (4) & 0.262 \\
      G333.0731$-$0.3996A & 16:20:37.76 & (3) & $-$50:36:26.66 & (8) & 2.13 & (2) & -52.639 & (3) & 0.676 & (7) & 1.084 \\
      G333.0699$-$0.3989A & 16:20:36.7 & (2) & $-$50:36:32.92 & (7) & 0.68 & (1) & -53.451 & (7) & 0.63 & (2) & 0.321 \\
      G333.0708$-$0.3990A & 16:20:37.0 & (1) & $-$50:36:30.80 & (9) & 0.68 & (3) & -53.75 & (2) & 0.81 & (4) & 0.415 \\
      G333.1040$-$0.5027A & 16:21:13.55 & (1) & $-$50:39:31.22 & (5) & 5.4 & (2) & -57.356 & (9) & 0.51 & (2) & 2.077 \\
      G333.1037$-$0.5036A & 16:21:13.7 & (1) & $-$50:39:34.49 & (6) & 1.66 & (7) & -58.22 & (1) & 0.54 & (3) & 0.677 \\
      G333.1045$-$0.5025A & 16:21:13.65 & (2) & $-$50:39:29.68 & (3) & 0.89 & (4) & -56.89 & (1) & 0.55 & (3) & 0.371 \\
      G333.1003$-$0.4985A & 16:21:11.43 & (5) & $-$50:39:30.1 & (2) & 0.56 & (6) & -55.15 & (2) & 0.45 & (5) & 0.192 \\
      G333.1210$-$0.4325A & 16:20:59.39 & (3) & $-$50:35:48.95 & (6) & 4.4 & (1) & -50.780 & (7) & 0.45 & (2) & 1.505 \\
      G333.1205$-$0.4337A & 16:20:59.6 & (2) & $-$50:35:53.4 & (4) & 1.30 & (6) & -53.17 & (1) & 0.56 & (3) & 0.551 \\
      G333.1206$-$0.4330A & 16:20:59.42 & (5) & $-$50:35:51.3 & (1) & 2.23 & (6) & -52.661 & (6) & 0.45 & (1) & 0.759 \\
      &  &  &  &  & 1.01 & (7) & -51.86 & (2) & 0.70 & (6) & 0.344 \\
      G333.1214$-$0.4334A & 16:20:59.75 & (2) & $-$50:35:50.46 & (4) & 0.8 & (2) & -53.51 & (3) & 0.3 & (1) & 0.181 \\
      &  &  &  &  & 0.62 & (1) & -53.908 & (4) & 0.372 & (9) & 0.141 \\
      G333.1254$-$0.4412A & 16:21:02.90 & (8) & $-$50:36:00.1 & (1) & 3.5 & (3) & -47.16 & (4) & 1.06 & (9) & 2.804 \\
      G333.1254$-$0.4412B & 16:21:02.87 & (3) & $-$50:36:00.1 & (1) & 55 & (1) & -49.983 & (5) & 0.52 & (1) & 21.447 \\
      &  &  &  &  & 13 & (2) & -49.12 & (8) & 1.4 & (2) & 5.069 \\
      G333.1254$-$0.4412C & 16:21:02.88 & (3) & $-$50:36:00.0 & (2) & 6.5 & (4) & -48.12 & (2) & 0.48 & (4) & 2.339 \\
      &  &  &  &  & 3.1 & (9) & -47.61 & (6) & 0.4 & (1) & 1.115 \\
      G333.1300$-$0.4373A & 16:21:03.06 & (1) & $-$50:35:38.51 & (2) & 4.35 & (2) & -51.528 & (1) & 0.358 & (2) & 1.172 \\
      G333.1310$-$0.4389A & 16:21:03.75 & (1) & $-$50:35:40.05 & (7) & 16.7 & (2) & -50.238 & (5) & 0.69 & (1) & 8.702 \\
      &  &  &  &  & 10 & (1) & -49.704 & (4) & 0.26 & (3) & 5.211 \\
      G333.1256$-$0.4398A & 16:21:02.58 & (7) & $-$50:35:56.2 & (2) & 11.8 & (5) & -49.09 & (1) & 0.56 & (3) & 4.978 \\
      G333.1257$-$0.4378A & 16:21:02.0 & (1) & $-$50:35:50.7 & (1) & 12 & (1) & -48.633 & (6) & 0.30 & (3) & 2.680 \\
      &  &  &  &  & 3.6 & (3) & -49.04 & (3) & 0.53 & (6) & 0.804 \\
      G333.1256$-$0.4378A & 16:21:02.03 & (4) & $-$50:35:50.78 & (6) & 10 & (1) & -48.69 & (2) & 0.47 & (5) & 3.545 \\
      G333.1255$-$0.4379A & 16:21:02.03 & (6) & $-$50:35:51.6 & (1) & 16.7 & (8) & -48.05 & (1) & 0.52 & (3) & 6.571 \\
      G333.1260$-$0.4398A & 16:21:02.7 & (3) & $-$50:35:55.12 & (9) & 11.6 & (7) & -47.40 & (2) & 0.56 & (4) & 4.893 \\
      G333.1254$-$0.4387A & 16:21:02.20 & (3) & $-$50:35:53.79 & (8) & 15.3 & (3) & -46.730 & (6) & 0.67 & (1) & 7.756 \\
      &  &  &  &  & 5.8 & (4) & -45.81 & (3) & 0.96 & (8) & 2.940 \\
      G333.1381$-$0.4247A & 16:21:01.89 & (3) & $-$50:34:45.58 & (8) & 10.8 & (2) & -56.070 & (3) & 0.464 & (7) & 3.771 \\
      G333.1319$-$0.4310A & 16:21:01.92 & (1) & $-$50:35:17.60 & (5) & 2.33 & (9) & -55.586 & (8) & 0.40 & (2) & 0.702 \\
      G333.1398$-$0.4260A & 16:21:02.69 & (6) & $-$50:34:44.92 & (4) & 1.28 & (5) & -53.770 & (9) & 0.54 & (2) & 0.522 \\
      G333.1629$-$0.1010A & 16:19:42.69 & (5) & $-$50:19:54.74 & (7) & 5.63 & (6) & -91.480 & (4) & 0.730 & (9) & 3.093 \\
      G333.1631$-$0.1011A & 16:19:42.75 & (7) & $-$50:19:54.56 & (4) & 1.4 & (2) & -90.83 & (4) & 0.7 & (1) & 0.690 \\
      G333.1610$-$0.1009A & 16:19:42.15 & (2) & $-$50:19:59.57 & (5) & 1.12 & (3) & -92.337 & (5) & 0.45 & (1) & 0.379 \\
      &  &  &  &  & 0.62 & (6) & -92.83 & (3) & 0.54 & (6) & 0.210 \\
      \hline
    \end{tabular}
  \end{center}
\end{table*}

\setcounter{table}{0}
\begin{table*}
  \begin{center}
    \caption{{\em - continued.}}
    \begin{tabular}{ l ll ll d{1}l d{1}l d{1}l d{5} }
      \hline
      Spot name & \multicolumn{2}{c}{$\alpha_{2000}$} & \multicolumn{2}{c}{$\delta_{2000}$}                        & \multicolumn{2}{c}{Peak flux} & \multicolumn{2}{c}{Peak}     & \multicolumn{2}{c}{FWHM}   & \multicolumn{1}{c}{Integrated} \\
                & \multicolumn{2}{c}{(h:m:s)}         & \multicolumn{2}{c}{($^\circ$:$^\prime$:$^{\prime\prime}$)} & \multicolumn{2}{c}{density}   & \multicolumn{2}{c}{velocity} & \multicolumn{2}{c}{(\kms{})} & \multicolumn{1}{c}{flux density}  \\
                & &                                   & &                                                          & \multicolumn{2}{c}{(Jy)}      & \multicolumn{2}{c}{(\kms{})}   & &                          & \multicolumn{1}{c}{(Jy\,\kms{})} \\
      \hline
      G333.1850$-$0.0929A & 16:19:46.47 & (2) & $-$50:18:38.1 & (2) & 0.97 & (4) & -86.18 & (1) & 0.73 & (3) & 0.536 \\
      G333.1839$-$0.0904A & 16:19:45.53 & (3) & $-$50:18:34.63 & (4) & 2.0 & (1) & -86.69 & (2) & 0.90 & (5) & 1.361 \\
      G333.1828$-$0.0880A & 16:19:44.59 & (6) & $-$50:18:31.04 & (2) & 1.41 & (7) & -86.83 & (1) & 0.33 & (2) & 0.355 \\
      G333.2302$-$0.0583A & 16:19:49.41 & (7) & $-$50:15:15.4 & (1) & 280 & (3) & -87.298 & (6) & 1.07 & (1) & 224.818 \\
      &  &  &  &  & 7 & (2) & -89.2 & (3) & 1.6 & (6) & 5.620 \\
      G333.2344$-$0.0628A & 16:19:51.71 & (2) & $-$50:15:16.24 & (4) & 1.51 & (3) & -91.542 & (7) & 0.65 & (2) & 0.733 \\
      G333.2323$-$0.0568A & 16:19:49.59 & (1) & $-$50:15:06.13 & (3) & 2.4 & (1) & -90.44 & (1) & 0.70 & (3) & 1.259 \\
      G333.2305$-$0.0582A* & 16:19:49.5 & (2) & $-$50:15:14.2 & (2) & 35 & (4) & -87.87 & (3) & 0.59 & (7) & 15.509 \\
      &  &  &  &  & 8 & (3) & -88.3 & (2) & 1.4 & (5) & 3.545 \\
      G333.2325$-$0.0628A & 16:19:51.22 & (2) & $-$50:15:21.18 & (2) & 3.76 & (3) & -89.012 & (2) & 0.429 & (5) & 1.213 \\
      G333.2332$-$0.0627A & 16:19:51.37 & (8) & $-$50:15:19.10 & (4) & 6.3 & (6) & -88.10 & (2) & 0.44 & (5) & 2.077 \\
      G333.2332$-$0.0627B & 16:19:51.38 & (7) & $-$50:15:19.0 & (1) & 2.1 & (5) & -88.5 & (1) & 0.8 & (2) & 1.329 \\
      G333.2307$-$0.0585A & 16:19:49.6 & (2) & $-$50:15:14.54 & (9) & 5.3 & (8) & -86.56 & (4) & 0.6 & (1) & 2.583 \\
      G333.2307$-$0.0585B & 16:19:49.61 & (7) & $-$50:15:14.4 & (1) & 10.2 & (7) & -88.16 & (2) & 0.47 & (4) & 3.634 \\
      G333.2345$-$0.0640A & 16:19:52.10 & (1) & $-$50:15:19.02 & (9) & 8.0 & (3) & -86.098 & (9) & 0.44 & (2) & 2.666 \\
      G333.2336$-$0.0632A & 16:19:51.61 & (7) & $-$50:15:19.14 & (6) & 2.9 & (3) & -85.42 & (5) & 1.0 & (1) & 2.107 \\
      G333.2337$-$0.0625A & 16:19:51.45 & (5) & $-$50:15:17.1 & (1) & 12.1 & (2) & -85.556 & (4) & 0.377 & (7) & 3.431 \\
      &  &  &  &  & 5.2 & (7) & -84.99 & (4) & 0.7 & (1) & 1.475 \\
      G333.2343$-$0.0610A & 16:19:51.24 & (8) & $-$50:15:11.92 & (8) & 8.62 & (9) & -85.492 & (3) & 0.506 & (7) & 3.285 \\
      G333.2198$-$0.4027A & 16:21:17.90 & (2) & $-$50:30:22.21 & (8) & 21.5 & (2) & -51.675 & (2) & 0.422 & (5) & 6.821 \\
      &  &  &  &  & 1.58 & (7) & -51.86 & (4) & 1.54 & (8) & 0.501 \\
      G333.2204$-$0.4030A & 16:21:18.13 & (8) & $-$50:30:21.58 & (9) & 0.92 & (5) & -54.65 & (2) & 0.82 & (5) & 0.566 \\
      G333.2204$-$0.4030B & 16:21:18.14 & (5) & $-$50:30:21.57 & (4) & 1.32 & (7) & -55.82 & (3) & 1.15 & (7) & 1.146 \\
      G333.2200$-$0.4008A & 16:21:17.43 & (4) & $-$50:30:16.9 & (2) & 2.9 & (2) & -53.74 & (3) & 0.90 & (7) & 1.964 \\
      G333.2203$-$0.4018A & 16:21:17.8 & (2) & $-$50:30:18.6 & (3) & 1.8 & (1) & -53.61 & (2) & 0.75 & (4) & 1.018 \\
      G333.2203$-$0.4019A & 16:21:17.8 & (1) & $-$50:30:19.0 & (2) & 2.5 & (2) & -52.72 & (2) & 0.56 & (4) & 1.050 \\
      &  &  &  &  & 1.6 & (2) & -53.43 & (4) & 0.8 & (1) & 0.672 \\
      G333.2845$-$0.3732A & 16:21:27.31 & (4) & $-$50:26:22.48 & (8) & 0.78 & (5) & -52.09 & (3) & 0.90 & (6) & 0.530 \\
      G333.3008$-$0.3516A & 16:21:25.92 & (8) & $-$50:24:46.3 & (1) & 0.53 & (3) & -49.45 & (2) & 0.70 & (4) & 0.279 \\
      G333.3117$+$0.1033A & 16:19:28.67 & (5) & $-$50:04:54.82 & (2) & 3.34 & (6) & -45.161 & (4) & 0.480 & (9) & 1.208 \\
      G333.3137$+$0.1071A & 16:19:28.21 & (4) & $-$50:04:39.96 & (6) & 1.72 & (6) & -48.22 & (1) & 0.69 & (3) & 0.887 \\
      G333.3123$+$0.1063A & 16:19:28.05 & (2) & $-$50:04:45.64 & (5) & 1.51 & (4) & -47.186 & (7) & 0.58 & (2) & 0.658 \\
      G333.3146$+$0.1054A & 16:19:28.90 & (1) & $-$50:04:42.00 & (2) & 2.5 & (2) & -44.43 & (5) & 1.0 & (1) & 1.941 \\
      G333.3358$-$0.3620A & 16:21:38.02 & (8) & $-$50:23:43.51 & (6) & 19.6 & (4) & -51.927 & (8) & 0.81 & (2) & 11.951 \\
      G333.3358$-$0.3620B & 16:21:38.05 & (7) & $-$50:23:43.35 & (5) & 12.5 & (2) & -52.43 & (8) & 0.2 & (2) & 1.883 \\
      &  &  &  &  & 3 & (3) & -52.87 & (5) & 0.2 & (2) & 0.452 \\
      G333.3376$-$0.3621A & 16:21:38.54 & (4) & $-$50:23:39.2 & (1) & 2.0 & (9) & -49.96 & (4) & 0.2 & (1) & 0.369 \\
      &  &  &  &  & 1.0 & (3) & -49.52 & (1) & 0.25 & (7) & 0.184 \\
      G333.3376$-$0.3621B & 16:21:38.5 & (1) & $-$50:23:39.3 & (1) & 6.6 & (3) & -50.29 & (1) & 0.41 & (2) & 2.047 \\
      G333.3376$-$0.3621C & 16:21:38.5 & (1) & $-$50:23:39.3 & (2) & 0.81 & (5) & -54.10 & (2) & 0.60 & (4) & 0.365 \\
      G333.3284$-$0.3643A & 16:21:36.67 & (5) & $-$50:24:08.16 & (2) & 4.9 & (1) & -50.867 & (9) & 0.68 & (2) & 2.519 \\
      G333.3285$-$0.3645A & 16:21:36.73 & (2) & $-$50:24:08.35 & (5) & 3.2 & (3) & -50.22 & (3) & 0.63 & (8) & 1.509 \\
      G333.3359$-$0.3616A & 16:21:37.95 & (7) & $-$50:23:42.20 & (7) & 4.2 & (1) & -51.098 & (7) & 0.41 & (1) & 1.303 \\
      &  &  &  &  & 1.92 & (8) & -50.54 & (2) & 1.01 & (5) & 0.596 \\
      G333.3384$-$0.3623A & 16:21:38.79 & (3) & $-$50:23:37.62 & (2) & 0.56 & (3) & -47.60 & (2) & 0.82 & (5) & 0.347 \\
      G333.3874$+$0.0322A & 16:20:07.52 & (3) & $-$50:04:45.95 & (4) & 3.1 & (1) & -70.22 & (2) & 0.96 & (4) & 2.231 \\
      G333.3860$+$0.0311A & 16:20:07.42 & (5) & $-$50:04:52.28 & (5) & 0.97 & (9) & -72.57 & (3) & 0.65 & (7) & 0.473 \\
      G333.3862$+$0.0311A & 16:20:07.48 & (6) & $-$50:04:51.84 & (4) & 1.65 & (5) & -72.03 & (1) & 0.83 & (3) & 1.027 \\
      G333.3872$+$0.0316A & 16:20:07.62 & (3) & $-$50:04:48.0 & (2) & 1.44 & (6) & -68.44 & (1) & 0.49 & (2) & 0.528 \\
      G333.3745$-$0.2023A & 16:21:05.93 & (4) & $-$50:15:17.91 & (7) & 9.1 & (4) & -61.31 & (1) & 0.44 & (2) & 3.000 \\
      G333.3773$-$0.2013A & 16:21:06.42 & (7) & $-$50:15:08.52 & (7) & 6.5 & (1) & -57.979 & (7) & 0.93 & (2) & 4.574 \\
      G333.4679$-$0.1601A & 16:21:19.62 & (2) & $-$50:09:33.1 & (2) & 17.3 & (1) & -43.007 & (2) & 0.558 & (5) & 7.267 \\
      G333.4679$-$0.1601B & 16:21:19.61 & (2) & $-$50:09:32.95 & (2) & 0.6 & (1) & -43.94 & (9) & 0.9 & (2) & 0.405 \\
      G333.4669$-$0.1621A & 16:21:19.88 & (4) & $-$50:09:40.75 & (6) & 6.1 & (1) & -45.275 & (5) & 0.55 & (1) & 2.508 \\
      &  &  &  &  & 0.8 & (1) & -45.2 & (1) & 1.4 & (3) & 0.329 \\
      G333.4658$-$0.1660A & 16:21:20.62 & (3) & $-$50:09:53.44 & (7) & 2.1 & (1) & -44.10 & (1) & 0.52 & (3) & 0.815 \\
      G333.4658$-$0.1660B & 16:21:20.62 & (1) & $-$50:09:53.47 & (3) & 0.6 & (1) & -44.79 & (8) & 0.8 & (2) & 0.364 \\
      G333.4664$-$0.1633A & 16:21:20.05 & (4) & $-$50:09:45.2 & (1) & 5.5 & (2) & -43.467 & (7) & 0.41 & (2) & 1.716 \\
      G333.4681$-$0.1613A & 16:21:19.97 & (9) & $-$50:09:35.64 & (7) & 1.1 & (1) & -42.11 & (2) & 0.53 & (5) & 0.435 \\
      G333.4681$-$0.1616A & 16:21:20.1 & (2) & $-$50:09:36.55 & (4) & 3.69 & (8) & -41.518 & (6) & 0.55 & (1) & 1.530 \\
      \hline
    \end{tabular}
  \end{center}
\end{table*}

\setcounter{table}{0}
\begin{table*}
  \begin{center}
    \caption{{\em - continued.}}
    \begin{tabular}{ l ll ll d{1}l d{1}l d{1}l d{5} }
      \hline
      Spot name & \multicolumn{2}{c}{$\alpha_{2000}$} & \multicolumn{2}{c}{$\delta_{2000}$}                        & \multicolumn{2}{c}{Peak flux} & \multicolumn{2}{c}{Peak}     & \multicolumn{2}{c}{FWHM}   & \multicolumn{1}{c}{Integrated} \\
                & \multicolumn{2}{c}{(h:m:s)}         & \multicolumn{2}{c}{($^\circ$:$^\prime$:$^{\prime\prime}$)} & \multicolumn{2}{c}{density}   & \multicolumn{2}{c}{velocity} & \multicolumn{2}{c}{(\kms{})} & \multicolumn{1}{c}{flux density}  \\
                & &                                   & &                                                          & \multicolumn{2}{c}{(Jy)}      & \multicolumn{2}{c}{(\kms{})}   & &                          & \multicolumn{1}{c}{(Jy\,\kms{})} \\
      \hline
      G333.4684$-$0.1649A & 16:21:21.00 & (6) & $-$50:09:44.19 & (3) & 1.31 & (6) & -38.91 & (1) & 0.51 & (3) & 0.502 \\
      G333.4973$+$0.1431A & 16:20:07.59 & (4) & $-$49:55:23.7 & (1) & 2.7 & (2) & -112.39 & (3) & 0.75 & (7) & 1.531 \\
      &  &  &  &  & 0.71 & (7) & -113.60 & (4) & 0.85 & (9) & 0.403 \\
      G333.4967$+$0.1424A & 16:20:07.61 & (3) & $-$49:55:26.86 & (8) & 0.70 & (5) & -113.53 & (2) & 0.64 & (5) & 0.339 \\
      G333.4967$+$0.1421A & 16:20:07.69 & (3) & $-$49:55:28.0 & (1) & 0.75 & (3) & -113.060 & (8) & 0.49 & (2) & 0.277 \\
      G333.5228$-$0.2747A & 16:22:04.49 & (1) & $-$50:12:05.3 & (1) & 7.2 & (2) & -50.220 & (9) & 0.78 & (2) & 4.211 \\
      G333.5507$-$0.2915A & 16:22:16.36 & (6) & $-$50:11:36.91 & (4) & 1.96 & (6) & -46.232 & (7) & 0.44 & (2) & 0.653 \\
      G333.5650$-$0.2952A & 16:22:21.11 & (4) & $-$50:11:09.8 & (1) & 1.16 & (9) & -46.65 & (3) & 0.72 & (6) & 0.631 \\
      G333.5618$-$0.0246A & 16:21:08.74 & (4) & $-$49:59:48.85 & (8) & 60 & (1) & -39.98 & (1) & 0.85 & (2) & 38.498 \\
      G333.5687$+$0.0284A & 16:20:56.61 & (3) & $-$49:57:15.91 & (7) & 7.1 & (1) & -84.796 & (4) & 0.504 & (9) & 2.693 \\
      G333.5934$-$0.2122A & 16:22:06.63 & (4) & $-$50:06:26.6 & (1) & 27.7 & (3) & -49.561 & (3) & 0.617 & (7) & 12.863 \\
      G333.5950$-$0.2108A & 16:22:06.70 & (2) & $-$50:06:19.2 & (1) & 1.2 & (1) & -48.75 & (5) & 1.1 & (1) & 1.021 \\
      G333.5950$-$0.2108B & 16:22:06.7 & (1) & $-$50:06:18.8 & (2) & 2.3 & (1) & -50.25 & (3) & 0.79 & (6) & 1.374 \\
      G333.5950$-$0.2108C & 16:22:06.7 & (2) & $-$50:06:18.77 & (6) & 0.8 & (1) & -51.34 & (5) & 0.7 & (1) & 0.407 \\
      G333.5941$-$0.2116A & 16:22:06.66 & (3) & $-$50:06:23.07 & (8) & 1.26 & (6) & -48.60 & (2) & 0.81 & (4) & 0.768 \\
      G333.6919$-$0.1955A & 16:22:28.19 & (2) & $-$50:01:32.9 & (1) & 3.29 & (9) & -50.610 & (5) & 0.358 & (9) & 0.886 \\
      &  &  &  &  & 0.46 & (9) & -50.7 & (1) & 1.1 & (2) & 0.124 \\
      G333.6958$-$0.1986A & 16:22:30.06 & (5) & $-$50:01:30.8 & (1) & 0.94 & (3) & -51.43 & (1) & 0.76 & (3) & 0.535 \\
      G333.7123$-$0.1158A & 16:22:12.50 & (4) & $-$49:57:18.37 & (5) & 6.2 & (1) & -31.495 & (6) & 0.73 & (1) & 3.407 \\
      G333.7098$-$0.1152A & 16:22:11.69 & (2) & $-$49:57:23.17 & (8) & 0.45 & (5) & -28.12 & (4) & 0.70 & (9) & 0.236 \\
      G333.7733$-$0.2578A & 16:23:06.16 & (4) & $-$50:00:43.31 & (2) & 2.55 & (5) & -49.405 & (5) & 0.51 & (1) & 0.985 \\
      G333.7722$-$0.0096A & 16:22:00.29 & (3) & $-$49:50:15.79 & (5) & 17.7 & (8) & -89.33 & (2) & 0.88 & (4) & 11.671 \\
      G333.8181$-$0.3026A & 16:23:29.83 & (2) & $-$50:00:42.0 & (1) & 20.4 & (8) & -48.31 & (1) & 0.58 & (3) & 8.895 \\
      G333.8187$-$0.3031A & 16:23:30.11 & (4) & $-$50:00:41.93 & (3) & 2.3 & (2) & -47.70 & (3) & 0.70 & (7) & 1.219 \\
      G333.8993$-$0.0977A & 16:22:56.86 & (8) & $-$49:48:35.3 & (1) & 1.5 & (2) & -63.65 & (4) & 0.8 & (1) & 0.885 \\
      &  &  &  &  & 1.5 & (1) & -62.83 & (3) & 0.65 & (7) & 0.885 \\
      G333.9007$-$0.0993A & 16:22:57.65 & (8) & $-$49:48:35.9 & (2) & 1.36 & (8) & -64.36 & (2) & 0.84 & (5) & 0.861 \\
      G333.9307$-$0.1342A & 16:23:14.73 & (2) & $-$49:48:47.5 & (1) & 3.8 & (2) & -42.10 & (2) & 0.69 & (4) & 1.960 \\
      G333.9305$-$0.1343A & 16:23:14.71 & (2) & $-$49:48:48.46 & (4) & 3.0 & (2) & -42.55 & (2) & 0.60 & (5) & 1.356 \\
      G333.9303$-$0.1316A & 16:23:13.95 & (2) & $-$49:48:41.8 & (1) & 0.89 & (6) & -41.67 & (2) & 0.52 & (4) & 0.345 \\
      G333.9744$+$0.0737A & 16:22:31.40 & (3) & $-$49:38:08.44 & (4) & 6.7 & (2) & -58.30 & (1) & 0.72 & (3) & 3.610 \\
      G333.9746$+$0.0736A & 16:22:31.49 & (2) & $-$49:38:08.19 & (2) & 3.2 & (3) & -59.67 & (4) & 0.93 & (9) & 2.246 \\
      G334.0266$-$0.0465A & 16:23:16.67 & (1) & $-$49:41:00.19 & (2) & 10.4 & (1) & -84.040 & (4) & 0.721 & (9) & 5.641 \\
      G334.7452$+$0.5068A & 16:23:57.70 & (1) & $-$48:46:59.92 & (4) & 3.52 & (7) & -64.554 & (5) & 0.51 & (1) & 1.360 \\
      G334.7459$+$0.5063A & 16:23:57.99 & (2) & $-$48:46:59.41 & (8) & 1.60 & (5) & -63.49 & (1) & 0.70 & (3) & 0.845 \\
      G334.7469$+$0.5058A & 16:23:58.40 & (7) & $-$48:46:58.0 & (1) & 2.3 & (2) & -61.15 & (2) & 0.46 & (4) & 0.803 \\
      &  &  &  &  & 0.45 & (5) & -60.22 & (4) & 0.8 & (1) & 0.157 \\
      \hline
    \end{tabular}
  \end{center}
\end{table*}

\onecolumn
\begin{landscape}
\begin{table*}
  \section{Gaussian spectral fits to auto-correlation thermal line emission}
  \label{app:thermal_detail}
  \begin{center}
    \caption{Gaussian fits for the spectra of thermal lines detected towards class~I~\choh{} maser sites. Column 1 lists the maser site name. Columns 2 to 5 contain information for \cs{}, columns 6 to 9 for \sio{}, columns 10 to 13 for thermal \choh{}. The columns for each type of emission list the fitted Gaussian parameters, as well as the integrated intensity bounded by these Gaussians. Note that the uncertainty for each parameter is quoted in parentheses, in units of the least significant figure. Rows without site names correspond to a second Gaussian fitted to the site directly above. If no Gaussian information is listed for an emission type, it was not detected.}
    \begin{tabular}{ l | d{1}l d{1}l d{1}l d{5} | d{1}l d{1}l d{1}l d{5} | d{1}l d{1}l d{1}l d{5} }
      \hline
      Site name   & \multicolumn{7}{c}{\cs{} (1--0) parameters} & \multicolumn{7}{c}{\sio{} (1--0) $v=0$ parameters} & \multicolumn{7}{c}{\choh{} 1$_0$-0$_0$ parameters} \\
                  & \multicolumn{2}{c}{Peak}      & \multicolumn{2}{c}{Peak}     & \multicolumn{2}{c}{FWHM}   & \multicolumn{1}{c}{Integrated} & \multicolumn{2}{c}{Peak}      & \multicolumn{2}{c}{Peak}     & \multicolumn{2}{c}{FWHM}   & \multicolumn{1}{c}{Integrated} & \multicolumn{2}{c}{Peak}      & \multicolumn{2}{c}{Peak}     & \multicolumn{2}{c}{FWHM}   & \multicolumn{1}{c}{Integrated} \\
                  & \multicolumn{2}{c}{intensity} & \multicolumn{2}{c}{velocity} & \multicolumn{2}{c}{(\kms{})} & \multicolumn{1}{c}{intensity} & \multicolumn{2}{c}{intensity} & \multicolumn{2}{c}{velocity} & \multicolumn{2}{c}{(\kms{})} & \multicolumn{1}{c}{intensity} & \multicolumn{2}{c}{intensity} & \multicolumn{2}{c}{velocity} & \multicolumn{2}{c}{(\kms{})} & \multicolumn{1}{c}{intensity} \\
                  & \multicolumn{2}{c}{(K)}       & \multicolumn{2}{c}{(\kms{})}   & &                          & \multicolumn{1}{c}{(K\,\kms{})} & \multicolumn{2}{c}{(K)}       & \multicolumn{2}{c}{(\kms{})}   & &                          & \multicolumn{1}{c}{(K\,\kms{})} & \multicolumn{2}{c}{(K)}       & \multicolumn{2}{c}{(\kms{})}   & &                          & \multicolumn{1}{c}{(K\,\kms{})} \\
      \hline
      G330.294$-$0.393 & 0.507 & (3) & -80.68 & (1) & 5.62 & (4) & 2.143 & & & & & & & & 0.02 & (2) & -80.7 & (3) & 9.7 & (7) & 0.147 \\
      G330.678$-$0.402 & 0.298 & (4) & -63.87 & (3) & 5.32 & (7) & 1.194 & 0.013 & (1) & -64.0 & (2) & 8.6 & (4) & 0.084 & 0.05 & (3) & -63.53 & (6) & 3.4 & (2) & 0.127 \\
      G330.779$+$0.249 & 0.336 & (5) & -43.51 & (1) & 2.72 & (3) & 0.687 & 0.027 & (1) & -43.81 & (9) & 6.7 & (2) & 0.137 & 0.10 & (4) & -43.48 & (3) & 2.55 & (7) & 0.192 \\
      G330.876$-$0.362 & 1.24 & (1) & -62.79 & (2) & 4.58 & (4) & 4.279 & 0.044 & (1) & -63.35 & (6) & 7.1 & (1) & 0.236 & 0.09 & (4) & -62.91 & (5) & 4.0 & (1) & 0.272 \\
      G330.871$-$0.383 & 1.39 & (1) & -62.94 & (2) & 4.88 & (4) & 5.102 & 0.060 & (1) & -63.63 & (5) & 7.1 & (1) & 0.322 & 0.10 & (4) & -63.25 & (5) & 4.9 & (1) & 0.370 \\
      G330.927$-$0.408 & 0.47 & (2) & -41.38 & (4) & 2.9 & (1) & 1.040 & 0.016 & (1) & -42.5 & (1) & 5.4 & (3) & 0.065 & 0.05 & (4) & -41.49 & (7) & 2.8 & (2) & 0.106 \\
      G330.931$-$0.260 & 0.158 & (4) & -90.18 & (6) & 5.2 & (1) & 0.622 & & & & & & & & 0.02 & (4) & -89.1 & (2) & 3.2 & (4) & 0.048 \\
      G330.955$-$0.182 & 1.483 & (8) & -91.16 & (1) & 7.01 & (4) & 7.820 & 0.174 & (1) & -90.01 & (3) & 7.96 & (7) & 1.043 & 0.12 & (5) & -90.59 & (6) & 5.5 & (1) & 0.496 \\
      G331.131$-$0.470 & 0.870 & (7) & -67.50 & (1) & 4.55 & (3) & 2.979 & 0.023 & (1) & -67.2 & (1) & 4.2 & (2) & 0.072 & 0.09 & (3) & -67.78 & (4) & 4.2 & (1) & 0.286 \\
      G331.134$-$0.488 & 0.526 & (6) & -66.23 & (2) & 5.43 & (5) & 2.152 & 0.016 & (1) & -66.5 & (2) & 7.7 & (4) & 0.093 & 0.03 & (2) & -66.1 & (1) & 5.4 & (2) & 0.123 \\
      G331.132$-$0.244 & 0.897 & (5) & -86.94 & (1) & 5.76 & (3) & 3.889 & 0.087 & (1) & -86.74 & (5) & 8.5 & (1) & 0.554 & 0.19 & (4) & -86.33 & (3) & 5.35 & (8) & 0.765 \\
      G331.134$+$0.156 & 0.277 & (6) & -76.96 & (3) & 3.05 & (6) & 0.637 & 0.017 & (1) & -76.2 & (1) & 5.1 & (3) & 0.065 & 0.06 & (3) & -76.52 & (5) & 4.4 & (1) & 0.198 \\
      G331.279$-$0.189 & 0.808 & (5) & -88.29 & (1) & 5.25 & (3) & 3.195 & 0.046 & (1) & -87.2 & (1) & 10.9 & (3) & 0.376 & 0.11 & (3) & -88.06 & (3) & 3.61 & (7) & 0.299 \\
      G331.341$-$0.347 & 0.519 & (8) & -65.90 & (2) & 3.06 & (5) & 1.197 & 0.019 & (1) & -66.5 & (1) & 6.9 & (3) & 0.098 & 0.04 & (2) & -66.46 & (6) & 4.2 & (1) & 0.126 \\
      G331.370$-$0.399 & 0.255 & (6) & -64.80 & (3) & 3.62 & (8) & 0.695 & 0.009 & (1) & -66.3 & (4) & 12.3 & (8) & 0.083 & 0.02 & (2) & -66.5 & (2) & 8.2 & (5) & 0.124 \\
	  & 0.051 & (5) & -69.9 & (1) & 3.1 & (3) & 0.121 & & & & & & & & & & & & & & \\
      G331.371$-$0.145 & 0.167 & (3) & -87.06 & (4) & 4.16 & (9) & 0.523 & & & & & & & & 0.03 & (3) & -87.7 & (1) & 6.1 & (3) & 0.138 \\
	  & 0.052 & (3) & -91.0 & (1) & 3.7 & (3) & 0.145 & & & & & & & & & & & & & & \\
      G331.380$+$0.149 & 0.344 & (8) & -45.05 & (5) & 4.8 & (1) & 1.255 & 0.069 & (1) & -44.63 & (6) & 6.7 & (1) & 0.350 & 0.20 & (6) & -45.03 & (3) & 4.32 & (7) & 0.650 \\
	  & 0.07 & (1) & -50.3 & (1) & 1.9 & (3) & 0.098 & & & & & & & & & & & & & & \\
      G331.409$-$0.164 & 0.348 & (4) & -86.36 & (4) & 5.73 & (8) & 1.502 & 0.028 & (1) & -85.05 & (9) & 5.0 & (2) & 0.105 & 0.11 & (3) & -85.99 & (4) & 4.48 & (8) & 0.371 \\
      G331.44$-$0.14 & 0.169 & (5) & -86.6 & (1) & 8.2 & (3) & 1.049 & 0.017 & (1) & -85.6 & (1) & 5.0 & (3) & 0.064 & 0.05 & (3) & -85.89 & (8) & 5.7 & (2) & 0.213 \\
      G331.440$-$0.187 & 0.240 & (7) & -88.0 & (1) & 8.9 & (3) & 1.605 & 0.060 & (1) & -87.97 & (6) & 7.1 & (1) & 0.320 & 0.18 & (4) & -88.36 & (3) & 5.51 & (7) & 0.746 \\
      G331.442$-$0.158 & 0.203 & (3) & -87.10 & (6) & 7.1 & (1) & 1.090 & 0.021 & (1) & -87.0 & (1) & 6.4 & (3) & 0.101 & 0.09 & (4) & -86.75 & (5) & 4.5 & (1) & 0.306 \\
      G331.492$-$0.082 & 0.586 & (7) & -88.83 & (2) & 4.33 & (6) & 1.908 & 0.035 & (1) & -88.8 & (1) & 7.7 & (2) & 0.202 & 0.06 & (5) & -88.72 & (7) & 3.3 & (2) & 0.149 \\
      G331.503$-$0.109 & 0.63 & (2) & -88.14 & (7) & 5.3 & (2) & 2.511 & 0.037 & (1) & -88.0 & (1) & 7.4 & (3) & 0.206 & 0.08 & (7) & -87.9 & (1) & 4.0 & (2) & 0.240 \\
	  & 0.269 & (6) & -101.38 & (6) & 5.2 & (1) & 1.055 & 0.032 & (1) & -100.8 & (2) & 8.5 & (4) & 0.206 & 0.08 & (4) & -101.30 & (6) & 4.0 & (1) & 0.238 \\
      G331.519$-$0.082 & 0.785 & (4) & -89.35 & (1) & 4.73 & (3) & 2.797 & 0.027 & (1) & -89.3 & (2) & 8.3 & (4) & 0.169 & 0.04 & (5) & -89.3 & (1) & 3.3 & (3) & 0.100 \\
      G331.530$-$0.099 & 1.196 & (6) & -88.58 & (1) & 5.07 & (3) & 4.566 & 0.063 & (1) & -88.63 & (6) & 7.4 & (1) & 0.353 & 0.14 & (5) & -89.16 & (4) & 4.21 & (9) & 0.443 \\
      G331.544$-$0.067 & 0.842 & (9) & -88.27 & (2) & 4.79 & (6) & 3.036 & 0.038 & (1) & -88.50 & (7) & 5.3 & (2) & 0.151 & 0.10 & (5) & -88.66 & (5) & 3.8 & (1) & 0.287 \\
      G331.555$-$0.122 & 0.558 & (7) & -100.37 & (3) & 4.51 & (6) & 1.894 & 0.024 & (1) & -98.2 & (2) & 16.5 & (6) & 0.299 & 0.09 & (5) & -100.56 & (5) & 3.6 & (1) & 0.247 \\
	  & 0.113 & (5) & -88.8 & (2) & 8.8 & (4) & 0.747 & & & & & & & & & & & & & & \\
      G331.72$-$0.20 & 0.464 & (6) & -47.44 & (2) & 2.88 & (4) & 1.007 & 0.020 & (1) & -46.5 & (1) & 4.3 & (2) & 0.064 & 0.05 & (4) & -46.7 & (1) & 6.0 & (3) & 0.225 \\
      G331.853$-$0.129 & 0.452 & (4) & -50.57 & (2) & 3.50 & (4) & 1.191 & 0.056 & (1) & -50.69 & (5) & 5.4 & (1) & 0.229 & 0.13 & (6) & -50.84 & (7) & 5.4 & (2) & 0.531 \\
      G331.887$+$0.063 & 0.343 & (3) & -87.57 & (2) & 4.95 & (5) & 1.278 & 0.081 & (1) & -87.39 & (3) & 4.89 & (7) & 0.298 & 0.20 & (4) & -87.44 & (2) & 4.06 & (5) & 0.611 \\
      G331.921$-$0.083 & 0.297 & (8) & -51.59 & (4) & 3.3 & (1) & 0.742 & 0.015 & (1) & -50.9 & (2) & 9.9 & (5) & 0.112 & 0.06 & (5) & -51.56 & (8) & 3.6 & (2) & 0.162 \\
      G332.092$-$0.420 & 0.776 & (6) & -56.73 & (2) & 4.47 & (4) & 2.613 & 0.018 & (1) & -57.0 & (2) & 6.1 & (4) & 0.082 & 0.06 & (4) & -56.97 & (7) & 4.1 & (2) & 0.185 \\
      G332.240$-$0.044 & 0.462 & (5) & -47.97 & (2) & 3.93 & (4) & 1.365 & 0.053 & (1) & -48.67 & (8) & 9.2 & (2) & 0.368 & 0.19 & (6) & -48.18 & (4) & 4.8 & (1) & 0.693 \\
	  & 0.080 & (3) & -48.6 & (2) & 14.1 & (5) & 0.847 & & & & & & & & & & & & & & \\
      \hline
    \end{tabular}
  \end{center}
\end{table*}

\setcounter{table}{0}
\begin{table*}
  \begin{center}
    \caption{{\em - continued.}}
    \begin{tabular}{ l | d{1}l d{1}l d{1}l d{5} | d{1}l d{1}l d{1}l d{5} | d{1}l d{1}l d{1}l d{5} }
      \hline
      Site name   & \multicolumn{7}{c}{\cs{} (1--0) parameters} & \multicolumn{7}{c}{\sio{} (1--0) $v=0$ parameters} & \multicolumn{7}{c}{\choh{} 1$_0$-0$_0$ parameters} \\
                  & \multicolumn{2}{c}{Peak}      & \multicolumn{2}{c}{Peak}     & \multicolumn{2}{c}{FWHM}   & \multicolumn{1}{c}{Integrated} & \multicolumn{2}{c}{Peak}      & \multicolumn{2}{c}{Peak}     & \multicolumn{2}{c}{FWHM}   & \multicolumn{1}{c}{Integrated} & \multicolumn{2}{c}{Peak}      & \multicolumn{2}{c}{Peak}     & \multicolumn{2}{c}{FWHM}   & \multicolumn{1}{c}{Integrated} \\
                  & \multicolumn{2}{c}{intensity} & \multicolumn{2}{c}{velocity} & \multicolumn{2}{c}{(\kms{})} & \multicolumn{1}{c}{intensity} & \multicolumn{2}{c}{intensity} & \multicolumn{2}{c}{velocity} & \multicolumn{2}{c}{(\kms{})} & \multicolumn{1}{c}{intensity} & \multicolumn{2}{c}{intensity} & \multicolumn{2}{c}{velocity} & \multicolumn{2}{c}{(\kms{})} & \multicolumn{1}{c}{intensity} \\
                  & \multicolumn{2}{c}{(K)}       & \multicolumn{2}{c}{(\kms{})}   & &                          & \multicolumn{1}{c}{(K\,\kms{})} & \multicolumn{2}{c}{(K)}       & \multicolumn{2}{c}{(\kms{})}   & &                          & \multicolumn{1}{c}{(K\,\kms{})} & \multicolumn{2}{c}{(K)}       & \multicolumn{2}{c}{(\kms{})}   & &                          & \multicolumn{1}{c}{(K\,\kms{})} \\
      \hline
      G332.295$-$0.094 & 0.564 & (7) & -48.86 & (2) & 3.68 & (5) & 1.561 & 0.022 & (1) & -50.3 & (1) & 9.2 & (3) & 0.153 & 0.06 & (3) & -49.38 & (8) & 5.3 & (2) & 0.239 \\
	  & 0.095 & (4) & -49.3 & (2) & 9.2 & (5) & 0.656 & & & & & & & & & & & & & & \\
      G332.318$+$0.179 & 0.197 & (5) & -48.30 & (3) & 2.50 & (7) & 0.370 & 0.022 & (1) & -47.9 & (1) & 5.2 & (3) & 0.085 & 0.06 & (4) & -48.23 & (6) & 4.1 & (1) & 0.183 \\
	  & 0.059 & (3) & -48.2 & (2) & 5.9 & (4) & 0.264 & & & & & & & & & & & & & & \\
      G332.355$-$0.114 & 0.523 & (5) & -49.97 & (1) & 2.73 & (3) & 1.076 & & & & & & & & 0.03 & (3) & -50.20 & (5) & 2.1 & (2) & 0.048 \\
      G332.583$+$0.147 & 0.124 & (9) & -45.03 & (8) & 2.2 & (2) & 0.209 & & & & & & & & & & & & & & \\
      G332.604$-$0.167 & 0.40 & (1) & -46.66 & (5) & 3.5 & (1) & 1.068 & 0.025 & (1) & -46.5 & (1) & 7.5 & (3) & 0.142 & 0.07 & (4) & -46.51 & (9) & 6.1 & (2) & 0.320 \\
      G332.716$-$0.048 & 0.140 & (5) & -39.54 & (7) & 4.2 & (2) & 0.443 & & & & & & & & 0.03 & (4) & -39.8 & (1) & 3.5 & (3) & 0.080 \\
      G333.002$-$0.437 & 0.860 & (6) & -55.62 & (1) & 3.31 & (3) & 2.140 & 0.019 & (1) & -54.5 & (1) & 6.1 & (3) & 0.087 & 0.02 & (3) & -54.6 & (2) & 6.5 & (5) & 0.097 \\
	  & 0.358 & (7) & -51.63 & (3) & 3.08 & (7) & 0.831 & & & & & & & & & & & & & & \\
      G333.029$-$0.063 & 0.49 & (1) & -41.24 & (3) & 2.87 & (7) & 1.058 & & & & & & & & 0.03 & (4) & -41.3 & (1) & 3.4 & (4) & 0.077 \\
      G333.029$-$0.024 & 0.45 & (1) & -41.73 & (3) & 2.16 & (7) & 0.731 & 0.016 & (1) & -41.3 & (2) & 4.9 & (4) & 0.058 & 0.02 & (7) & -41.77 & (9) & 2.1 & (2) & 0.031 \\
      G333.014$-$0.466 & 0.818 & (8) & -54.33 & (2) & 4.40 & (5) & 2.707 & 0.027 & (1) & -53.8 & (1) & 8.5 & (3) & 0.173 & 0.01 & (6) & -54.9 & (4) & 5.9 & (9) & 0.045 \\
	  & 0.10 & (1) & -60.6 & (2) & 3.2 & (4) & 0.239 & & & & & & & & & & & & & & \\
      G333.068$-$0.446 & 1.43 & (1) & -53.48 & (2) & 4.68 & (4) & 5.039 & 0.064 & (1) & -53.27 & (6) & 6.2 & (1) & 0.298 & 0.10 & (8) & -53.73 & (5) & 4.1 & (1) & 0.306 \\
	  & 0.190 & (9) & -59.9 & (1) & 4.7 & (3) & 0.671 & & & & & & & & & & & & & & \\
      G333.071$-$0.399 & 0.842 & (7) & -53.59 & (2) & 5.07 & (5) & 3.212 & 0.047 & (2) & -52.5 & (1) & 6.0 & (2) & 0.213 & 0.06 & (7) & -53.68 & (8) & 4.6 & (2) & 0.209 \\
		 & & & & & & & & 0.020 & (1) & -58.4 & (4) & 18 & (1) & 0.278 & & & & & & & \\
G333.103$-$0.502 & 1.156 & (9) & -55.87 & (1) & 3.63 & (3) & 3.159 & 0.027 & (1) & -55.4 & (1) & 9.1 & (4) & 0.184 & 0.04 & (7) & -56.37 & (8) & 3.5 & (2) & 0.106 \\
G333.121$-$0.433 & 2.24 & (1) & -51.50 & (2) & 5.65 & (4) & 9.533 & 0.096 & (2) & -51.54 & (7) & 8.8 & (2) & 0.634 & 0.07 & (8) & -51.23 & (8) & 5.7 & (2) & 0.302 \\
		 & 0.207 & (8) & -60.8 & (1) & 6.8 & (3) & 1.063 & & & & & & & & & & & & & & \\
G333.126$-$0.439 & 2.54 & (2) & -50.98 & (2) & 5.74 & (5) & 10.976 & 0.139 & (2) & -50.99 & (6) & 8.9 & (1) & 0.927 & 0.1 & (1) & -50.49 & (7) & 6.0 & (2) & 0.455 \\
		 & 0.30 & (1) & -59.5 & (1) & 6.7 & (3) & 1.521 & & & & & & & & & & & & & & \\
G333.137$-$0.427 & 2.84 & (1) & -52.32 & (1) & 6.48 & (4) & 13.843 & 0.171 & (2) & -52.80 & (5) & 6.8 & (1) & 0.876 & 0.13 & (9) & -52.82 & (6) & 6.4 & (1) & 0.625 \\
		 & & & & & & & & 0.077 & (2) & -44.9 & (1) & 11.9 & (3) & 0.688 & & & & & & & \\
G333.162$-$0.101 & 0.35 & (1) & -92.06 & (4) & 2.47 & (9) & 0.651 & & & & & & & & 0.02 & (4) & -88.8 & (4) & 16 & (1) & 0.242 \\
		 & 0.151 & (8) & -87.0 & (1) & 5.1 & (3) & 0.580 & & & & & & & & & & & & & & \\
G333.184$-$0.090 & 0.522 & (9) & -86.14 & (3) & 3.09 & (6) & 1.216 & 0.018 & (1) & -86.4 & (2) & 5.1 & (4) & 0.069 & 0.08 & (8) & -86.23 & (5) & 3.6 & (1) & 0.216 \\
		 & 0.134 & (9) & -90.4 & (1) & 3.7 & (3) & 0.370 & & & & & & & & & & & & & & \\
G333.233$-$0.061 & 0.157 & (6) & -88.7 & (2) & 8.3 & (4) & 0.976 & 0.091 & (2) & -87.85 & (5) & 5.9 & (1) & 0.407 & 0.15 & (7) & -87.93 & (3) & 4.52 & (6) & 0.510 \\
G333.220$-$0.402 & 0.842 & (9) & -52.00 & (2) & 3.19 & (4) & 2.024 & 0.027 & (1) & -52.3 & (1) & 4.9 & (3) & 0.100 & 0.04 & (8) & -52.24 & (7) & 2.7 & (2) & 0.081 \\
		 & 0.101 & (9) & -47.1 & (1) & 2.9 & (3) & 0.220 & & & & & & & & & & & & & & \\
G333.24$+$0.02 & 0.348 & (8) & -69.87 & (5) & 4.7 & (1) & 1.240 & 0.019 & (1) & -68.4 & (2) & 11.9 & (5) & 0.170 & 0.03 & (5) & -69.1 & (1) & 6.8 & (3) & 0.153 \\
G333.284$-$0.373 & 1.983 & (7) & -51.947 & (9) & 5.46 & (2) & 8.154 & 0.098 & (2) & -51.52 & (4) & 4.84 & (9) & 0.357 & 0.2 & (1) & -51.90 & (3) & 4.44 & (7) & 0.668 \\
G333.301$-$0.352 & 1.623 & (8) & -50.27 & (1) & 4.80 & (3) & 5.868 & 0.066 & (1) & -50.42 & (4) & 4.4 & (1) & 0.217 & 0.18 & (9) & -50.48 & (3) & 4.06 & (6) & 0.551 \\
		 & 0.191 & (8) & -58.7 & (1) & 4.8 & (2) & 0.690 & & & & & & & & & & & & & & \\
G333.313$+$0.106 & 0.468 & (8) & -46.41 & (4) & 4.80 & (9) & 1.691 & 0.039 & (1) & -46.8 & (1) & 7.8 & (3) & 0.228 & 0.11 & (9) & -46.25 & (6) & 5.5 & (1) & 0.459 \\
G333.335$-$0.363 & 1.112 & (8) & -50.70 & (2) & 4.70 & (4) & 3.936 & 0.067 & (2) & -49.93 & (9) & 6.0 & (2) & 0.302 & 0.16 & (8) & -50.29 & (2) & 3.80 & (5) & 0.457 \\
		 & & & & & & & & 0.019 & (1) & -59.3 & (6) & 21 & (1) & 0.293 & & & & & & & \\
G333.387$+$0.031 & 0.324 & (9) & -69.87 & (5) & 3.6 & (1) & 0.890 & 0.025 & (1) & -69.3 & (2) & 11.7 & (5) & 0.220 & 0.05 & (8) & -69.88 & (7) & 3.6 & (2) & 0.136 \\
G333.376$-$0.202 & 0.357 & (8) & -59.94 & (4) & 3.08 & (8) & 0.827 & 0.021 & (1) & -55.5 & (3) & 15.4 & (7) & 0.244 & 0.03 & (6) & -59.0 & (1) & 4.9 & (3) & 0.111 \\
G333.467$-$0.163 & 0.35 & (1) & -43.81 & (4) & 3.0 & (1) & 0.802 & 0.041 & (1) & -44.1 & (1) & 9.6 & (3) & 0.298 & 0.07 & (7) & -43.49 & (6) & 4.6 & (1) & 0.244 \\
		 & 0.284 & (6) & -44.52 & (9) & 8.8 & (2) & 1.890 & 0.011 & (1) & -58.2 & (6) & 14 & (1) & 0.117 & & & & & & & \\
      \hline
    \end{tabular}
  \end{center}
\end{table*}

\setcounter{table}{0}
\begin{table*}
  \begin{center}
    \caption{{\em - continued.}}
    \begin{tabular}{ l | d{1}l d{1}l d{1}l d{5} | d{1}l d{1}l d{1}l d{5} | d{1}l d{1}l d{1}l d{5} }
      \hline
      Site name   & \multicolumn{7}{c}{\cs{} (1--0) parameters} & \multicolumn{7}{c}{\sio{} (1--0) $v=0$ parameters} & \multicolumn{7}{c}{\choh{} 1$_0$-0$_0$ parameters} \\
                  & \multicolumn{2}{c}{Peak}      & \multicolumn{2}{c}{Peak}     & \multicolumn{2}{c}{FWHM}   & \multicolumn{1}{c}{Integrated} & \multicolumn{2}{c}{Peak}      & \multicolumn{2}{c}{Peak}     & \multicolumn{2}{c}{FWHM}   & \multicolumn{1}{c}{Integrated} & \multicolumn{2}{c}{Peak}      & \multicolumn{2}{c}{Peak}     & \multicolumn{2}{c}{FWHM}   & \multicolumn{1}{c}{Integrated} \\
                  & \multicolumn{2}{c}{intensity} & \multicolumn{2}{c}{velocity} & \multicolumn{2}{c}{(\kms{})} & \multicolumn{1}{c}{intensity} & \multicolumn{2}{c}{intensity} & \multicolumn{2}{c}{velocity} & \multicolumn{2}{c}{(\kms{})} & \multicolumn{1}{c}{intensity} & \multicolumn{2}{c}{intensity} & \multicolumn{2}{c}{velocity} & \multicolumn{2}{c}{(\kms{})} & \multicolumn{1}{c}{intensity} \\
                  & \multicolumn{2}{c}{(K)}       & \multicolumn{2}{c}{(\kms{})}   & &                          & \multicolumn{1}{c}{(K\,\kms{})} & \multicolumn{2}{c}{(K)}       & \multicolumn{2}{c}{(\kms{})}   & &                          & \multicolumn{1}{c}{(K\,\kms{})} & \multicolumn{2}{c}{(K)}       & \multicolumn{2}{c}{(\kms{})}   & &                          & \multicolumn{1}{c}{(K\,\kms{})} \\
      \hline
      G333.497$+$0.143 & 0.18 & (4) & -113.3 & (3) & 2.8 & (6) & 0.379 & 0.013 & (1) & -114.6 & (3) & 11.4 & (8) & 0.111 & 0.03 & (6) & -113.28 & (7) & 2.4 & (2) & 0.053 \\
      G333.523$-$0.275 & 0.936 & (8) & -49.68 & (1) & 3.51 & (3) & 2.474 & 0.037 & (1) & -49.3 & (1) & 6.4 & (2) & 0.178 & 0.05 & (8) & -49.64 & (6) & 2.9 & (1) & 0.110 \\
      G333.558$-$0.293 & 0.589 & (9) & -45.83 & (2) & 2.87 & (5) & 1.272 & 0.013 & (1) & -47.8 & (3) & 12.5 & (8) & 0.122 & 0.03 & (7) & -45.95 & (7) & 2.4 & (2) & 0.055 \\
      G333.562$-$0.025 & 0.145 & (9) & -46.7 & (1) & 3.2 & (2) & 0.345 & 0.016 & (1) & -38.9 & (2) & 7.8 & (5) & 0.094 & 0.03 & (7) & -39.5 & (1) & 4.2 & (3) & 0.096 \\
      G333.569$+$0.028 & 0.256 & (9) & -84.67 & (9) & 5.3 & (2) & 1.013 & 0.031 & (1) & -84.4 & (1) & 7.5 & (3) & 0.174 & 0.08 & (6) & -84.64 & (4) & 4.2 & (1) & 0.254 \\
      G333.595$-$0.211 & 1.96 & (1) & -47.56 & (1) & 5.33 & (3) & 7.858 & 0.069 & (1) & -47.93 & (6) & 6.5 & (1) & 0.338 & 0.0 & (1) & -47.9 & (1) & 3.8 & (2) & 0.000 \\
      G333.694$-$0.197 & 0.289 & (9) & -50.53 & (4) & 2.8 & (1) & 0.606 & 0.026 & (1) & -49.9 & (2) & 8.0 & (4) & 0.157 & 0.08 & (7) & -50.47 & (3) & 2.98 & (8) & 0.180 \\
      G333.711$-$0.115 & 0.071 & (7) & -29.0 & (2) & 4.6 & (5) & 0.247 & & & & & & & & & & & & & & \\
      G333.773$-$0.258 & 0.377 & (6) & -47.69 & (4) & 5.5 & (1) & 1.563 & 0.048 & (1) & -48.64 & (8) & 5.8 & (2) & 0.211 & 0.06 & (6) & -48.66 & (7) & 4.9 & (2) & 0.220 \\
      G333.772$-$0.010 & 0.121 & (8) & -86.0 & (1) & 3.5 & (3) & 0.319 & & & & & & & & 0.01 & (5) & -86.8 & (4) & 7.5 & (9) & 0.057 \\
      G333.818$-$0.303 & 0.141 & (6) & -46.6 & (1) & 5.4 & (3) & 0.574 & & & & & & & & 0.01 & (3) & -46.6 & (4) & 9 & (1) & 0.069 \\
      G333.900$-$0.098 & 0.134 & (7) & -65.1 & (1) & 4.0 & (2) & 0.401 & 0.014 & (1) & -65.1 & (2) & 6.6 & (5) & 0.070 & 0.03 & (6) & -64.9 & (1) & 4.6 & (3) & 0.104 \\
      G333.930$-$0.133 & 0.198 & (7) & -40.01 & (9) & 5.2 & (2) & 0.773 & 0.020 & (1) & -39.4 & (1) & 5.2 & (3) & 0.078 & 0.02 & (6) & -39.9 & (2) & 5.1 & (4) & 0.077 \\
      G333.974$+$0.074 & 0.22 & (1) & -59.93 & (7) & 2.4 & (2) & 0.406 & & & & & & & & 0.01 & (5) & -58.0 & (5) & 10 & (1) & 0.072 \\
      G334.027$-$0.047 & 0.366 & (8) & -84.93 & (4) & 3.67 & (9) & 1.011 & & & & & & & & 0.03 & (5) & -84.72 & (7) & 3.3 & (2) & 0.075 \\
	  & 0.06 & (1) & -89.0 & (2) & 2.0 & (4) & 0.089 & & & & & & & & & & & & & & \\
      G334.746$+$0.506 & 0.255 & (7) & -62.45 & (6) & 4.0 & (1) & 0.771 & 0.023 & (1) & -62.0 & (2) & 12.2 & (5) & 0.212 & 0.06 & (7) & -62.36 & (6) & 4.4 & (2) & 0.201 \\
      \hline
    \end{tabular}
  \end{center}
\end{table*}
\end{landscape}

\setcounter{table}{1}
\begin{table*}
  \begin{center}
    \caption{Gaussian fits for the spectra of \hfta{} radio recombination lines detected towards class~I~\choh{} maser sites. If a maser site is not listed, then \hfta{} emission was not detected. Column 1 lists the maser site name. Columns 2 to 5 list the fitted Gaussian parameters, as well as the integrated intensity bounded by these Gaussians. Note that the uncertainty for each parameter is quoted in parentheses, in units of the least significant figure.}
    \begin{tabular}{ l | d{1}l d{1}l d{1}l d{5} }
      \hline
      Site name   & \multicolumn{2}{c}{Peak}      & \multicolumn{2}{c}{Peak}     & \multicolumn{2}{c}{FWHM}     & \multicolumn{1}{c}{Integrated}  \\
                  & \multicolumn{2}{c}{intensity} & \multicolumn{2}{c}{velocity} & \multicolumn{2}{c}{(\kms{})} & \multicolumn{1}{c}{intensity}   \\
                  & \multicolumn{2}{c}{(K)}       & \multicolumn{2}{c}{(\kms{})} & &                            & \multicolumn{1}{c}{(K\,\kms{})} \\
      \hline
G330.294$-$0.393 & 0.054 & (1) & -83.1 & (3) & 27.8 & (6) & 1.130 \\
G330.678$-$0.402 & 0.026 & (1) & -62.2 & (4) & 21 & (1) & 0.405 \\
G330.876$-$0.362 & 0.082 & (1) & -51.3 & (2) & 32.8 & (4) & 2.032 \\
G330.871$-$0.383 & 0.070 & (1) & -54.8 & (2) & 32.8 & (5) & 1.729 \\
G330.955$-$0.182 & 0.174 & (1) & -90.50 & (9) & 33.6 & (2) & 4.411 \\
G331.134$-$0.488 & 0.019 & (1) & -70.3 & (4) & 15 & (1) & 0.218 \\
G331.279$-$0.189 & 0.070 & (1) & -80.1 & (2) & 24.7 & (4) & 1.302 \\
G331.341$-$0.347 & 0.041 & (1) & -61.5 & (3) & 28.5 & (8) & 0.878 \\
G331.492$-$0.082 & 0.038 & (1) & -90.3 & (3) & 21.7 & (7) & 0.620 \\
G331.503$-$0.109 & 0.057 & (1) & -89.1 & (2) & 19.2 & (4) & 0.824 \\
G331.519$-$0.082 & 0.225 & (1) & -91.72 & (6) & 26.2 & (1) & 4.445 \\
G331.530$-$0.099 & 0.151 & (1) & -89.52 & (8) & 22.9 & (2) & 2.600 \\
G331.544$-$0.067 & 0.099 & (1) & -90.8 & (1) & 21.6 & (3) & 1.611 \\
G331.555$-$0.122 & 0.041 & (1) & -99.2 & (3) & 32.2 & (8) & 0.988 \\
G332.295$-$0.094 & 0.017 & (1) & -46.3 & (8) & 29 & (2) & 0.372 \\
G333.002$-$0.437 & 0.085 & (2) & -53.9 & (2) & 22.2 & (5) & 1.421 \\
G333.014$-$0.466 & 0.154 & (1) & -53.09 & (6) & 21.8 & (1) & 2.537 \\
G333.068$-$0.446 & 0.062 & (1) & -56.3 & (2) & 27.3 & (4) & 1.279 \\
G333.071$-$0.399 & 0.022 & (1) & -54.1 & (5) & 34 & (1) & 0.560 \\
G333.103$-$0.502 & 0.052 & (1) & -53.7 & (2) & 26.0 & (5) & 1.021 \\
G333.121$-$0.433 & 0.286 & (1) & -47.75 & (4) & 26.53 & (9) & 5.704 \\
G333.126$-$0.439 & 0.250 & (1) & -50.34 & (5) & 28.1 & (1) & 5.294 \\
G333.137$-$0.427 & 0.300 & (1) & -50.44 & (4) & 29.50 & (9) & 6.671 \\
G333.162$-$0.101 & 0.040 & (1) & -92.0 & (2) & 22.0 & (5) & 0.668 \\
G333.184$-$0.090 & 0.034 & (1) & -92.4 & (3) & 19.1 & (6) & 0.490 \\
G333.220$-$0.402 & 0.056 & (1) & -51.5 & (2) & 17.5 & (4) & 0.736 \\
G333.284$-$0.373 & 0.168 & (1) & -51.81 & (7) & 28.1 & (2) & 3.560 \\
G333.301$-$0.352 & 0.051 & (1) & -53.2 & (2) & 27.9 & (5) & 1.070 \\
G333.335$-$0.363 & 0.031 & (1) & -46.9 & (3) & 26.7 & (8) & 0.630 \\
G333.467$-$0.163 & 0.055 & (1) & -43.1 & (2) & 20.7 & (4) & 0.863 \\
G333.523$-$0.275 & 0.027 & (1) & -51.7 & (4) & 28.3 & (9) & 0.571 \\
G333.558$-$0.293 & 0.018 & (1) & -52.3 & (5) & 18 & (1) & 0.242 \\
G333.595$-$0.211 & 0.837 & (1) & -45.38 & (2) & 31.78 & (5) & 20.020 \\
G333.694$-$0.197 & 0.014 & (1) & -46.9 & (8) & 24 & (2) & 0.241 \\
      \hline
    \end{tabular}
  \end{center}
\end{table*}
\twocolumn

\twocolumn[{
    \vspace{-8.75mm}
    \section{GLIMPSE images}
    \label{app:glimpse}
    \begin{minipage}{\textwidth}
      \centering
      \includegraphics[height=0.30\textheight]{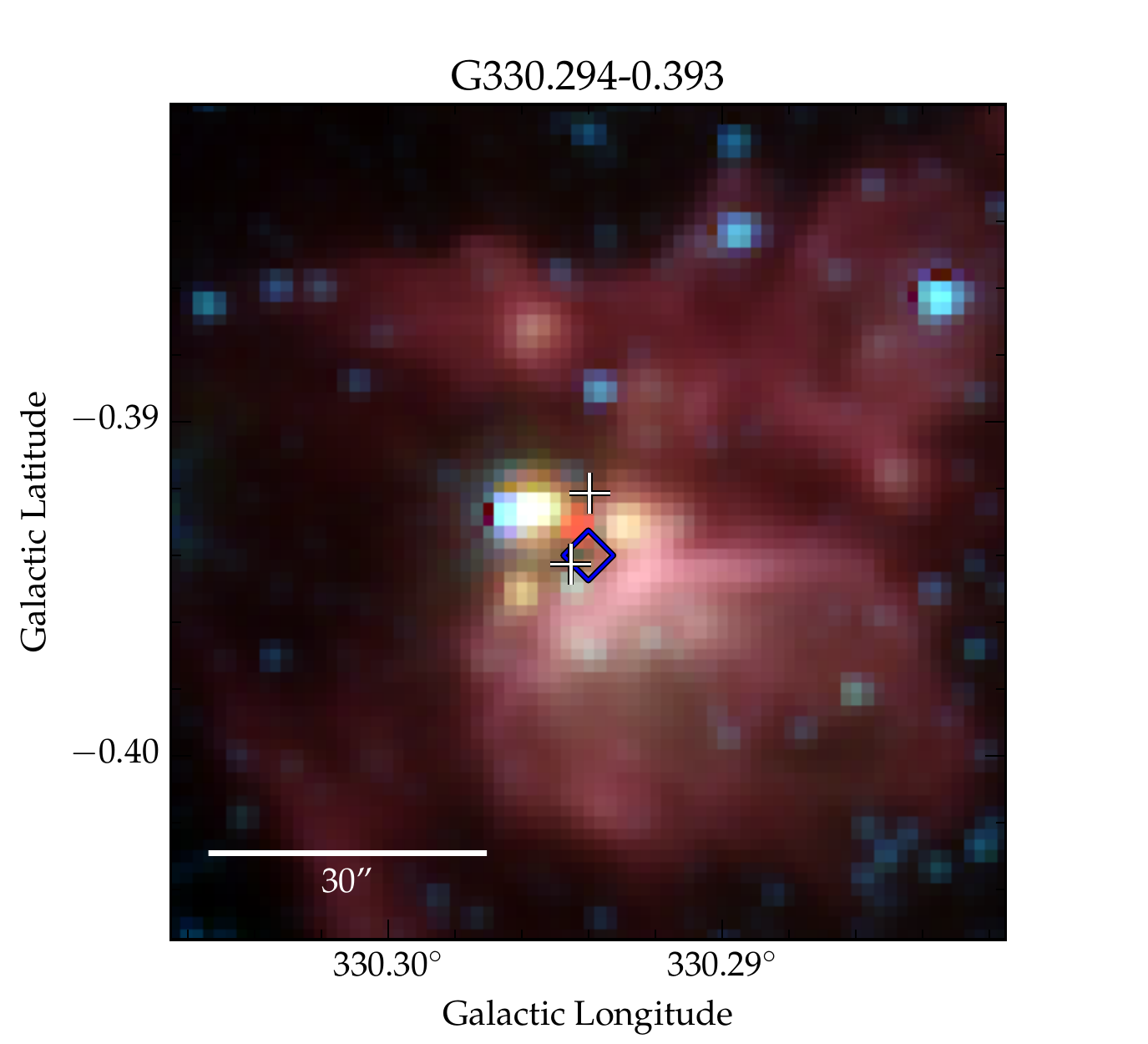}
      \includegraphics[height=0.30\textheight]{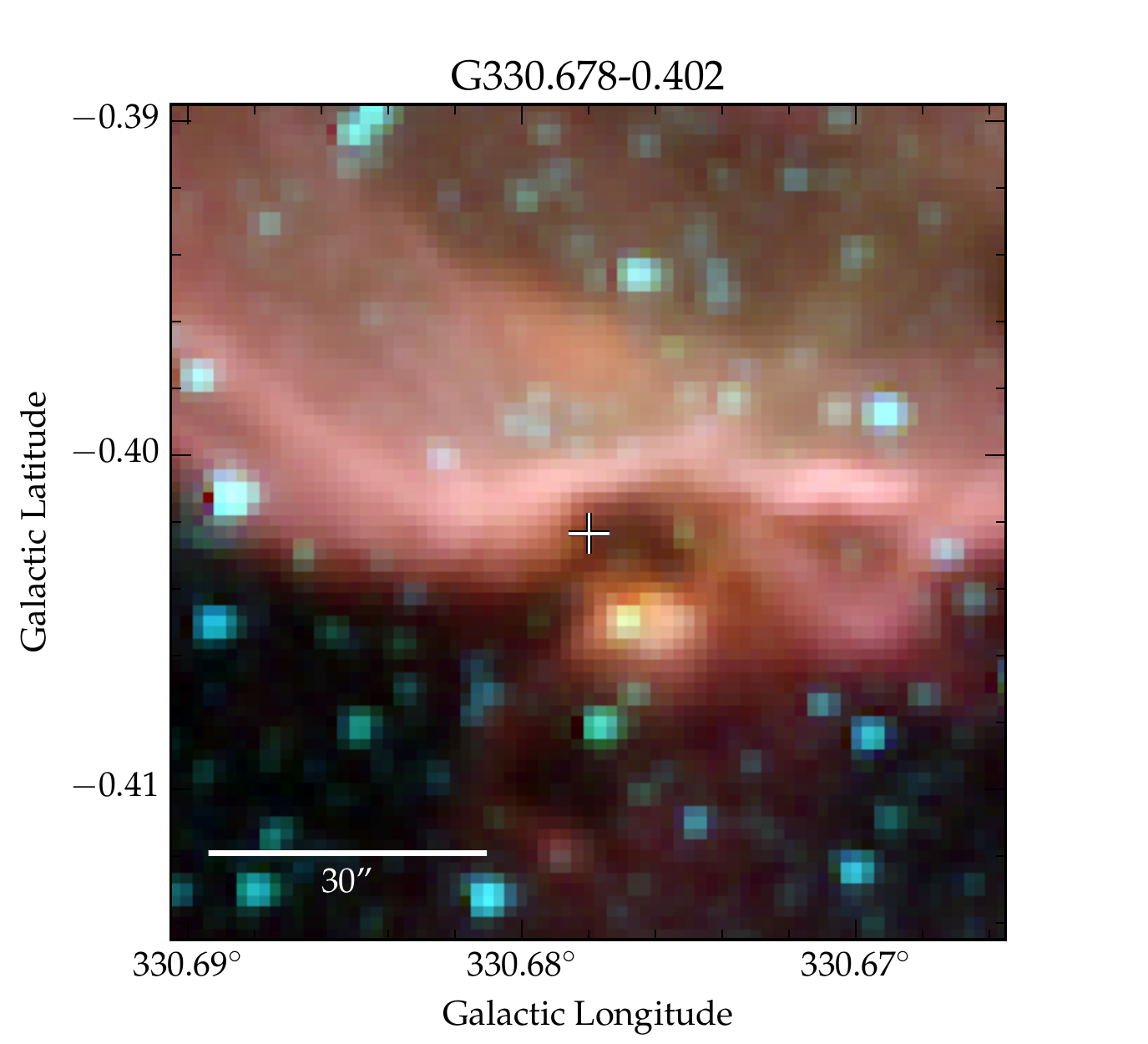}
      \includegraphics[height=0.30\textheight]{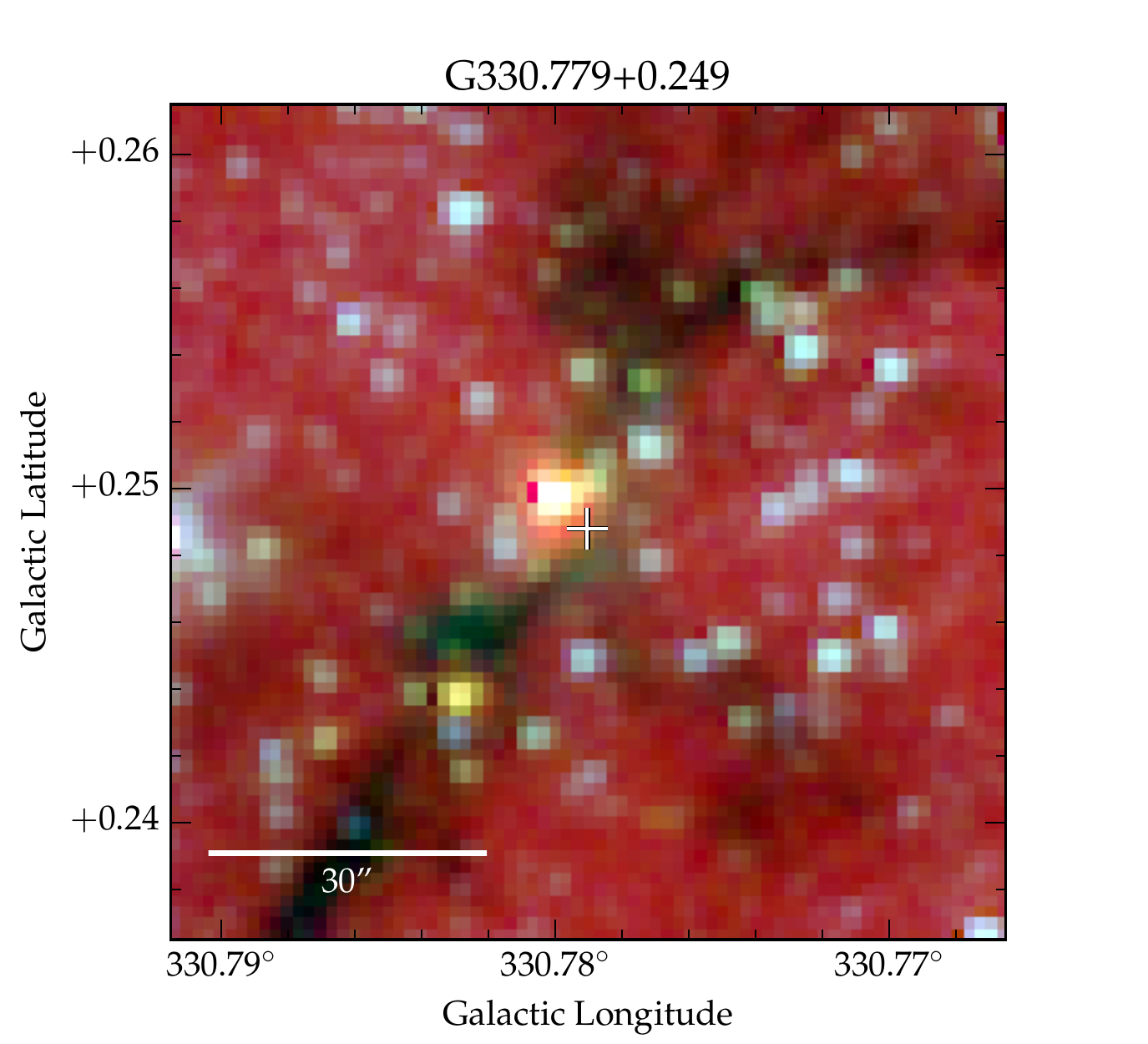}
      \includegraphics[height=0.30\textheight]{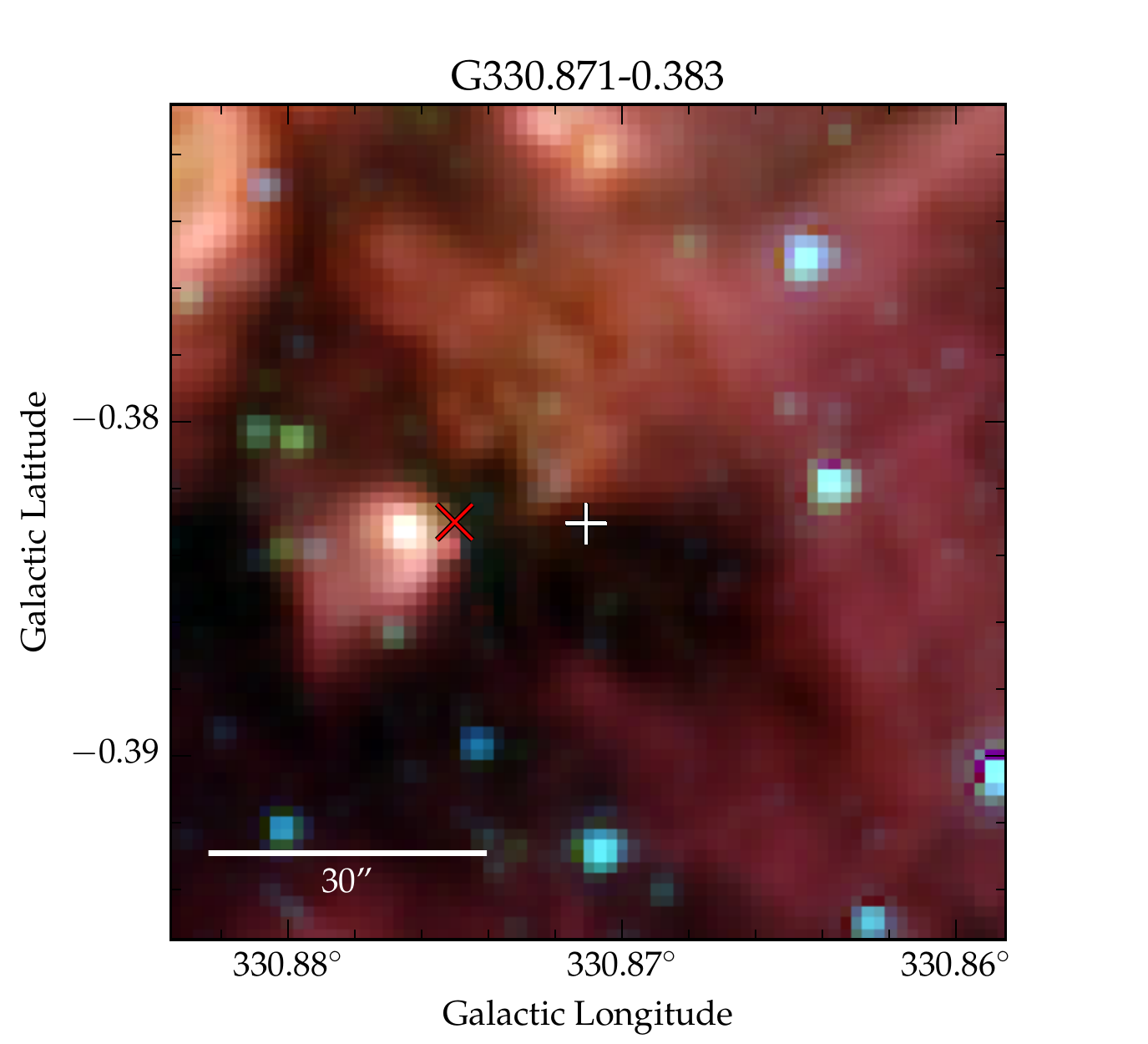}
      \includegraphics[height=0.30\textheight]{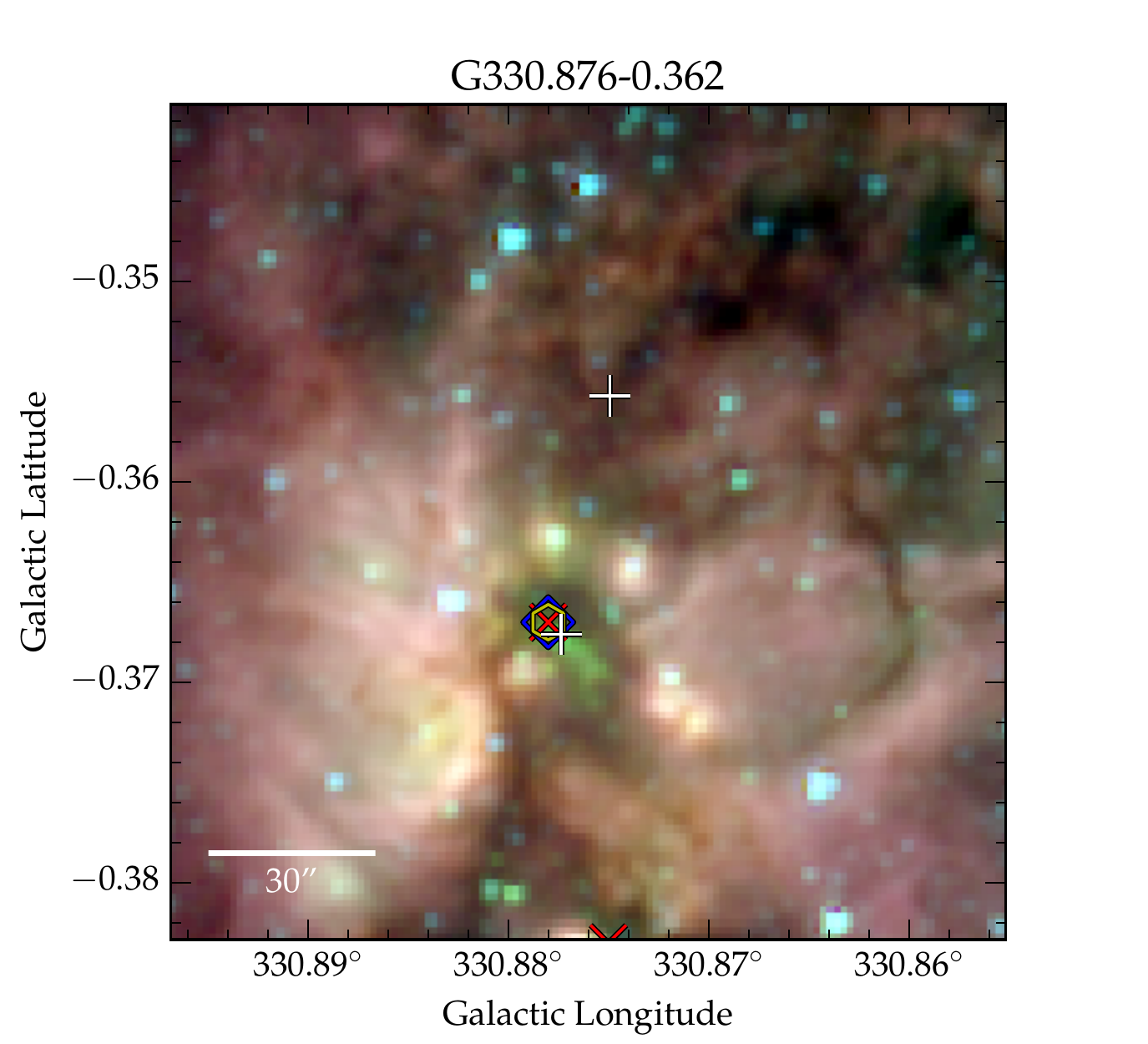}
      \includegraphics[height=0.30\textheight]{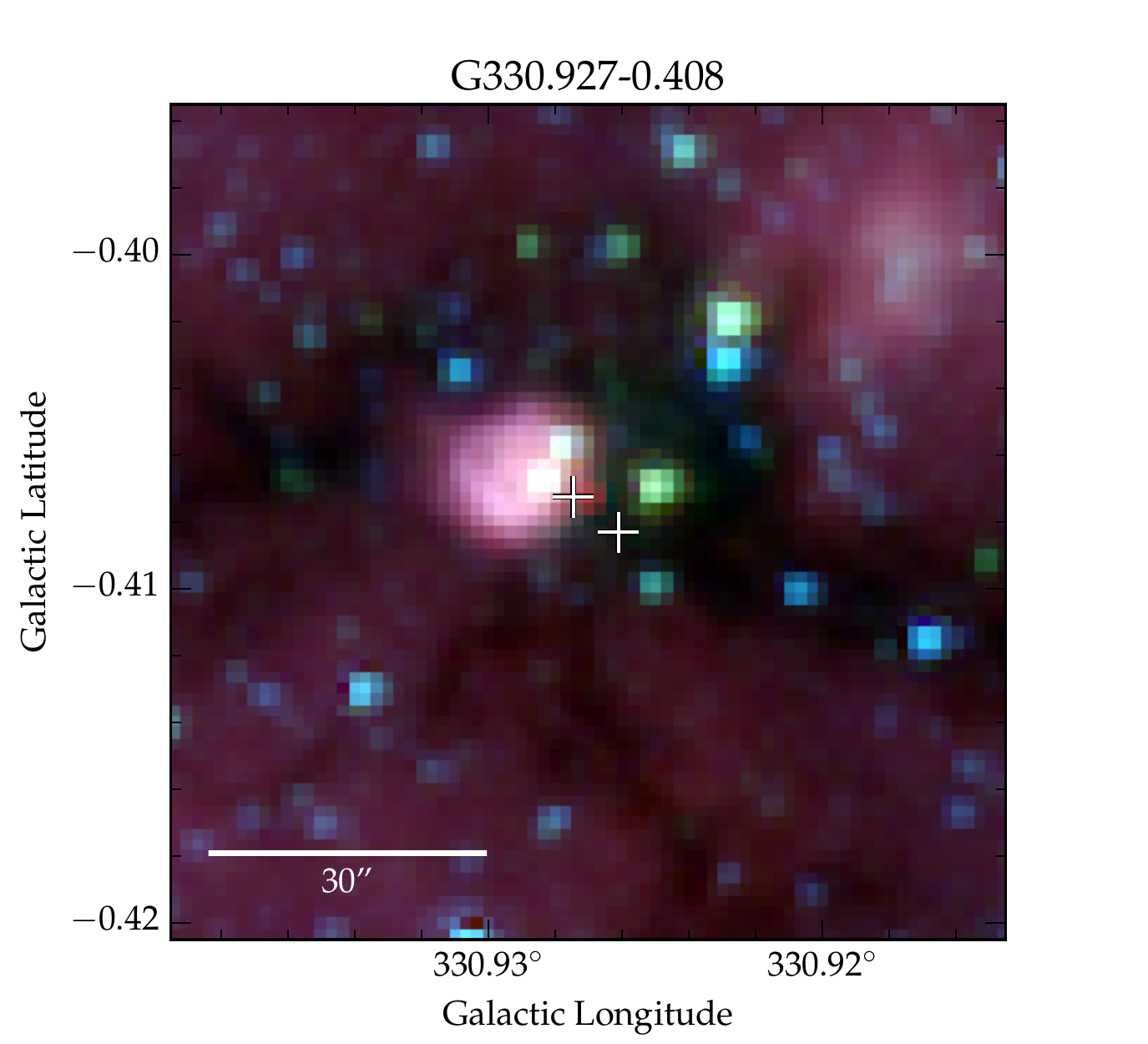}
      \captionof{figure}{\emph{Spitzer} GLIMPSE three-colour images (RGB = 3.6, 4.5 and 8.0\,$\mu$m) overlaid with the positions of class~I~\choh{} masers (white plus symbols). The size of the symbols reflects the maximum positional uncertainty for a maser spot. The published positions of class~II~\choh{}, \water{} and \oh{} masers are overlaid (red cross, blue diamond and yellow hexagon symbols, respectively; \citealt{caswell11,breen10,walsh11,walsh14,sevenster97,caswell98}).}
    \end{minipage}
}]
\setcounter{figure}{0}
\twocolumn[{
    \begin{minipage}{\textwidth}
      \centering
      \includegraphics[height=0.30\textheight]{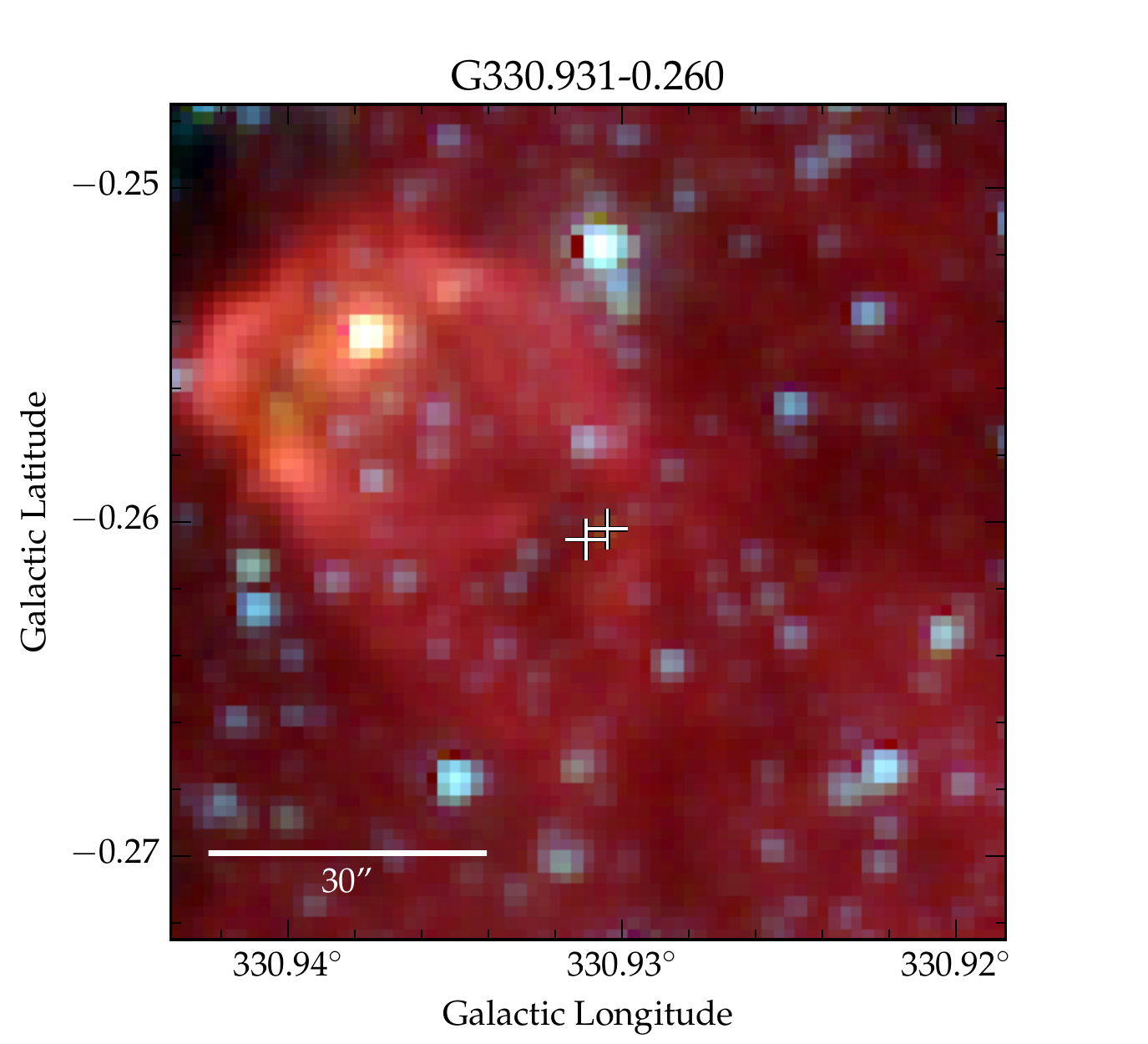}
      \includegraphics[height=0.30\textheight]{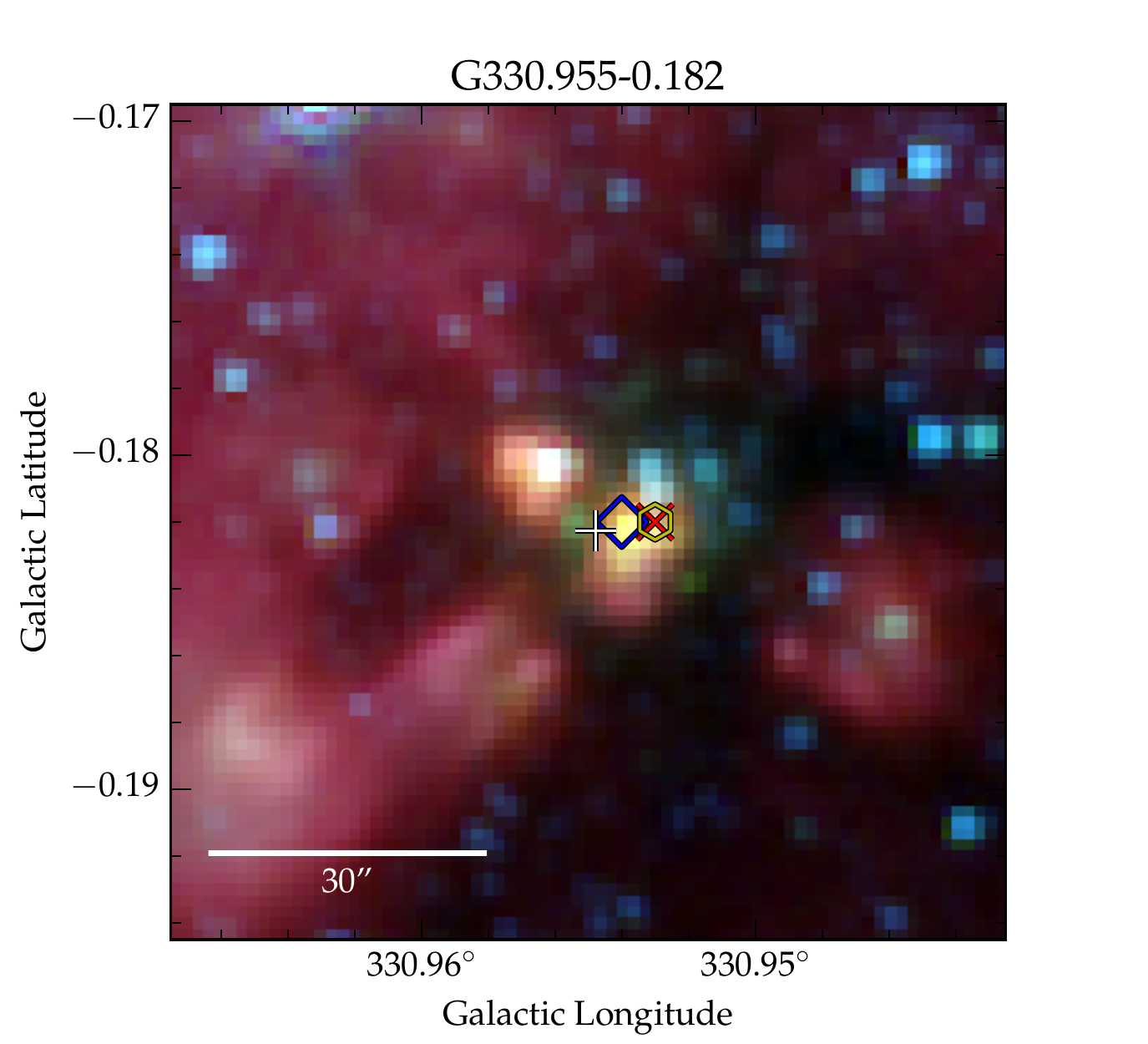}
      \includegraphics[height=0.30\textheight]{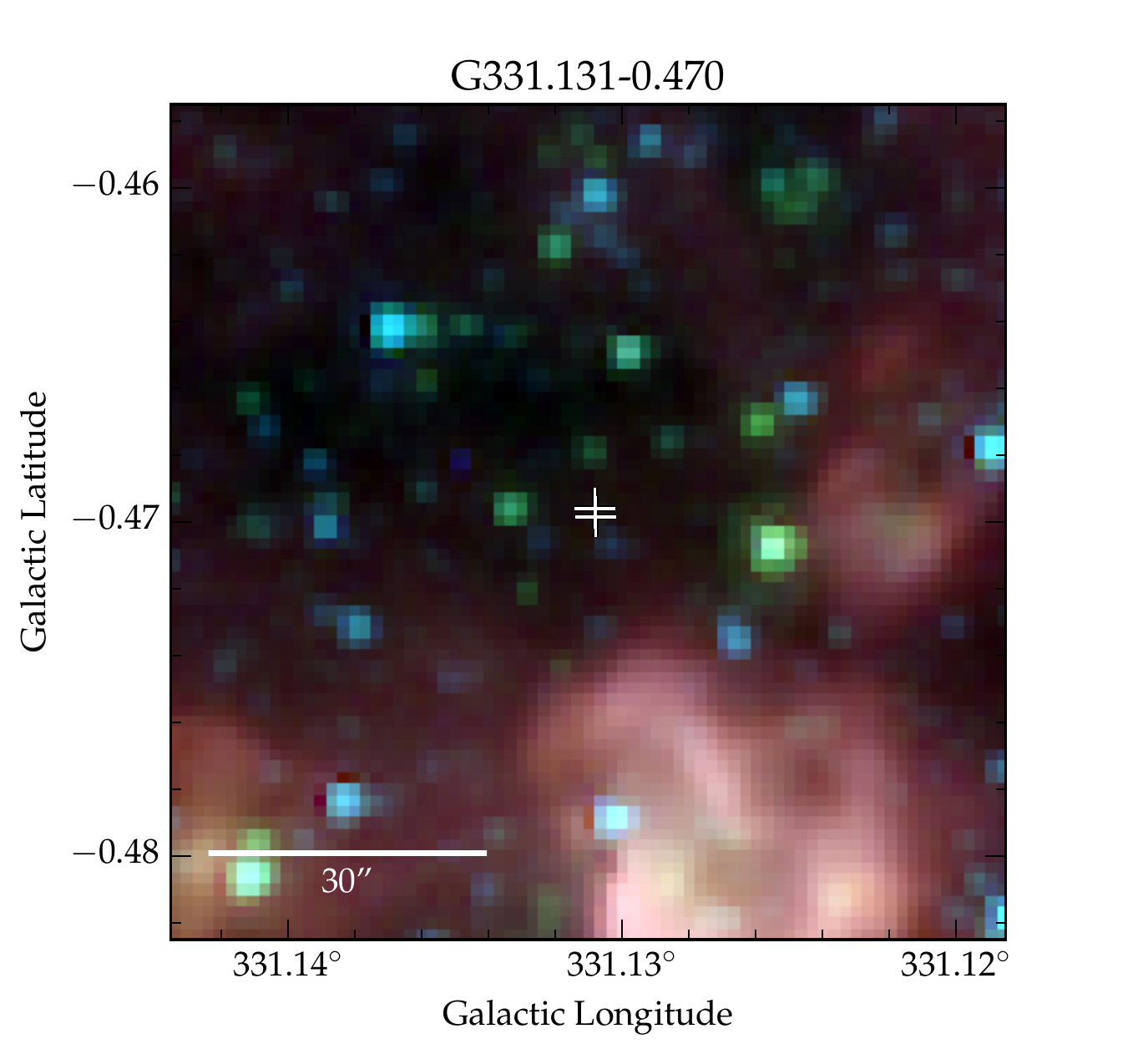}
      \includegraphics[height=0.30\textheight]{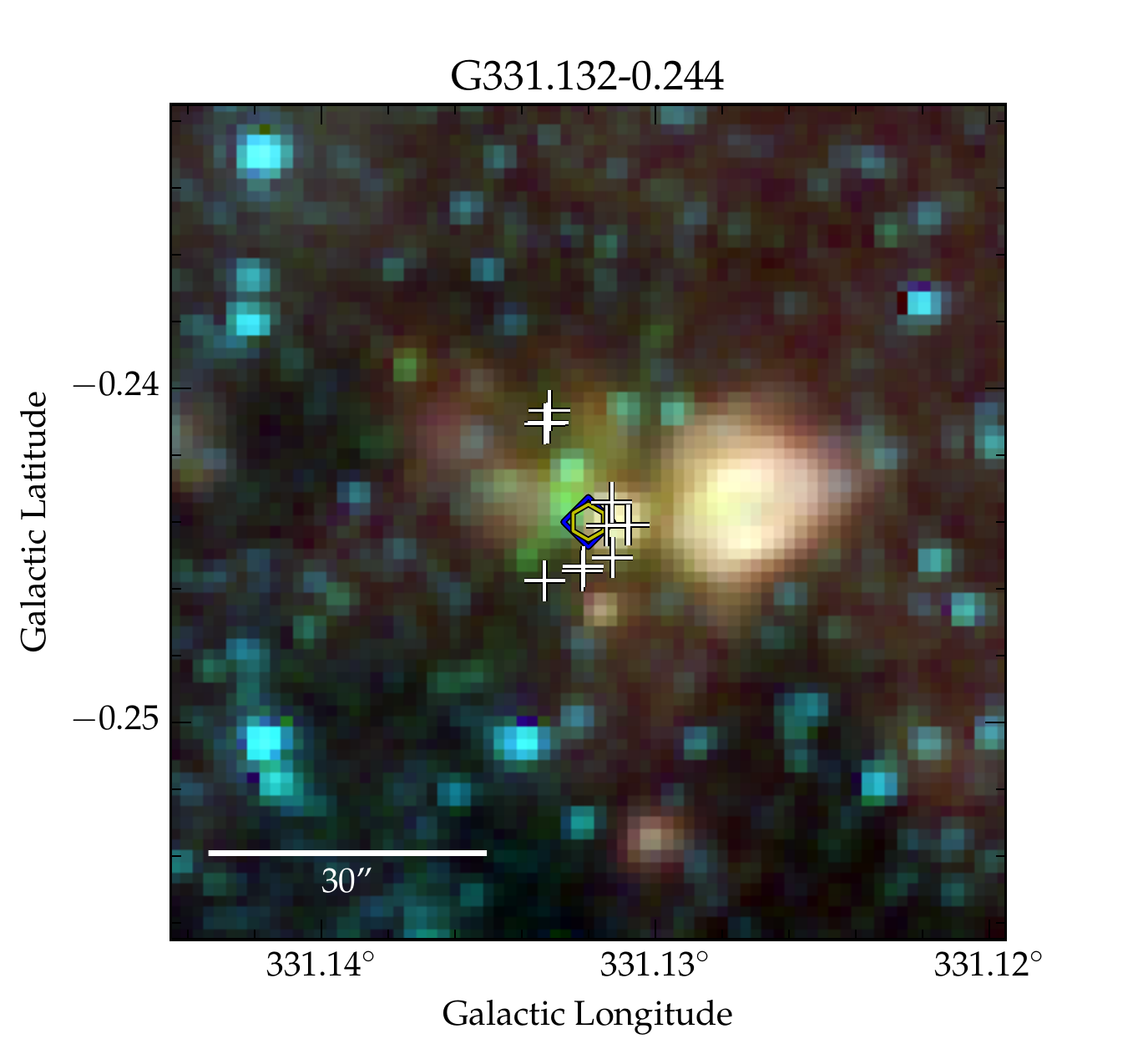}
      \includegraphics[height=0.30\textheight]{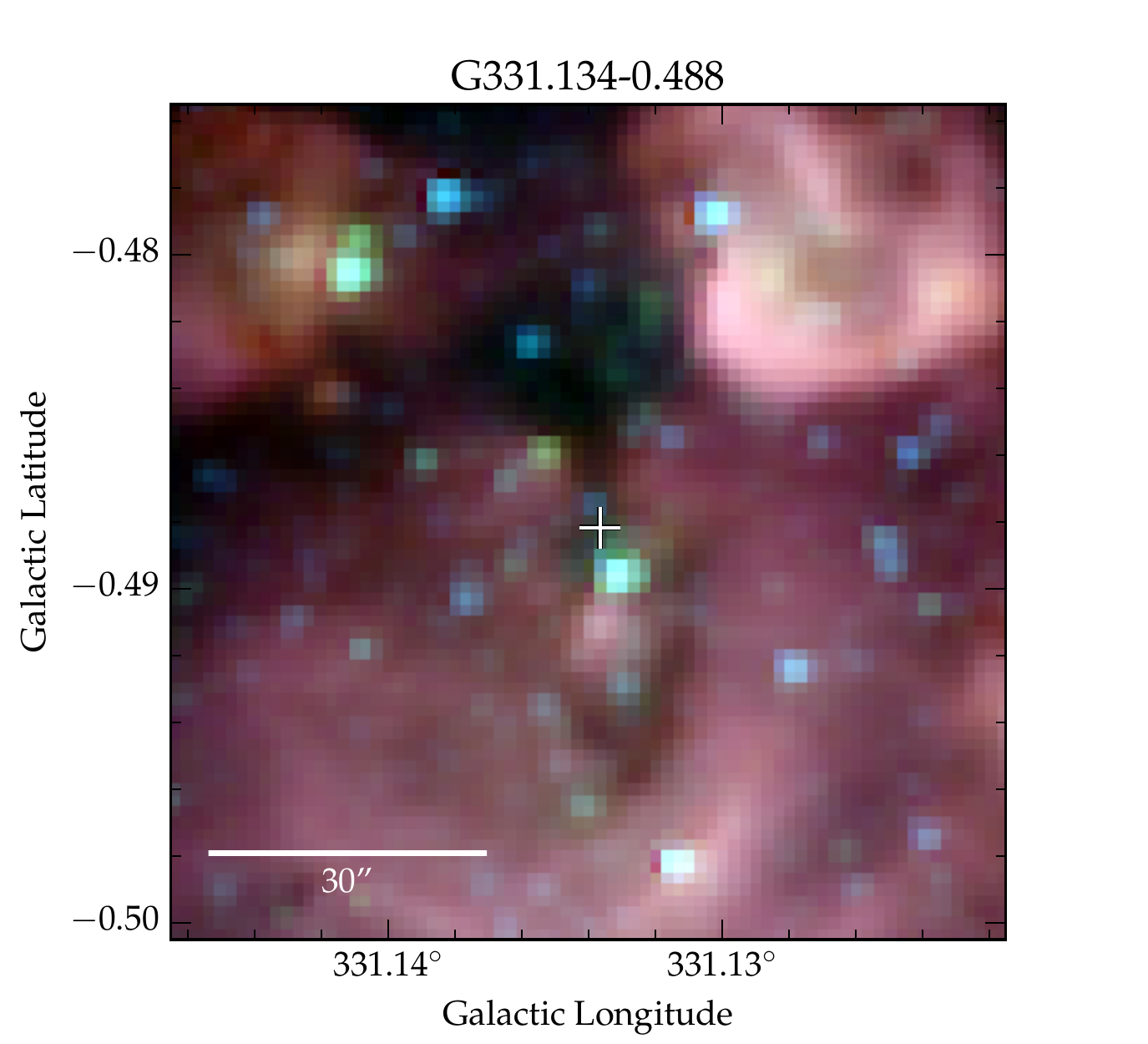}
      \includegraphics[height=0.30\textheight]{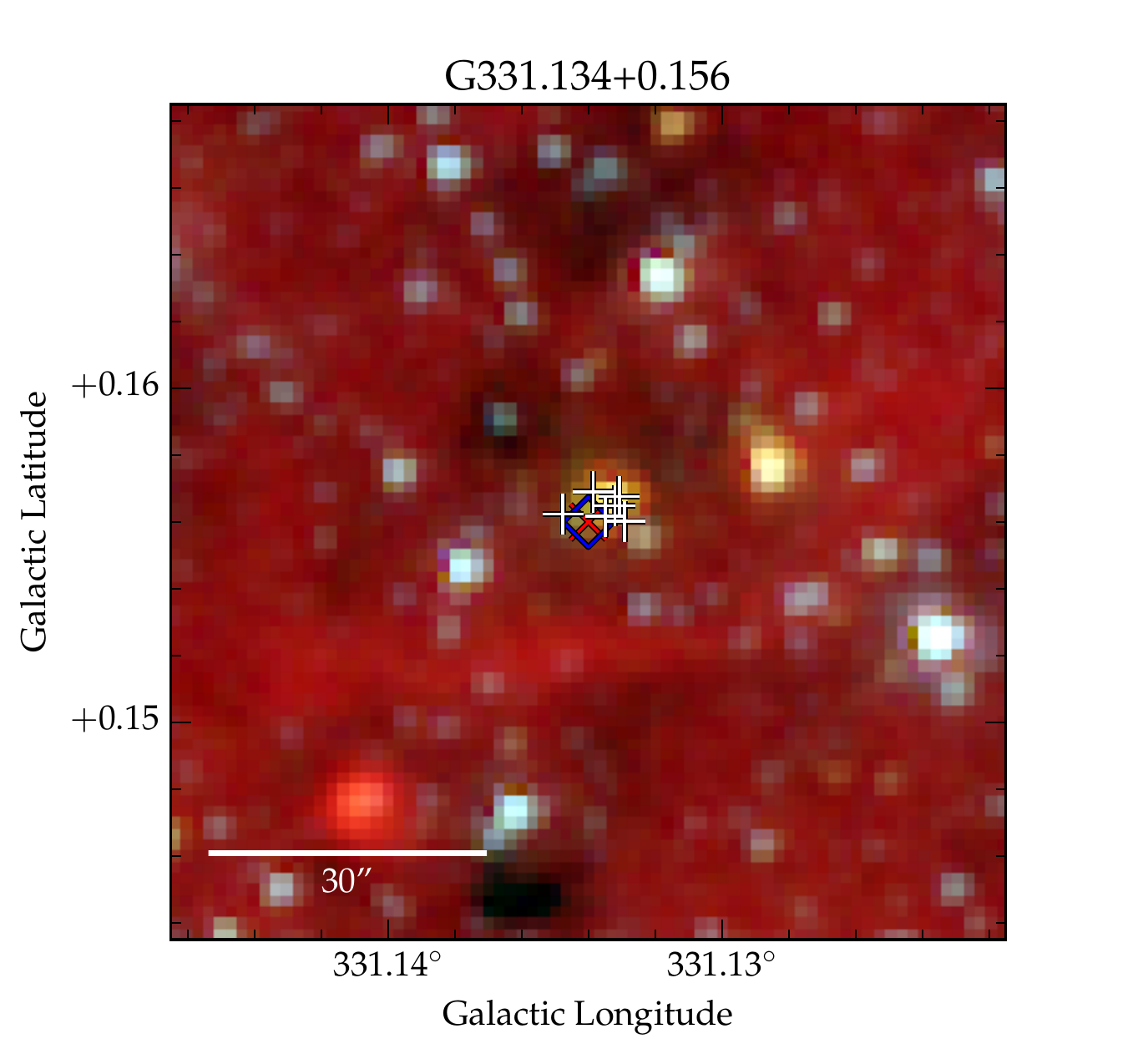}
      \captionof{figure}{\emph{continued}}
    \end{minipage}
}]
\setcounter{figure}{0}
\twocolumn[{
    \begin{minipage}{\textwidth}
      \centering
      \includegraphics[height=0.30\textheight]{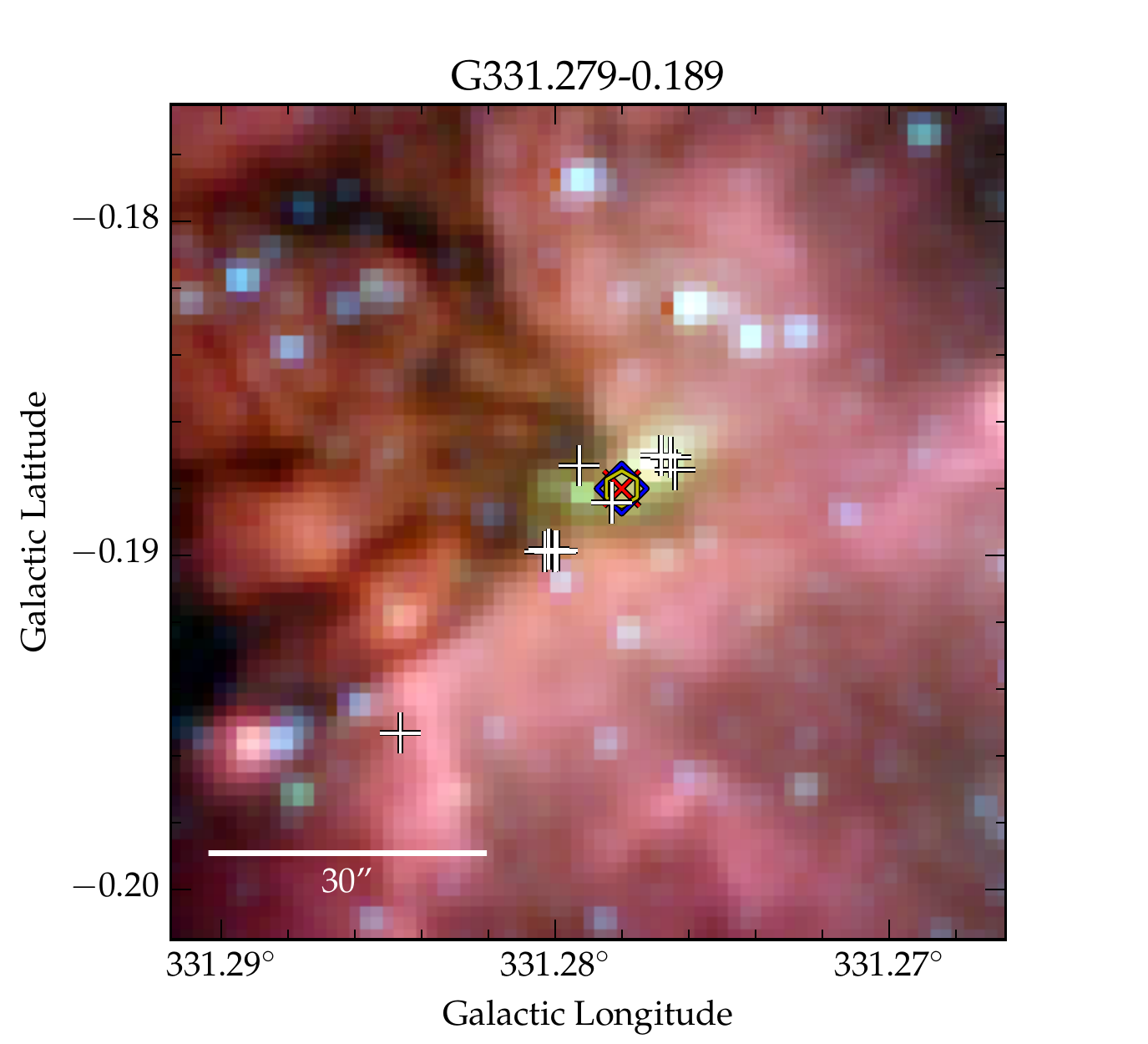}
      \includegraphics[height=0.30\textheight]{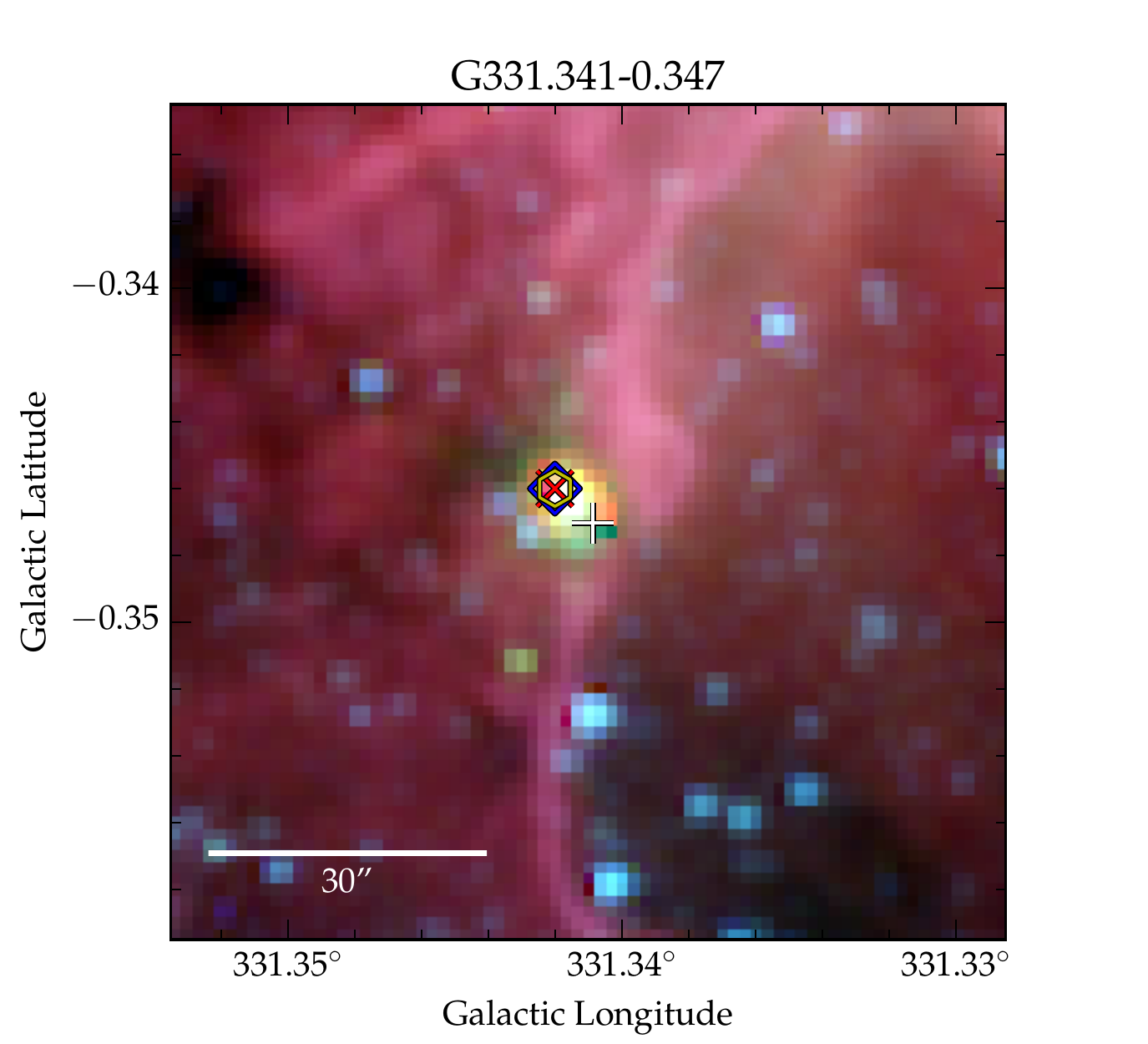}
      \includegraphics[height=0.30\textheight]{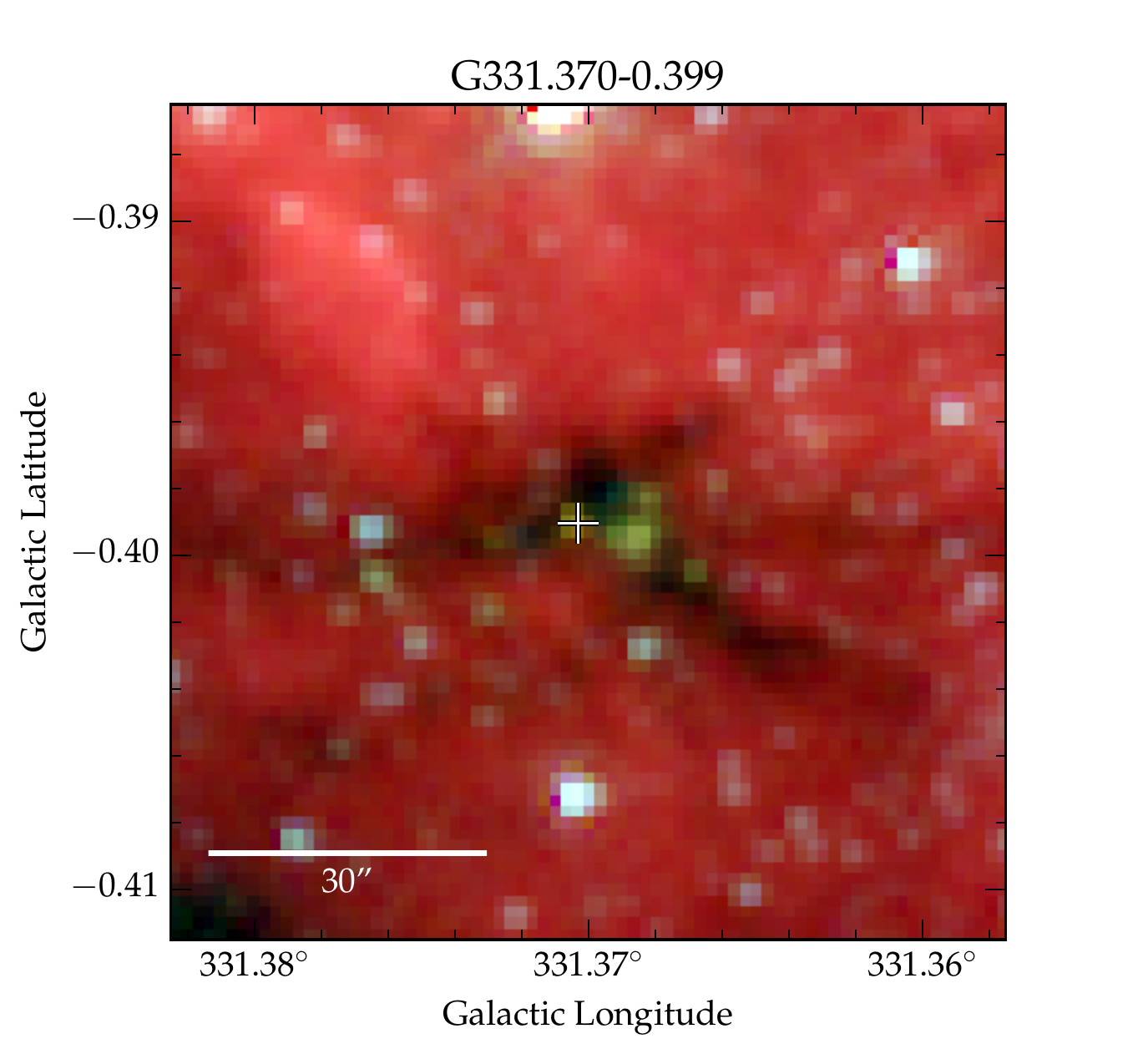}
      \includegraphics[height=0.30\textheight]{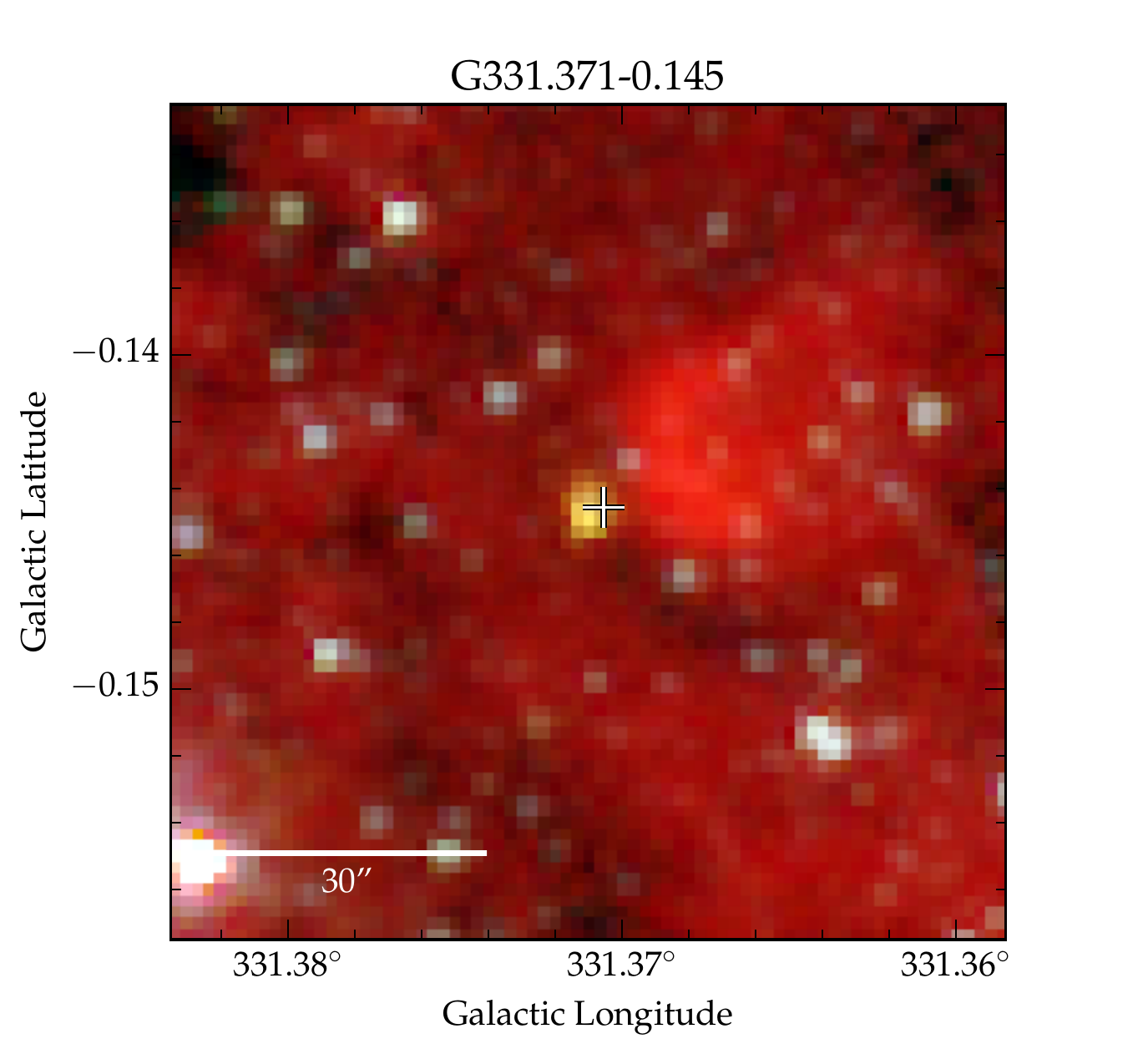}
      \includegraphics[height=0.30\textheight]{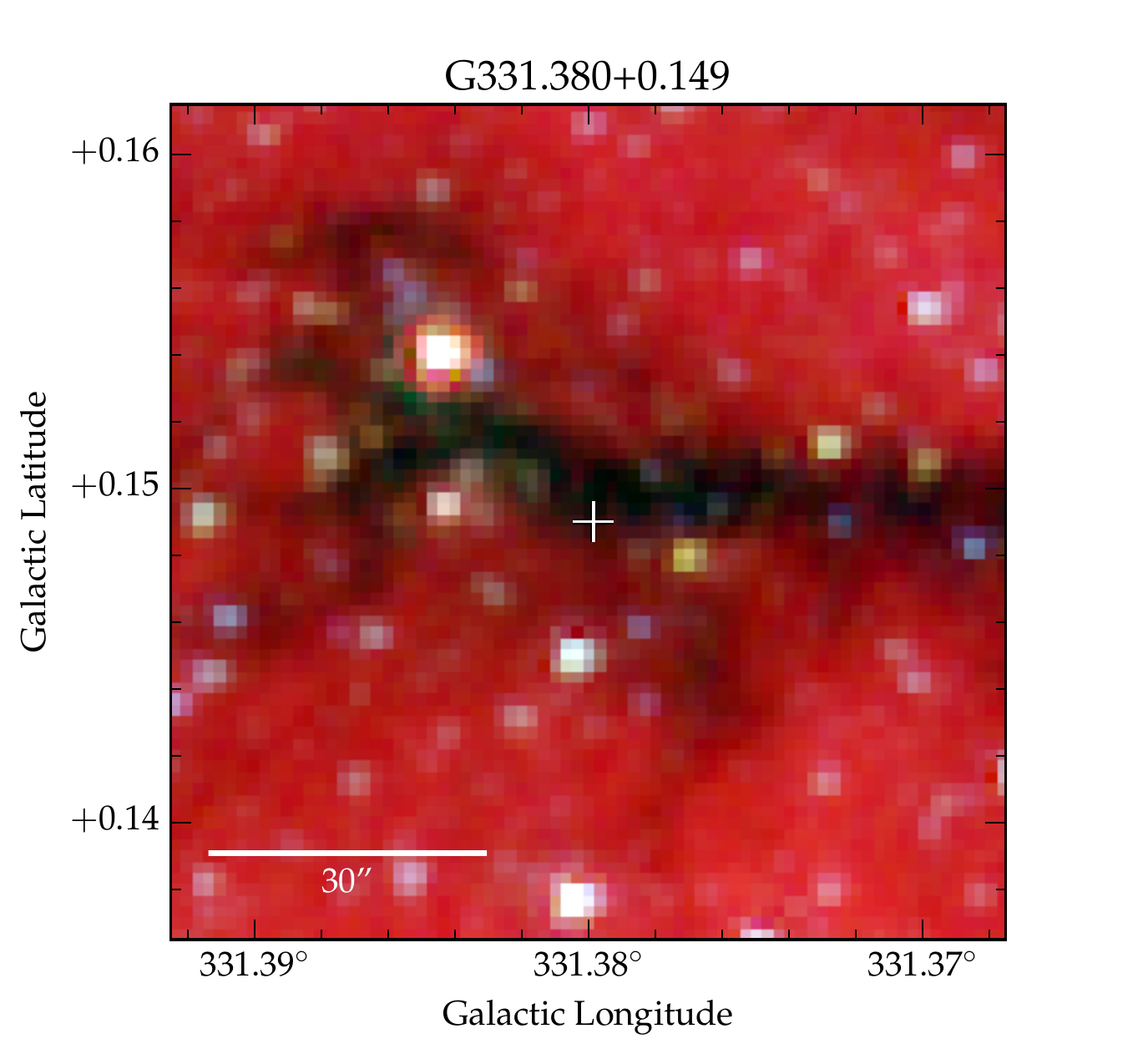}
      \includegraphics[height=0.30\textheight]{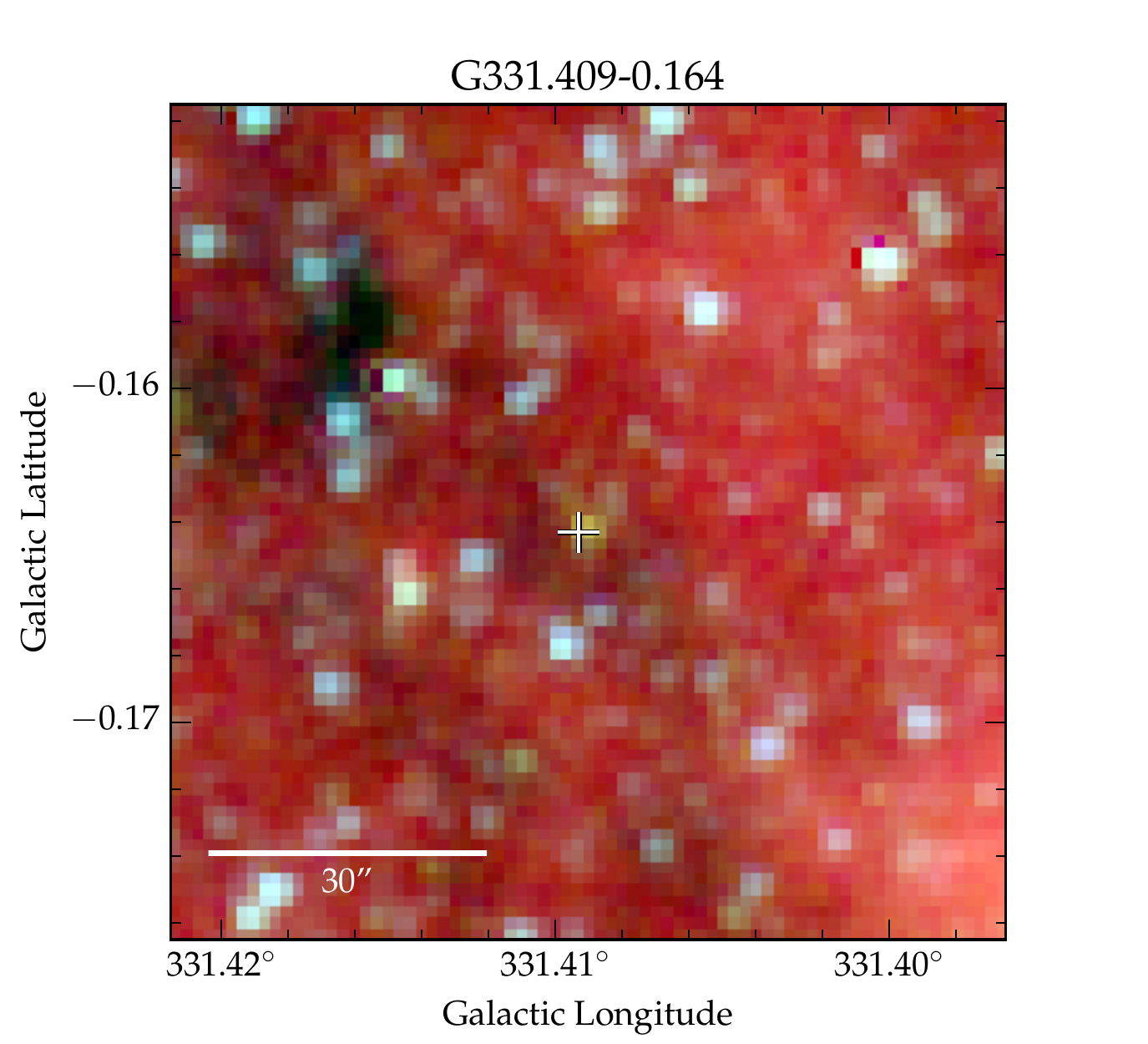}
      \captionof{figure}{\emph{continued}}
    \end{minipage}
}]
\setcounter{figure}{0}
\twocolumn[{
    \begin{minipage}{\textwidth}
      \centering
      \includegraphics[height=0.30\textheight]{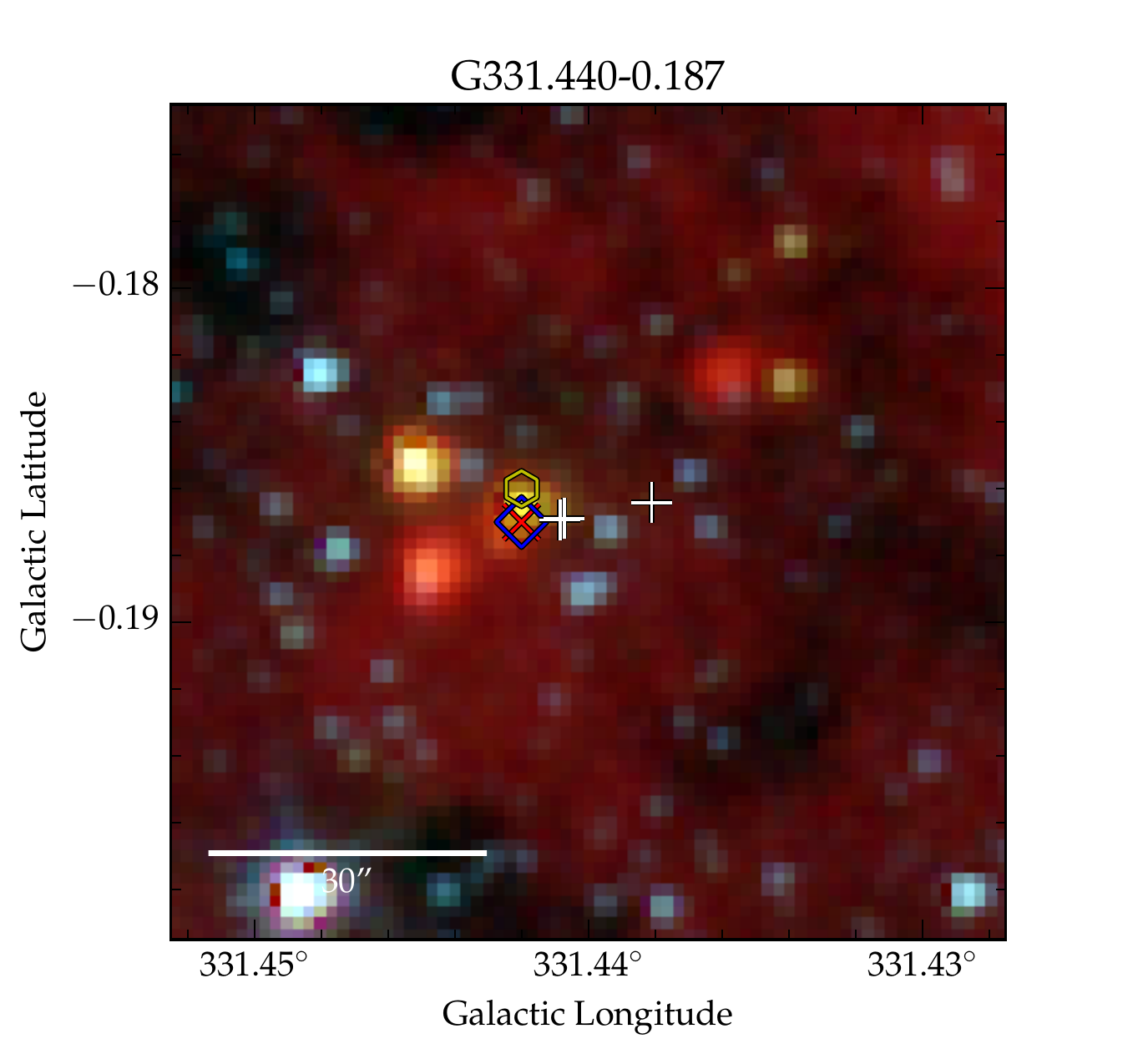}
      \includegraphics[height=0.30\textheight]{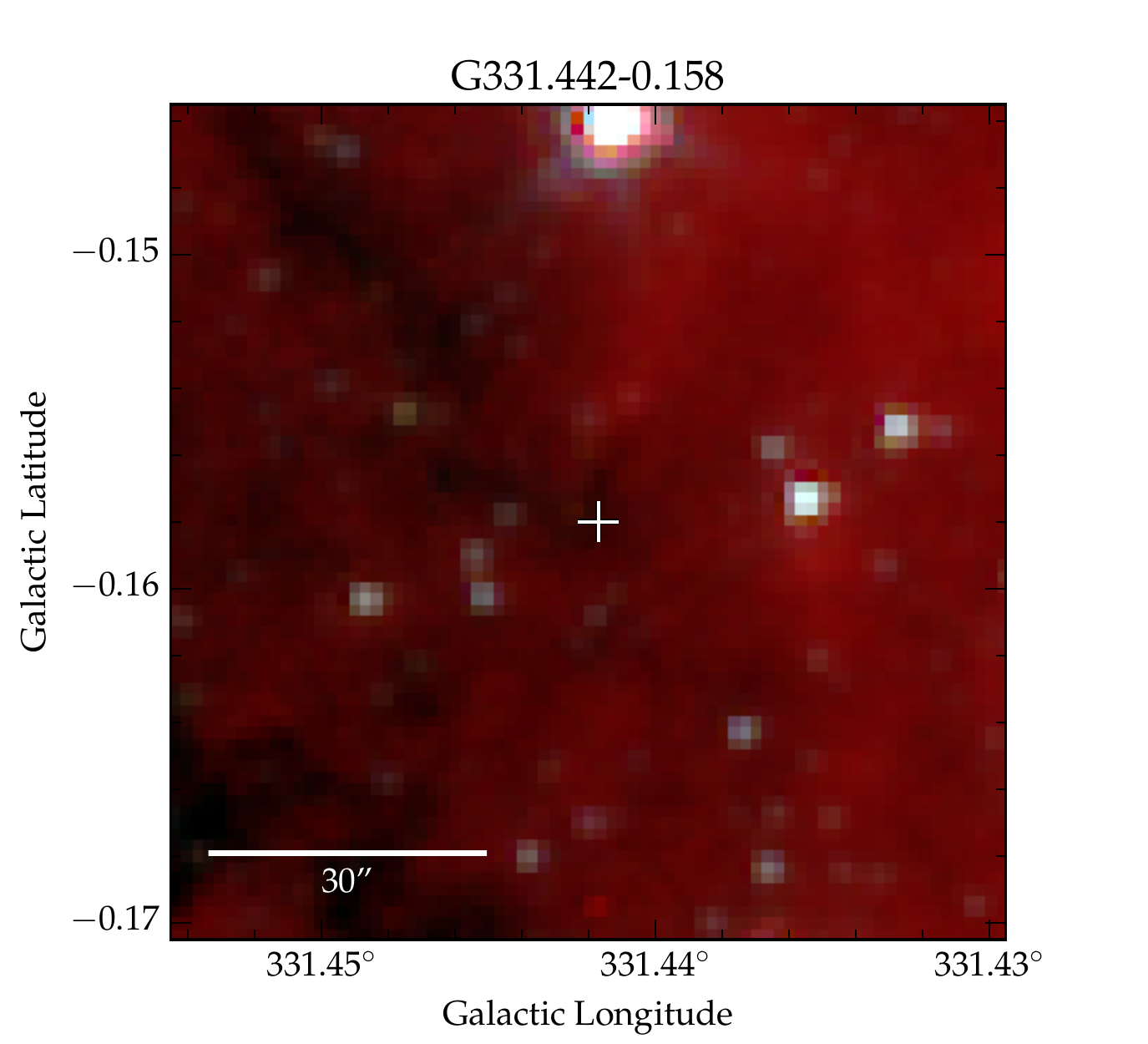}
      \includegraphics[height=0.30\textheight]{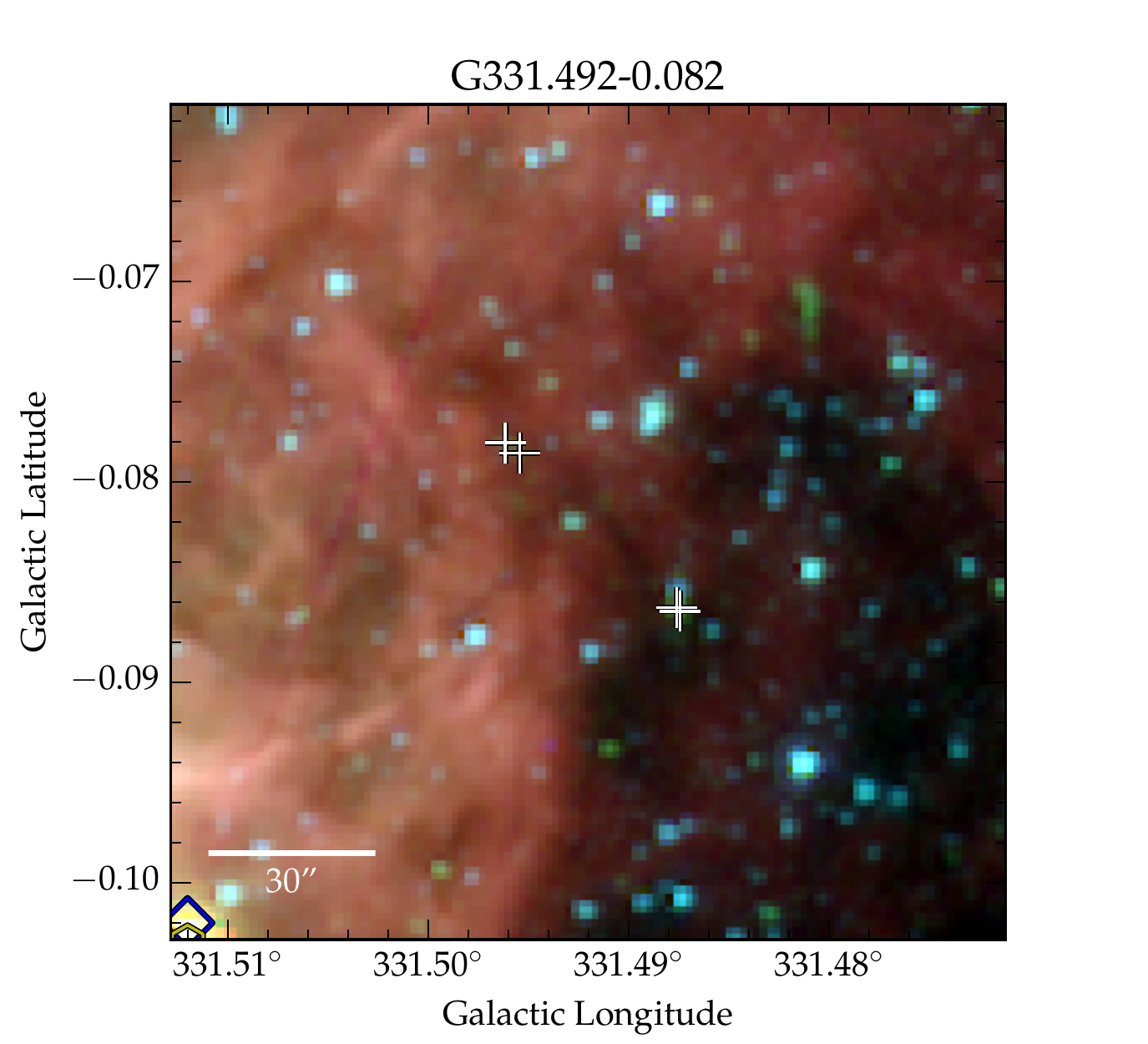}
      \includegraphics[height=0.30\textheight]{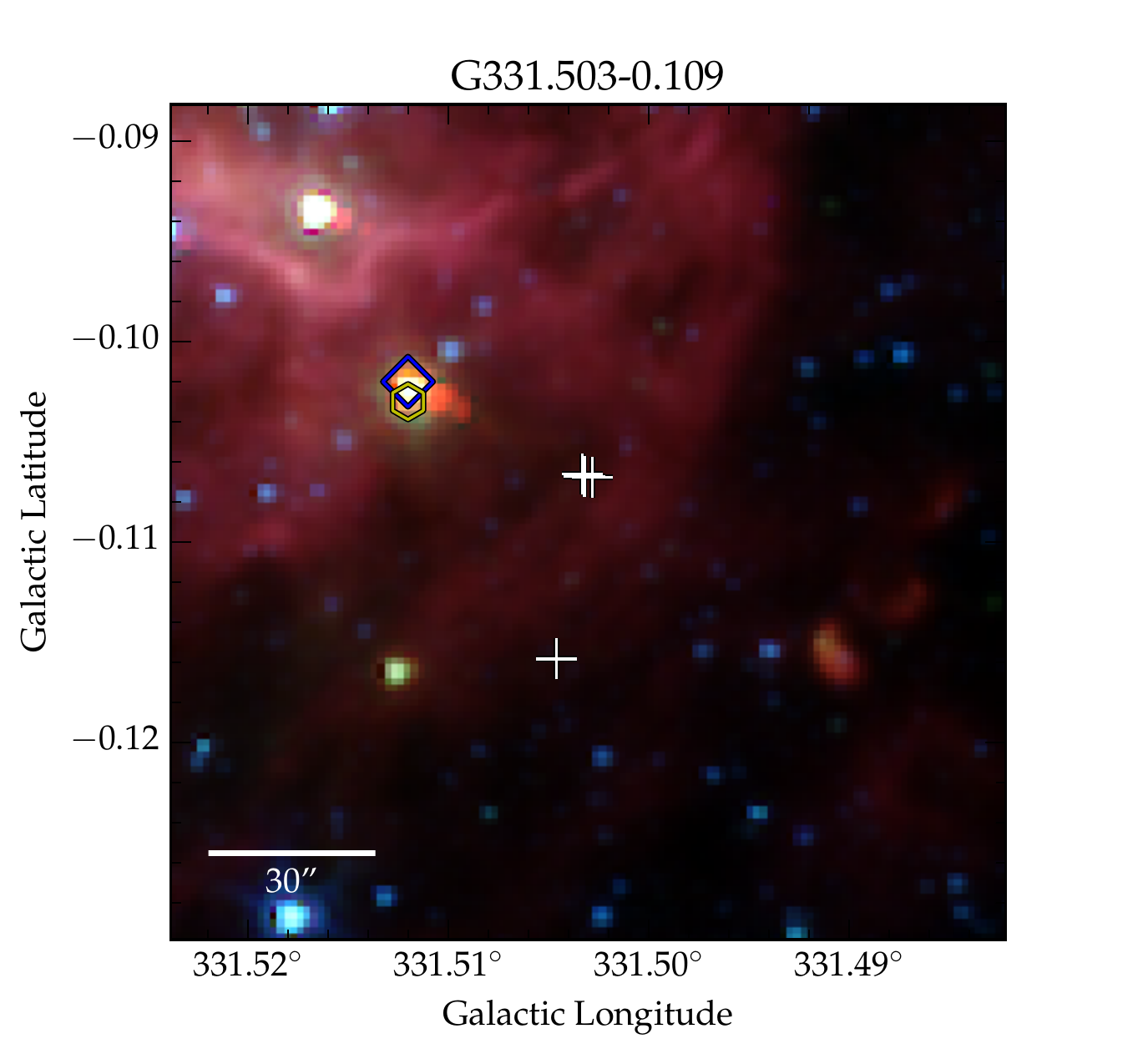}
      \includegraphics[height=0.30\textheight]{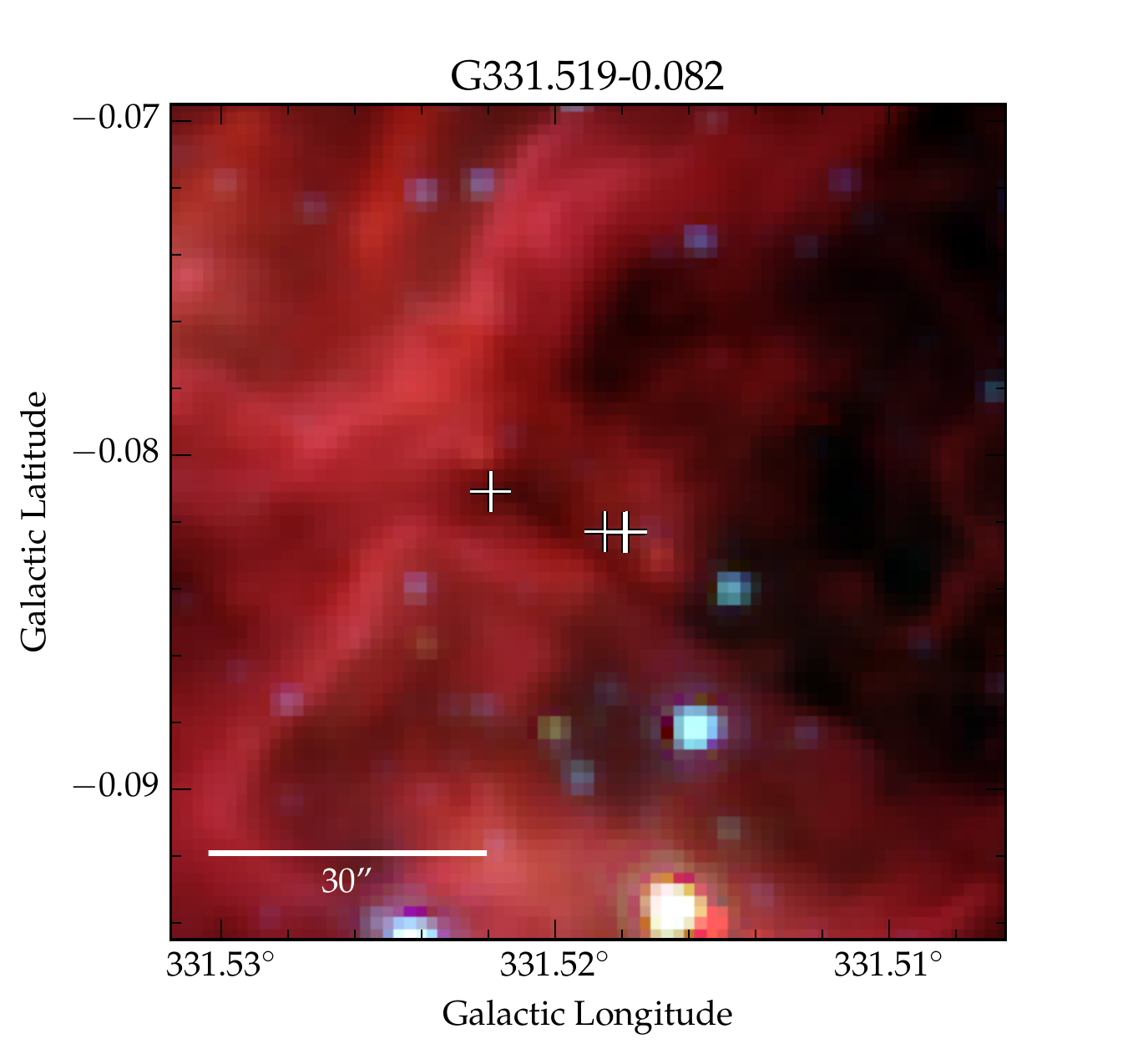}
      \includegraphics[height=0.30\textheight]{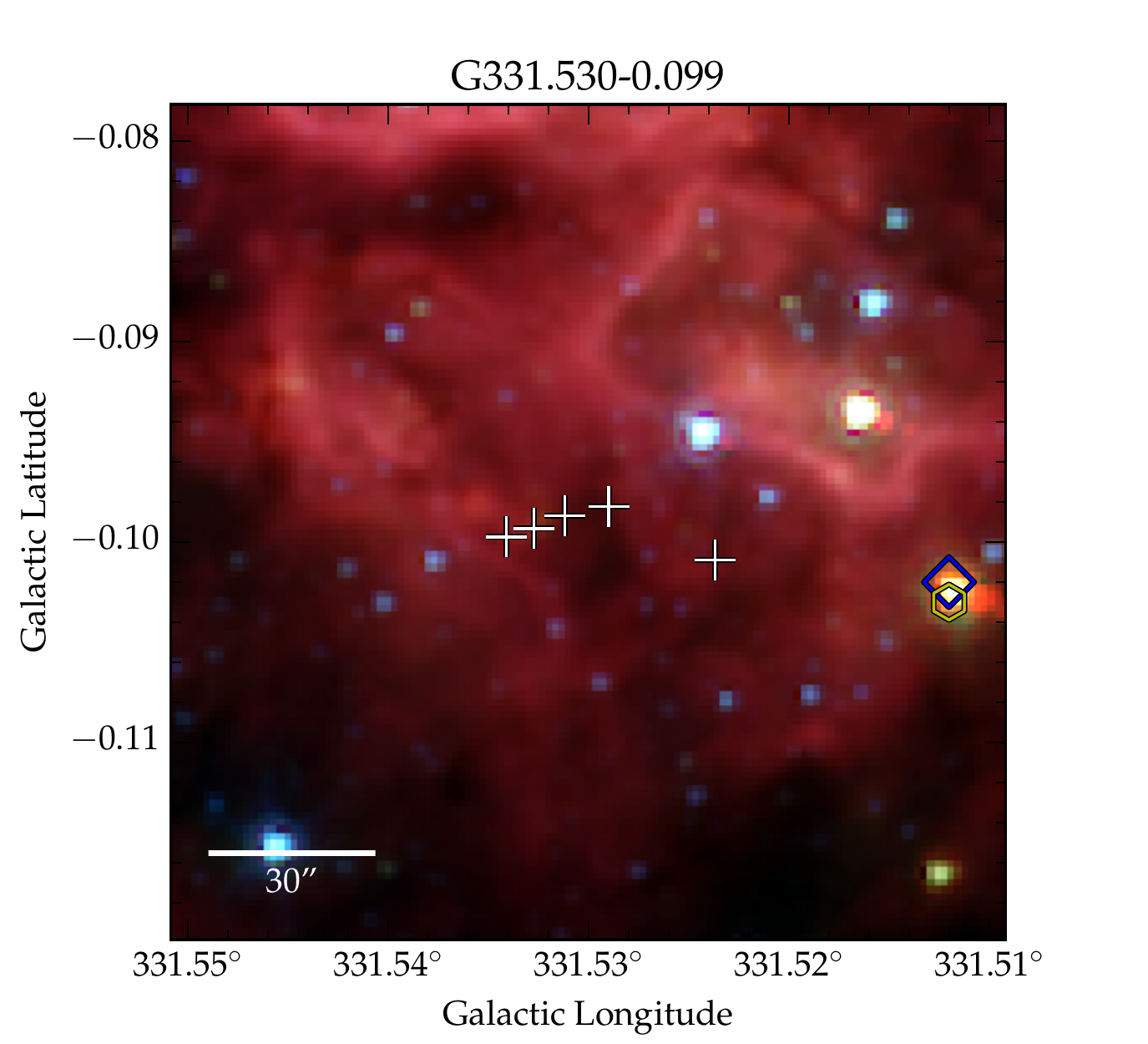}
      \captionof{figure}{\emph{continued}}
    \end{minipage}
}]
\setcounter{figure}{0}
\twocolumn[{
    \begin{minipage}{\textwidth}
      \centering
      \includegraphics[height=0.30\textheight]{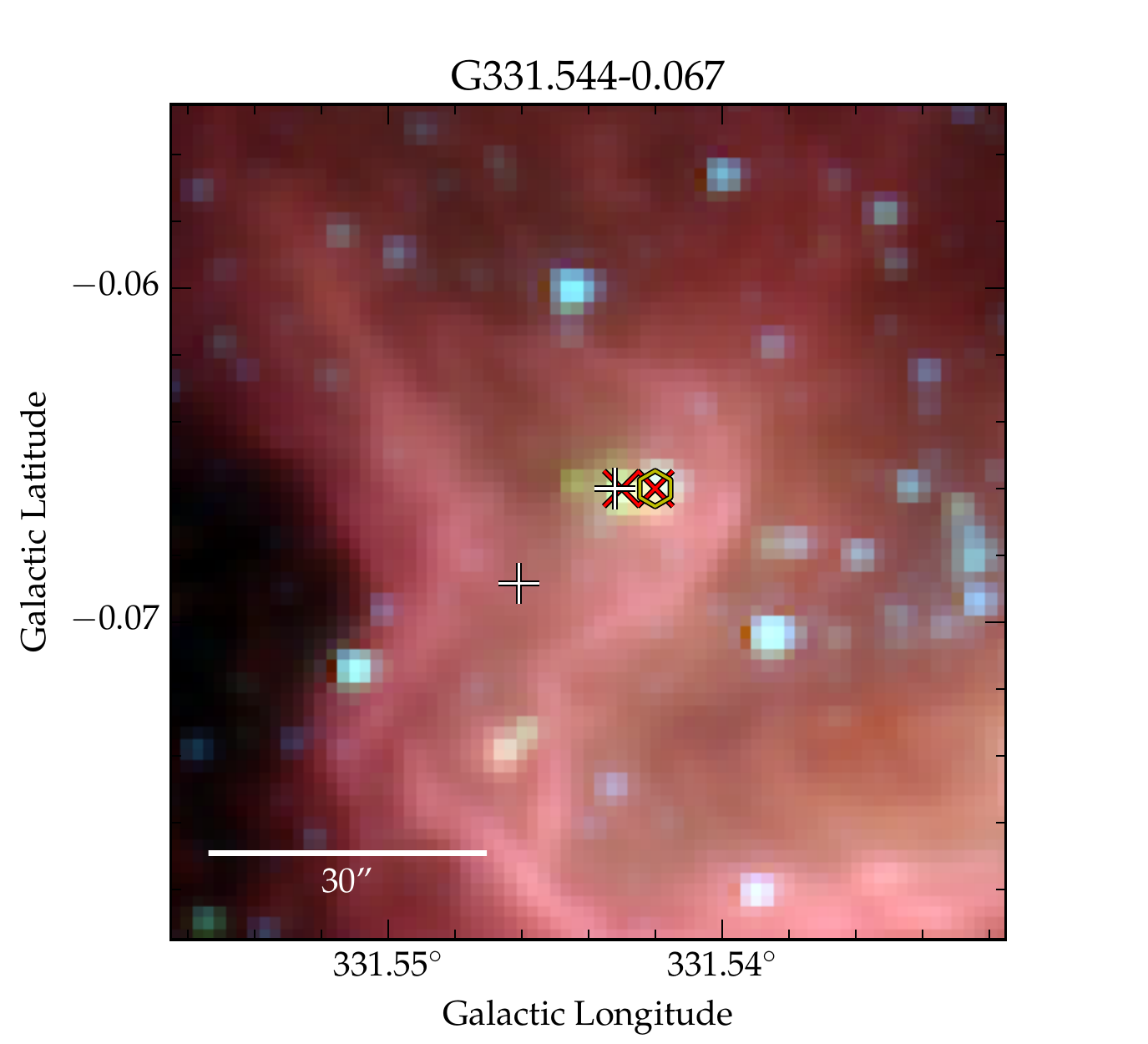}
      \includegraphics[height=0.30\textheight]{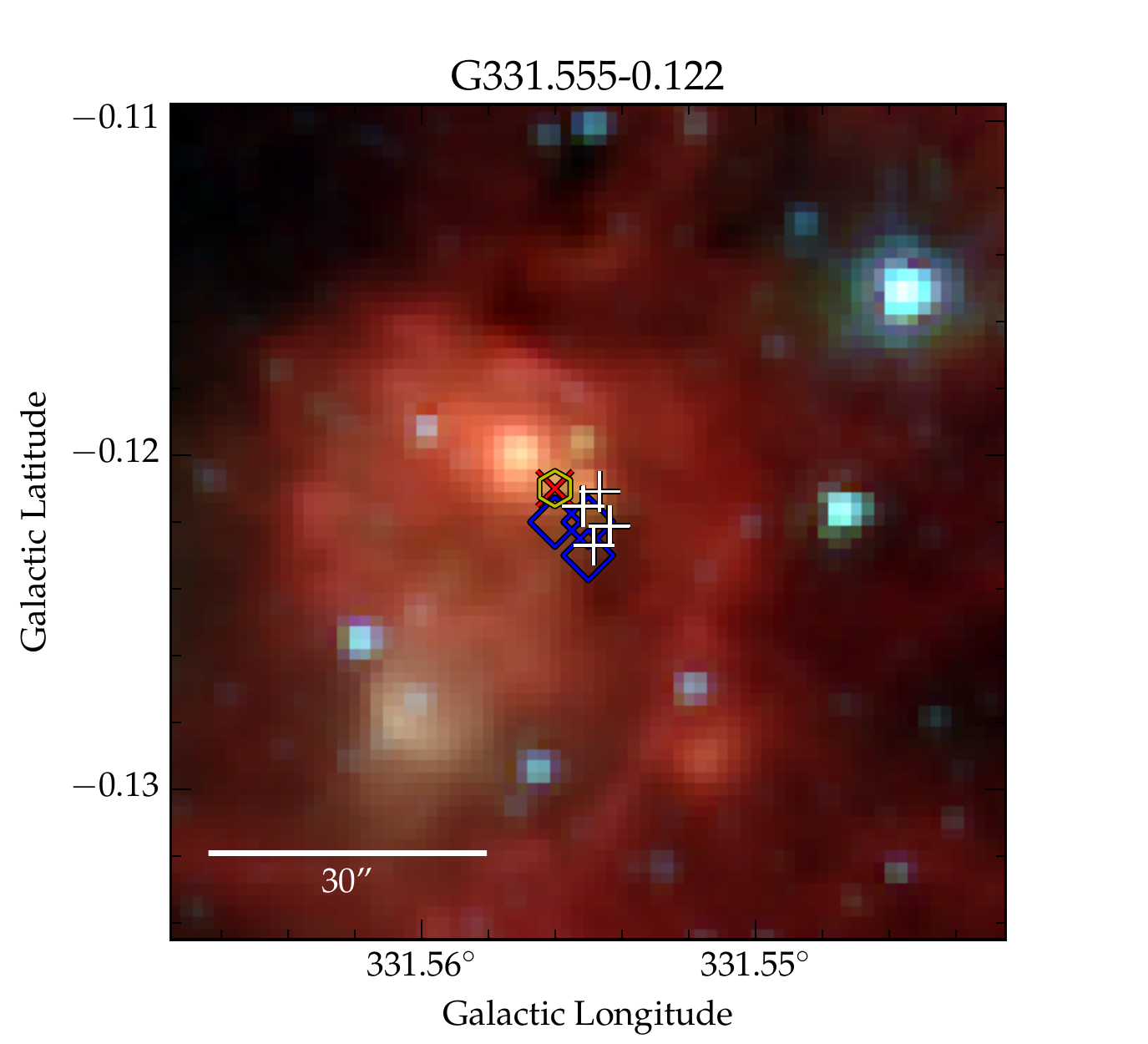}
      \includegraphics[height=0.30\textheight]{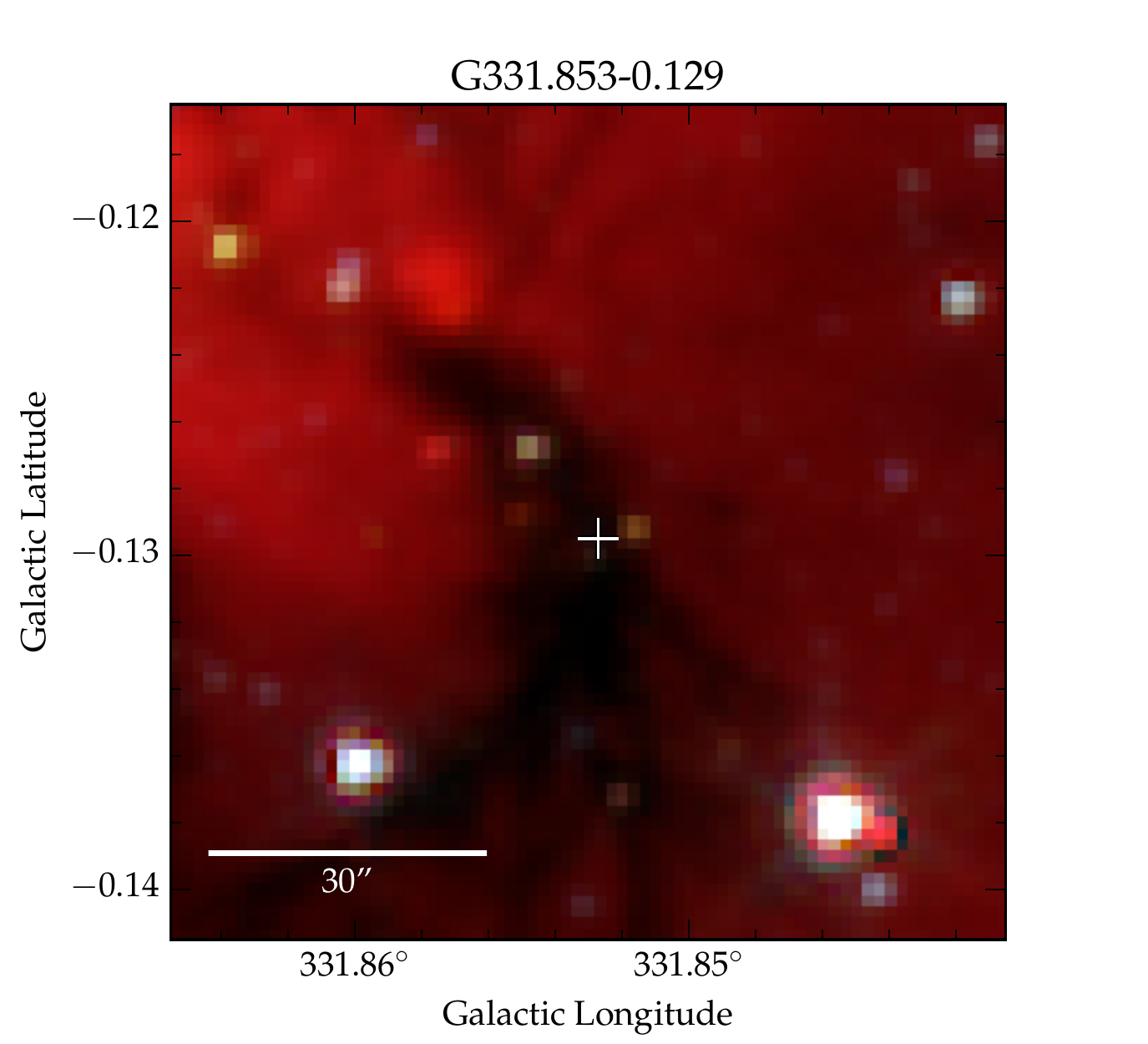}
      \includegraphics[height=0.30\textheight]{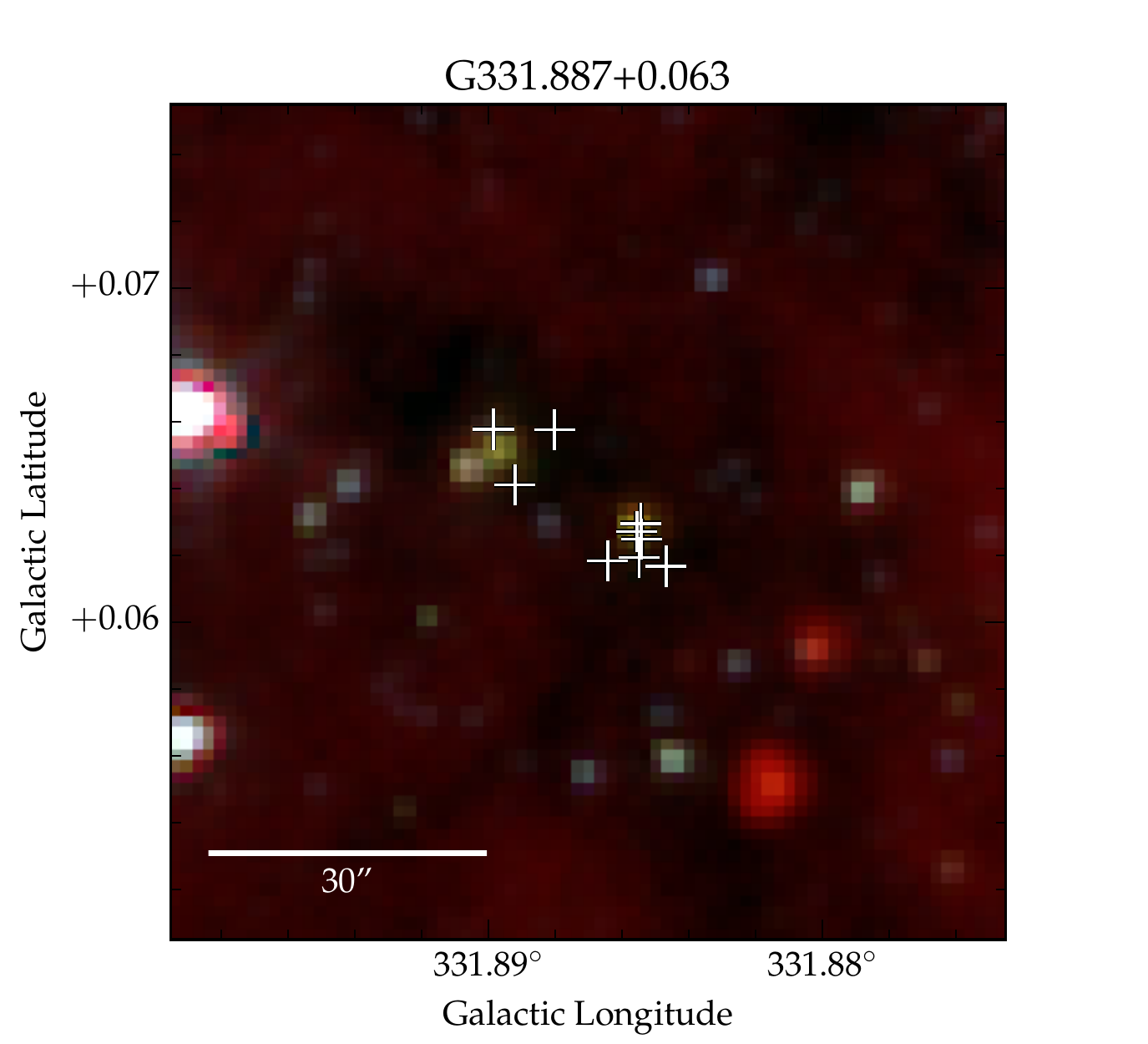}
      \includegraphics[height=0.30\textheight]{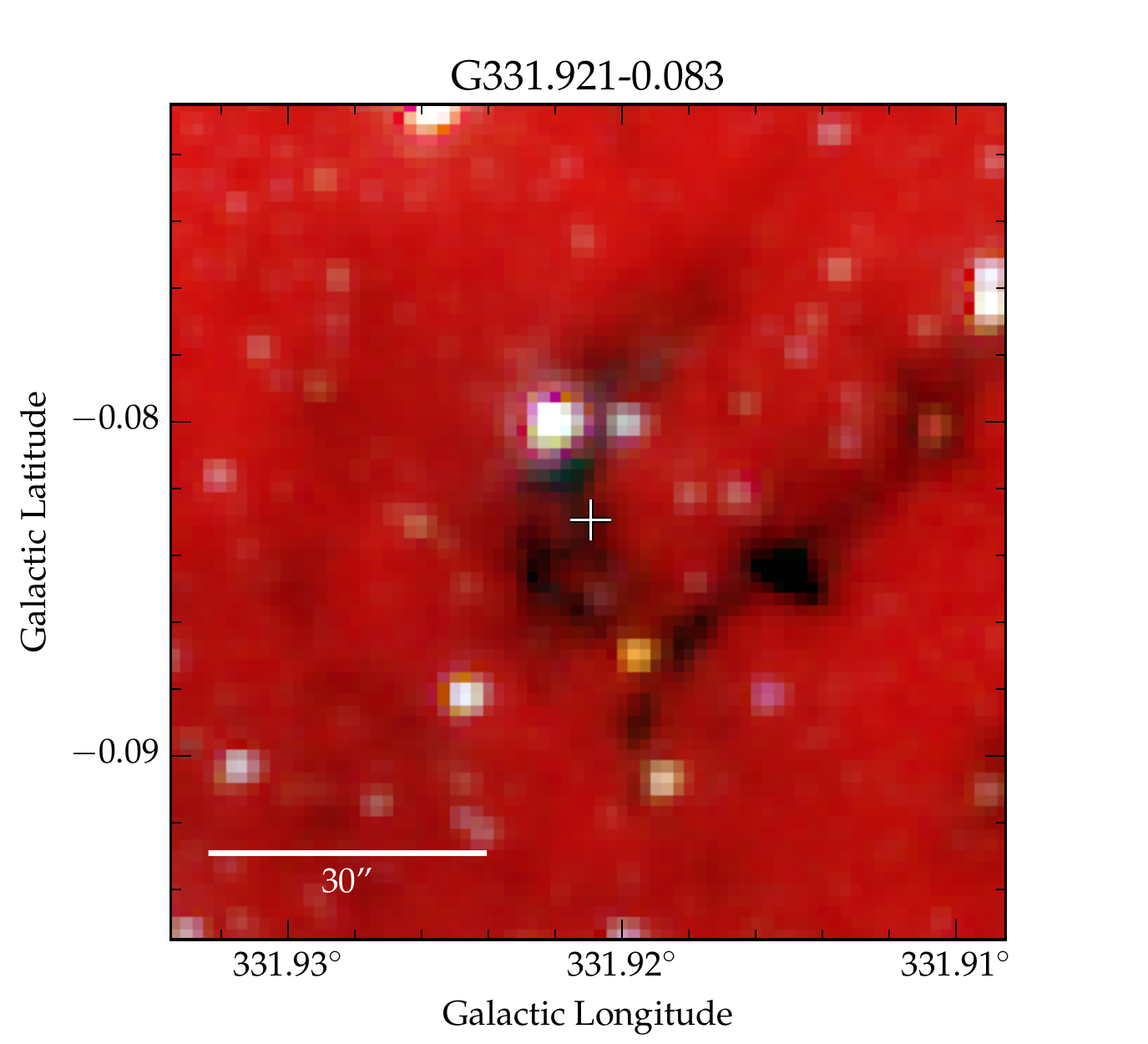}
      \includegraphics[height=0.30\textheight]{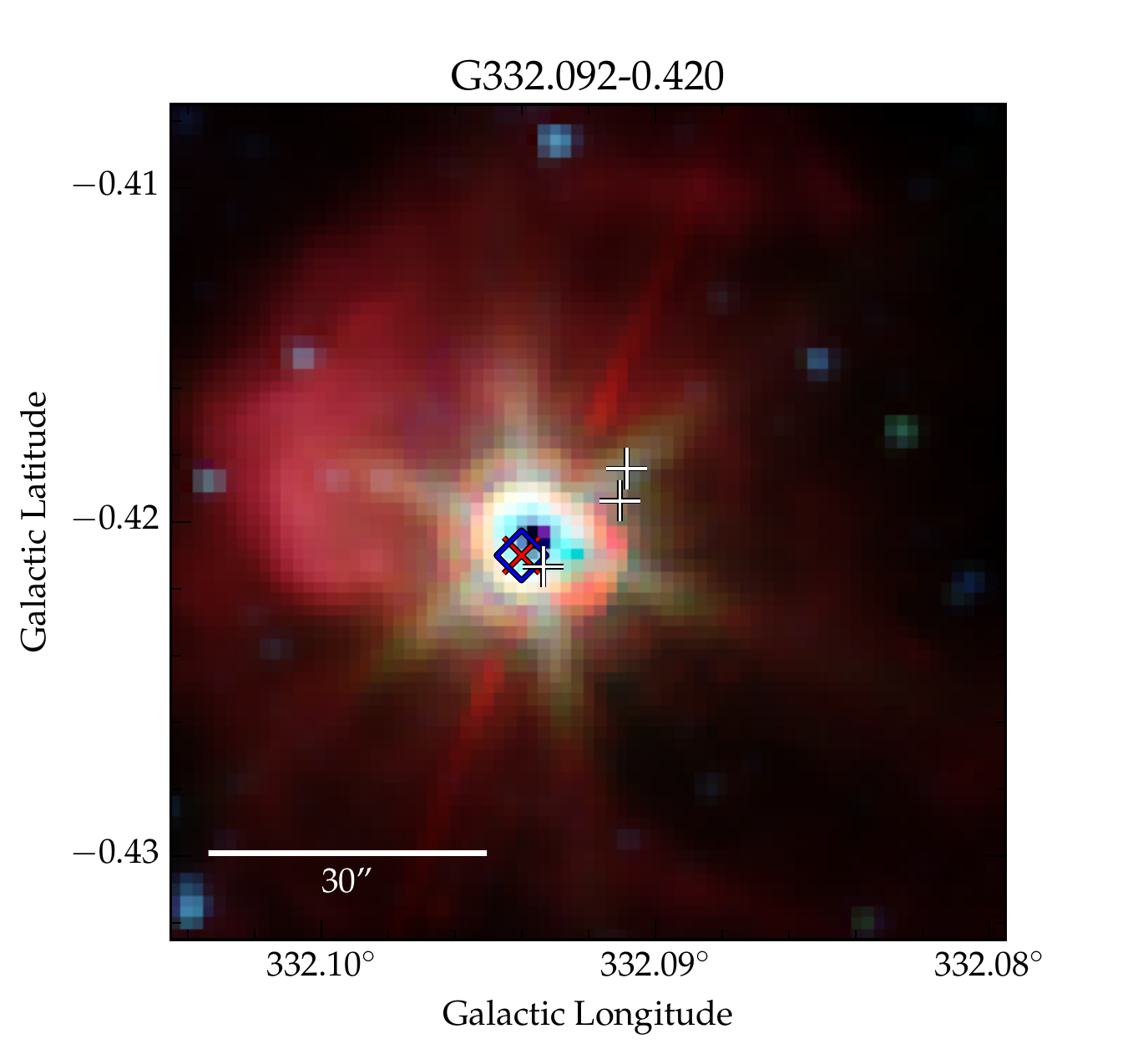}
      \captionof{figure}{\emph{continued}}
    \end{minipage}
}]
\setcounter{figure}{0}
\twocolumn[{
    \begin{minipage}{\textwidth}
      \centering
      \includegraphics[height=0.30\textheight]{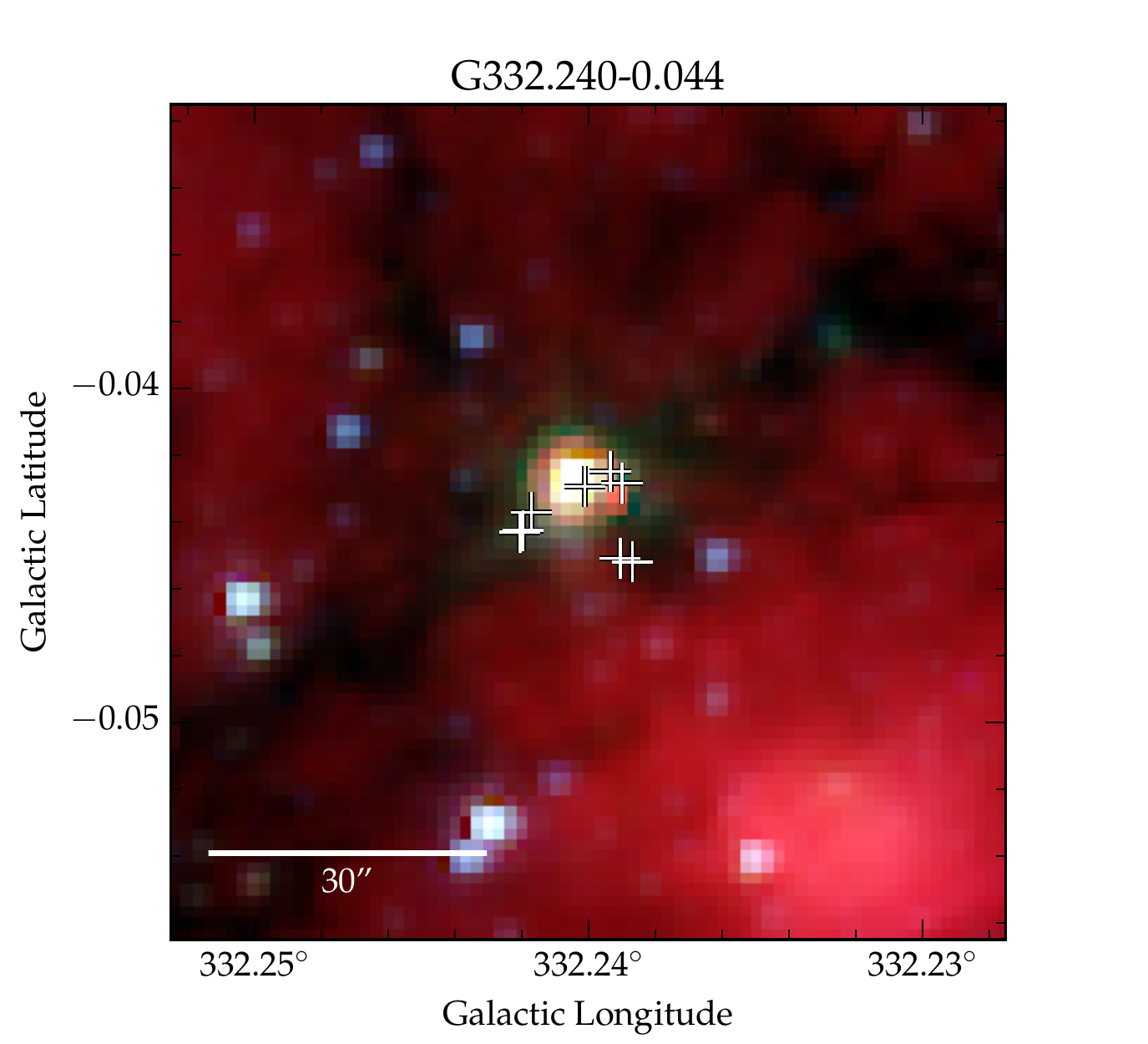}
      \includegraphics[height=0.30\textheight]{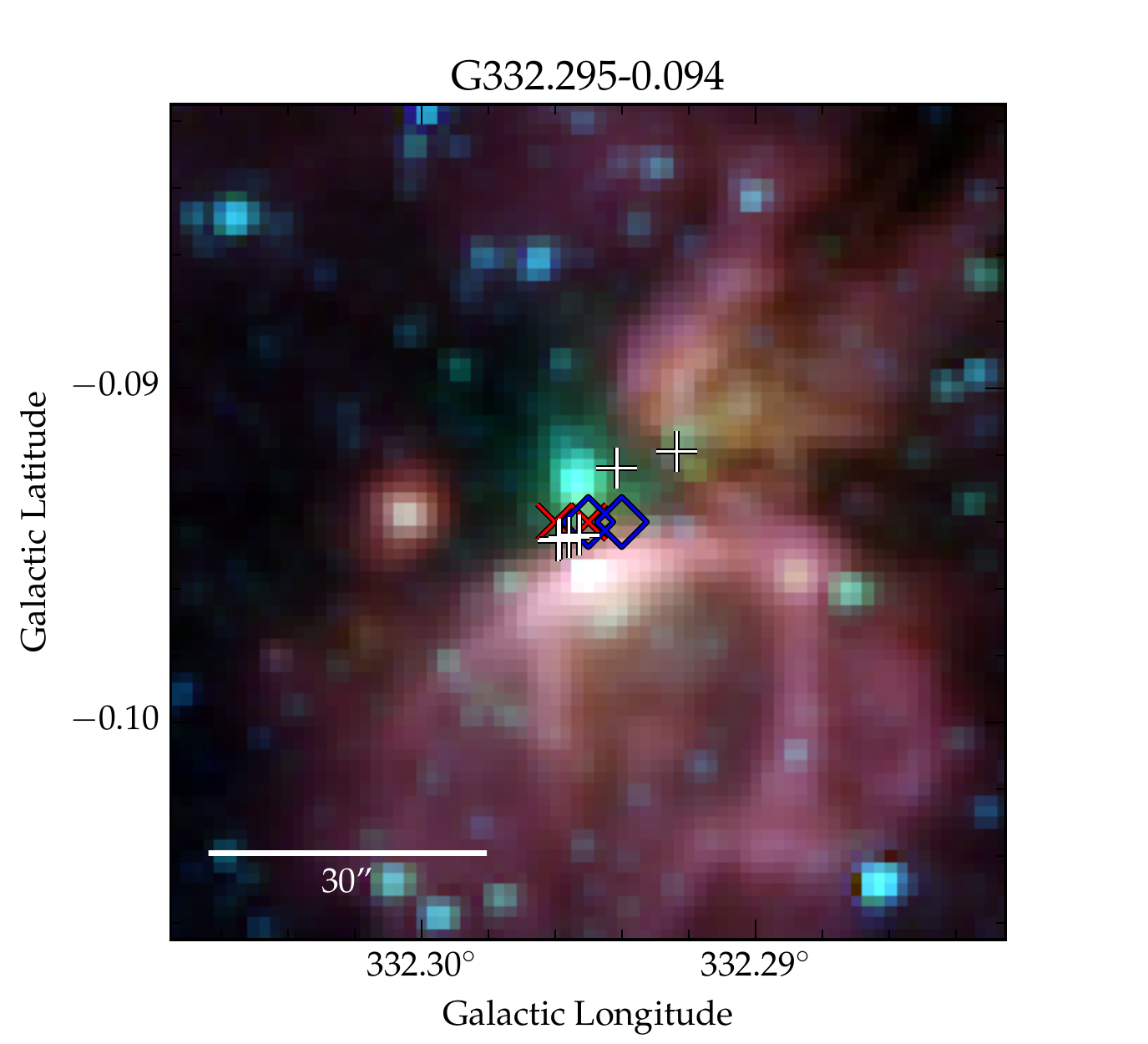}
      \includegraphics[height=0.30\textheight]{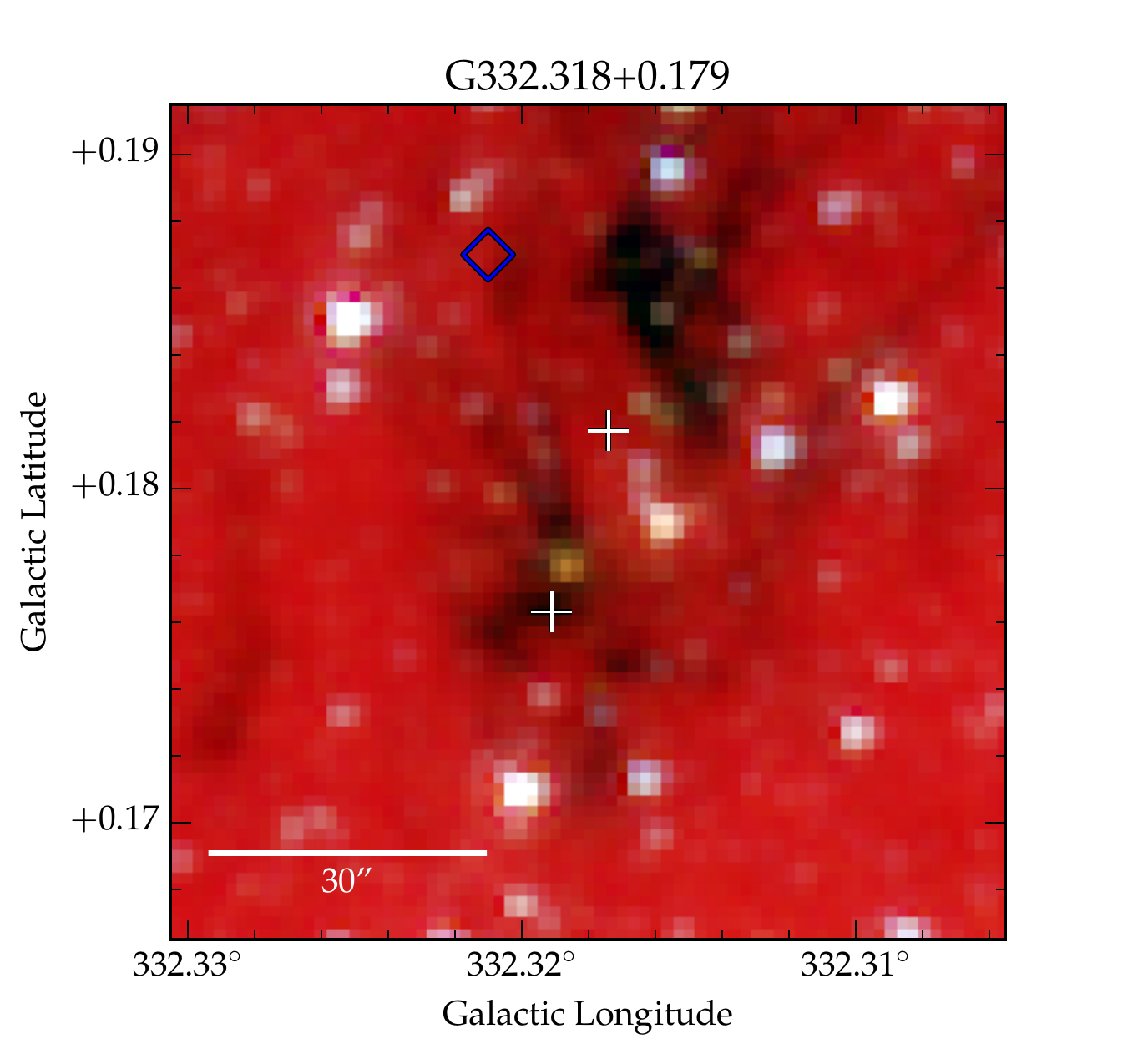}
      \includegraphics[height=0.30\textheight]{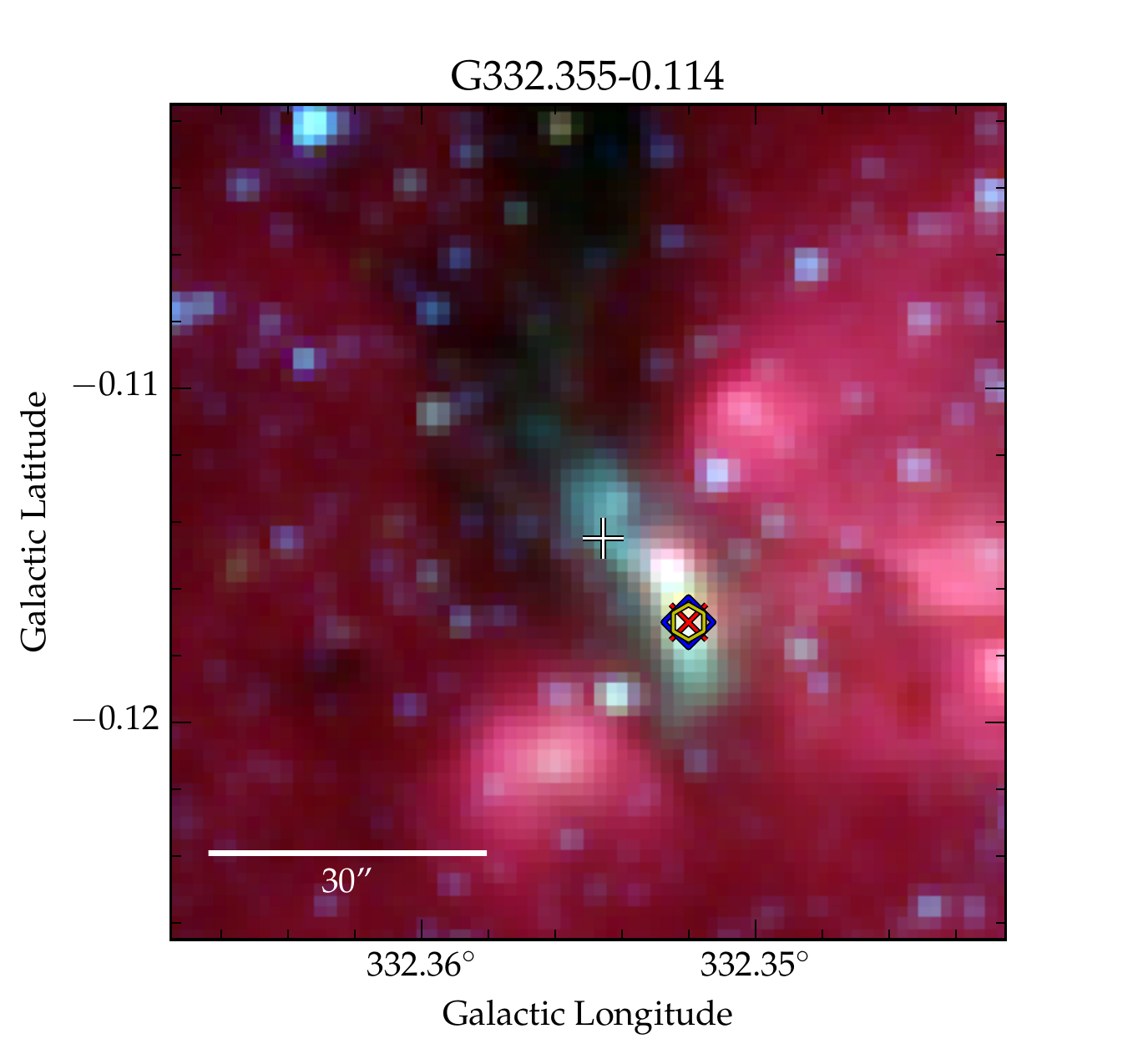}
      \includegraphics[height=0.30\textheight]{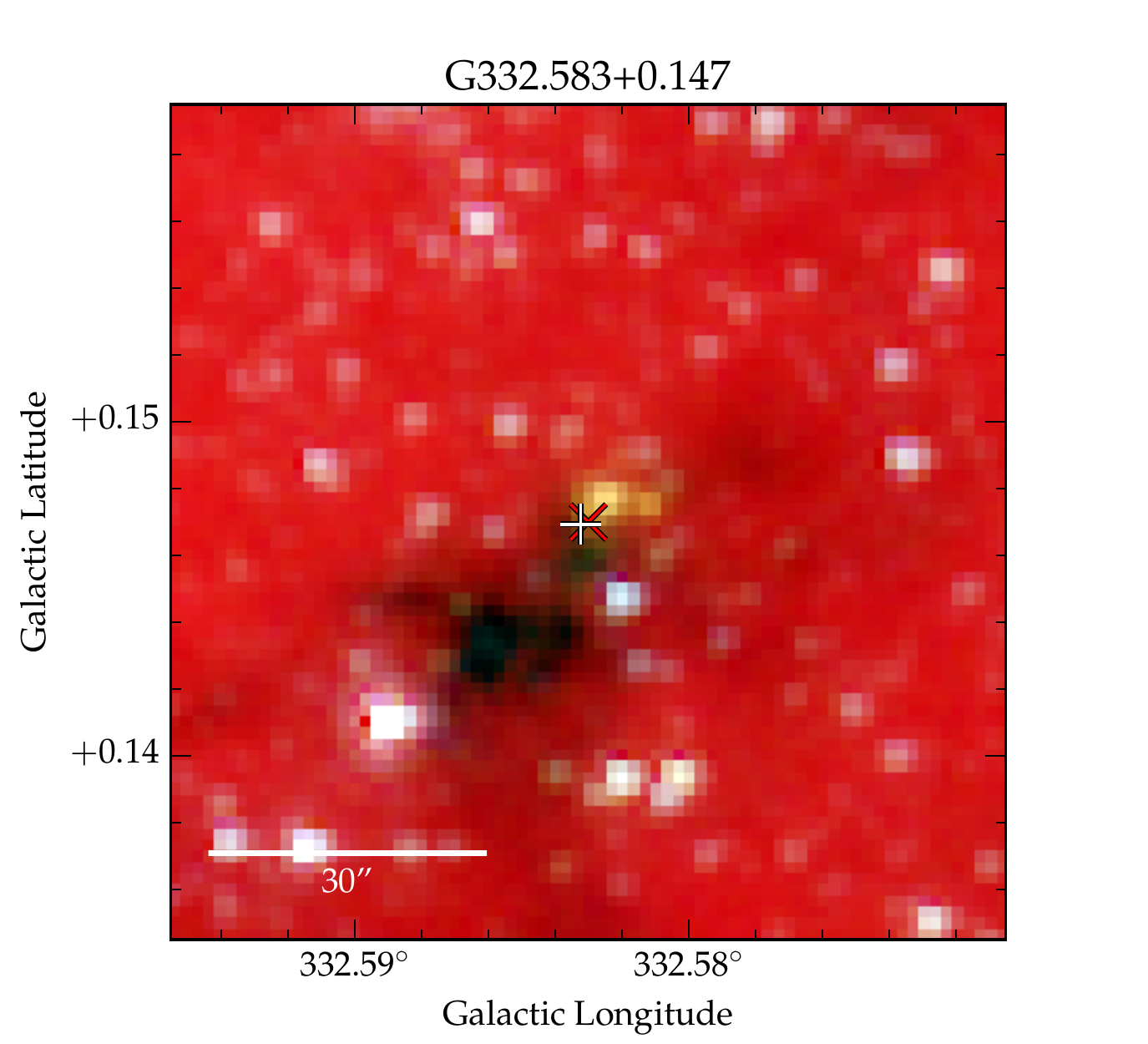}
      \includegraphics[height=0.30\textheight]{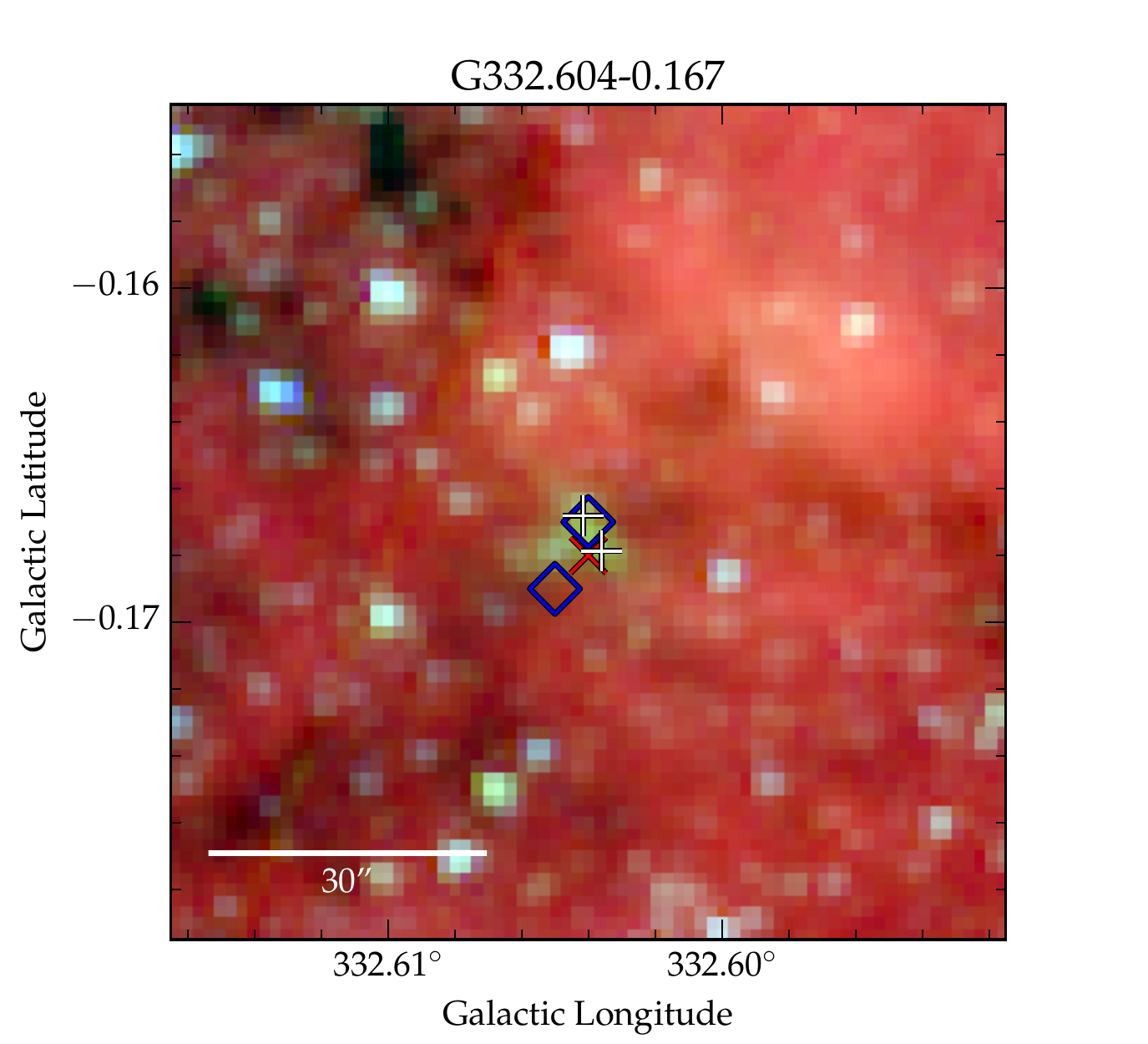}
      \captionof{figure}{\emph{continued}}
    \end{minipage}
}]
\setcounter{figure}{0}
\twocolumn[{
    \begin{minipage}{\textwidth}
      \centering
      \includegraphics[height=0.30\textheight]{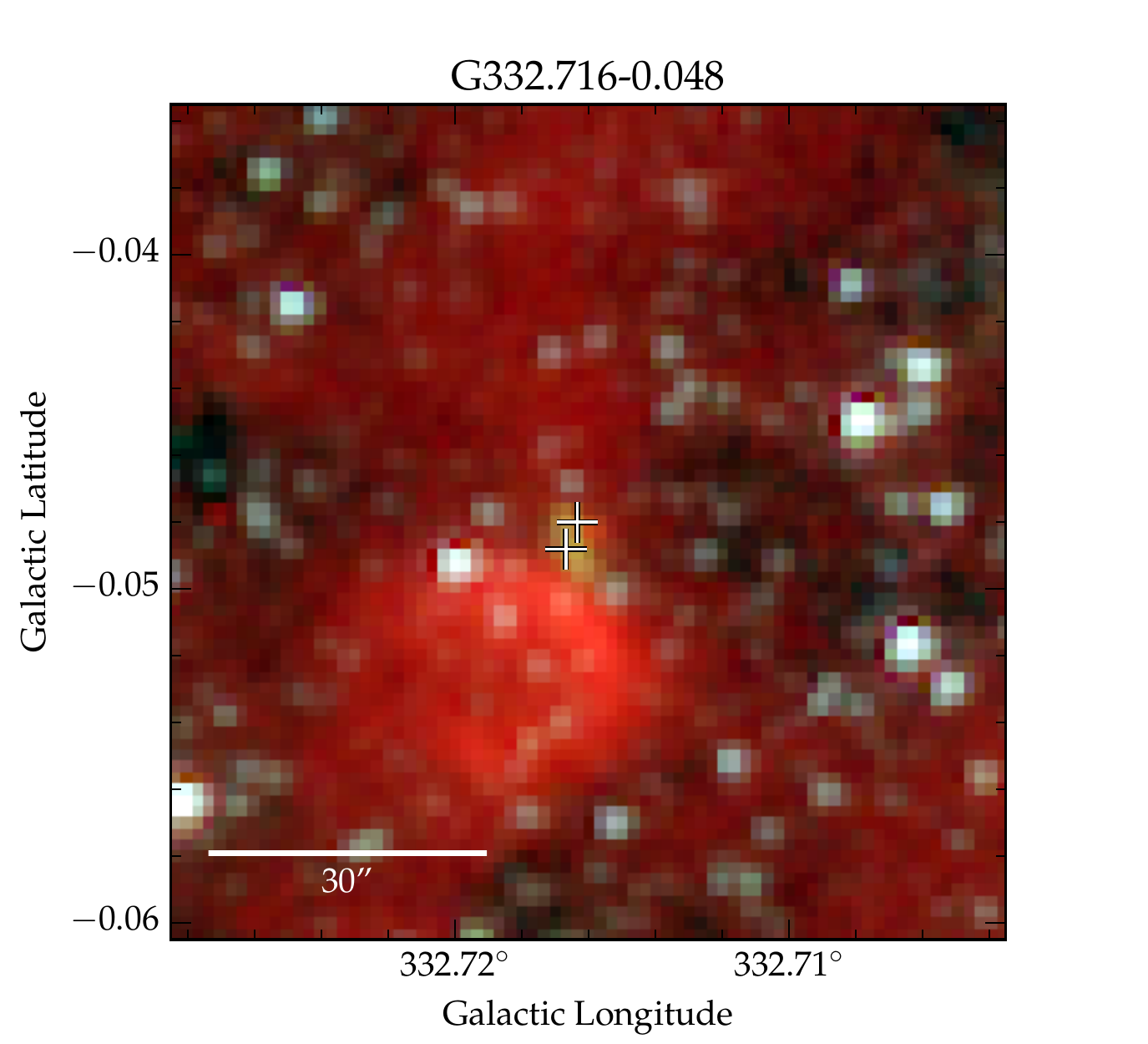}
      \includegraphics[height=0.30\textheight]{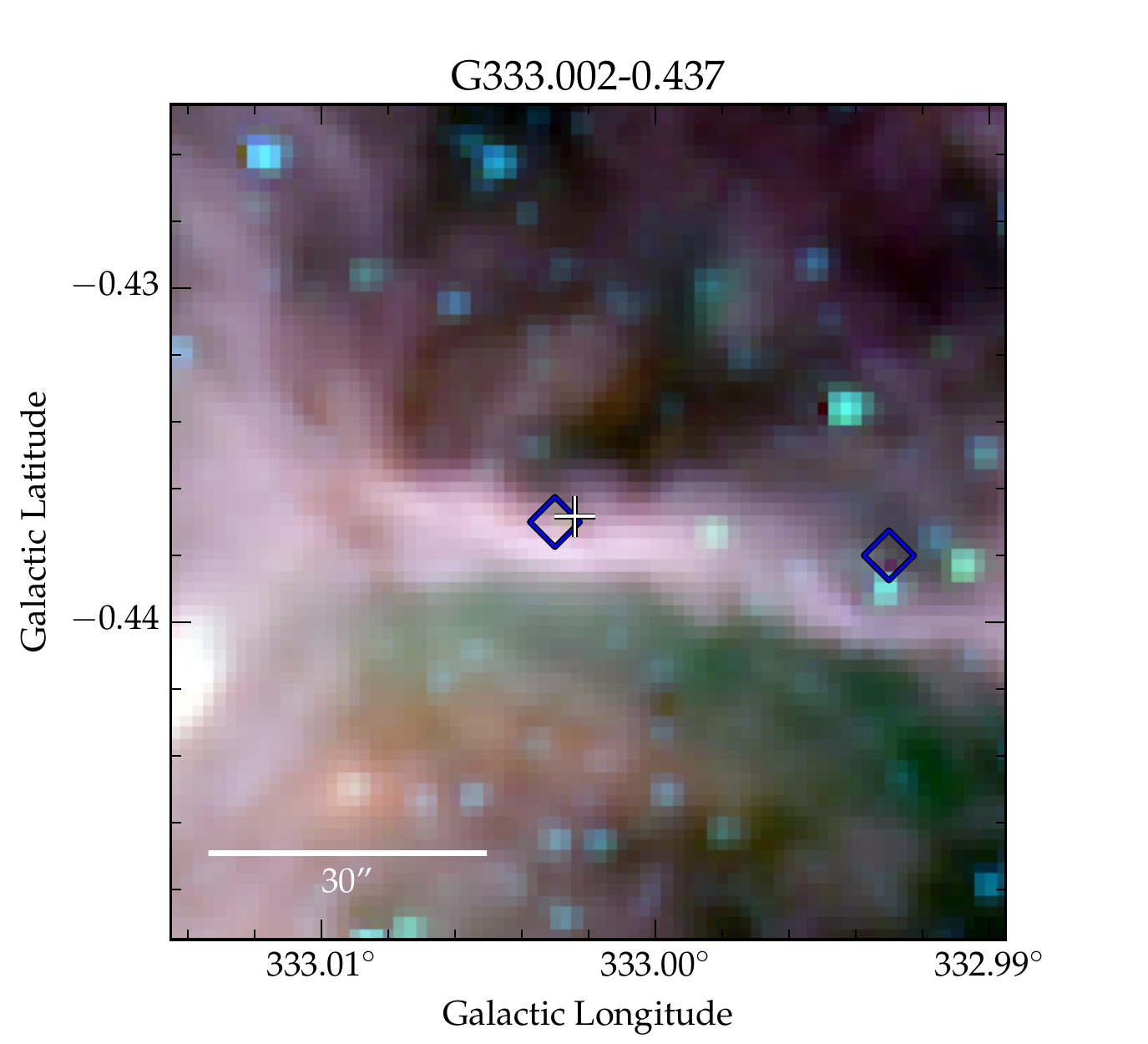}
      \includegraphics[height=0.30\textheight]{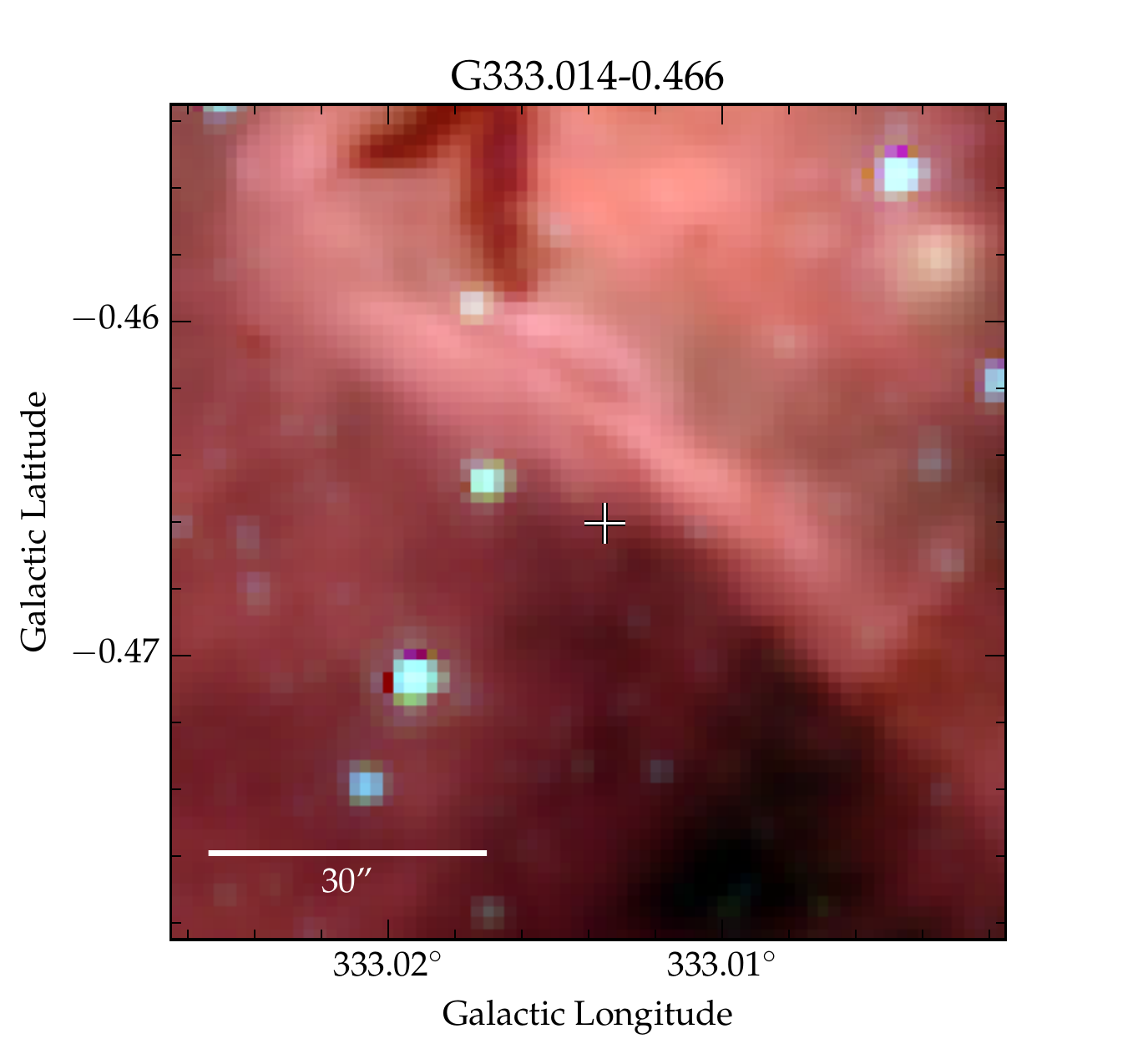}
      \includegraphics[height=0.30\textheight]{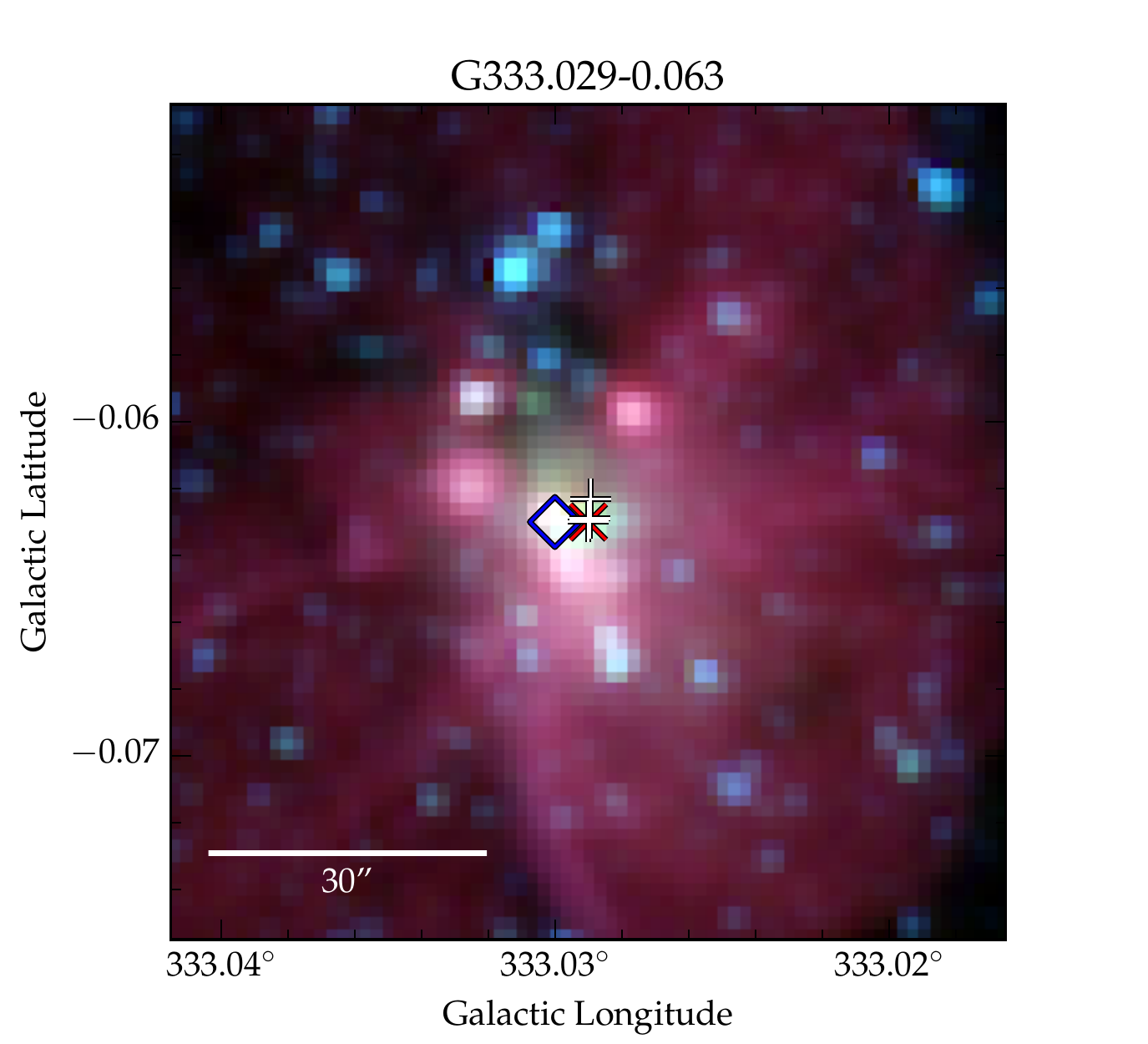}
      \includegraphics[height=0.30\textheight]{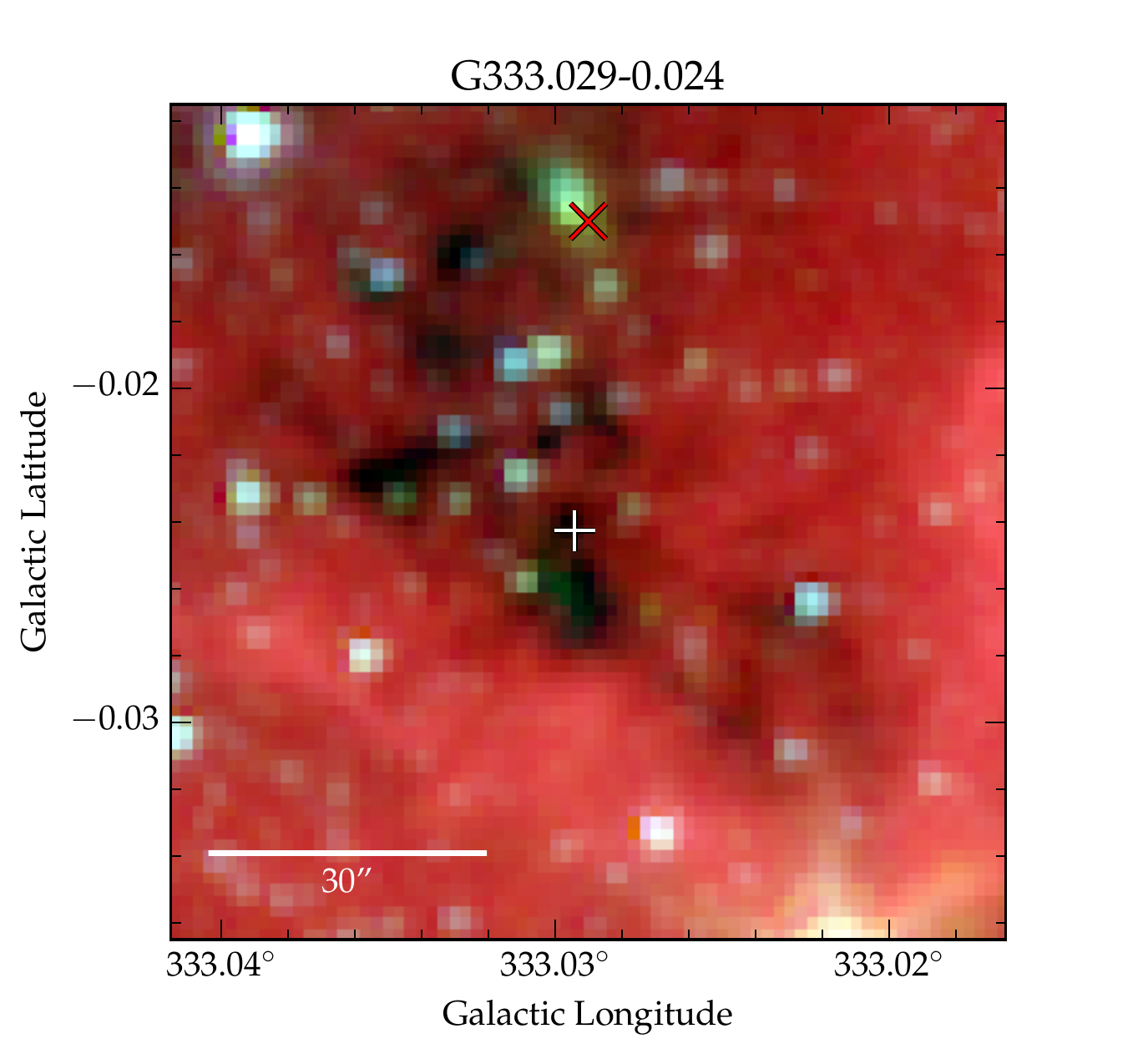}
      \includegraphics[height=0.30\textheight]{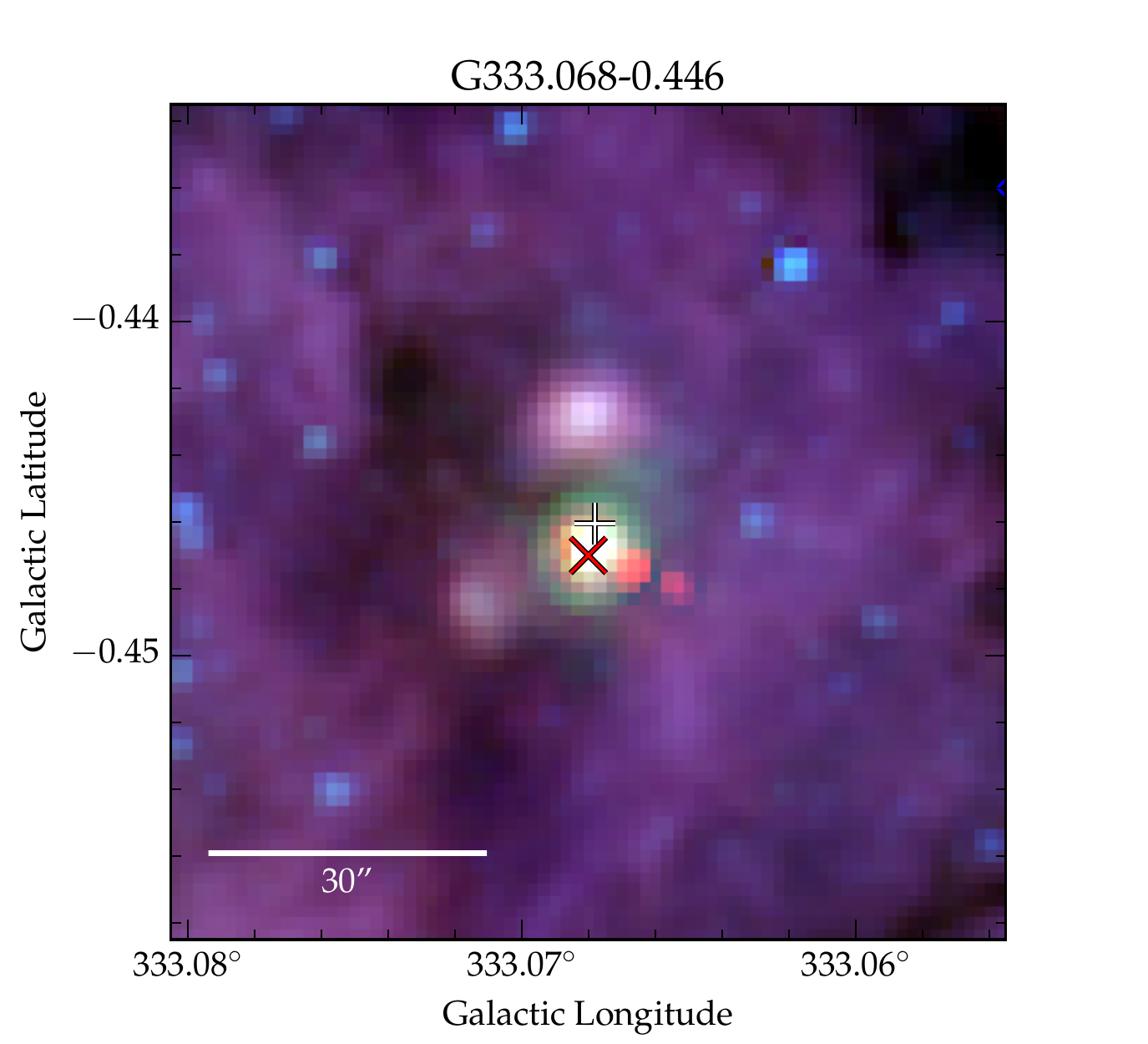}
      \captionof{figure}{\emph{continued}}
    \end{minipage}
}]
\setcounter{figure}{0}
\twocolumn[{
    \begin{minipage}{\textwidth}
      \centering
      \includegraphics[height=0.30\textheight]{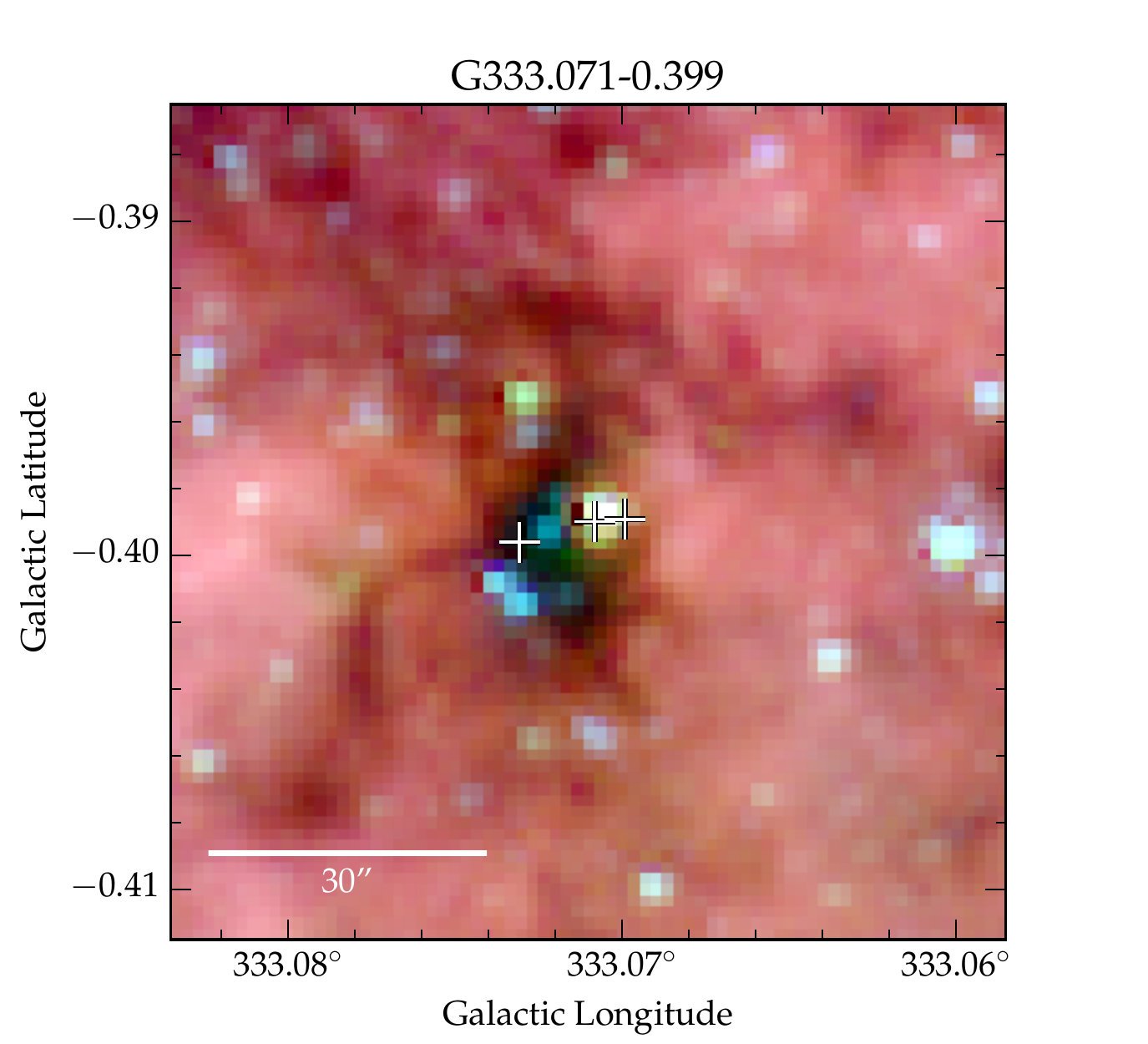}
      \includegraphics[height=0.30\textheight]{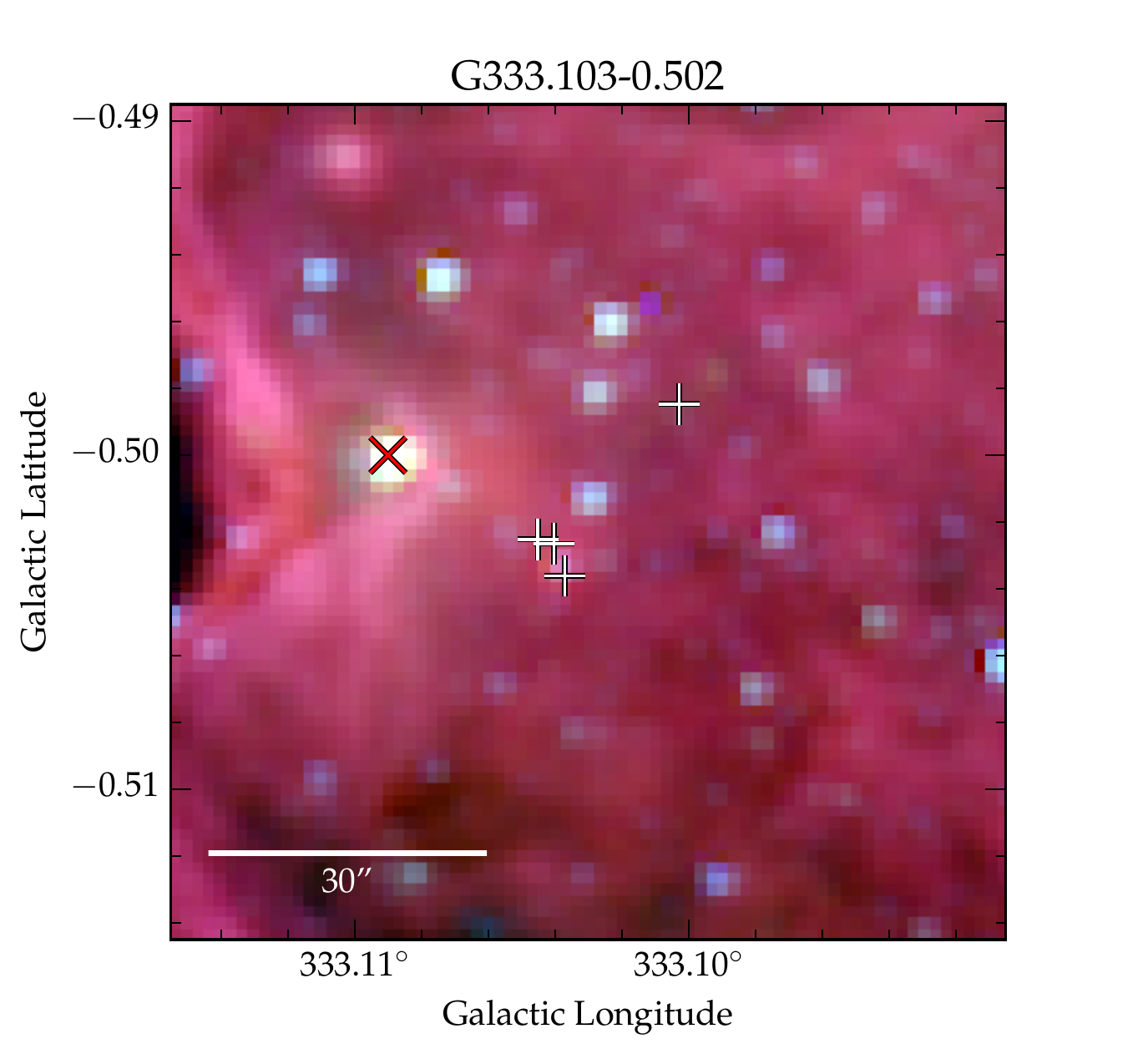}
      \includegraphics[height=0.30\textheight]{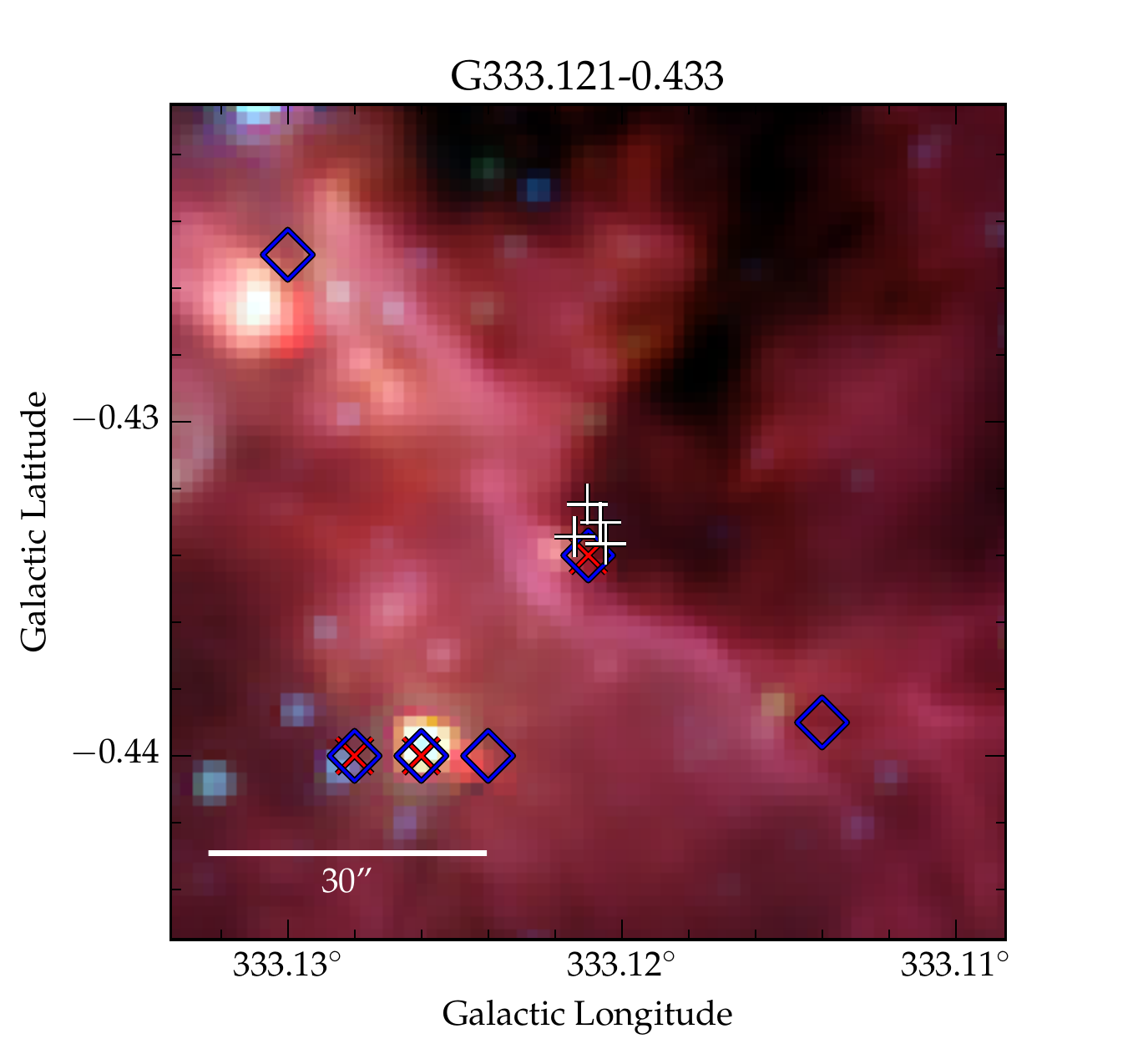}
      \includegraphics[height=0.30\textheight]{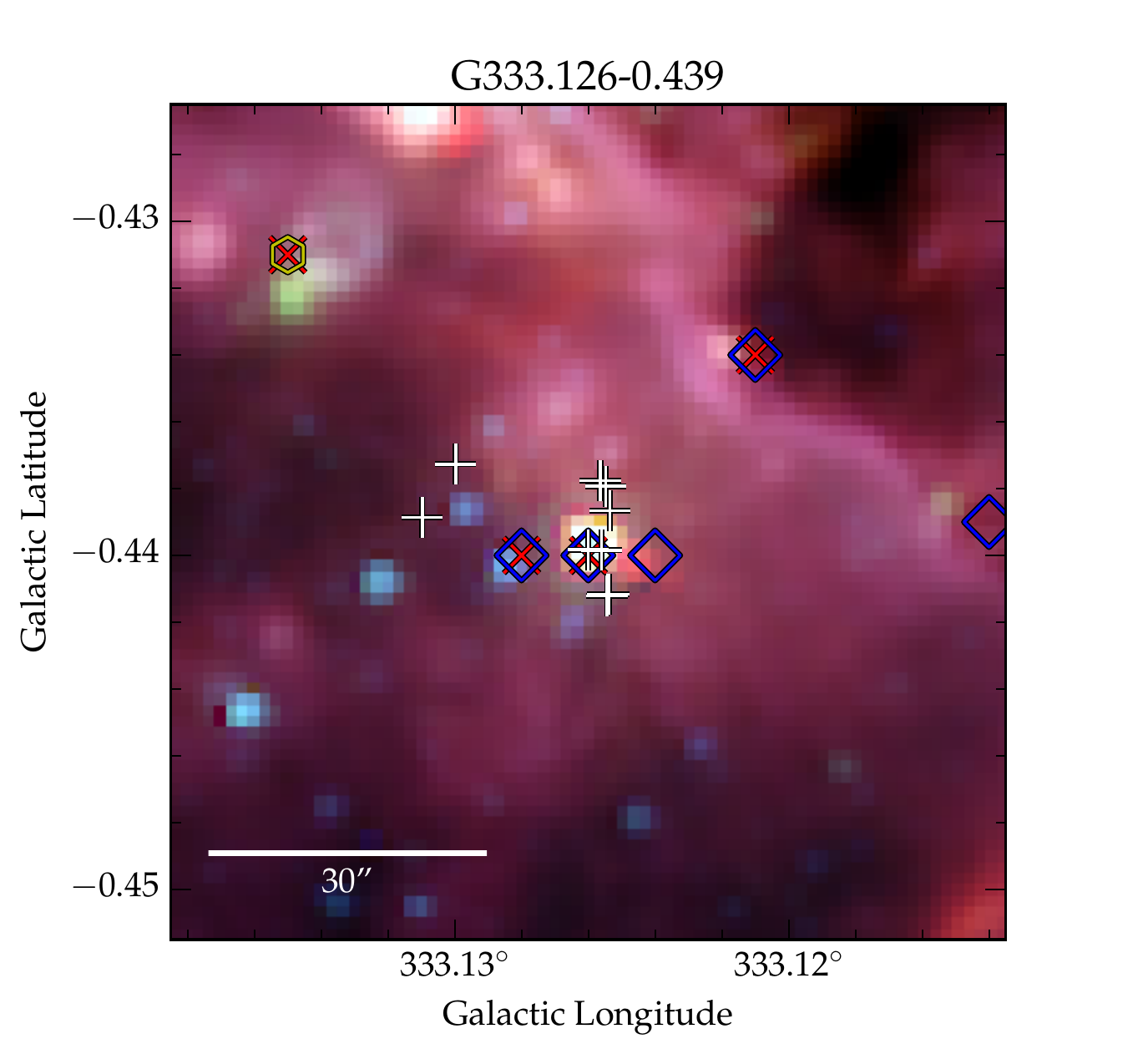}
      \includegraphics[height=0.30\textheight]{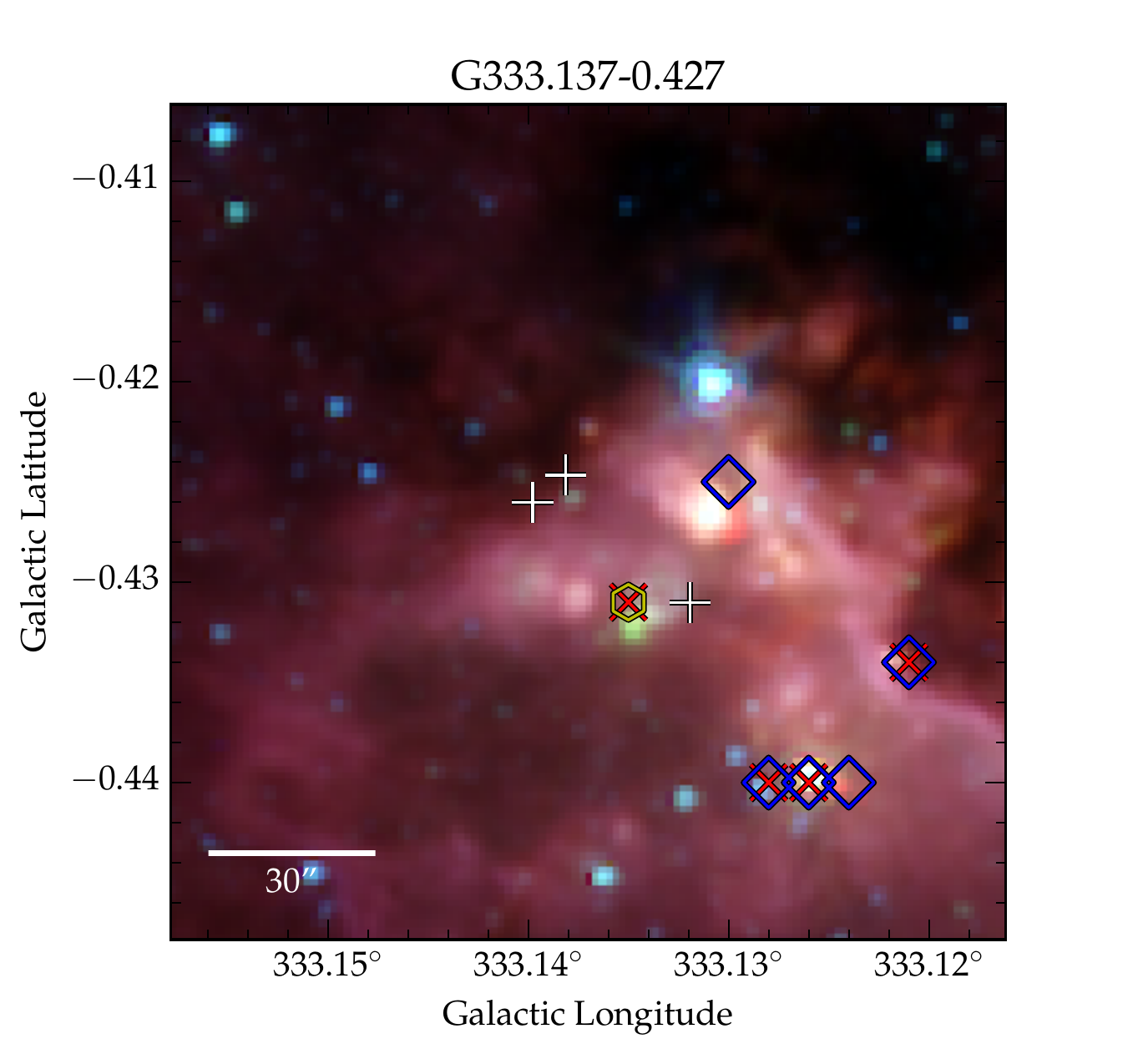}
      \includegraphics[height=0.30\textheight]{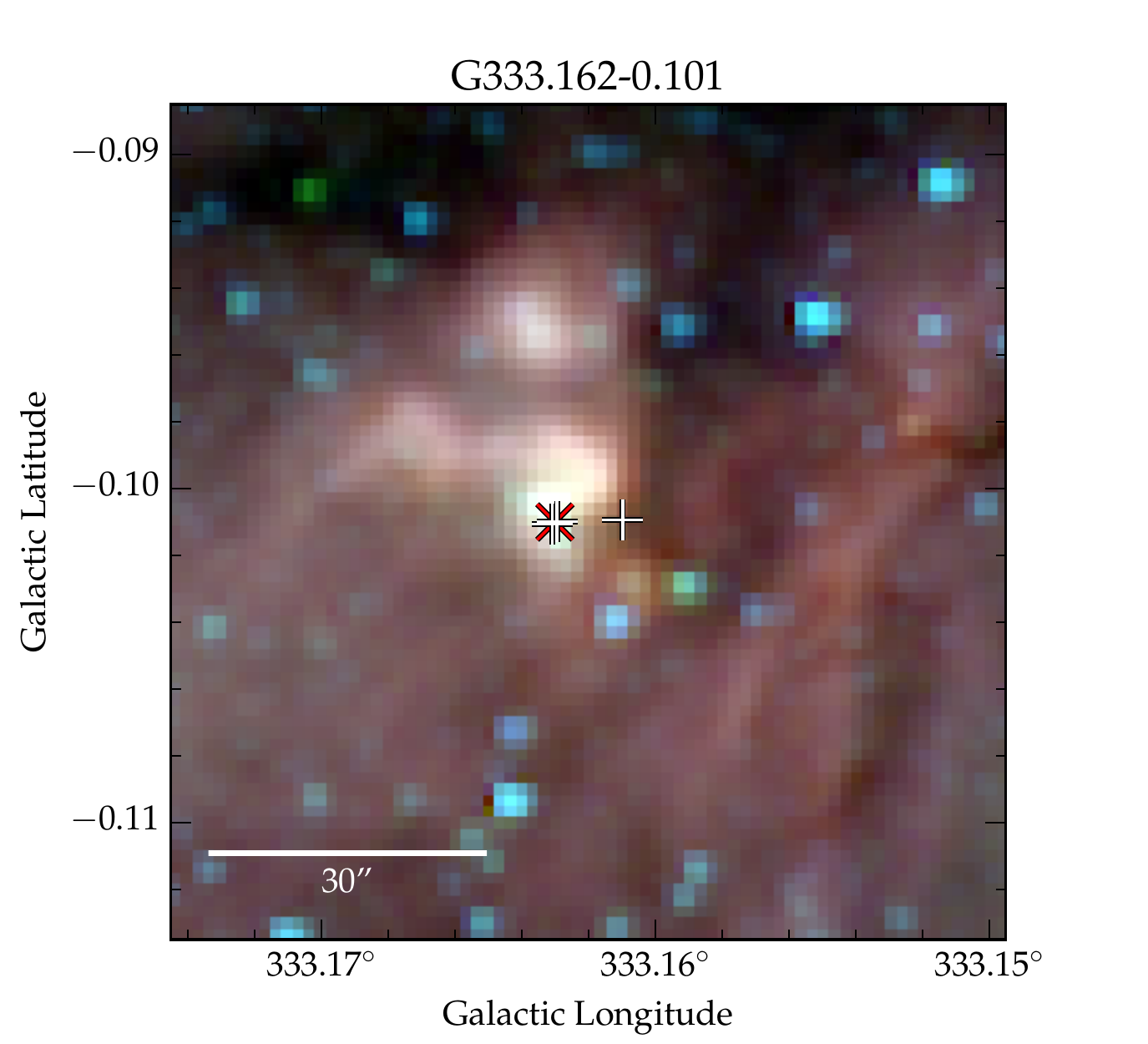}
      \captionof{figure}{\emph{continued}}
    \end{minipage}
}]
\setcounter{figure}{0}
\twocolumn[{
    \begin{minipage}{\textwidth}
      \centering
      \includegraphics[height=0.30\textheight]{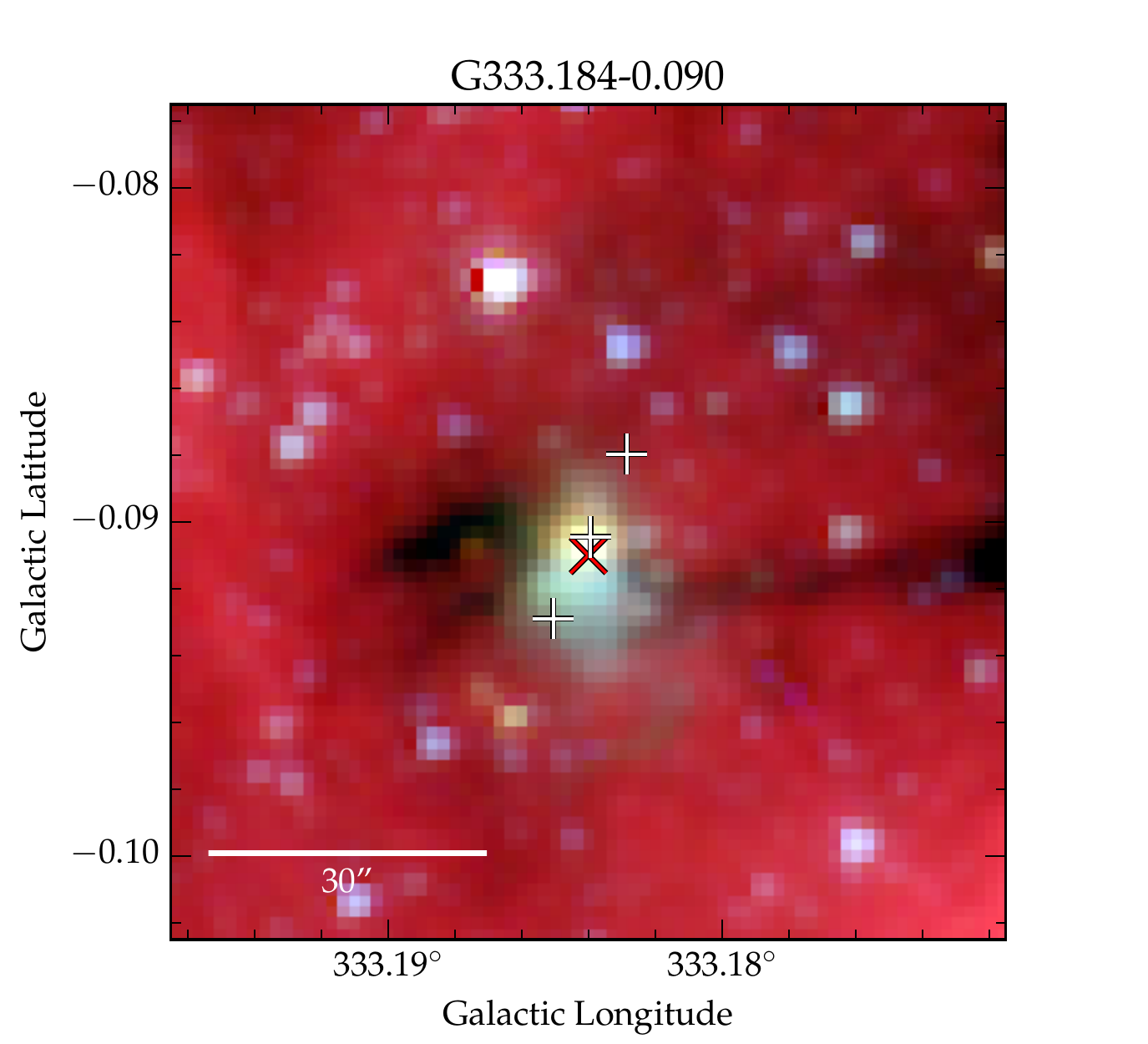}
      \includegraphics[height=0.30\textheight]{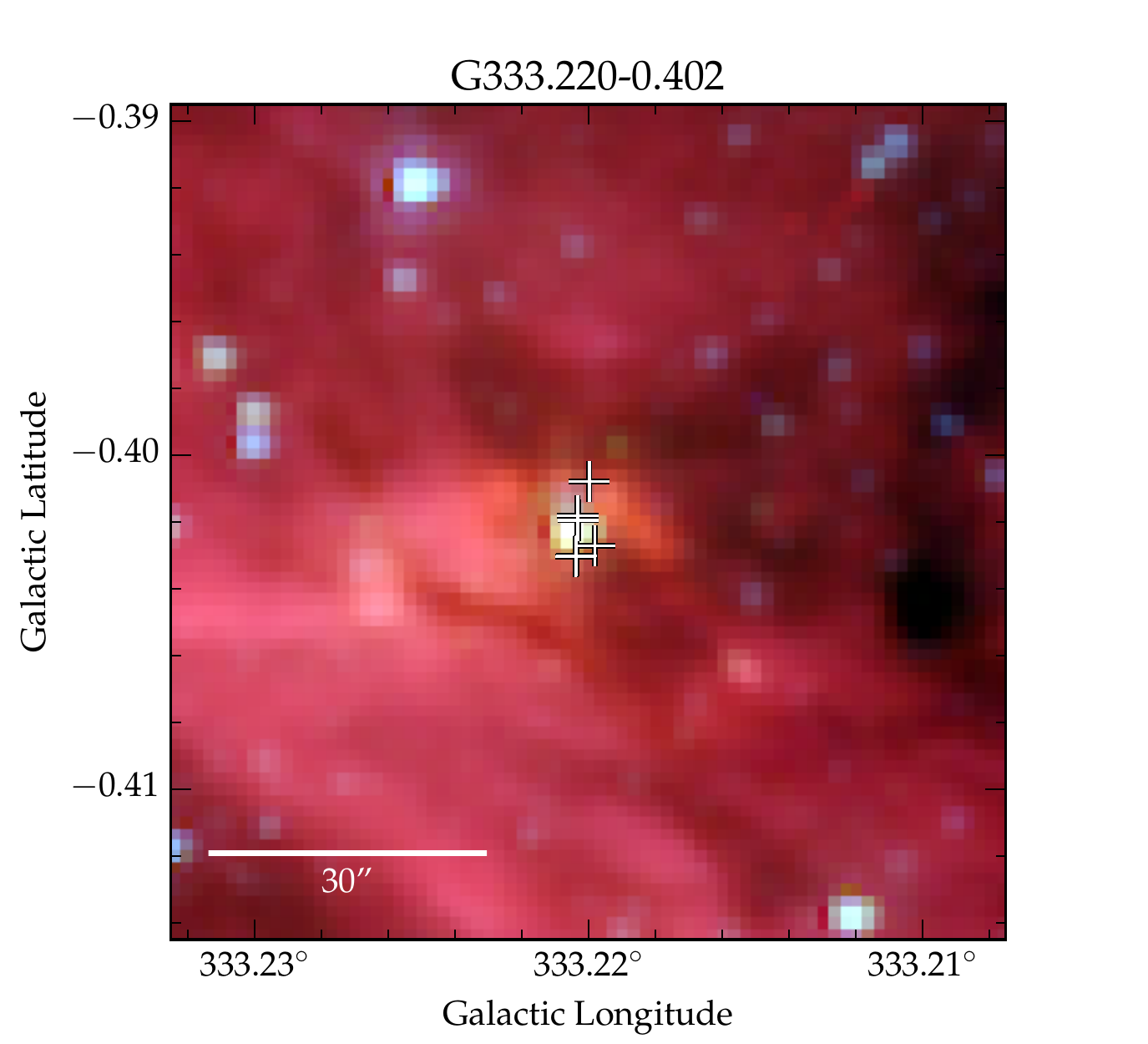}
      \includegraphics[height=0.30\textheight]{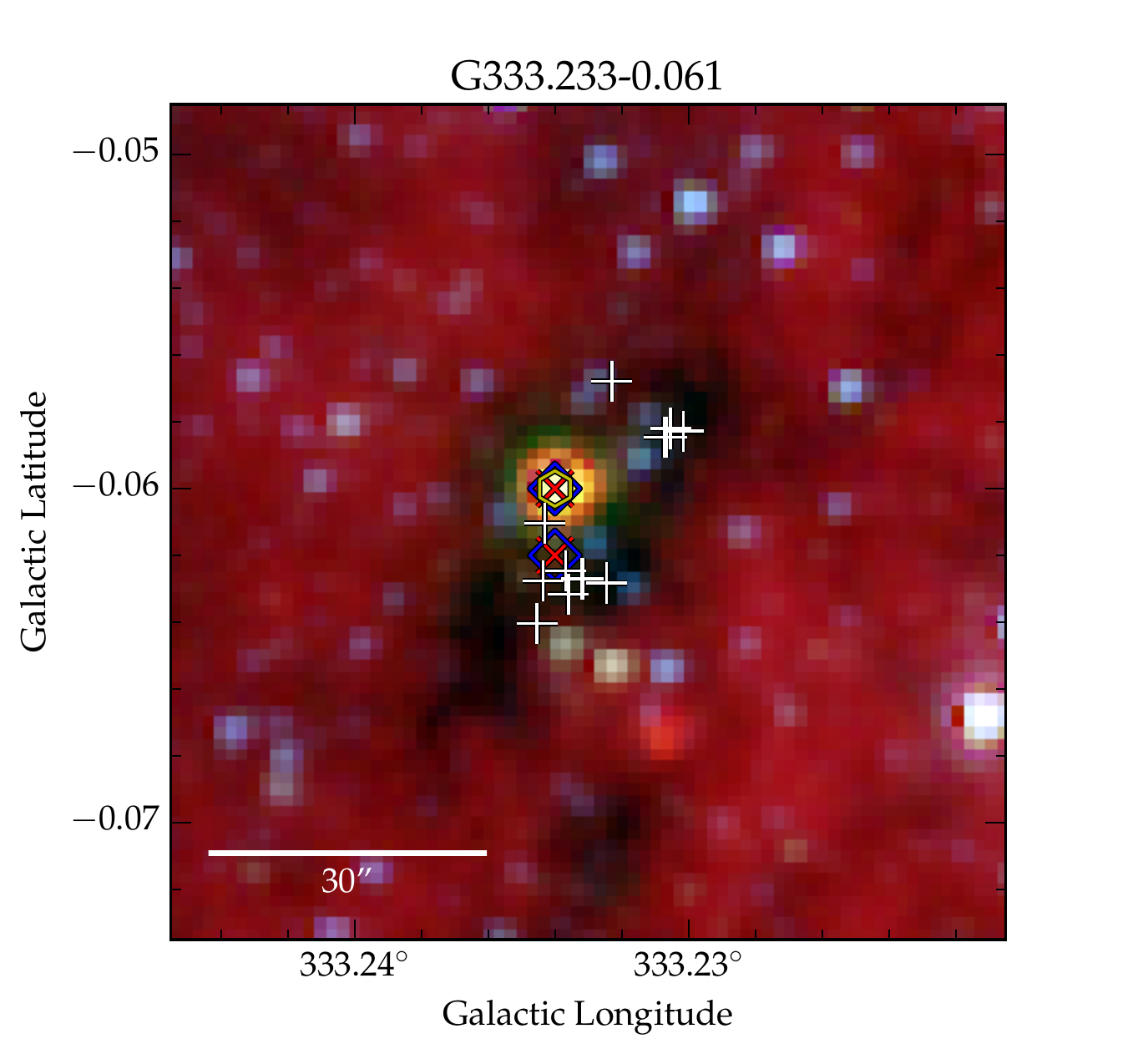}
      \includegraphics[height=0.30\textheight]{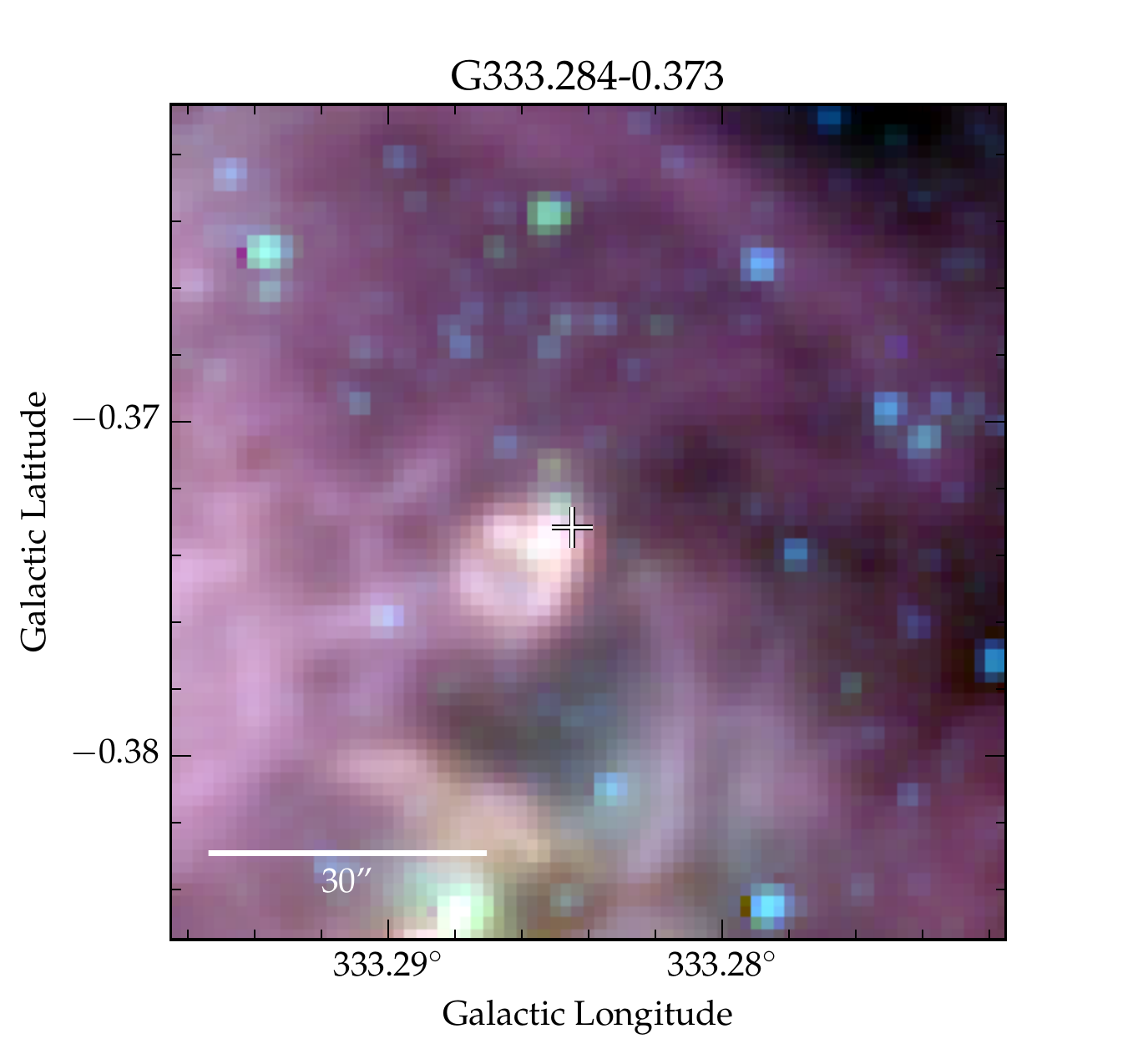}
      \includegraphics[height=0.30\textheight]{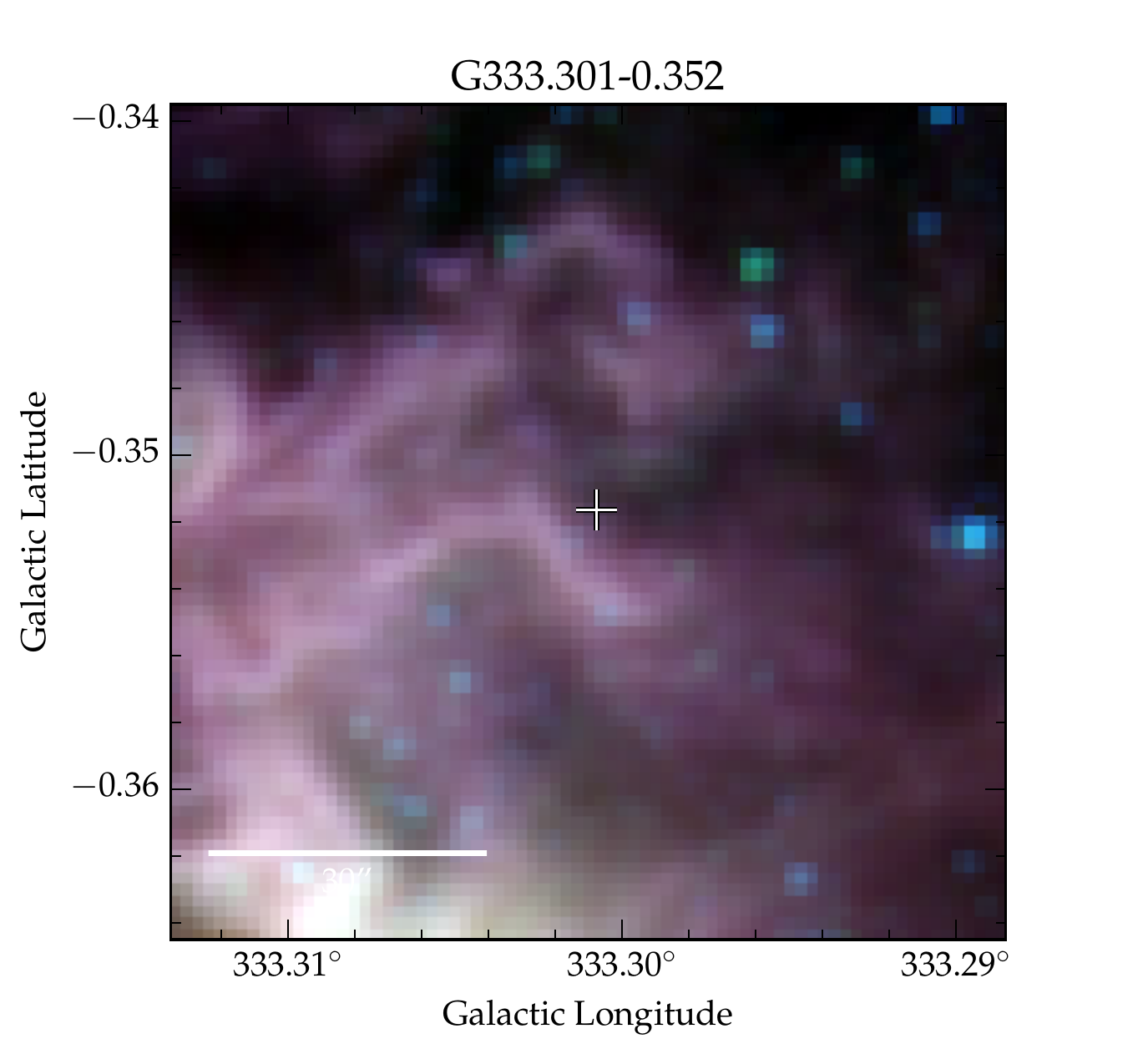}
      \includegraphics[height=0.30\textheight]{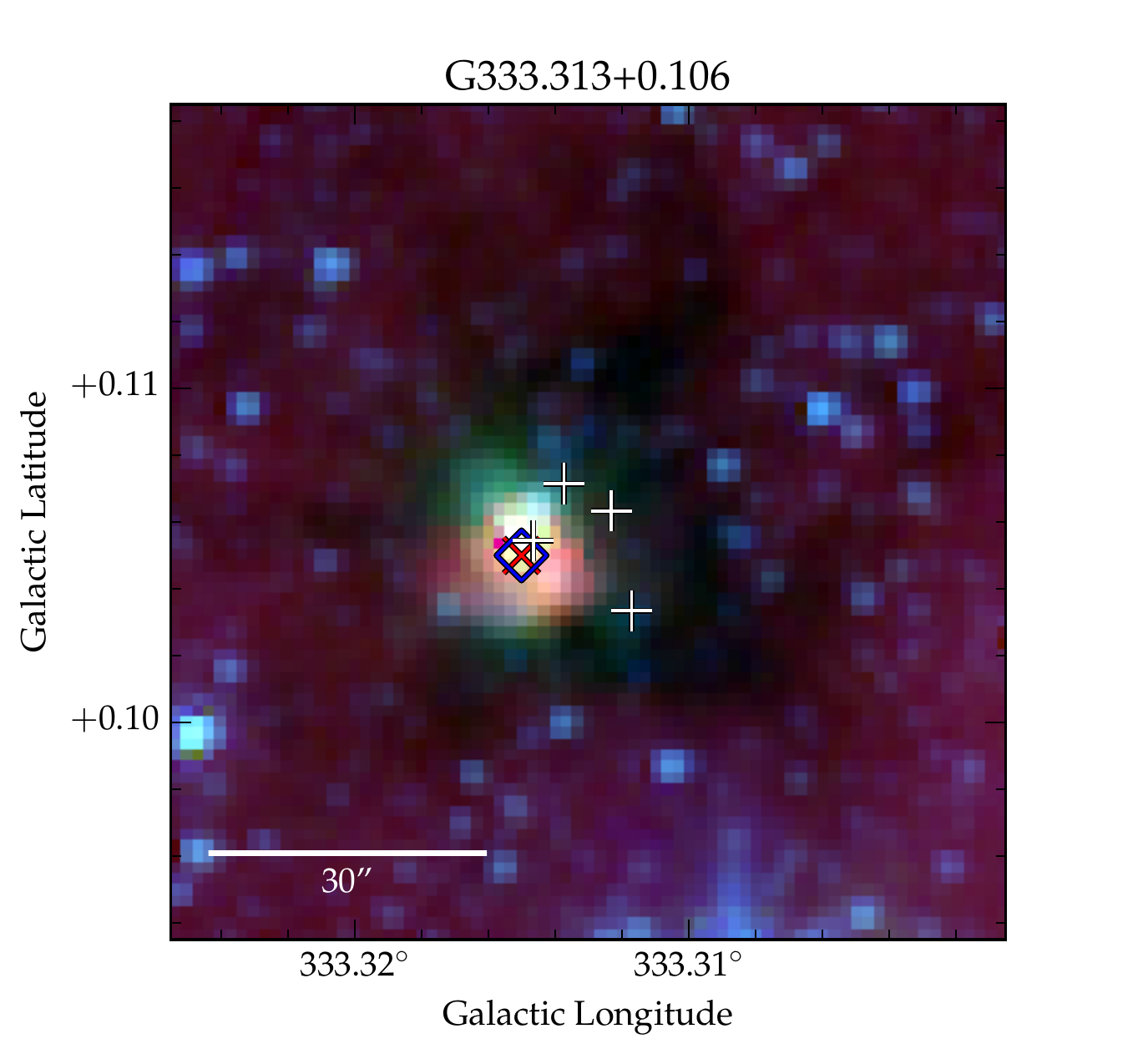}
      \captionof{figure}{\emph{continued}}
    \end{minipage}
}]
\setcounter{figure}{0}
\twocolumn[{
    \begin{minipage}{\textwidth}
      \centering
      \includegraphics[height=0.30\textheight]{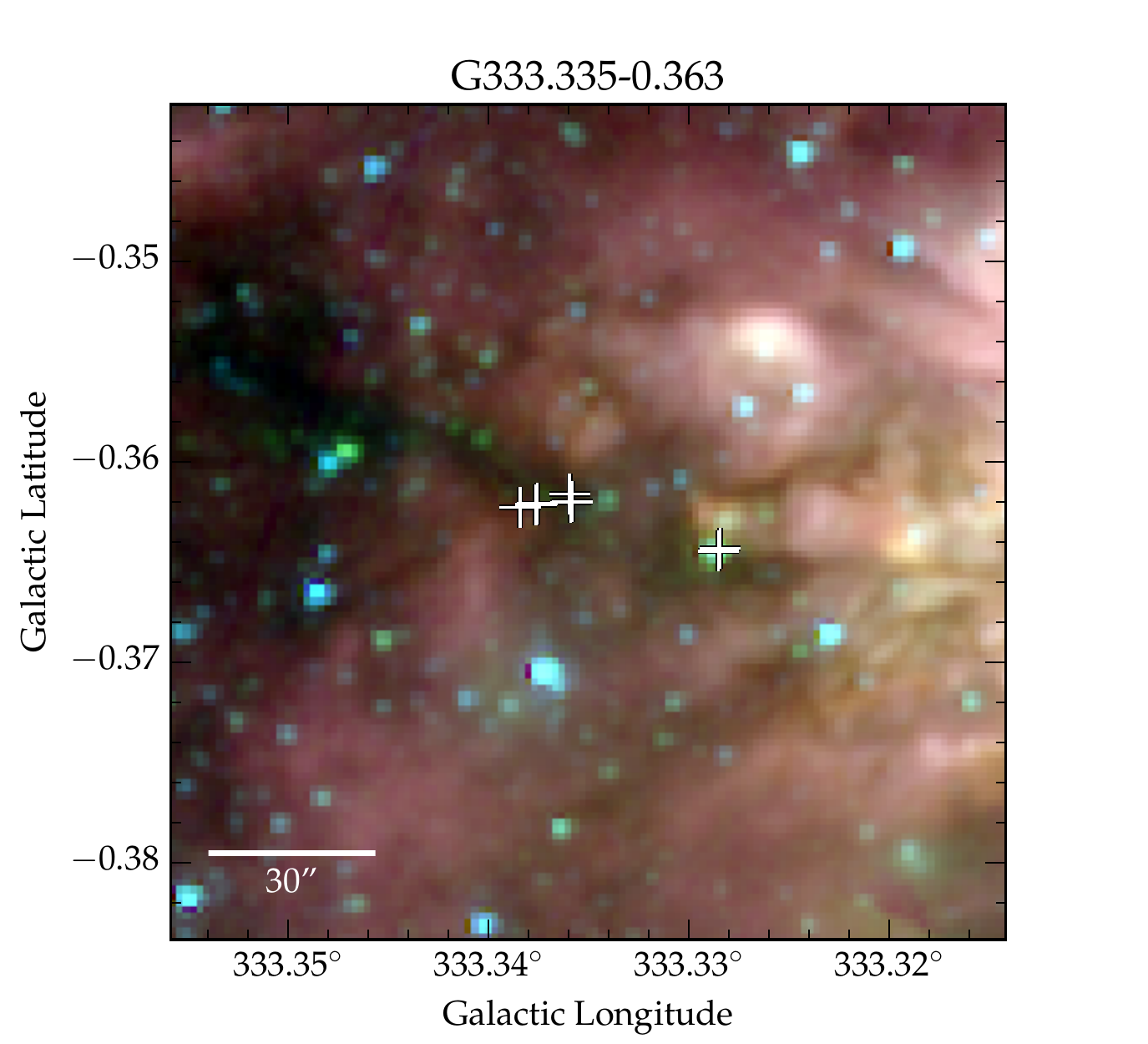}
      \includegraphics[height=0.30\textheight]{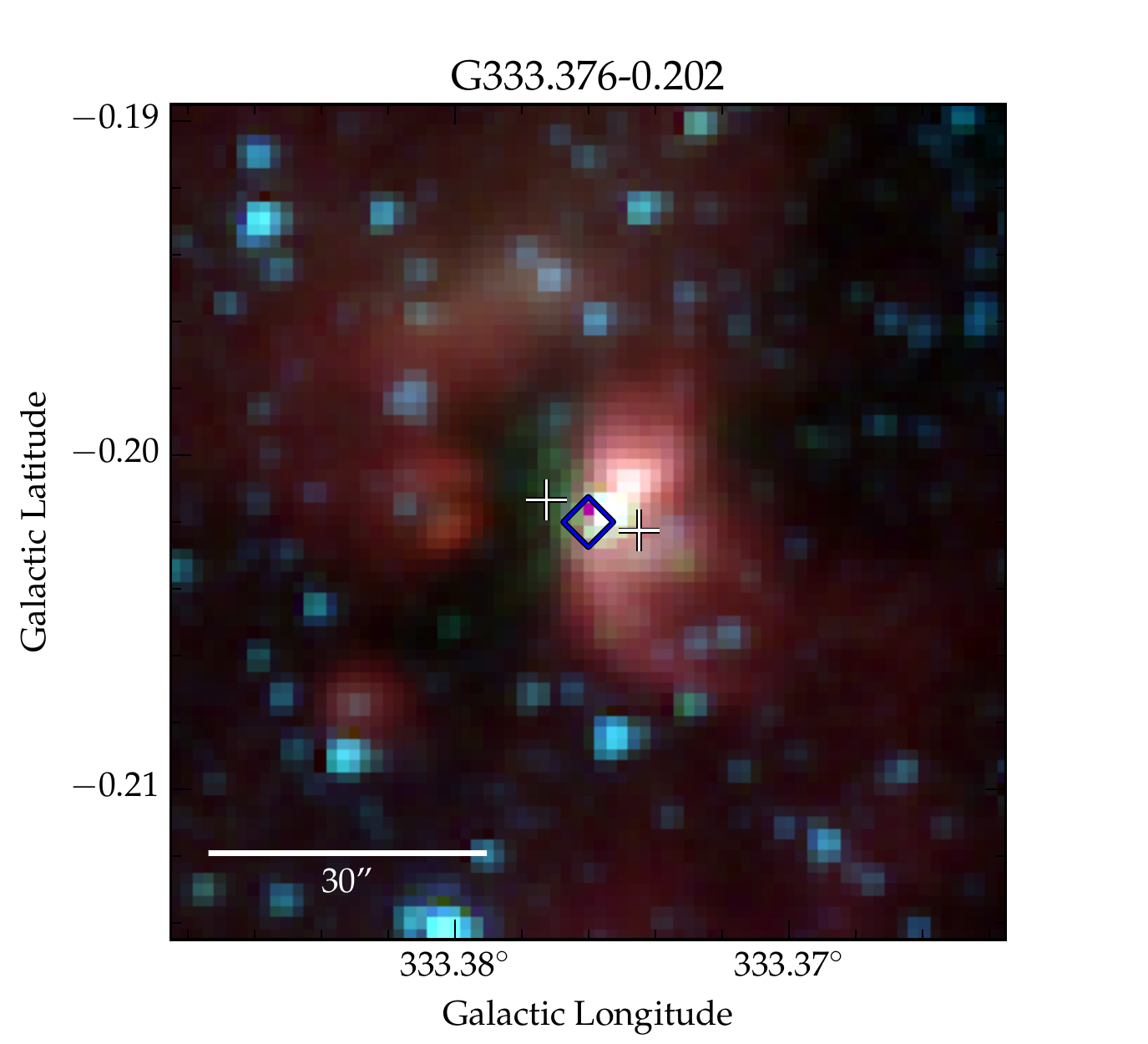}
      \includegraphics[height=0.30\textheight]{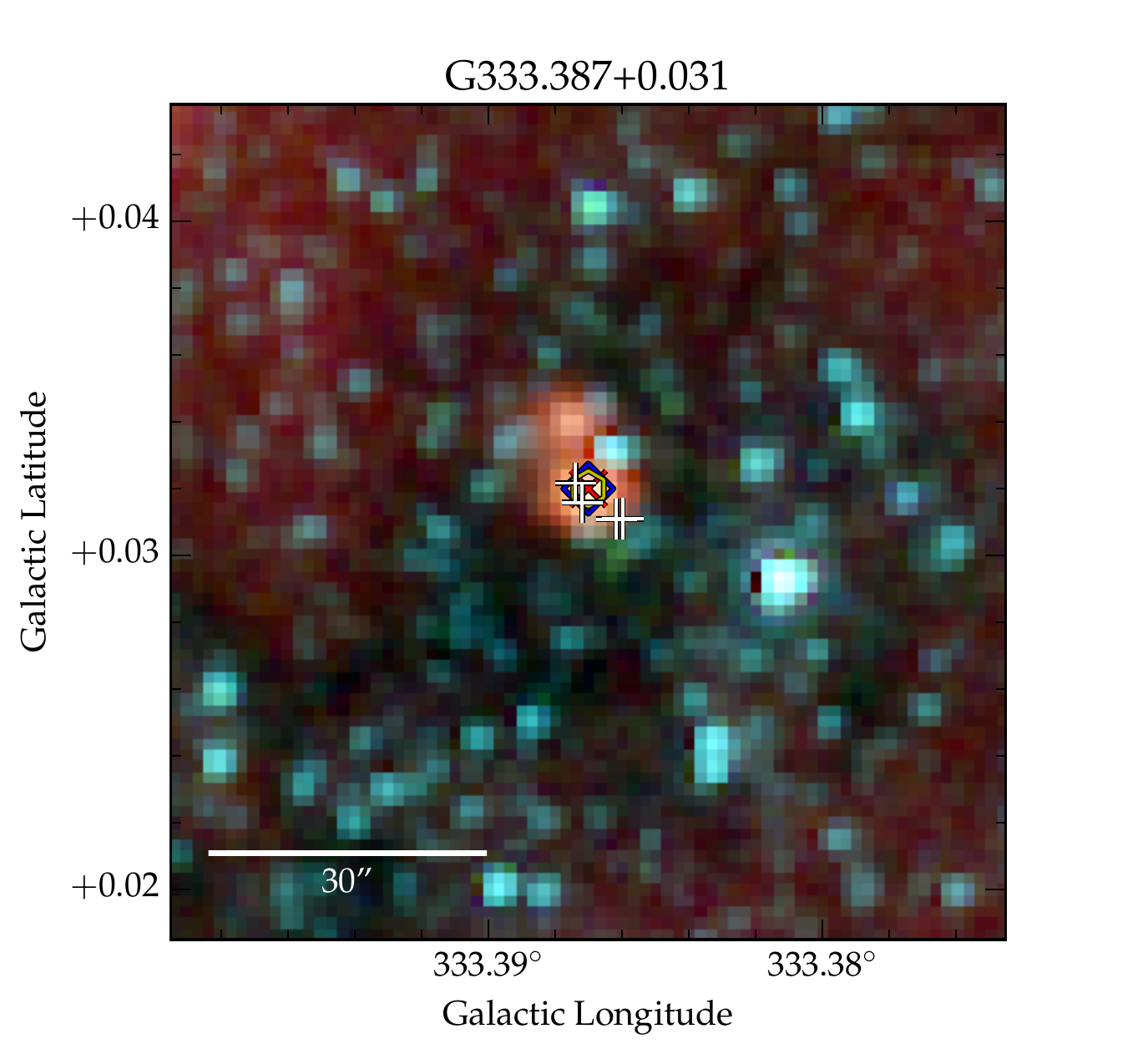}
      \includegraphics[height=0.30\textheight]{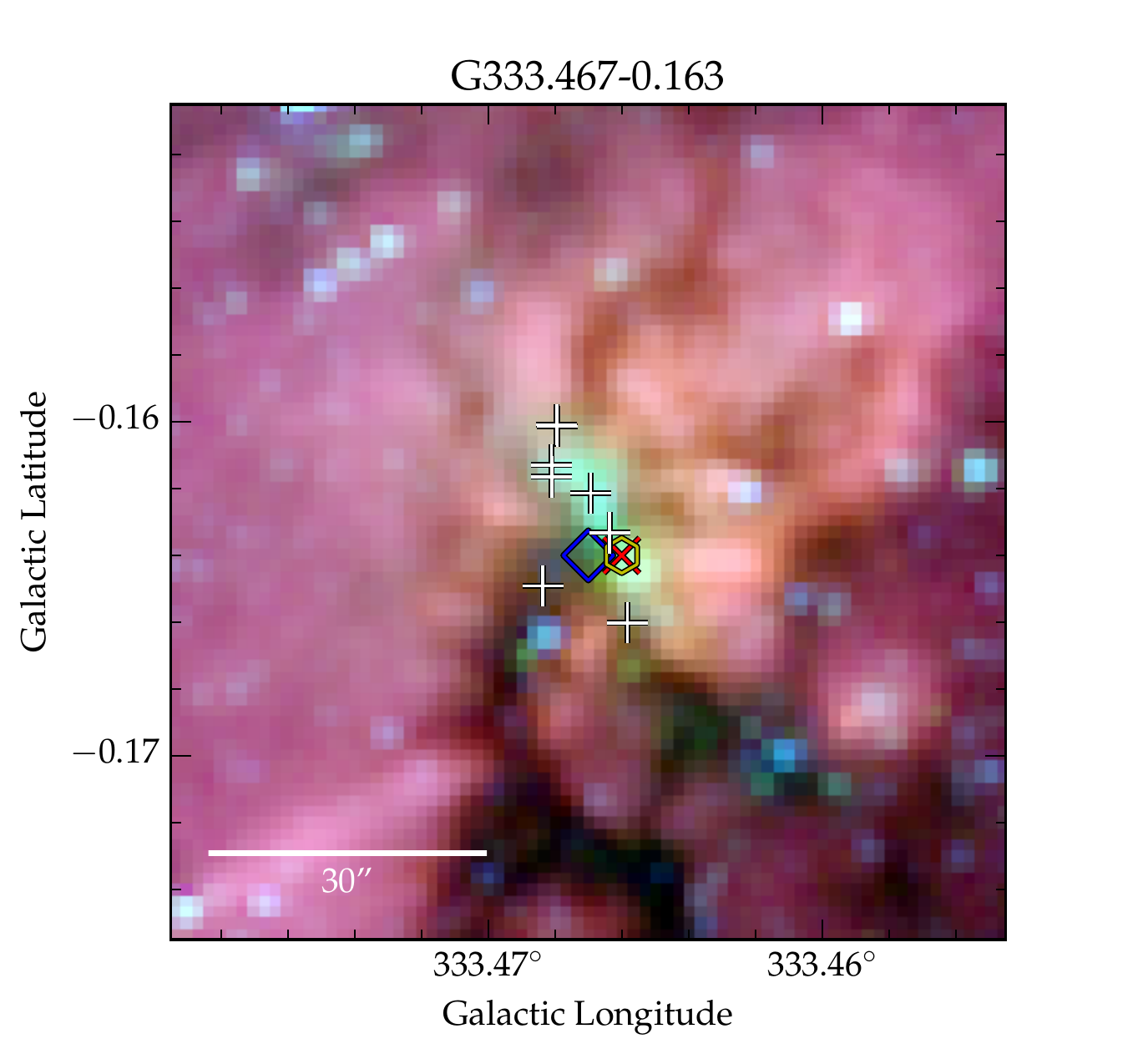}
      \includegraphics[height=0.30\textheight]{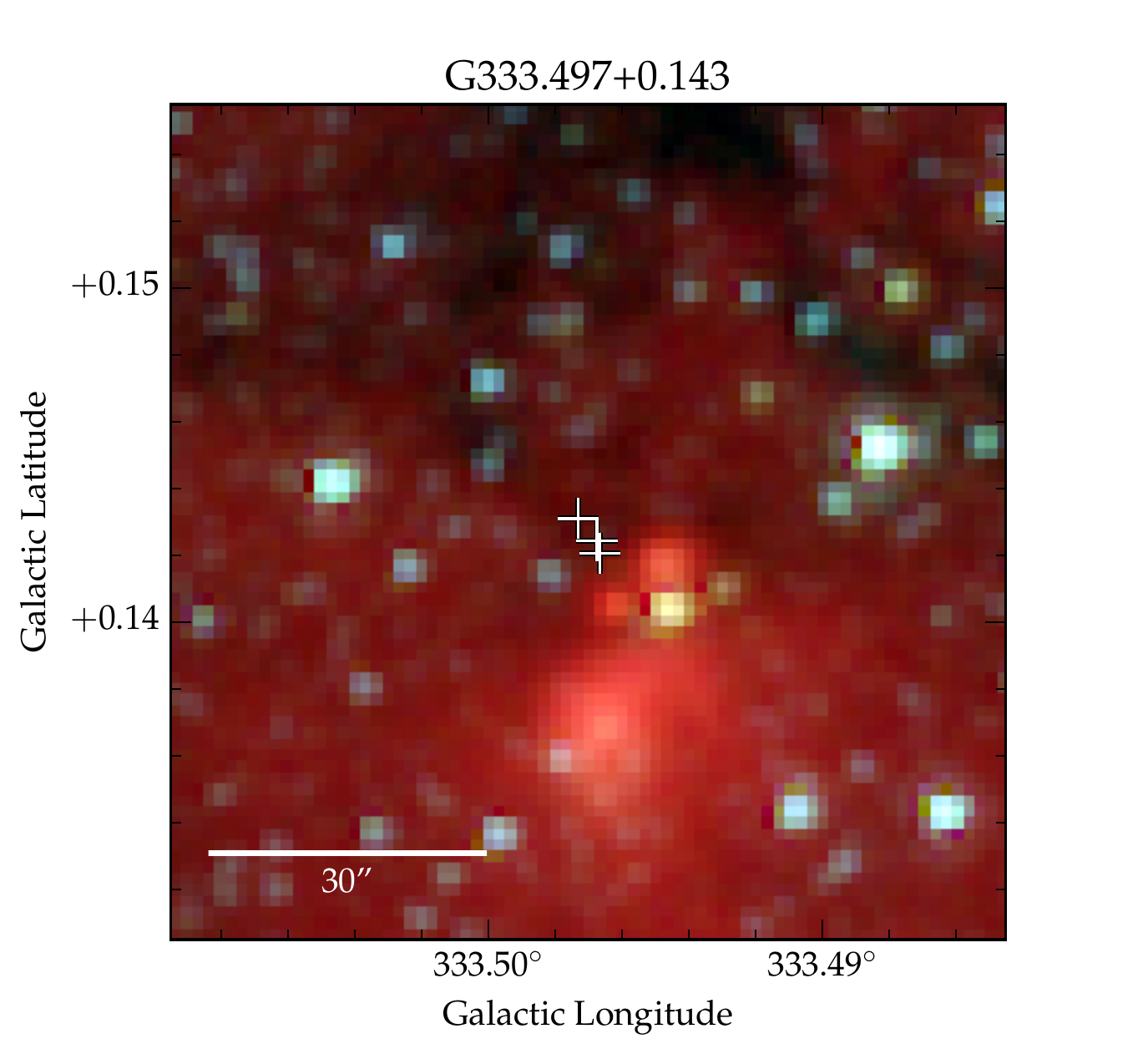}
      \includegraphics[height=0.30\textheight]{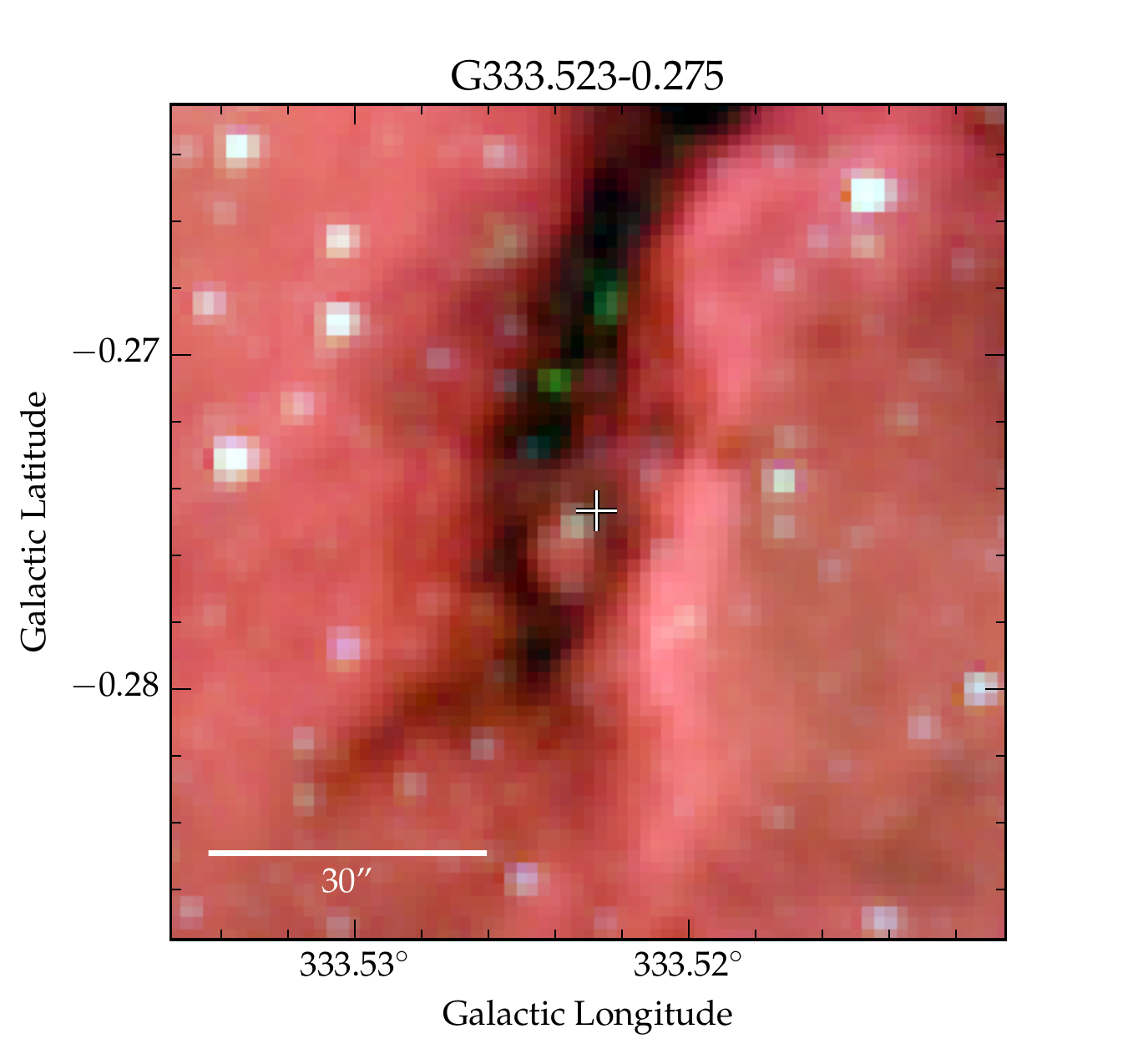}
      \captionof{figure}{\emph{continued}}
    \end{minipage}
}]
\setcounter{figure}{0}
\twocolumn[{
    \begin{minipage}{\textwidth}
      \centering
      \includegraphics[height=0.30\textheight]{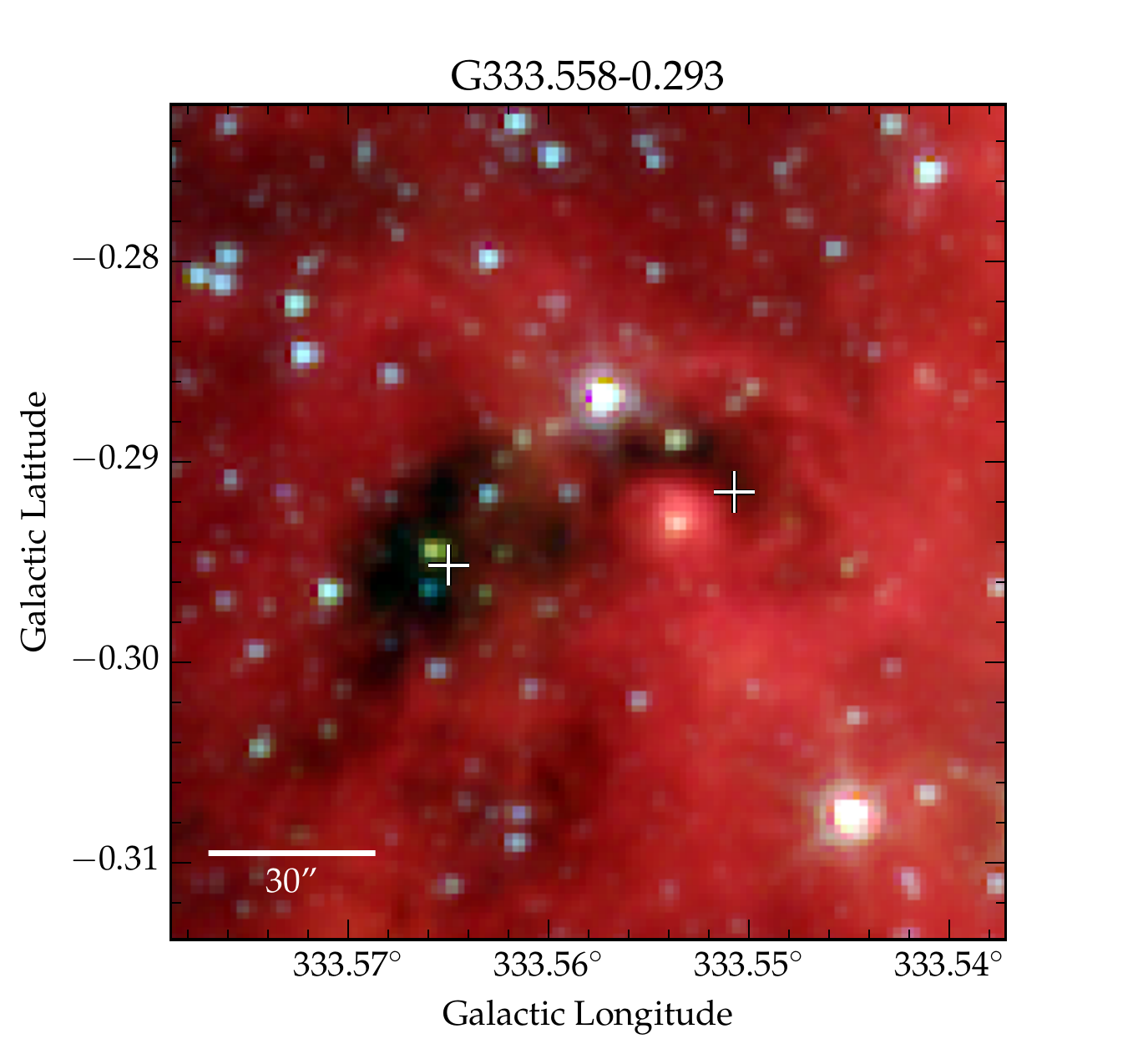}
      \includegraphics[height=0.30\textheight]{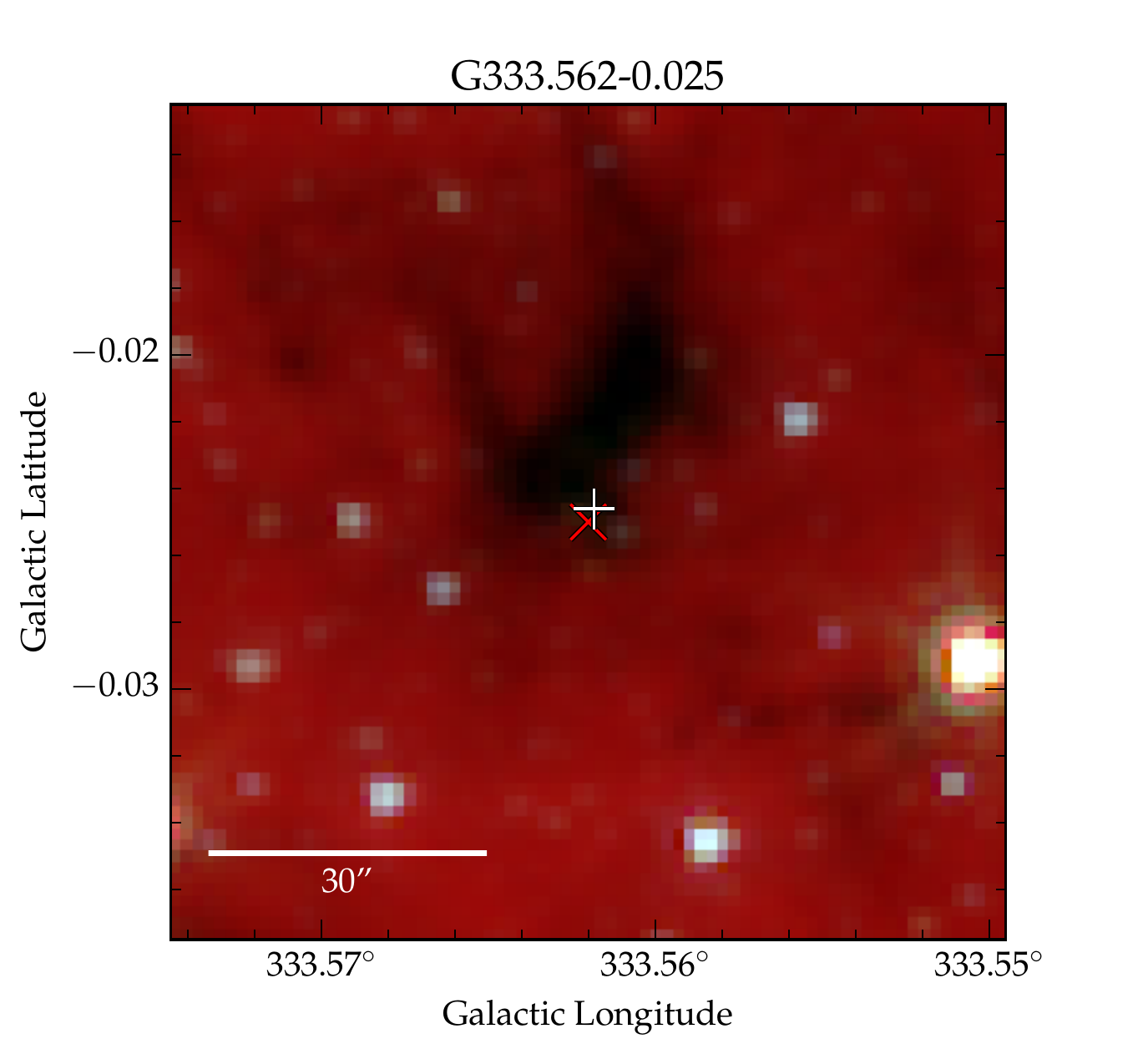}
      \includegraphics[height=0.30\textheight]{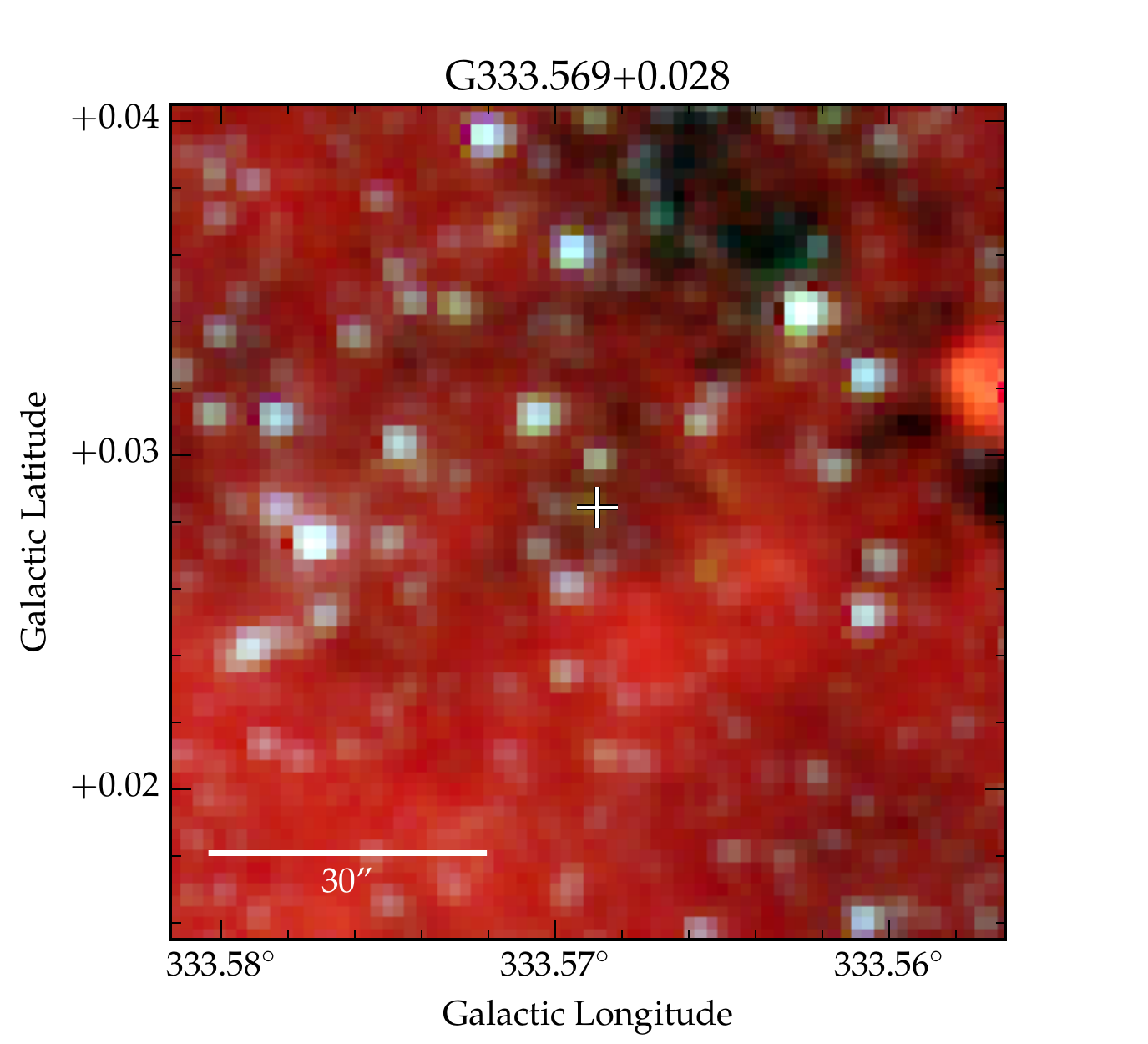}
      \includegraphics[height=0.30\textheight]{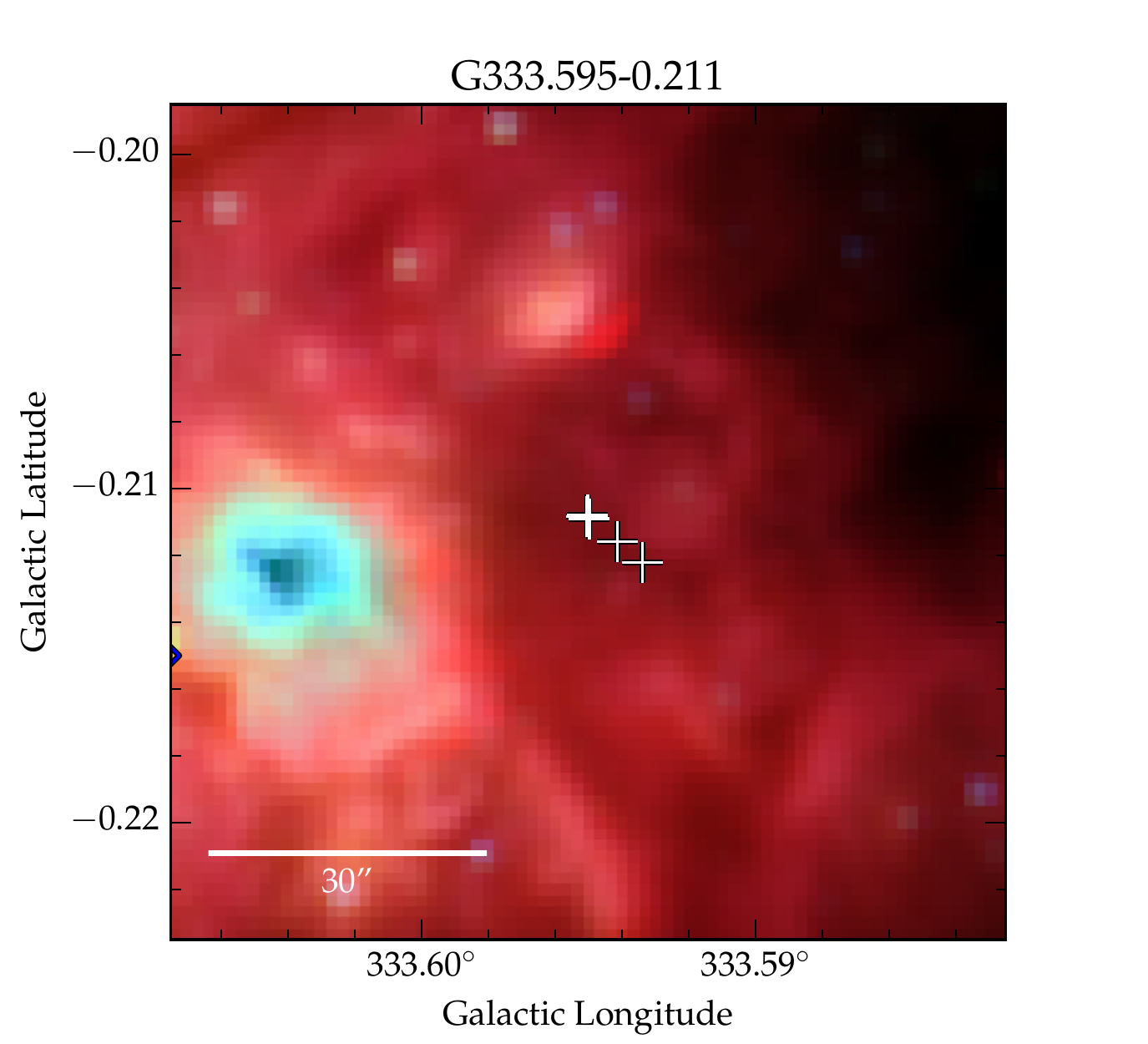}
      \includegraphics[height=0.30\textheight]{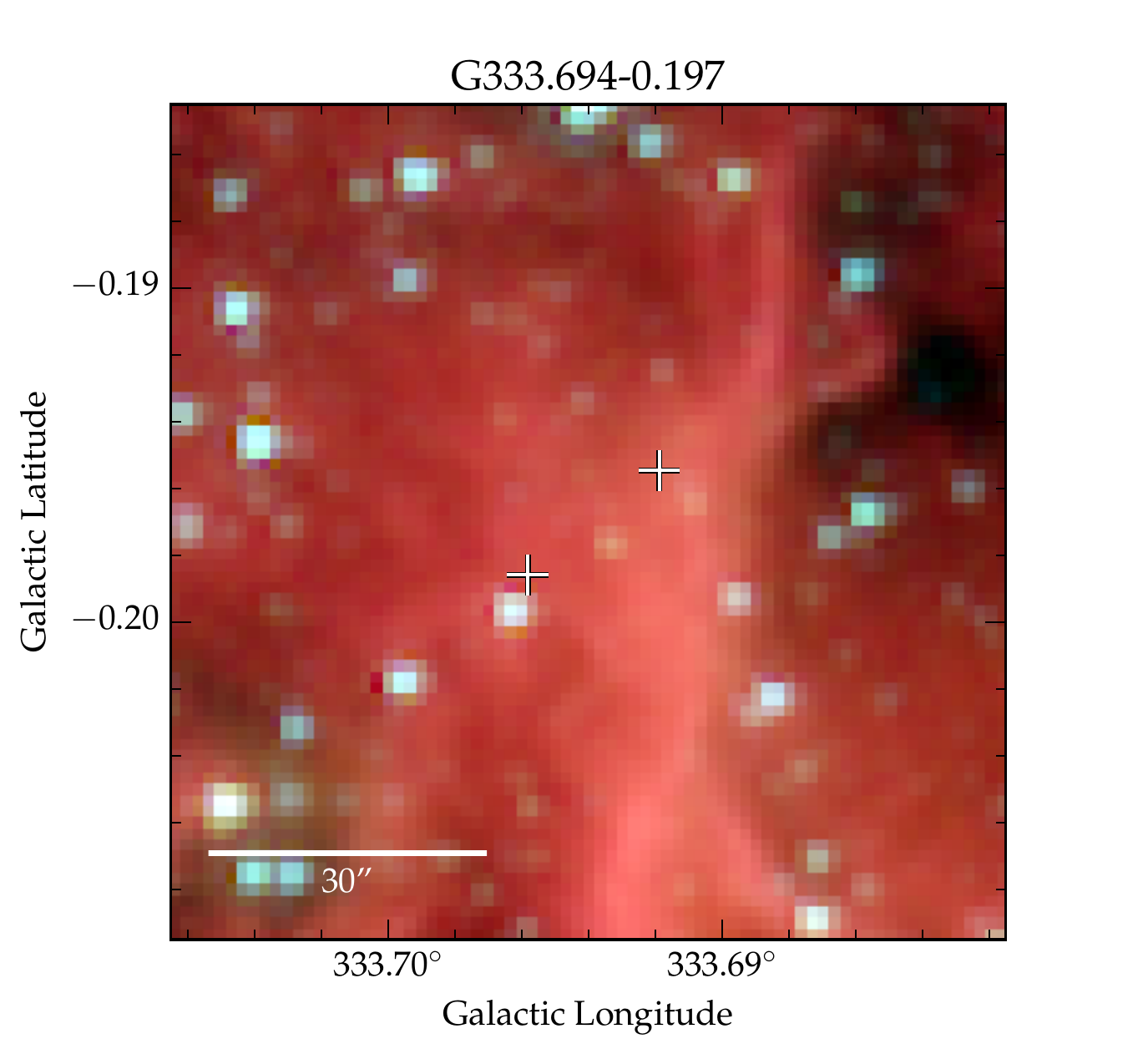}
      \includegraphics[height=0.30\textheight]{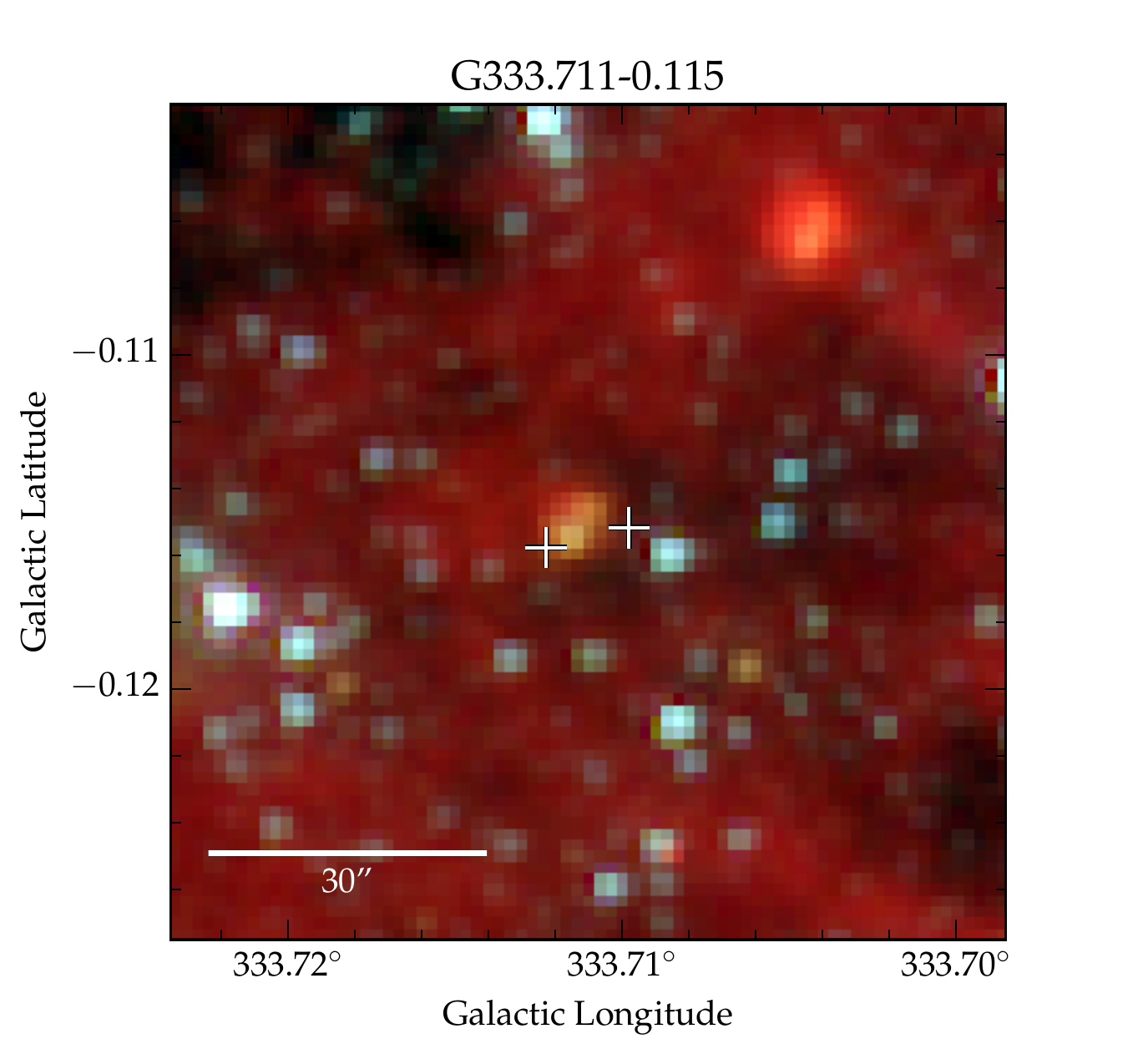}
      \captionof{figure}{\emph{continued}}
    \end{minipage}
}]
\setcounter{figure}{0}
\twocolumn[{
    \begin{minipage}{\textwidth}
      \centering
      \includegraphics[height=0.30\textheight]{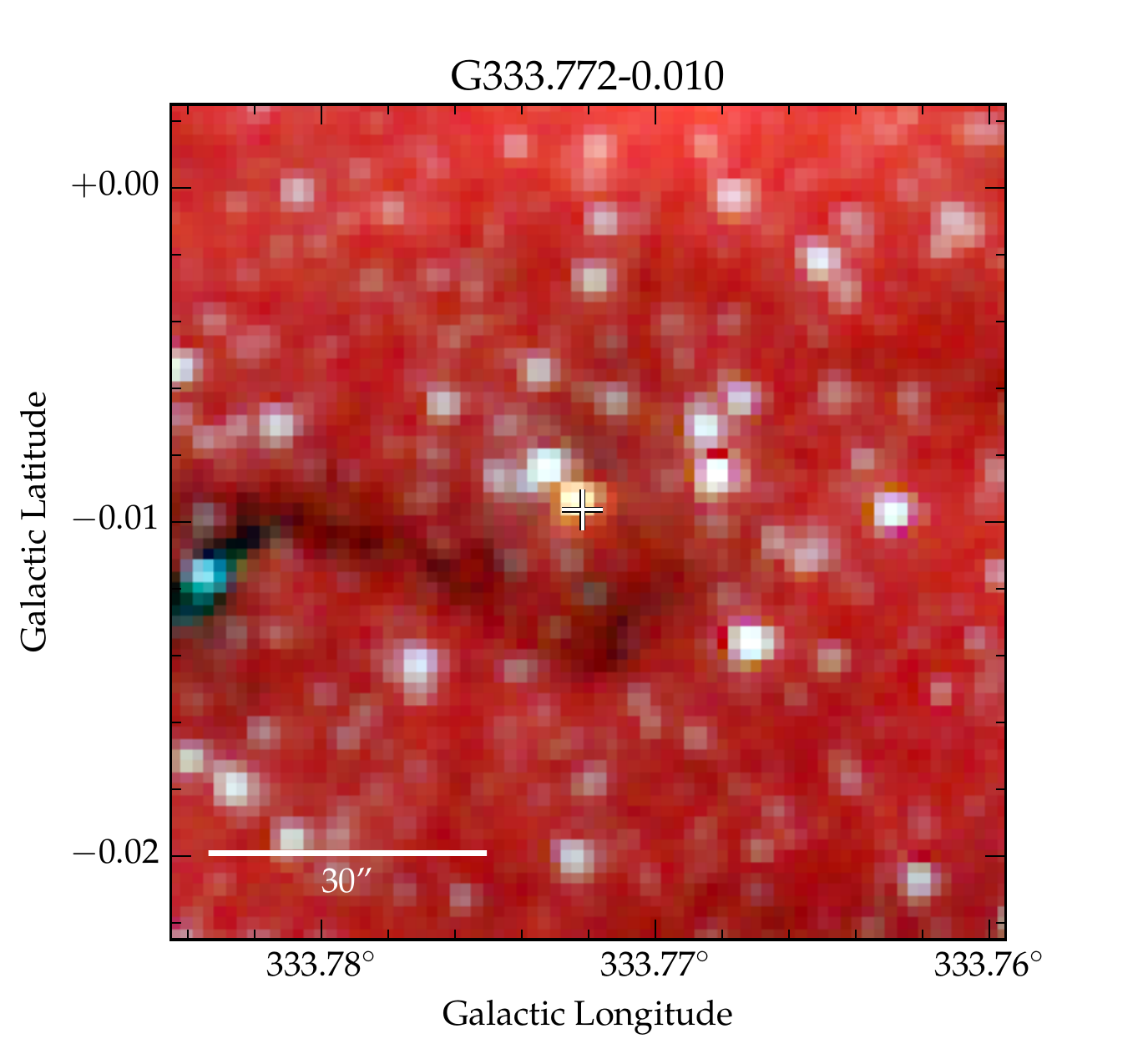}
      \includegraphics[height=0.30\textheight]{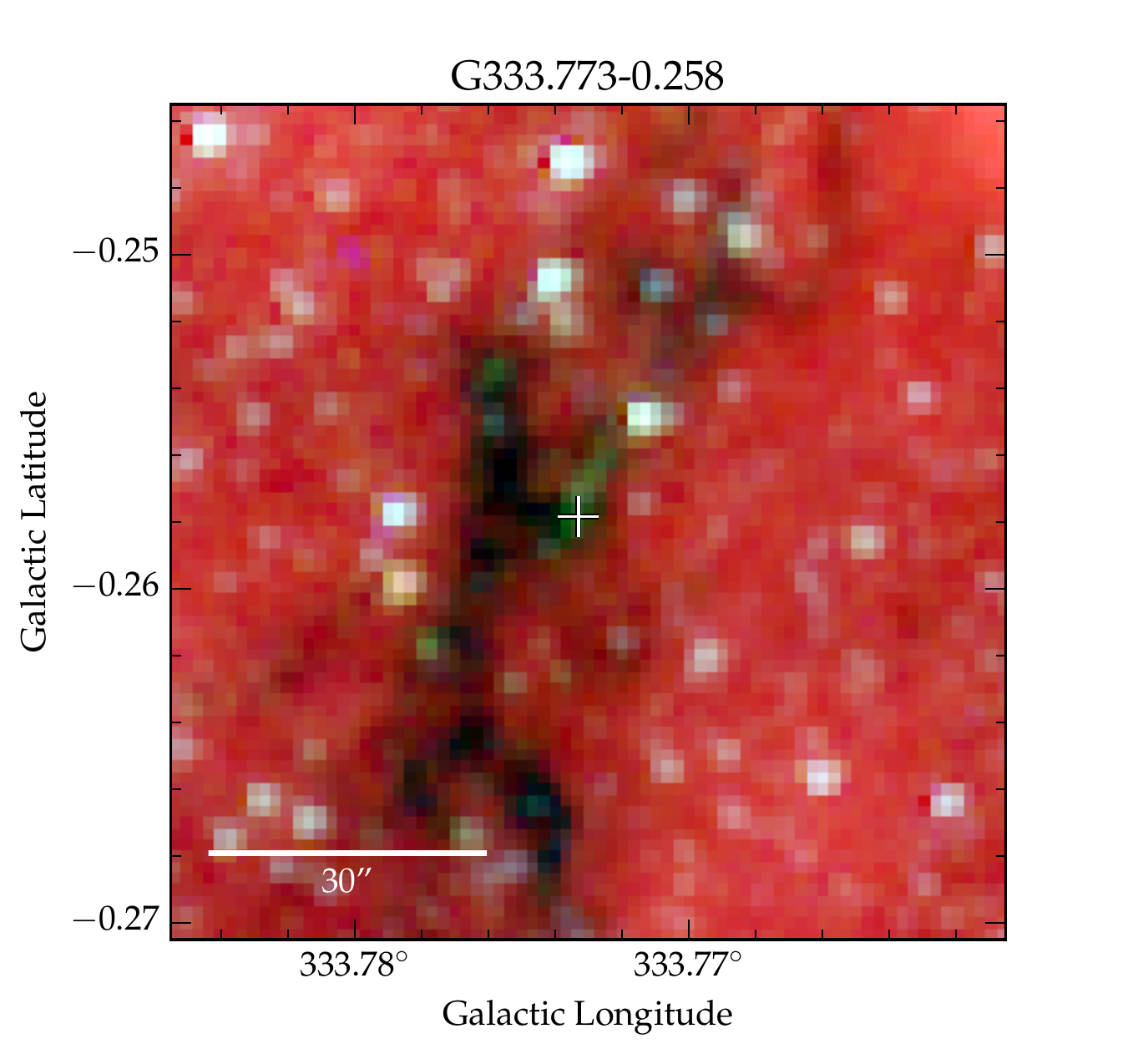}
      \includegraphics[height=0.30\textheight]{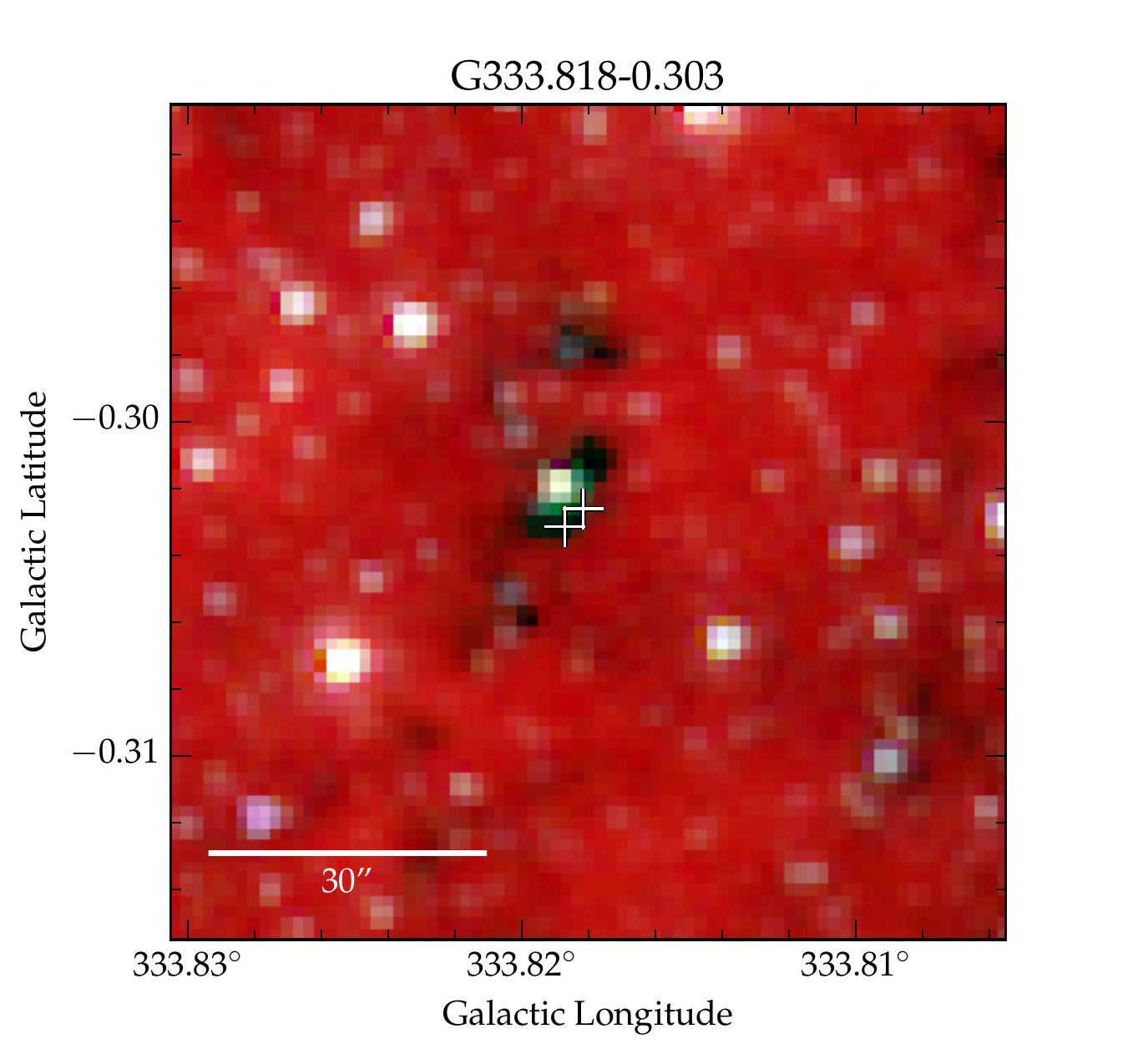}
      \includegraphics[height=0.30\textheight]{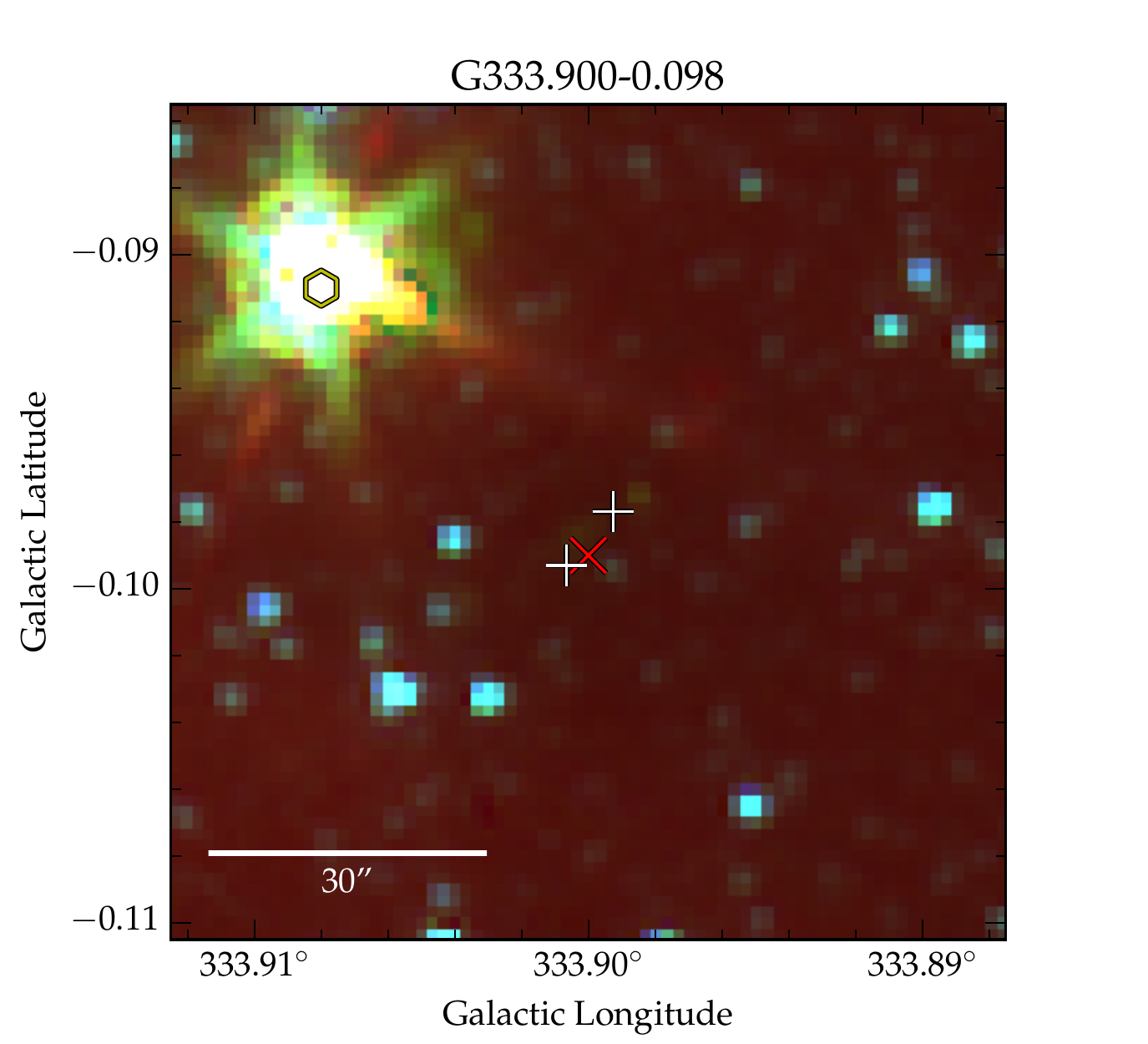}
      \includegraphics[height=0.30\textheight]{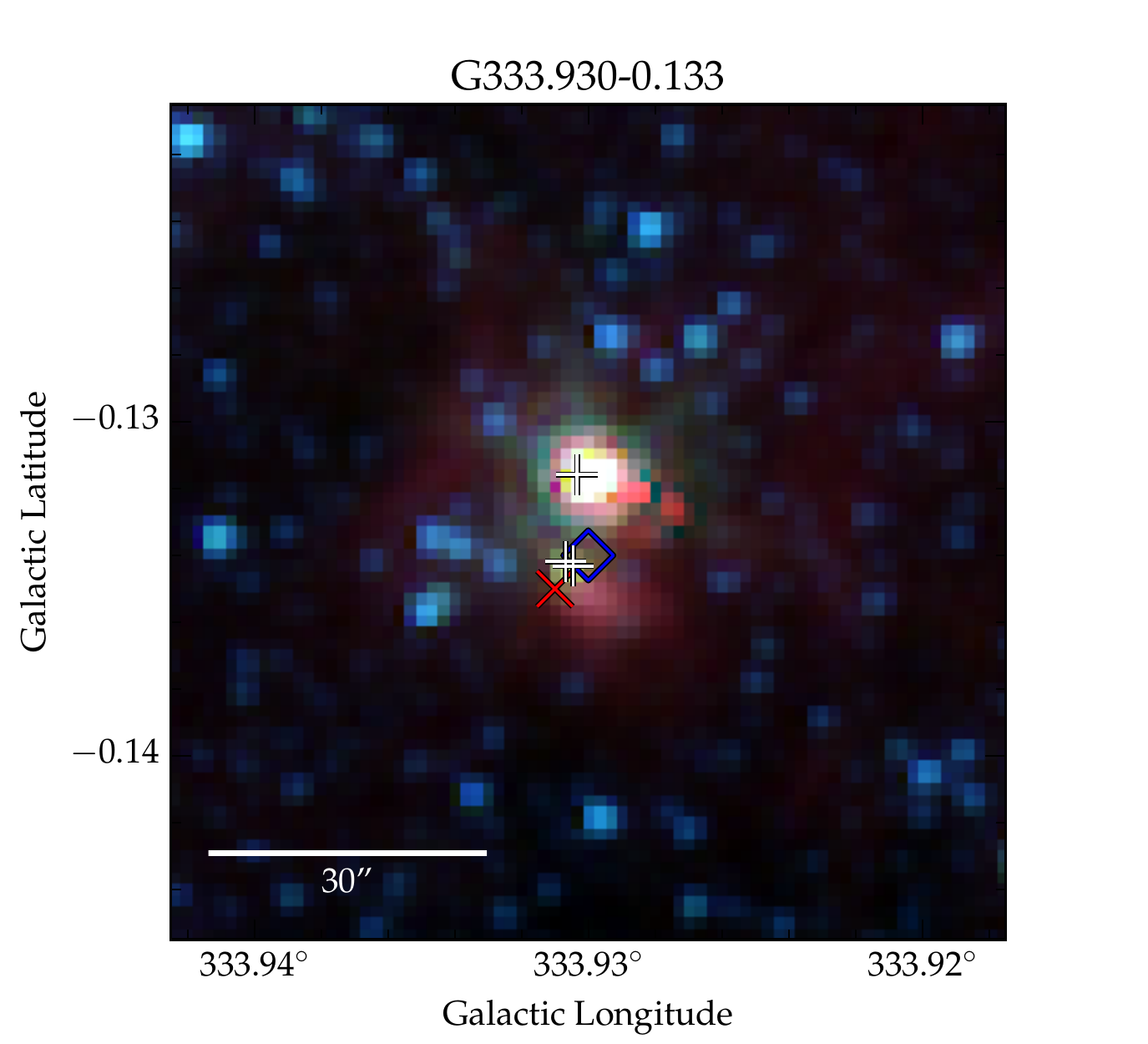}
      \includegraphics[height=0.30\textheight]{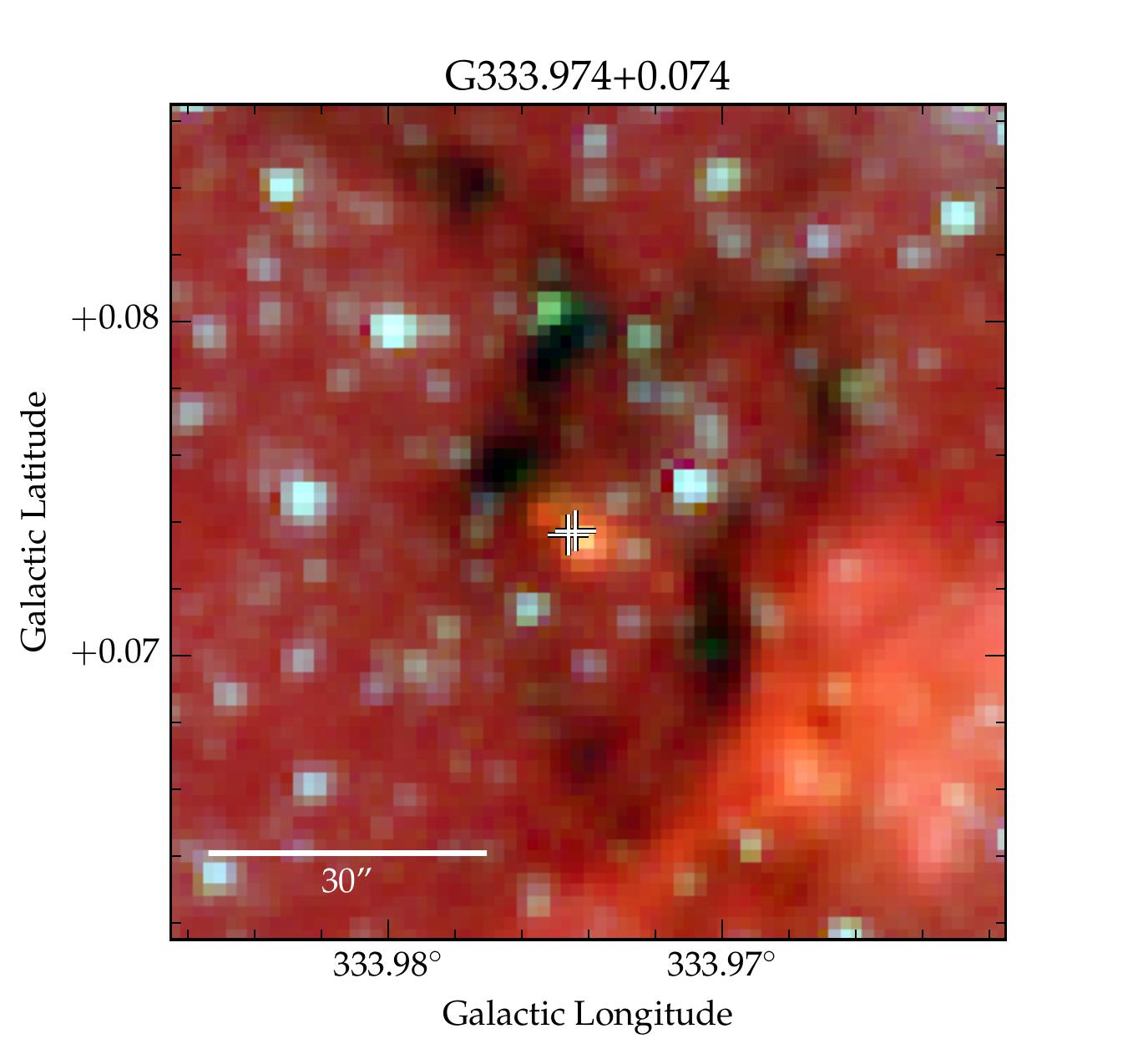}
      \captionof{figure}{\emph{continued}}
    \end{minipage}
}]
\setcounter{figure}{0}
\twocolumn[{
    \begin{minipage}{\textwidth}
      \centering
      \includegraphics[height=0.30\textheight]{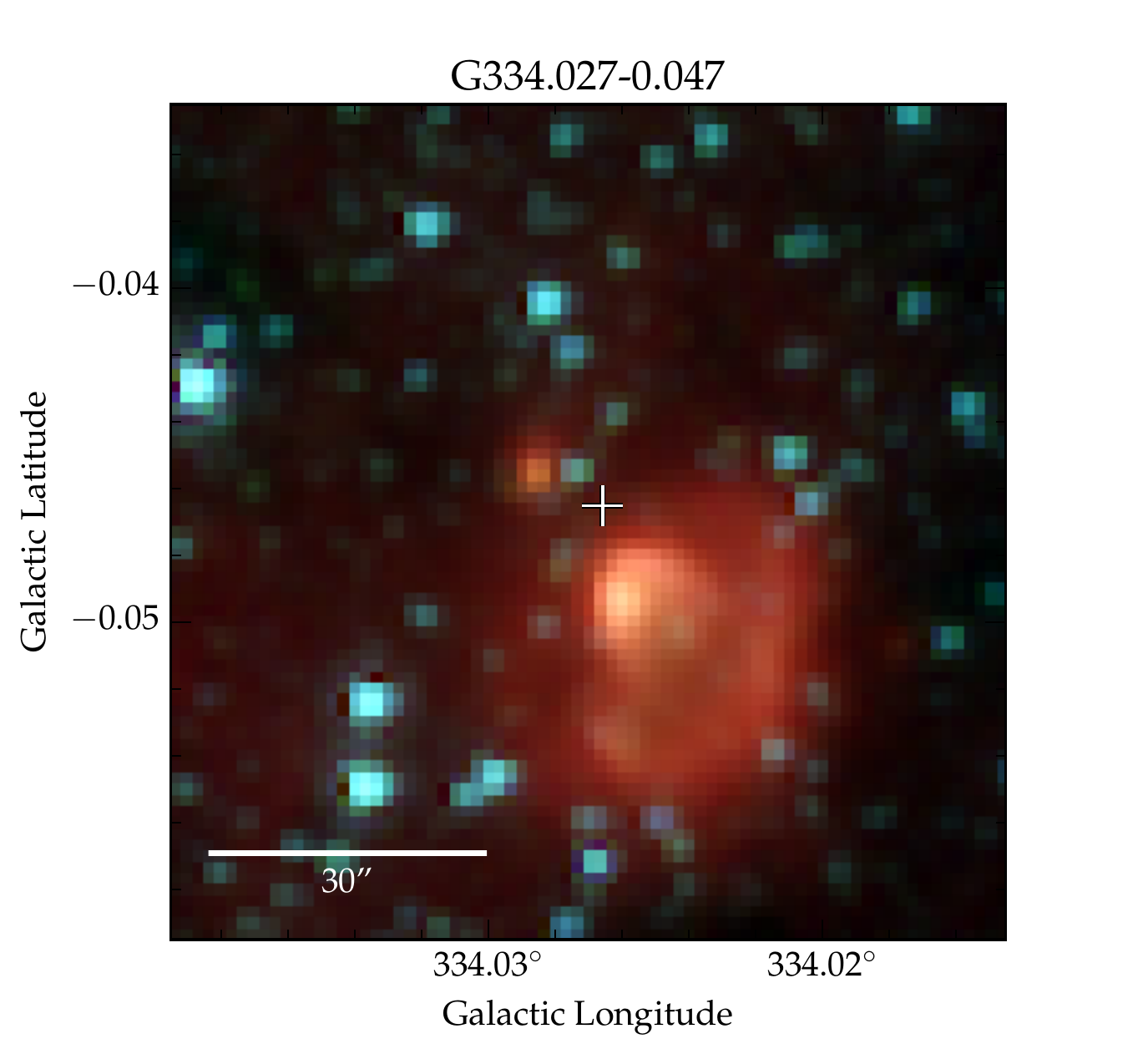}
      \includegraphics[height=0.30\textheight]{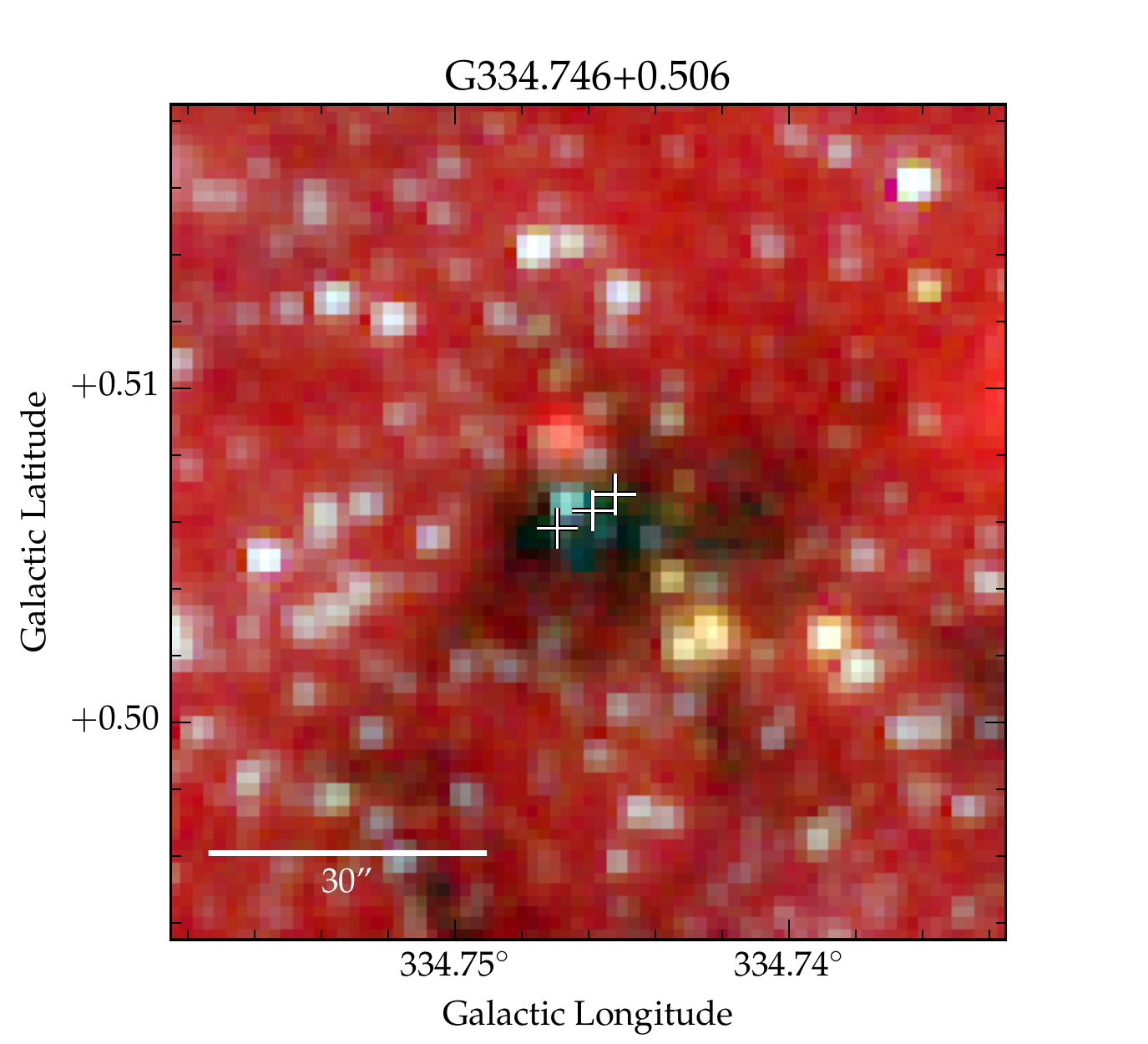}
      \captionof{figure}{\emph{continued}}
    \end{minipage}
}]

\bsp	
\label{lastpage}
\end{document}